\documentclass[a4paper,11pt,twoside]{book}

\usepackage{a4wide}
\usepackage[nottoc]{tocbibind}
\usepackage{bbm}
\usepackage[T1]{fontenc}
\usepackage[latin1]{inputenc}
\usepackage[english]{babel}
\usepackage{graphicx}
\usepackage{palatino}
\usepackage{axodraw,tabularx,array,titlesec,titletoc}
\usepackage{pstricks,pst-coil,graphics}
\usepackage{epsfig,feynmf}
\usepackage[final]{pdfpages}

\usepackage{amsmath}
\usepackage{amssymb}
\usepackage{psfrag}
\usepackage{setspace}
\usepackage[leftbars]{changebar}
\usepackage[footnotesize]{caption}
\usepackage{multirow}
\usepackage{verbatim}

\usepackage{fancyhdr}
\fancypagestyle{plain}{%
\fancyhead{} 
}

\newcommand{\bea}{\begin{eqnarray}}
\newcommand{\eea}{\end{eqnarray}}
\newcommand{\beq}{\begin{equation}}
\newcommand{\eeq}{\end{equation}}

\newcommand{\pdir}{p\kern -5.2pt\raise 0.2ex\hbox {/}}
\newcommand{\vdir}{v\kern -5.75pt\raise 0.15ex\hbox {/}}
\newcommand{\kdir}{k\kern -5.75pt\raise 0.15ex\hbox {/}}
\newcommand{\epsdir}{\epsilon\kern -5.0pt\raise 0.15ex\hbox {/}}
\newcommand{\bvdir}{\bar{v}\kern -5.75pt\raise 0.15ex\hbox {/}}
\newcommand{\Ddir}{D\kern -7.75pt\raise 0.20ex\hbox {/}}
\newcommand{\Adir}{A\kern -7.75pt\raise 0.20ex\hbox {/}}
\newcommand{\ldir}{l\kern -5.0pt\raise 0.2ex\hbox{/}}
\newcommand{\varepsdir}{\varepsilon\kern -5.5pt\raise 0.15ex\hbox{/}}

\newcommand{\beqns}{\begin{eqnarray*}}
\newcommand{\eeqns}{\end{eqnarray*}}

\newcommand{\Prvp}{\frac{1+\!\not\!v}{2}}

\newcommand{\Prvm}{\frac{1-\!\not\!v}{2}}

\newcommand{\gc}{\gamma^5}

\newcommand{\lgl}{\langle}
\newcommand{\rgl}{\rangle}

\newcommand{\ga}{\gamma}

\newcommand{\m}{\mu}
\newcommand{\n}{\nu}
\newcommand{\p}{\psi}

\newcommand{\nn}{\nonumber}
\newcommand{\msb}{\overline{\rm{MS}}}

\def\elematrice#1#2#3{\lgl#1|#2|#3\rgl}

\def\Journal#1#2#3#4{{#1} {\bf #2}, #3 (#4)}

\def\PLB{{\em Phys. Lett.}  B}

\def\PRD{{\em Phys. Rev.} D}

\definecolor{Red}{rgb}{1,0,0}
 \definecolor{PineGreen}{rgb}{0.5,0.65,0}
 \definecolor{Blue}{rgb}{0,0,1}
 \definecolor{Magenta}{cmyk}{0,1,0,0}
 \definecolor{Black}{gray}{0}
 \definecolor{Gray}{gray}{0}
 \definecolor{Lavender}{rgb}{1,0,1}
\newcommand{\mgt}{\cmag}

\newcommand{\cred}{\color{Red}}
\newcommand{\cblu}{\color{Blue}}
\newcommand{\cmag}{\color{Magenta}}

\newcommand{\cbla}{\color{Black}}

\newcommand{\bla}{\cbla}
\renewcommand{\red}{\cred}

\newcommand{\blu}{\cblu}

\newcommand{\statpropp}[1]  
{\unitlength #1
\begin{picture}(200,30)
  \linethickness{0.15 mm}
  \blu\put(0,0){\vector( 1, 0){100}}
  \blu\put(100,0){\line( 1, 0){100}}
\end{picture}
}

\newcommand{\statpropm}[1]  
{\unitlength #1
\begin{picture}(200,30)
  \linethickness{0.15 mm}
  \blu\put(0,0){\vector( -1, 0){100}}
  \blu\put(-100,0){\line( -1, 0){100}}
\end{picture}
}

\newcommand{\wavef}[1]  
{\unitlength #1
\begin{picture}(200,30)
  \linethickness{0.3 mm}
  \bla\put(0,0){\line( 0, 1){20}}
\end{picture}
}

\newcommand{\lproppp}[1]  
{\unitlength #1
\begin{picture}(200,30)
  \linethickness{0.1 mm}
  \red\put(0,0){\vector( 4, 1){100}}
  \red\put(100,25){\line( 4, 1){100}}
\end{picture}
}
\newcommand{\lproppm}[1]  
{\unitlength #1
\begin{picture}(200,30)
  \linethickness{0.1 mm}
  \red\put(0,0){\vector( 4, -1){100}}
  \red\put(100,-25){\line( 4, -1){100}}
\end{picture}
}
\newcommand{\lpropmp}[1]  
{\unitlength #1
\begin{picture}(200,30)
  \linethickness{0.1 mm}
  \red\put(0,0){\vector( -4, 1){100}}
  \red\put(-100,25){\line( -4, 1){100}}
\end{picture}
}
\newcommand{\lpropmm}[1]  
{\unitlength #1
\begin{picture}(200,30)
  \linethickness{0.1 mm}
  \red\put(0,0){\vector(-4, -1){100}}
  \red\put(-100,-25){\line(-4, -1){100}}
\end{picture}
}

\newcommand{\testv}[1]  
{             \unitlength #1
              \begin{picture}(200,30)                
                \put(230,0 ){\statpropm{#1}}
                \put(30,30){\statpropp{#1}}
                \put(30,-100){\wavef{#1}}
                \put(30,-50){\lproppp{#1}}
              \end{picture}
}

\newcommand{\vbbdag}[1]  
{             \unitlength #1
              \begin{picture}(200,125)                
                \put(0,25){\wavef{#1}}
                \put(0,50){\wavef{#1}}
                \put(200,75 ){\statpropm{#1}}
                \put(0,25 ){\statpropp{#1}}
                \put(200,0 ){\wavef{#1}}
                \put(200,75){\wavef{#1}}
                \put(200,0){\lproppp{#1}}
                \put(400,50){\lpropmp{#1}}
                \put(0,105){$x_0$}
                \put(200,105){$y_0$}
                \put(400,55){$z_0'$}
              \end{picture}
}

\newcommand{\vbbdagt}[1]  
{             \unitlength #1
              \begin{picture}(200,125)                
                \put(0,25){\wavef{#1}}
                \put(0,50){\wavef{#1}}
                \put(205,75 ){\statpropm{#1}}
                \put(5,25 ){\statpropp{#1}}
                \put(210,0 ){\wavef{#1}}
                \put(210,75){\wavef{#1}}
                \put(210,0){\lpropmp{#1}}
                \put(10,50){\lproppp{#1}}
                \put(0,105){$x_0$}
                \put(210,105){$y_0$}
              \end{picture}
}

\newcommand{\vbbdaggg}[1]  
{             \unitlength #1
              \begin{picture}(200,90)                
                \put(200,0 ){\statpropm{#1}}
                \put(0,0){\wavef{#1}}
                \put(200,0){\wavef{#1}}
                \put(0,20){\lproppp{#1}}
                \put(0,70){\lproppm{#1}}
                \put(0,70){\wavef{#1}}
                \put(200,70){\wavef{#1}}
                \put(200,90 ){\statpropm{#1}}
                \put(0,100){$x_0$}
                \put(200,100){$y_0$}
              \end{picture}
}

\newcommand{\vbb}[1]  
{             \unitlength #1
              \begin{picture}(200,120)                
                \put(0,0){\statpropp{#1}}
                \put(0,0){\wavef{#1}}
                \put(200,0){\wavef{#1}}
                \put(200,70){\lpropmm{#1}}
                \put(200,20){\lpropmp{#1}}
                \put(0,70){\wavef{#1}}
                \put(200,70){\wavef{#1}}
                \put(0,90 ){\statpropp{#1}}
                \put(0,100){$x_0$}
                \put(200,100){$y_0$}
              \end{picture}
}

\newcommand{\lprop}[1]  
{\unitlength #1
\begin{picture}(200,30)
  \linethickness{0.3 mm}
  \mgt \qbezier(0,0)(40,30)(100,30)     
     \qbezier(100,30)(160,30)(200,0)  
\end{picture}
}
\newcommand{\lpropp}[1]  
{\unitlength #1
\begin{picture}(200,30)
  \linethickness{0.3 mm}
  \mgt \qbezier(0,0)(40,30)(100,30)     
     \qbezier(100,30)(160,30)(200,-15)  
\end{picture}
}
\newcommand{\hprop}[1]  
{\unitlength #1
\begin{picture}(200,30)
  \linethickness{0.3 mm}
  \blu \qbezier(0,0)(40,-10)(100,-10)     
     \qbezier(100,-10)(160,-10)(200,0)  
\end{picture}
}

\newcommand{\statprop}[1]  
{\unitlength #1
\begin{picture}(200,30)
  \linethickness{0.3 mm}
  \blu\put(0,0){\line( 1, 0){200.0}}
\end{picture}
}

\newcommand{\kinprop}[1]  
{\unitlength #1
\begin{picture}(200,30)
  \linethickness{0.3 mm}
  \blu\put(0,0){\line( 1, 0){60.0}}
  \blu\put(60,0){\line( 0, -1){15.0}}
  \blu\put(60,-15){\line( 1, 0){140.0}}
  \put(40,-20){$y$}
\end{picture}
}

\newcommand{\spinprop}[1]  
{\unitlength #1
\begin{picture}(200,30)
  \linethickness{0.3 mm}
  \blu\put(0,0){\line( 1, 0){200.0}}
  \red\put(60,0){\circle*{8.0}}
  \put(40,-20){$\sigma\cdot{\mathbf B}(y)$}
\end{picture}
}

\newcommand{\kinpropb}[1]  
{\unitlength #1
\begin{picture}(200,30)
  \linethickness{0.3 mm}
  \blu\put(0,0){\line( 1, 0){200.0}}
  \red\put(60,0){\circle*{8.0}}
  \put(30,-20){${\mathbf D^2}(y)$}
\end{picture}
}

\newcommand{\hlcft}[1]  
{             \unitlength #1
              \begin{picture}(200,30)                
                \put(30,0){\lprop{#1}}
                \put(30,0){\hprop{#1}}
              \end{picture}
}

\newcommand{\statcft}[1]  
{             \unitlength #1
              \begin{picture}(200,30)                
                \put(30,0){\lprop{#1}}
                \put(30,0){\statprop{#1}}
              \end{picture}
}

\newcommand{\kincft}[1]  
{             \unitlength #1
              \begin{picture}(200,30)                
                \put(30,0){\lpropp{#1}}
                \put(30,0){\kinprop{#1}}
              \end{picture}
}

\newcommand{\kincftb}[1]  
{             \unitlength #1
              \begin{picture}(200,30)                
                \put(30,0){\lprop{#1}}
                \put(30,0){\kinpropb{#1}}
              \end{picture}
}

\newcommand{\spincft}[1]  
{             \unitlength #1
              \begin{picture}(200,30)                
                \put(30,0){\lprop{#1}}
                \put(30,0){\spinprop{#1}}
              \end{picture}
}

\begin{document}
\thispagestyle{empty}

\begin{center}
\Huge{UNIVERSIT\'E PARIS-SUD XI}\\
\vspace{0.4cm}
\Huge{UFR SCIENTIFIQUE D'ORSAY}\\
\vspace{2cm}
\Large{TH\`ESE}\\
pr\'esent\'ee pour obtenir\\
\vspace{0.7cm}
\Large{L'HABILITATION A DIRIGER DES RECHERCHES}\\
\vspace{0.7cm}
par\\
\vspace{0.4cm}
Beno\^\i t BLOSSIER\\
\vspace{0.6cm}
sujet:\\
\vspace{0.3cm}
{\bf On the first-principles determination}\\ 
{\bf of the Standard Model fundamental parameters in the quark sector}\\

\vspace{1cm}
Soutenue le 3 Avril 2014 devant la Commission d'examen compos\'ee de\\
\vspace{0.8cm}
\begin{tabular}{ll}
M. Ulrich ELLWANGER&(Pr\'esident)\\
Mme Svjetlana FAJFER&(Rapporteur)\\
M. Vicente\, GIMENEZ-GOMEZ&\\
M. Christopher\, SACHRAJDA&(Rapporteur)\\
M. Achille STOCCHI&\\
M. Hartmut\, WITTIG&(Rapporteur)\\
\end{tabular}
\end{center}

\chapter*{Acknowledgements}

\noindent Before developing my discussion in this "Th\`ese d'Habilitation \`a Diriger des Recherches", 
I have the pleasure to thank Ulrich Ellwanger for being the president of my jury,  
Svjetlana Fajfer, Christopher Sachrajda and Hartmut Wittig to have referred on my manuscript 
and Vicente Gimenez-Gomez and Achille Stocchi, who kindly accepted to be examiners in the jury.

\noindent My research career is still quite short but I had the great opportunity to meet passionate, 
fascinating people who are also full of humanity. Olivier P\`ene was my PhD advisor, he let me mature my 
reflexions and my thoughts without any pressure but he was always on my side quickly, each time that I needed 
any support. During my postdoctoral stay in DESY Zeuthen, Rainer Sommer and Karl Jansen were also
particularly careful to give me the freedom to collaborate and exchange ideas with both of them, opening
the door to a fruitful and continuous work with 2 international teams. I could start and lead projects with Vittorio Lubicz, 
Hubert Simma, Silvano Simula, Cecilia Tarantino, Marc Wagner. The atmosphere was made nice and 
studious thanks to young colleagues and visitors: Stefano Actis, 
Isabella Birenbaum, Guillaume Bossard, Nicolas Garron, Gregorio Herdoiza, Georg von Hippel, Tereza Mendes, Misha Rogal, Andrea Shindler.
Initiated by James Zanotti, the "S-Bahn Party", that consisted in coming back to Berlin each Friday evening with chips and beers, was a very 
pleasant moment: sometimes I miss it when I sit in the famous RER B. It is with a lot of emotion that I walk along 
Frankfurter Allee during my yearly visit in Berlin, "+/- 0" in Simon-Dach Strasse is still one of my favorite restaurants, "Palm-Beach" is a hearty caf\'e.
I really enjoyed to become a permanent researcher in LPT Orsay; Damir Becirevic, Philippe Boucaud and Alain Le Yaouanc are extraordinary persons, Jean-Pierre Leroy finds very often witticisms to make a situation funny, Asmaa Abada and S\'ebastien Descotes-Genon 
give me regularly the chance to explain why lattice QCD is still relevant for phenomenology, I have long and enriching conversations with Martial Mazars and Samuel Wallon about teaching. I thank my first PhD student, Antoine G\'erardin, for his kindness and his reliability: I am pretty sure that he will know great successes in academic research. Outside the laboratory, Mariane Mangin-Brinet, Vincent Mor\'enas  
and Pepe Rodriguez Quintero are colleagues with whom I feel excited to have long-term projects. I lived the very interesting experience of preparing a series of pedagogy documents together with teachers and inspectors: Nicole Audoin, V\'eronique Bancel, Sonia Duval, Thierry Gozzi, Florence Janssens, Olivier Launay, Marie-Christine L\'evi, Nicolas Magnin, Marl\`ene N\'eel, Corinne Neuhart and Myriam Vial helped me to write down what I had in mind; they accepted that a physicist managed their demanding task, I could observe the working of a central administration in a ministry and I tend to pursue my effort to make the French educative system more accessible to disabled pupils and students.

\noindent Finally, I have a special thought to my family, they were a bit worry when they learnt that I had to live abroad a couple of years but they are all the time close to me in good and in more complicated moments, especially to organize my movings or to arrange my daily life.

\tableofcontents

\chapter*{Introduction}

\fancyhead[LO]{\bfseries \leftmark}
\fancyhead[RE]{\bfseries \rightmark}

\noindent During the write-up of this report, the Nobel prize in physics 2013 was awarded to Fran\c{c}ois Englert and Peter Higgs in recognition of their prediction (together or independently with their colleagues Brout, Guralnik, Hagen and Kibble) that a Goldstone boson was at the origin of masses of other fundamental
particles, through the spontaneous breaking of electroweak symmetry. It was an appealing  explanation of the short range  behaviour of the weak interaction. This theoretical work performed in the 1960's was experimentally confirmed 50 years later at Large Hadron Collider (CERN), after a huge effort by the world-wide community of particle physicists, both theorists and experimentalists, as well as engineer and technical staff. This magnificient achievement illustrates how much tenacious is Humanity to understand better our Universe and track fundamental forces that rule it. The Higgs boson is the last missing brick of the Standard Model. Experiments tell us that what is observed at 126 GeV is most probably a spin-0, parity even, particle. The question whether it is composite or not is not completely solved yet; however there are quite strong clues in favour of no internal structure. The Higgs field $h$ interacts with matter fields $\psi$ (charged leptons and quarks) through Yukawa couplings $\bar{\psi} \psi h$.  A non zero vacuum expectation value $\langle v \rangle$ of $h$ induces a Dirac mass term $m \bar{\psi} \psi$.\\
Of course, although the SM describes very well the microscopic world up to the electroweak scale, a lot of problems are still there. The mechanism of neutrino mass generation seems to be completely different and is under deep investigation. The weak phase appearing in the quark flavour mixing, source of CP violation through the Cabibbo-Kobayashi-Maskawa mechanism, is not sufficient to modelise the matter-antimatter asymmetry observed in our Universe. Charged leptons are characterized by a strong hierarchy among their mass: 3 orders of magnitude between $m_e$ and $m_\mu$ and 1 order of magnitude betwen $m_\mu$ and $m_\tau$. In the quark sector the same picture emerges: $m_t \approx 100 m_{c,b} \approx 1000 m_s \approx 10^5 m_{u,d}$. Is there any dynamics behind the disparity among Yukawa couplings? At which scale occurs the breaking of flavour symmetry? The Higgs field interacts with itself through a quartic term. This is known to generate a quadratic divergence of the mass: hence the theory is inconsistent from that point of view. Many extensions of the SM are proposed in the literature to cure that problem, with new degrees of freedom that are helpful to cancel the divergences in quantum loops. There is an intense experimental activity to detect those hypothetic new particles, without any success for the moment.\\
Probing new physics effects is nowadays a research topic of key importance in high energy physics. The direct search consists in detecting new particles, typically at the electroweak scale, but low energy processes are also quite attractive because they offer a complementary set of constraints. Those can be either new couplings, for instance mediated by right-handed currents, or high energy particles circulating in virtual loops, typically in box or penguin diagrams. In the operator product expansion formalism (OPE) it will translate in corrections of Wilson coefficients with respect to what is known in the Standard Model or to the contribution of non-forbidden form factors to decay amplitudes. The main theoretical difficulty of studying low-energy processes in the hadronic sector is how to appropriately treat the long-distance dynamics of Quantum Chromodynamics (QCD). Indeed, quarks are not directly observed at colliders: they are confined in hadrons. Quantifying precisely the confinement is an extremely hard task. In the language of perturbation theory and Feynman diagrams it would consist in resumming terms with soft or collinear gluons to correct subleading divergences, without any hope because the strong coupling is anyway too large in the infrared regime to control whatever convergence of the series at the target level of precision. Analytical methods can give under certain circumstances precious indications. QCD sum rules and dispersion relations, in application of the Cauchy theorem, are among the most popular ones and are intimately related to basic mathematical properties of correlation functions in quantum field theories. Systematics arises when assumptions, like quark-hadron duality, are not valid anymore, typically beyond the leading order of a given expansion. As the low energy spectrum of mesons roughly verifies an SU(3) flavour symmetry, it has been applied since long an effective field theory approach, known as chiral perturbation theory ($\chi$PT): the (pseudo) Goldstone bosons $\pi$, $K$, $\eta$ are put together in a common field and an effective Lagrangian is obtained in terms of derivative vertices, with couplings denoted as low-energy constants. As far as processes within the soft regime $\lesssim$ 1 GeV and light pseudoscalar mesons are considered, $\chi$PT reveals justified and nice predictions can be made without being spoiled by uncontrolled errors. It is not the same story when transitions with energetic light particles are emitted. A similar approach is at work for heavy flavour physics: Heavy Quark Effective Theory (HQET) is an expansion in $\Lambda_{\rm QCD}/m_Q$ around $m_Q \to \infty$. Heavy Quark Symmetry (HQS) is the symmetry of the effective Lagrangian: it helps to predict 
relations among amplitudes in the charm and the beauty sectors or write them in terms of a couple of universal form factors. Unfortunately describing the charm quark in that framework is not totally safe from the theoretical point of view.\\ 
At the end of the day, the best way to compute the hadronic part of $H_1 \to H_2$ and $H \to 0$ hadronic transitions and decays is lattice QCD. Since the beginning of 2000's, an incredible amount of progresses have been made thanks to improvements in computation science and theory. An acceleration has even been observed since 2005: simulations performed with ${\rm N_f}=2+1+1$ dynamical quarks ($u/d$, $s$ and $c$), or at a physical pion mass, or with automatically $O(a)$ improved fermion regularisations, a (quenched) $b$-quark treated in HQET and matched to QCD in a fully non perturbative pattern. Ways to control systematics from excited states on the computation of hadron masses and matrix elements have been extensively explored. Massively parallel high performance computers are now in our professionnal daily life: 200 TFlops-crate are made available to the French lattice community, which represents 3 orders of magnitude more than in 2005.\\ 
In this report we will discuss some work, done during the recent years, that was realized with a satisfying control on systematics and a competitive precision to experiments thanks to those improvements. It concerns the determination of SM fundamental parameters in the quark sector that are closely related to the Higgs boson: in the first chapter we will present our measurement of $u/d$, $s$ and $c$ charm quark masses, in the second chapter we will describe our extraction of the $b$ quark mass based on HQET regularisation and in the third chapter we will detail our effort to estimate by an \emph{ab-initio} method the strong coupling constant that, as it is well known, governs the Higgs production by gluon-gluon fusion.


\chapter{First and second families quark masses}

\fancyhead[LO]{\bfseries \leftmark}
\fancyhead[RE]{\bfseries \rightmark}

QCD is the quantum field theory that is known to describe the strong interaction of quarks and gluons. Experimental and phenomenological evidence are numerous, for instance the observation of tracks let by 
almost free partons in high energy scatterings or the numerical confirmation that 2 sources of static colour interact through a linear interaction potential with their distance of separation. Confinement and spontaneous breaking of chiral symmetry are two facts that make impossible any direct detection of quarks and gluons and perturbative computation with a satisfying control. As a consequence, quark masses, that appear in the QCD Lagrangian, cannot be fixed in any "laboratory" renormalization scheme, in the sense that one cannot measure them experimentally by letting fly the quarks through detectors as one does for the leptons. One has to use more "theoretical" renormalization schemes, or at least not much related to experiments: the $\overline{\mbox{MS}}$ scheme and, more marginally because it is affected by infrared divengences, the pole scheme.\\
In the framework of chiral perturbation theory, one is able to predict ratios of light quark masses. In the isospin symmetry limit and without incorporating any electromagnetic corrections, one has:
\beq\nonumber
m^2_\pi = 2 B \hat{m}, \quad m^2_K = B (m_s + \hat{m}), \quad m^2_\eta = \frac{B(2 \hat{m}+4m_s)}{3}, \quad
\hat{m}=\frac{m_u+m_d}{2},
\eeq
\beq\nonumber
\frac{m_s}{\hat{m}}=\frac{2 m^2_K - m^2_\pi}{m^2_\pi}.
\eeq
Of course, formulae change a bit when one includes electromagnetic effects and strong isospin symmetry breaking corrections \cite{LeutwylerQG} but the kind of results one gets is similar: at the end of the day one obtains an ellipse shape constraining the ratios of masses $m_u/m_d$ and $m_s/m_d$ \cite{KaplanRU}:
\beq\nonumber
\left(\frac{m_u}{m_d}\right)^2 + \frac{1}{Q^2}\left(\frac{m_s}{m_d}\right)^2=1, \quad Q^2=\frac{m^2_s - \hat{m}^2}{m^2_u - m^2_d}\,.
\eeq
Absolute mass scales are extracted from QCD sum rules. One uses dispersion relations, Cauchy theorem and OPE formalism and derives generic formulae: 
\beq\nonumber
\Psi(q^2)=\int ds \frac{\rho(s)}{s-q^2 -i0}.
\eeq
$\Psi$ are functions of quark masses, typically they are defined as products of 2 hadronic current divergences, while the spectral functions $\rho$ are computed after an analysis of experimental data. $K \pi$ scalar form factors, $e^+e^- \to$ hadrons and light pseudoscalar spectral functions with a suppression of the contribution from resonances are the most popular inputs to measure the light and strange quark masses \cite{JaminTJ} - \cite{DominguezTT}.\\
The charm mass is computed following the same approaches: $e^+ e^- \to c\bar{c}$ processes and 2-pt correlators of vector current analysed in QCD sum rules \cite{BoughezalPX} - \cite{NarisonXY}. Within non-relativistic QCD, there is also the determination of $m_c$ by studying the $c \bar{c}$ potential, expanding it in $1/m_c$ and performing a matching with a lattice result at a large distance scale \cite{LaschkaZR}. Deep inelastic scattering offers an alternative method: charm production is described by means of perturbation theory and confronted to experimental results \cite{AlekhinUN}, \cite{AlekhinVU}. analysing the moments of $B$ inclusive decays to charmed hadrons is also popular \cite{HoangZW} - \cite{AubertQDA}.\\
Lattice QCD is particularly appealing to measure the quark masses. Indeed, as they enter as free parameters of the simulations, it is \emph{a priori} straightforward to establish the corresponding hadron masses dependence: $\chi$ PT or polynomials in $1/m_Q$ will be guides in the extrapolations/interpolations to the physical point, that is determined by imposing matching conditions with experiment. However there are a couple of subtle points to keep in mind. First, some of the largely used quark regularisations, like the Wilson-Clover action \cite{SheikholeslamiIJ}, suffer from an explicit breaking of chiral symmetry by the cut-off: an additive renormalization of the bare mass is necessary. Then, one has to pay attention that the chiral fits are performed on data not affected by uncontrolled finite volume effects. Furthermore, absolute calibration of lattice spacings is a source of systematic error. Finally cut-off effects on heavy quarks need still a careful treatment, with highly improved or tuned fermion actions \cite{ElKhadraMP} - \cite{FollanaUV}.

\section{Quality criteria of lattice results}

In the recent past years the lattice community realised the importance of making clear to outsiders how reliable are the results included in global averages. Quark masses are a pedagogical example to discuss. Typically, $\hat{m}$ is obtained by analysing the ratio $m_\pi/f_\pi$, $m_s$ from $m^2_K/m^2_\pi$ and $m_c$ from $m_D$. When necessary, the dimensionful results are usually first rescaled by the Sommer parameter $r_0$ \cite{SommerCE}, or its American friend, the Bernard parameter $r_1$ \cite{BernardGD}, both corresponding to phenomenological distances characterizing the force derived from the static potential: $\left. r^2 F(r)\right|_{r=r_C}=C$, where $C=1.65$ for $r_0$ and $C=1$ for $r_1$. Expressed in lattice units, they are converted to physical units through $f_\pi$ or the splitting $\Upsilon(2S) - \Upsilon(1S)$. Then one gets the quark masses in physical units as well.\\ 
Those determinations involve extrapolations to the continuum and chiral limits. A certain number of quality criteria have been established by the flavour Lattice Averaging Group \cite{FLAG}:\\
-- as far as continuum limit extrapolation is concerned, "good" means 3 or more lattice spacings, with at least two of them smaller than 0.1 fm, $a^{2}_{\rm max}/a^{2}_{\rm min} \geq 2$, $D(a_{\rm min})\leq 2\%$, $\delta(a_{\rm min})\leq 1$, 
"soso" means 2 or more lattice spacings, with one of them smaller than 0.1 fm, $a^{2}_{\rm max}/a^{2}_{\rm min} \geq 1.4$, $D(a_{\rm min})\leq 10\%$, $\delta (a_{\rm min})\leq 2$, "bad" otherwise; here,  $D(a) = \frac{Q(a) - Q(0)}{Q(a)}$, $\delta(a)=\frac{Q(a) - Q(0)}{\sigma_{Q}}$ and $\sigma_Q$ is the total error (statistical and systematic) of the continuum result;\\
-- concerning renormalization and matching, "good" means that no renormalization is needed (correlators made of conserved currents) or it is done non-pertubatively, "soso" refers to computations in 1-loop perturbation theory or higher with an estimate of truncation error, "bad" in other cases;\\
-- the control on finite-volume effects is considered as "good" if $m_{\pi} L \gtrsim 4$ or at least 3 volumes at fixed parameters of the simulation, "soso" if $m_{\pi} L \gtrsim 3$ and at least 2 volumes, and "bad" otherwise;\\
-- the control on chiral extrapolation is "good" if $m_{\pi\,{\rm min}}\lesssim 250$ MeV, "soso" if 250 MeV $\lesssim m_{\pi\, {\rm min}}\lesssim 400$ MeV and "bad" otherwise.\\
Only "good" and "soso" results are put in global averages.

\section{An action with a twisted-mass term}

We have already mentionned that Wilson-Clover fermions break explicitly the chiral symmetry because of a cut-off effect. It translates into a critical $\kappa$ larger than 1/8, inducing actually an additive renormalization for the quark mass. The Clover term in the action helps to eliminate some of the ${\cal O}(a)$ effects but it is not sufficient: counterterms are also necessary to improve the currents, with both pertubative and non perturbative computations of coefficients of improvement \cite{LuscherUG} like $c_A$ in the case of the axial bilinear $\bar{\psi}\gamma^\mu\gamma^5\psi$. All in all it makes tedious the achievement of a rich physical program in a handable amount of time (meson form factors, bag parameters, structure functions,...). Noting that, for an isospin doublet $\psi=\left(\begin{array}{c} \psi_u\\\psi_d\end{array}\right)$, applying the traceless transformation 
\beq\nonumber
\psi \rightarrow \chi=e^{i \omega \gamma^{5}\tau^3/2} \psi, \quad \bar{\psi} \rightarrow \bar{\chi} = \bar{\psi} e^{i \omega\tau^{3}/2 \gamma^{5}},
\eeq
the QCD Lagrangian takes the form 
\beq\nonumber
{\cal L}_{\rm tmQCD} = \bar{\chi} [\not\!\!D + m + i \mu \tau^3 \gamma^5] \chi, \quad \tan \omega = \mu/m, \quad M=\sqrt{m^2+\mu^2}.
\eeq
The tmQCD Lagrangian shares the same symmetry properties as the standard QCD Lagrangian, after the following modification of transformation laws:
\beq\nonumber
{\cal P}_\omega: \begin{array}{l}\chi(x_0,\vec{x}) \longrightarrow \gamma^0 e^{i \omega  
\gamma^5\tau^{3}} \chi(x_0,-\vec{x})\\
\bar{\chi}(x_0,\vec{x}) \longrightarrow \bar{\chi}(x_0,-\vec{x}) e^{i \omega \tau^3 \gamma^5} \gamma^0
\end{array},\quad
{\cal T}_\omega: \begin{array}{l}
\chi(x_0,\vec{x}) \longrightarrow i\gamma^0 \gamma^5 e^{i\omega \gamma^5 \tau^{3}}\chi(-x_0,\vec{x}) \\
\bar{\chi}(x_0,\vec{x}) \longrightarrow  -i \bar{\chi}(-x_0,\vec{x}) e^{i\omega \tau^3 \gamma^5}
\end{array},
\eeq
\beq\nonumber
{\cal C}: \begin{array}{l}\chi(x) \longrightarrow  C^{-1} \bar{\chi}^T(x)\\
\bar{\chi}(x) \longrightarrow  -\chi^T(x) C
\end{array}, \quad
C\gamma^\mu C^{-1}=-\gamma^{\mu\,T}, \quad 
C\gamma^5 C^{-1}=\gamma^5.
\eeq
Linear combinations relate axial and vector currents of $\chi$ and $\psi$ fields. The tmQCD Lagrangian is invariant under an $SU_V(2)_\omega$ transformation 
\beq\nonumber
\begin{array}{l}
\chi \longrightarrow e^{-i \omega \gamma^5 \tau^3/2} e^{i\alpha^a_V \tau^a/2} e^{i \omega \gamma^5 \tau^3/2} \chi\\
\bar{\chi} \longrightarrow \bar{\chi}e^{i \omega \tau^{3}/2 \gamma^5 } e^{-i\alpha^a_V \tau^a/2} e^{-i \omega \tau^{3}/2 \gamma^5 }\end{array},
\eeq
while the mass term breaks the $SU_A(2)_\omega$ symmetry
\beq\nonumber
\begin{array}{l}\chi \longrightarrow e^{-i \omega \gamma^5 \tau^3/2} e^{i\alpha^a_A  \gamma^5\tau^{a}/2} e^{i \omega \gamma^5 \tau^3/2} \chi\\
\bar{\chi} \longrightarrow \bar{\chi}e^{i \omega\tau^{3}/2 \gamma^5 } e^{i\alpha^a_A \tau^a/2\gamma^5} e^{-i \omega\tau^{3}/2 \gamma^5 }\end{array}.
\eeq
The equivalence between QCD and tmQCD is true at the classical level, it remains also true in a regularisation that preseves the chiral symmetry of the massless theory \cite{FrezzottiNK}: it is the case for instance with Ginsparg-Wilson fermions \cite{GinspargBJ}. Referring to the universality class argument, that equivalence can be extended to regularisations that break the chiral symmetry,  once the theory is appropriately renormalized. For example, Wilson fermions with a twisted mass term is equivalent to standard QCD, up to cut-off effects. Those cut-off effects depend on the angle $\omega$ that defines the regularisation. A particularly interesting situation is at \emph{maximal twist}, i.e. $\omega=\pi/2$. indeed, it was proven that the cut-off effects come only at ${\cal O}(a^2)$ on physical quantities \cite{FrezzottiNI}. That discovery opened the door to a fruitful exploration of that regularisation. The fermion action reads
\beq\nonumber
S_F=a^4 \sum_{x,y} \bar{\chi}(y) D^{yx}_W(r) \chi(x) + a \bar{\chi} (x) [m_0 + i\mu \tau^3 \gamma^5] \chi(x),
\eeq
with $4r + am_0 =1/2\kappa$ and $D_W$ is the Wilson kernel. The maximal twist corresponds to a real mass term, \emph{including the Wilson term}, equal to zero. It means $\kappa = \kappa_c$; this condition is realised by imposing that the untwisted PCAC mass vanishes, that is an optimal condition for an automatic ${\cal O}(a)$ improvement even at small quark masses $\mu/\Lambda_{\rm QCD}\sim (a\Lambda_{\rm QCD})^2$ \cite{AokiTA}, \cite{SharpeNY}:
\beq\nonumber
m_{\rm PCAC}\equiv \frac{\langle \partial_0 A^a_0(x) P^a(0)\rangle}{\langle 2P^a(x) P^a(0)\rangle}=0, \; A_0=\bar{\chi} \tau^a \gamma^0 \gamma^5 \chi, \; P=\bar{\chi} \tau^a \gamma^5 \chi,\; a=1,2.
\eeq
We recall that in the twisted basis $\langle A_0 P \rangle$ is a parity-odd quantity, hence it vanishes in the continuum limit. One can choose to determine $\kappa_c$ at each twisted mass $\mu$ and extrapolate the results to $\mu \to 0$ to get $\kappa^{\rm 0}_c$ that is reinjected at all simulated $\mu$'s. However it revealed enough to compute $\kappa_c$ at the minimal simulated twisted mass $\mu^{\rm min}$ and use it for all other $\mu$'s: the residual cut-off effects are indeed of ${\cal O}(a^2 \mu^{\rm min} \Lambda_{\rm QCD})$ \cite{BoucaudXU}.\\
The Wilson term breaks parity ${\cal P}_{\pi/2}$ and time reversal ${\cal T}_{\pi/2}$, as well as $SU_V(2)_{\pi/2}$;
the main consequence is that the pion masses $m_{\pi^\pm}$ and $m_{\pi^0}$, extracted by studying the correlation functions\\
$\sum_{\vec x}\langle [\bar{\chi} \tau^\pm \gamma^5 \chi] (\vec{x},t) 
[\bar{\chi} \tau^\mp \gamma^5 \chi] (0)\rangle$ and $\sum_{\vec x}\langle [\bar{\chi}  \chi] (\vec{x},t) 
[\bar{\chi}  \chi] (0)\rangle$, are different. In simulations with dynamical quarks, it was even observed the Sharpe-Singleton scenario, with a massless neutral pion mass for the twisted mass $\mu \sim {\cal O}(a^2)$ \cite{SharpeXM}, \cite{SharpePS}.
However the tmQCD action at maximal twist is still invariant under modified parity and time reversal transformations:
\beq\nonumber
{\cal P}^{1,2}_F: \begin{array}{l}
U_0(x_0, \vec{x}) \rightarrow U_0(x_0, -\vec{x}), U_k(x_0, \vec{x}) \rightarrow U^\dag_k(x_0, -\vec{x}-a\hat{k}) \\
\chi(x_0,\vec{x})\rightarrow i\tau^{1,2}\gamma^0\chi(x_0,-\vec{x})\\
\bar{\chi}(x_0,\vec{x})\rightarrow -i\bar{\chi}(x_0,-\vec{x})\tau^{1,2}\gamma^0
\end{array},
\eeq
\beq\nonumber
{\cal T}^{1,2}_F: \begin{array}{l}
U_0(x_0, \vec{x}) \rightarrow U^\dag_0(-x_0-a\hat{0}, \vec{x}), U_k(x_0, \vec{x}) \rightarrow U_k(-x_0, \vec{x}) \\
\chi(x_0,\vec{x})\rightarrow i\tau^{1,2}\gamma^0 \gamma^5\chi(-x_0,\vec{x})\\
\bar{\chi}(x_0,\vec{x})\rightarrow -i\bar{\chi}(-x_0,\vec{x})\tau^{1,2}\gamma^5 \gamma^0
\end{array}.
\eeq
The charged axial tranformations are defined by 
\beq\nonumber
U^{1,2}_A(1)_{\pi/2}:
\begin{array}{l}\chi \longrightarrow e^{-i (\pi/4) \gamma^5 \tau^3} e^{i\alpha^{1,2}_A \tau^{1,2}/2\,\gamma^5} e^{i (\pi/4) \gamma^5 \tau^3} \chi = e^{\pm i \frac{\alpha^{1,2}_A}{2} \tau^{2,1}}\chi\\
\bar{\chi} \longrightarrow \bar{\chi}e^{i (\pi/4) \gamma^5 \tau^3} e^{i\alpha^{1,2}_A \tau^{1,2}/2\gamma^5} e^{-i (\pi/4) \gamma^5 \tau^3}=\bar{\chi}e^{\pm i \frac{\alpha^{1,2}_A}{2} \tau^{2,1}}\end{array}
\eeq
At zero twisted mass, the tmQCD action is invariant under $U^1_A(1)_{\pi/2} \otimes U^2_A(1)_{\pi/2}$. At finite mass, it is invariant under $U^3_V(1)_{\pi/2}$ defined by:
\beq\nonumber
U^{3}_V(1)_{\pi/2}:
\begin{array}{l}\chi \longrightarrow e^{i \frac{\alpha^{3}_V}{2} \tau^{3}}\chi\\
\bar{\chi} \longrightarrow \bar{\chi}e^{-i \frac{\alpha^{3}_V}{2} \tau^{3}}\chi\end{array}.
\eeq
A key consequence of those symmetries are that the twisted mass renormalizes only multiplicatively: indeed, the counterterms $\bar{\chi} \tau^{1,2} \chi$ and $\bar{\chi} \gamma^5 \tau^{1,2} \chi$ break $U^3_V(1)_{\pi/2}$, $\bar{\chi}\gamma^5 \chi$ and $\bar{\chi} \tau^3 \chi$ are forbidden because they break ${\cal P}^{1,2}_F$. Finally the operator $\epsilon_{\mu\nu\rho\sigma}F_{\mu\nu}F_{\rho\sigma}$ also breaks ${\cal P}^{1,2}_F$.

\section{Extraction of $u/d$, $s$ and $c$ quark masses from TmQCD lattice simulations}

After exploratory studies led in the quenched approximation by the $\chi$LF Collaboration to understand how the twisted-mass regularisation was working \cite{JansenIR} - \cite{JansenCG}, the European Twisted Mass Collaboration (ETMC) has decided since 2005 to realise a physics program from simulations with ${\rm N_f}=2$ dynamical twisted-mass fermions. In the gauge sector, the tree-level Symanzik improved action
\cite{WeiszZW} has been chosen and reads
\begin{equation*}
S_\mathrm{G}[U] = \frac{\beta}{6} \bigg(b_0 \sum_{x,\mu\neq\nu} 
\textrm{Tr}\Big(1 - P^{1 \times 1}(x;\mu,\nu)\Big) + b_1 \sum_{x,\mu\neq\nu} \textrm{Tr}
\Big(1 - P^{1 \times 2}(x;\mu,\nu)\Big)\bigg) ,
\end{equation*}
where $b_0 = 1 - 8 b_1$ and $b_1 = -1/12$. The light sector is unitary, i.e. valence and sea light quarks are
described by the same TmQCD action with a twisted mass $\mu_l$, while strange and charm quarks are quenched. We introduce two doublets $\chi^{\cal S}=\left(\begin{array}{c} \chi_s\\\chi_{s'}\end{array}\right)$ and $\chi^{\cal C}=\left(\begin{array}{c} \chi_c\\\chi_{c'}\end{array}\right)$ and write the corresponding TmQCD action for those fermions with twisted masses $\mu_s$ and $\mu_c$, respectively. Valence and sea sectors have the same untwisted mass $\kappa$ tuned to $\kappa_c$ through $m_{\rm PCAC}=0$. 
Parameters of the simulations are collected in Table \ref{tab:simul}. 4 lattice spacings are considered to deal with cut-off effects, pion masses in the range [270 -- 500] MeV help to control the chiral extrapolation, finite size effects are checked by analysing 2 volumes at the same lattice spacing and sea mass.\\
$\pi$, $K$ and $D$ meson masses are extracted by analysing the ``charged'' two-point correlation functions 
\begin{equation}
C_{f_1\,f_2}(t)=-\sum_{\vec{x}} \langle [\bar{\chi}_{f_1} \gamma^5 \tau^+ \chi_{f_2}](\vec{x},t)  [\bar{\chi}_{f_2} \gamma^5 \tau^- \chi_{f_1}](0,0) \rangle.
\end{equation}
\begin{table}[t]
\begin{center}
\begin{tabular}{|c|c|c|c|c|c|}
\hline
$\beta$&$a$ [fm]&$L^3\times T$&$m_\pi$ [MeV]&$L m_\pi$&\#cfgs\\
\hline
3.8&0.098&$24^3 \times 48$&410&5.0&240\\
&&&480&5.8&240\\
\hline
3.9 & 0.085 &$24^3 \times 48$& 315 & 3.3&480\\ 
       &         &                & 400 & 4.1&240\\ 
       &         &                & 450 & 4.7&240\\ 
       &         &                & 490 & 5.0&240\\ 
\hline
3.9 &0.0085&$32^3 \times 64$& 275 & 3.7&240 \\ 
       &         &                 & 315 & 4.3&240\\ 
\hline
4.05 & 0.067 &$32^3 \times 64$&  300 & 3.3&240\\ 
       &         &                 & 420 & 4.5&240\\ 
       &         &                 & 485 & 5.2&240\\ 
\hline
4.2 & 0.054 &$48^3 \times 96$ & 270 & 3.5&80\\ 
       & 0.054 &$32^3 \times 64$  & 495 & 4.3&240\\ 
\hline
\end{tabular}
\end{center}
\vspace{-0.4cm}
\caption{Details of the ensembles of gauge configurations used in the
present study: gauge coupling $\beta$, 
lattice spacing $a$, lattice size $V=L^3 \times T$ in lattice units, approximate pion mass; approximate $m_\pi \, L$ product;
statistical used. For each ensemble, the time separation between measurements, in molecular dynamics units, is larger than the autocorrelation time of the light-light pseudoscalar meson masses.}
\label{tab:simul}
\end{table}
The use of all to all propagators with stochastic sources $\eta[i]$ diluted in time
\cite{FoleyAC} to gain in statistics and the improvement of the variance to signal ratio thanks to the one-end trick \cite{FosterWU}, \cite{McNeileBZ} are nowadays very popular and it has been extensively employed by ETMC.\\
First, we define a \emph{volume source} $\eta[i]^a_\alpha(x) = \pm 1/\sqrt{2}\pm i/\sqrt{2}$\footnote{Another possibility is a $U(1)$ source.} with 
\beq\nonumber
\frac{1}{N}\sum_{i=1}^N \eta[i]^a_\alpha(x) (\eta[i]^{b}_\beta(y))^* = \delta_{ab}\delta_{\alpha\beta}\delta_{xy} + {\cal O}(1/\sqrt{N}).
\eeq
Solving the Dirac equation 
\begin{equation}
\sum_{y} D[f,r]^{ab}_{\alpha \beta}(x,y)\ 
\phi[i,f,r]^b_{\beta}(y)\ =\ \eta[i,f]^a_\alpha(x),\, i=1,...,N,
\end{equation}
with $\tau^{3} \chi = r \chi$, we get an unbiased estimate of the quark propagator ${\cal S}[f,r](x;y)$:
\begin{equation}
\frac{1}{N} \sum_{i=1}^N \phi[i,f,r]^a_{\alpha}(x) (\eta[i,f]^b_\beta(y))^*=
{\cal S}[f,r]^{ab}_{\alpha\beta}(x;y) + {\cal O}\left(\sqrt{V}/\sqrt{N}\right).
\end{equation}
The issue here is that the signal to noise ratio of the stochastic propagator is in $\sqrt{N/V}$. It can be improved by using a \emph{diluted} source, especially in time: $\eta[i,f,\tilde{t}]^a_\alpha(\vec{x})=
\eta[i,f]^{a}_{\alpha}(x) \delta_{t_{x} \tilde{t}}$. The signal to noise ratio is in $\sqrt{N/V_{\rm space}}$.
A further improvement is the \emph{spin dilution}: $\eta[i,f,\tilde{\alpha},\tilde{t}]^a(\vec{x})=\eta[i,f]^{a}_{\alpha}(x) \delta_{t_{x} \tilde{t}}\delta_{\alpha \tilde{\alpha}}$.
However, with the ``one-end trick'', the stochastic improvement is much more significant: it consists in an unbiased estimate of the correlator itself and not of the propagators, separately.
In other words, we define
\begin{eqnarray}
\nonumber
A&=&\frac{1}{N} \sum_{i=1}^N \sum_{\vec x}\phi^{*}[i,f_{1},r,\beta,\tilde{t}]^a_{\alpha}(\vec{x},t_{x}+\tilde{t})  \phi[i,f_{2},r,\Gamma(\beta),\tilde{t}]^a_{\Gamma(\alpha)}(\vec{x},t_{x}+\tilde{t})\\
\nonumber
&=&\frac{1}{N}\sum_{i=1}^{N}\sum_{\vec{x},\vec{y},\vec{z}}{\cal S}^{*}[f_{1},r]^{ab}_{\alpha\beta}(\vec{x},t+\tilde{t};\vec{y},\tilde{t})\, \eta^{*}[i,\beta,\tilde{t}]^b(\vec{y})
\times\eta[i,\Gamma(\beta),\tilde{t}]^c(\vec{z})\,
{\cal S}[f_{2},r]^{ac}_{\Gamma(\alpha)\Gamma(\beta)}(\vec{x},t+\tilde{t};\vec{z},\tilde{t})\\
&=& \sum_{\vec{x},\vec{y}} 
{\rm Tr} \left[\Gamma {\cal S}^{\dag}[f_{1},r](\vec{x},t+\tilde{t};\vec{y},\tilde{t}) 
\Gamma{\cal S}[f_{2},r](\vec{x},t+\tilde{t};\vec{y},\tilde{t}) \right] + {\cal O}(V/\sqrt{N}).
\end{eqnarray}
Implicit sums over colour and spinor indices are such that $\eta^{*}[\alpha] \Gamma_{\alpha\beta}
\eta[\beta]=1$, $\Gamma_{\alpha\beta} \propto  \delta_{\beta \Gamma(\alpha)}$. This time, the signal to noise ratio is in $\sqrt{N}$, the gain is huge. In practice, $N=1$ is sufficient. In the following, $\Gamma$ is the $4\times 4$ identity matrix.
Once averaged over a sample of gauge configurations, and using the hermiticity 
property of the quark propagator in TmQCD 
\beq\nonumber
{\cal S}[f,r](x;\,y)\ =\ \gamma_5\, 
{\cal S}^\dag[f,-r](y;\,x)\, \gamma_5,
\eeq 
$\langle A \rangle$ is actually the pseudoscalar two-point correlation function $C_{f_{1}\,f_{2}}(t)$.\\
The overlap of the interpolating fields with the ground states is improved by
building interpolating fields generically written as 
$\bar{\chi}_1 P \otimes \Gamma \chi_2$, where $P$ is a path of links and $\Gamma$
is any Dirac matrix. We use interpolating fields of the so-called Gaussian 
smeared-form~\cite{GuskenAD}
\beq\nonumber
P = \left(\frac{1 + \kappa_G a^2 \Delta}{1+6\kappa_G}\right)^{R},
\eeq
where $\kappa_G=0.15$ is a hopping parameter, $R=30$ is the number of 
applications of the operator $(1 + \kappa_G a^2 \Delta)/(1+6\kappa_G)$, and $\Delta$ the 
gauge-covariant 3-D Laplacian constructed from three-times APE-blocked 
links~\cite{AlbaneseDS}. 
The Dirac equation, which is then solved, reads:
\[
\begin{aligned}
&\sum_{y} D[f,r]^{ab}_{\alpha \beta}(x,y)\  
\phi[i,f,r,P,\alpha,\tilde{t}]^b_{\beta}(y)= (P\eta[i,\alpha])^a(\vec{x}).\\
\end{aligned}
\]
We compute the ``charged'' $\pi$, $K$ and $D$ two-point correlators 
$C^{(2)\,f_1\,f_2}_{P_1;P_2}(t)$ which read
\cite{FrezzottiDR}:
\begin{eqnarray}
\nonumber
C^{(2)\,f_1\,f_2}_{P_1;P_2}(t)&=&
\frac{1}{2}\sum_{r=\pm 1} \left\langle {\rm Tr} 
\left[\sum_{\vec{x},\vec{y}} \gamma^5\, 
{\cal S}^{P_1}[f_{1},r](\vec{y},\,\tilde{t};\,\vec{x},\,\tilde{t}+t)\,\gamma^5\,
{\cal S}^{P_2}[f_2,-r](\vec{x},\,\tilde{t}+t;\,\vec{y},\,\tilde{t})\right]\right\rangle\\
\nonumber
&=&
\frac{1}{2}\sum_{r=\pm 1}
\frac{1}{N} \sum_{i=1}^N \left\langle 
\left\{\sum_{\vec{x}}  
\phi^*[i,f_1,r,P_1,\beta,\tilde{t}]^{a}_{\alpha}(\vec{x},\tilde{t}+t)\right.\right.\\
&&\hspace{3cm}\left.\left.\times\,(P_2\, 
\phi[i,f_2,r,\beta,\tilde{t}])^a_{\alpha}(\vec{x},\tilde{t}+t)\right\}\right\rangle.
\end{eqnarray}
As already mentionned, extrapolations to the chiral and continuum limits are possible if meson and quark masses are converted in ``pseudo''-physical units.
\begin{figure}[t]
    \begin{center}
\vspace{0.5cm}
      \includegraphics[draft=false,angle=-90,width=0.5\textwidth]{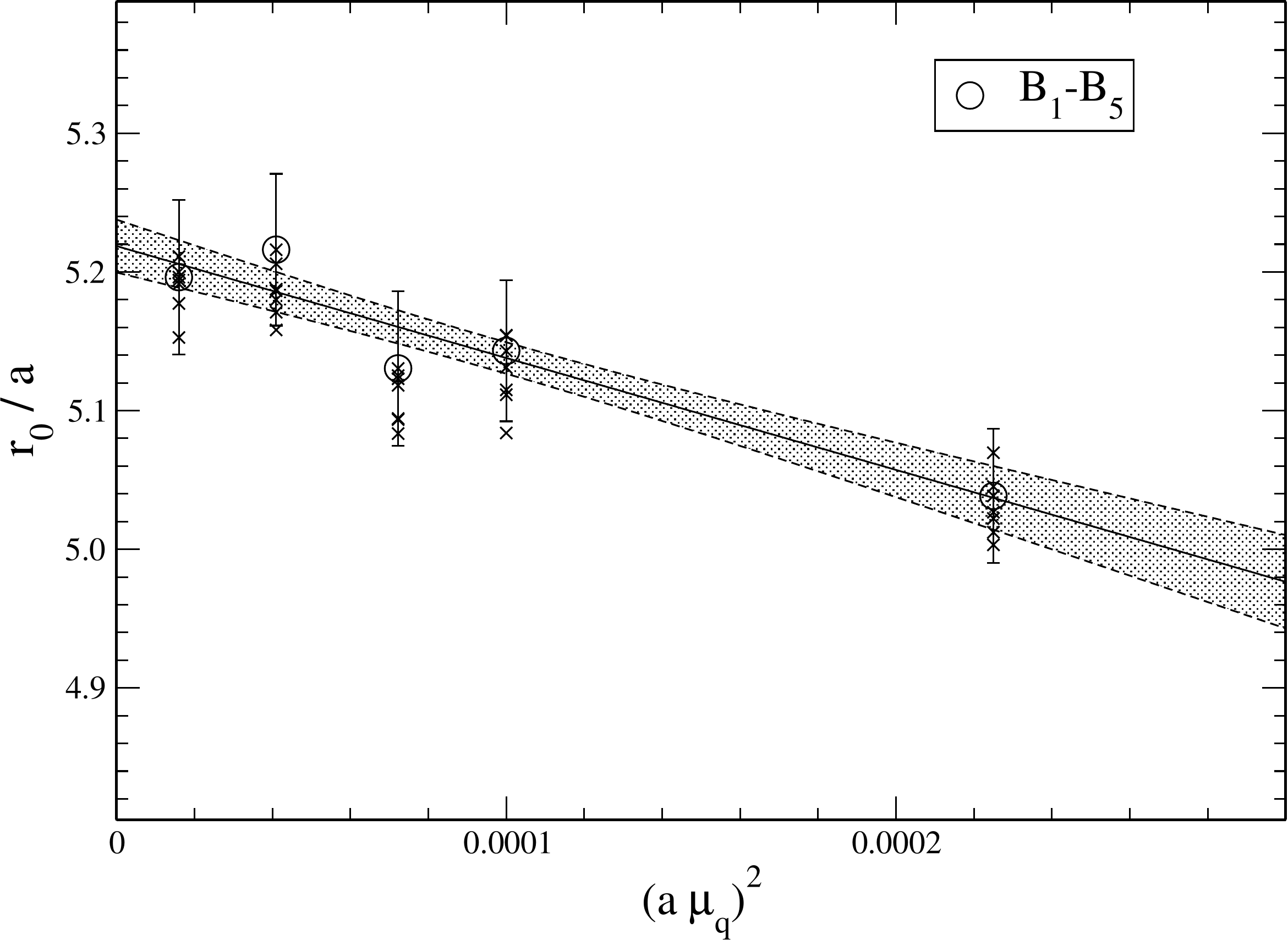}
    \end{center}
\caption{Mass dependence of $r_0/a$ for the ETMC ensembles ($\beta=3.9$) \cite{BoucaudXU}. 
The shaded area 
shows the error band of
  the quadratic fit in $\mu_{\rm sea}$ (full line) to the data (circles). The additional plus symbols
  are further determinations of $r_0/a$ corresponding to different values of the fit 
  parameters to Wilson loop $C(r,t)=Z(r)e^{-t(V_0+\alpha/r+\sigma r)}$. With respect to Table \ref{tab:simul} a heavier pion mass $\sim$ 600 MeV was included in that analysis.
\label{fig:r0etmc}}
\end{figure}
The choice made by ETMC is the phenomenological length $r_{0}$ that characterizes the potential between two static colour sources: 
\beq\nonumber
\left. r^{2} \frac{d V}{d r}\right |_{r=r_{0}}=1.65.
\eeq
It corresponds roughly to the distance where the linear behaviour of the potential dominates the coulombian part. $r_{0}/a$ is extracted by computing the static potential for each sea mass $\mu_{\rm sea}$, deducing an $r_{0}/a[\mu_{\rm sea}]$ and taking the chiral limit. As sketched in Figure \ref{fig:r0etmc} a very mild dependence 
on $\mu_{\rm sea}$ is observed: a quadratic polynomial describes well the data, a linear term (or rather, in $|\mu_{\rm sea}|$) comes in principle from the spontaneous breaking of chiral symmetry.\\
ETMC has always applied the RI-MOM renormalization scheme to obtain renormalized quark masses $\mu_R=Z_\mu \mu_q$. This mass-independent scheme is also independent of any regularisation; the renormalization constants (RC's) can be computed non perturbatively, which is a great advantage compared to $\overline{\rm MS}$ scheme. 
\begin{figure}[t]
\begin{center}
\includegraphics*[draft=false,width=0.2\textwidth]{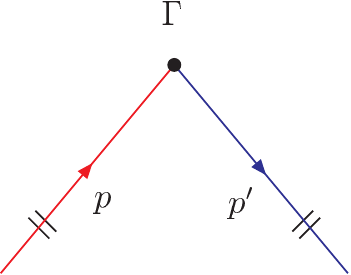}
\end{center}
\caption{Kinematical configuration of an amputated Green function $G_\Gamma(p,p')$.}
\label{fig:RIMOM}
\end{figure}
The steps are well established \cite{MartinelliTY}: we compute in the Landau gauge (we need to fix the gauge to get non-zero average values
of Green functions of quarks) the following Green functions:
\begin{eqnarray}
\nonumber
S_{u(d)}(p)&=&\sum_x e^{ip\cdot x} \lgl \chi^{u(d)}(x) \bar{\chi}^{u(d)}(0) \rgl,\\
\nonumber
G^c_\Gamma(p)&=&\sum_{x,y} e^{ip\cdot (x-y)} \lgl \chi^u(x) \bar{\chi}^u(0) 
\Gamma \chi^d(0) \bar{\chi}^d(y)\rgl.
\eea 
RI-MOM renormalization conditions read:
\beq\label{zq}
\frac{1}{Z_q} \lim_{\mu_q\to 0} \left[\frac{i}{12}{\rm Tr} \left(\frac{\sum_\mu \gamma_\mu \sin(a p_\mu)\, 
S^{-1}(p)}{\sum_\mu \sin^2(a p_\mu)}\right)_{p^2=\mu^2}\right]=1,
\quad \frac{Z_\Gamma}{Z_q} \lim_{\mu_q\to 0} \left[\frac{1}{12}{\rm Tr} 
(\Lambda^\Gamma(p) P_\Gamma)_{p^2=\mu^2}\right]=1,
\eeq
\beq\nonumber 
\Lambda^\Gamma(p) = S^{-1}_u(p) G^c_\Gamma(p)S^{-1}_d(p), \quad \Gamma P_\Gamma = 1.
\eeq
An averaged Green function over up and down quarks is computed to get the RC's, after an extrapolation to the chiral limit.\\ 
Using Ward identities of tmQCD, one can show that $Z_\mu=1/Z_P$. Indeed, applying an infinitesimal vectorial transformation 
\bea\nonumber
\delta \chi(x)&=&\left[\left(\omega^a_V \right) \frac{\tau^a}{2} \right] \chi(x),\\
\delta\bar{\chi}(x)&=&-\bar{\chi}(x)\left[\left(\omega^a_V \right) \frac{\tau^a}{2} 
\right],
\eea
with any operator ${\cal O}(x_1,...,x_n)$, $x_1\neq...\neq x_n$ (to avoid contact terms), we can derive the following equalities,
\beq
\left\lgl\frac{\delta S}{\delta \omega^a(x)}{\cal O}(x_1,...,x_n)\right\rgl=
\left\lgl\frac{\delta {\cal O}(x_1,...,x_n)}{\delta \omega^a(x)}\right\rgl,
\eeq
on the tmQCD action $S$. We obtain:
\bea\nonumber
\frac{\delta S}{\delta \omega^a_V(x)}&=&i\mu\bar{\chi}(x)\gamma^5\left[\tau^3,\frac{\tau^a}{2}\right]\chi(x)
-\frac{1}{2a}\sum_\m\left[\bar{\chi}(x+\m)(1+\ga_\m)\frac{\tau^a}{2} U^\dag_\m(x)\chi(x)
-\bar{\chi}(x)(1-\ga_\m)\frac{\tau^a}{2} U_\m(x)\chi(x+\m)\right.\\
\nonumber
&&\left.\hspace{3cm}-\bar{\chi}(x)(1+\ga_\m) \frac{\tau^a}{2}U^\dag_\m(x-\m)\chi(x-\m)
+\bar{\chi}(x-\m)(1-\ga_\m)\frac{\tau^a}{2} U_\m(x-\m)\chi(x)\right]\\
&=&i\mu\bar{\chi}(x)\gamma^5 \left[\tau^3,\frac{\tau^a}{2}\right]\chi(x)-\sum_\m \Delta_\m \tilde{V}^{a\m}(x),
\eea
\beq\nonumber
\tilde{V}^{a\m}(x)=\frac{1}{2}\left[\bar{\chi}(x+\m)(1+\ga_\m)\frac{\tau^a}{2} U^\dag_\m(x)\chi(x)
-\bar{\chi}(x)(1-\ga_\m)\frac{\tau^a}{2} U_\m(x)\p(x+\m)\right], \quad 
\Delta^x_\m f(x)=\frac{f(x)-f(x-\m)}{a}.
\eeq
\beq
\left\lgl\frac{\delta {\cal O}(x_1,...,x_n)}{\delta \omega^a_V(x)}\right\rgl=
-\sum_\m\Delta^x_\m \lgl{\cal O}(x_1,...,x_n)\tilde{V}^a_\m(x)\rgl
+i\mu \left\lgl {\cal O}(x_1,...,x_n)\bar{\chi}(x) \gamma^5 \left[\tau^3, \frac{\tau^a}{2}\right]\chi(x)\right\rgl.
\eeq
By imposing that it is still true in the chiral limit, we deduce that the non-local vector current $\tilde{V}^{a\m}$ is
conserved on the lattice and $Z_{\tilde{V}}=1$.
With ${\cal O}(x_1)=P^{1}(x_1)$, $P^{1}(y)=\bar{\chi}(y)\gc \tau^1 \chi(y)$, and $x\neq x_1$, we get
\beq
\sum_\m \Delta^x_\m \lgl P^{1}(x_1) \tilde{V}^{+}_\m(x)\rgl
=-\mu \lgl P^{1}(x_1) P^{+}(x) \rgl.
\eeq
Under the condition that this equation is also verified by renormalized correlators, we conclude that $Z_\m
Z_P=1$.\\
Thus, we need to examine in details the amputated charged pseudoscalar vertex to extract the quark mass renormalization constant. ETMC has chosen to work in an "exceptionnal" kinematical set-up: it means that the momentum of in-going and out-going quarks, $p$ and $p'$, sketched in Figure \ref{fig:RIMOM}, are the same. The three-point Green function $G^c_P(p,\mu_{\rm val})$ has in this case a coupling to a Goldstone pole that goes like $1/\mu_{\rm val}$ \cite{LaneBB}, \cite{PagelsBA}, where $\mu_{\rm val}$ is the valence mass: the associated power correction in the OPE analysis explodes in the chiral limit, as numerically observed \cite{MartinelliTY}, \cite{GockelerYE}, \emph{even at large $p^2$} \cite{CudellIC}. A subtraction of the pole to $\Gamma^c_P$ is required: 
\beq\nonumber
\Gamma^c_{P\,{\rm sub}}(p,\mu_{\rm val})=\Gamma^c_P(p,\mu_{\rm val}) 
- \frac{a^\Gamma(p) + b^\Gamma(p)\mu_{\rm val} + c^\Gamma(p) \mu^2_{\rm val}}{\mu_{\rm val}}.
\eeq 
\begin{figure}[!]
    \begin{center}
      \includegraphics*[angle=-90,width=6cm]{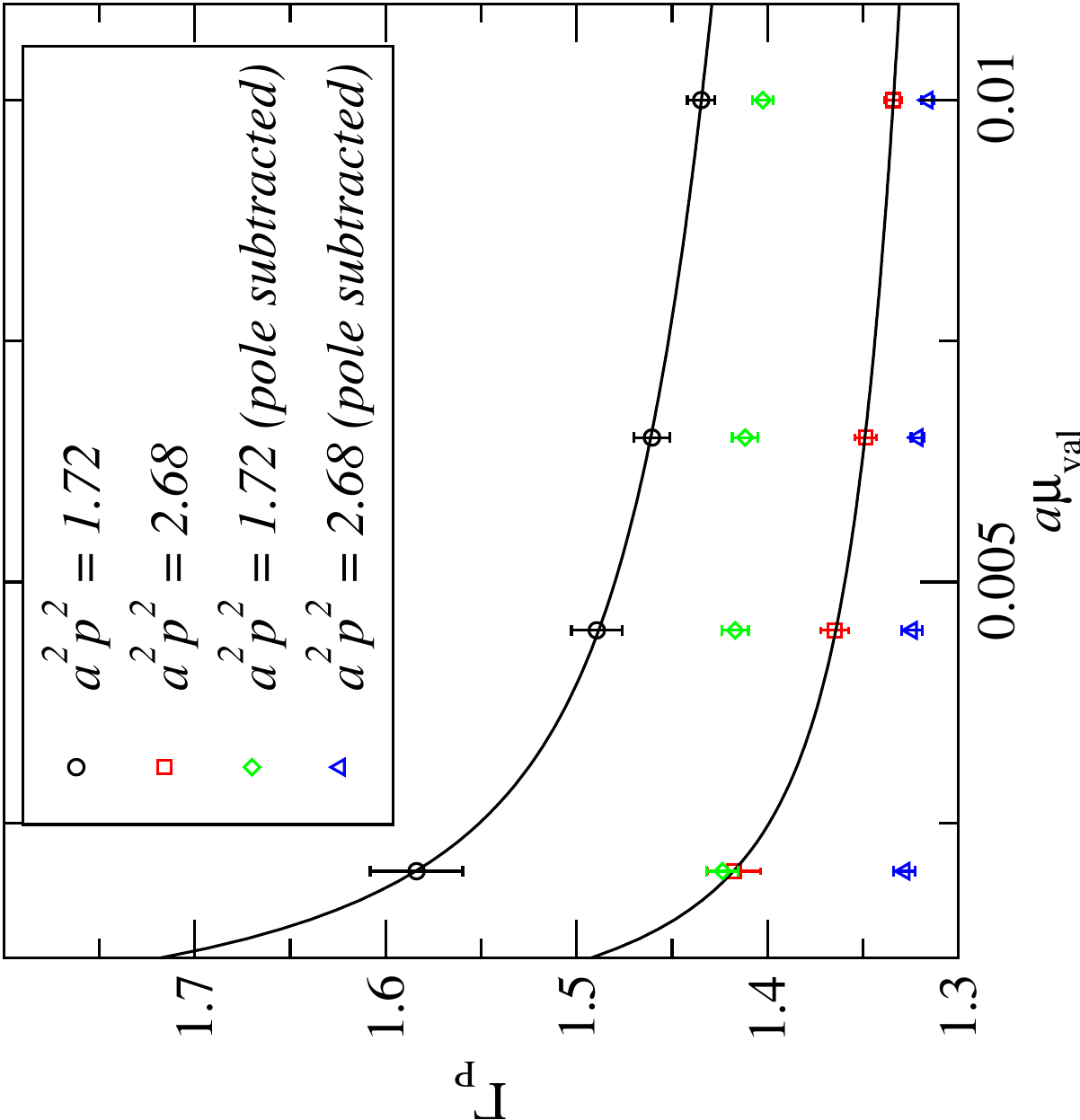}
    \end{center}
\caption{Subtraction of the Goldstone pole to the pseudoscalar amputated vertex by the fit 
$\Gamma_P(p,\mu_{\rm val},\mu_{\rm sea}) = \Gamma_{P\,{\rm sub}}(p,\mu_{\rm val},\mu_{\rm sea})
+ \frac{a^\Gamma(p) + b^\Gamma(p)\mu_{\rm val} + c^\Gamma(p) \mu^2_{\rm val}}{\mu_{\rm val}}$; the ETMC ensemble is ($\beta=4.2$, $\mu_{\rm sea}=0.002$, $L=24$) and has not been used in the extraction of hadronic quantities. \label{fig:vertPsub}}
\end{figure}
We have illustrated in Figure \ref{fig:vertPsub} how the subtraction is working for one of the ETMC ensembles.\\
Then, the standard RI-MOM procedure is applied: from each $p^2$ scale, a perturbative running \cite{ChetyrkinPQ} is performed to the scale $\mu=1/a$ to get $Z_P(\mu=1/a,p^2)$. It is equal to $Z_P(\mu=1/a)$ up to ${\cal O}(a^2 p^2)$ cut-off effects that are eliminated by subtracting the ${\cal O}(g^2_0)$ contribution from perturbation theory \cite{ConstantinouTR} and extrapolating the reminder at $a^2 p^2=0$ by a linear fit. The techniques is depicted in Figure \ref{fig:ZPMOM} that we took from \cite{ConstantinouGR}. Then the result is converted to the $\overline{\rm MS}$ scheme at 2 GeV.\\
\noindent We can finally analyse the (squared) light-light meson masses $r^2_0 m^2_{ll}$ and decay constants $r_0 f_{ll}$ in function of the renormalized light quark masses $r_0 \bar{m}_l$ ($\overline{\rm MS},\,2\,{\rm GeV}$). We perform a fit using the $\chi$PT formula at NLO
\begin{figure}[t]
    \begin{center}
      \includegraphics*[draft=false,angle=-90,width=0.5\textwidth]{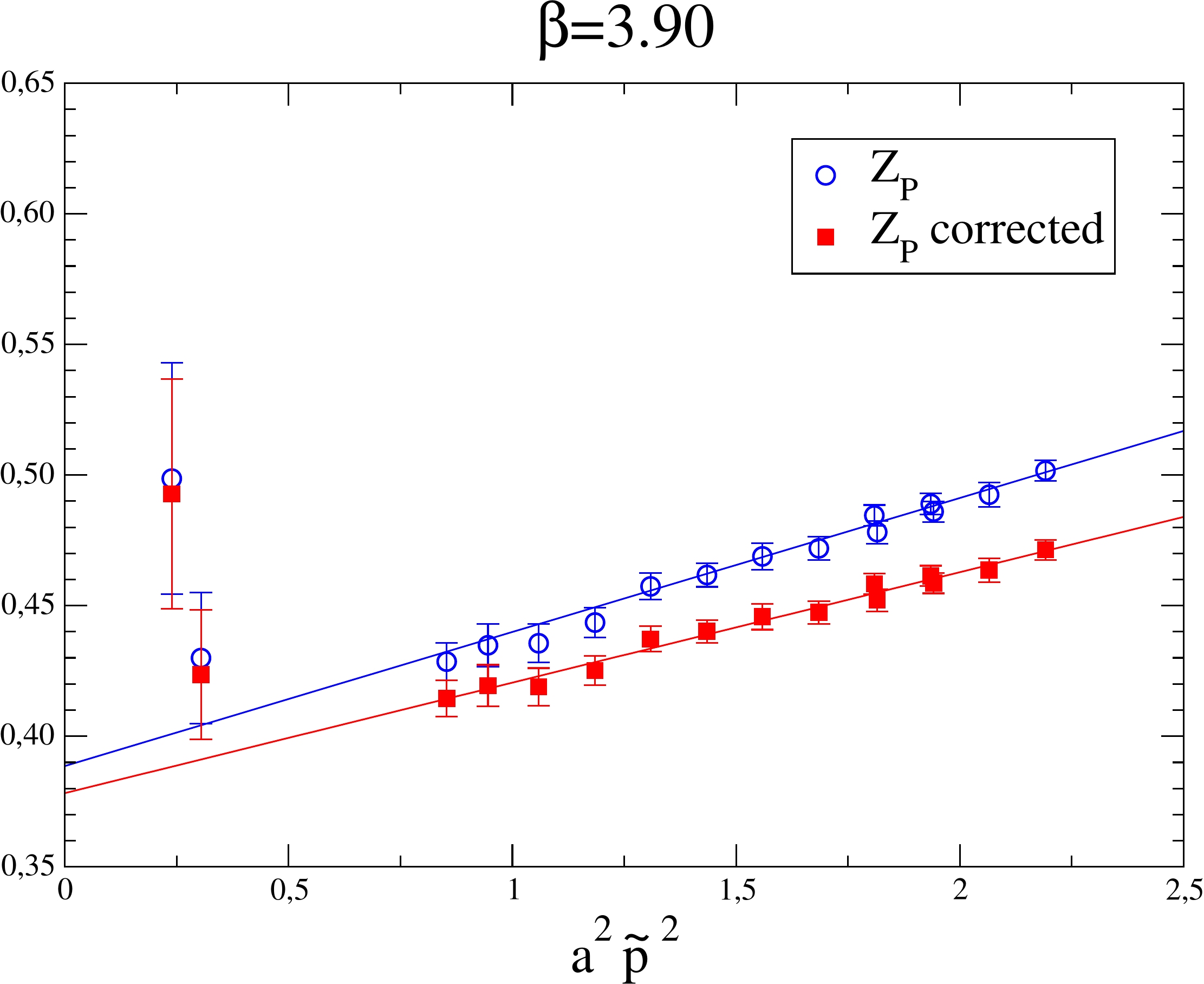}
    \end{center}
\caption{Renormalization constant $Z_P(\mu=1/a, p^2)$ at $\beta = 3.9$, run at $\mu=1/a$ from the initial scale $p^2$ defined by $a^2 \tilde{p}^2 \equiv \sum_i \sin^2(ap)_i$. Filled squares and empty circles are results obtained with and without the subtraction of the ${\cal O}(g^2a^2)$ discretization effects computed in perturbation theory \cite{ConstantinouTR}. The solid lines are linear fits to the data. The material comes from \cite{ConstantinouGR}. \label{fig:ZPMOM}}
\end{figure}
\begin{eqnarray}
\nonumber
r^2_0 m^2_{ll}&=& 2 r^2_0 B_0 \bar{m}_l \left\{1 + \frac{2 B_0 \bar{m}_l}{16 \pi^2 f^2_0} \ln \left(\frac{2 B_0 \bar{m}_l}{16 \pi^2 f^2_0}\right) + d_1 r_0 \bar{m}_l + \frac{a^2}{r^2_0} \left[d_2 + d_3 \ln\left(\frac{2 B_0 \bar{m}_l}{16 \pi^2 f^2_0}\right)
\right]\right\},\\
r_0 f_{ll}&=& r_0 f_0 \left\{1 - 2\frac{2 B_0 \bar{m}_l}{16 \pi^2 f^2_0} \ln \left(\frac{2 B_0 \bar{m}_l}{16 \pi^2 f^2_0}\right) + d_4 r_0 \bar{m}_l + \frac{a^2}{r^2_0} \left[d_5 + d_6 \ln\left(\frac{2 B_0 \bar{m}_l}{16 \pi^2 f^2_0}\right)\right]\right\}\,,
\end{eqnarray}
where $f_0$ and $B_0$ are usual leading order chiral Lagrangian low energy constants. The logarithmic contributions to the ${\cal O}(a^2)$ cut-off effect term stand for the mass splitting between charge and neutral pions in TmQCD and, at LO, are known as $d_3=-d_6=\frac{4c_2r^2_0}{(4\pi f_0)^2}$ \cite{BarJK}, \cite{HerdoizaSLA}, where $c_2$ is a linear combination of LEC's of the Wilson $\chi$PT theory \cite{BarMH}, that combines expansion in light quark masses and cut-off effects for Wilson fermions. Lattice data are obtained, of course, by simulating QCD in a finite volume of Euclidean space. Thus one has to correct for finite size effects the masses $m^2_{ll,L}$ and decay constants $f_{ll,L}$ extracted on the lattice: $m^2_{ll,L}=m^2_{ll} K^{\rm FSE}_M$ and $f_{ll,L}=f_{ll} K^{\rm FSE}_f$, where $K^{\rm FSE}_M$ and $K^{\rm FSE}_f$ are known in the literature \cite{ColangeloCU} for TmQCD, again taking into account the mass splitting.
\begin{figure}[t]
    \begin{center}
      \includegraphics*[draft=false,angle=-90,width=0.5\textwidth]{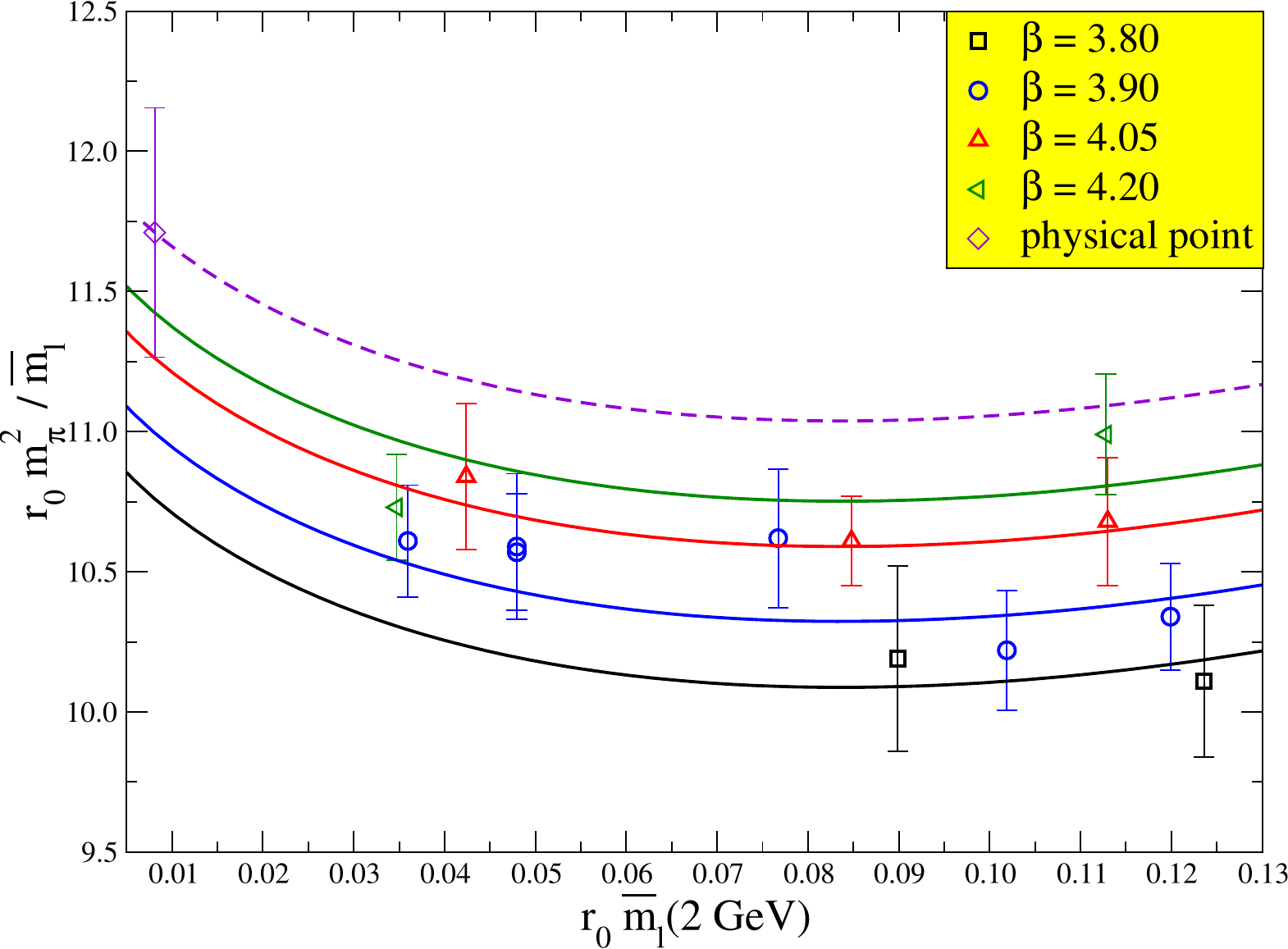}
    \end{center}
\caption{Dependence of $r_0 m^2_\pi/\bar{m}_l$ on the renormalized light quark mass $\bar{m}_l$ (${\overline{\rm MS}}$ scheme, $2\,{\rm GeV}$) at the four lattice spacings considered by ETMC.
\label{fig:mll}}
\end{figure}
We plot in Figure \ref{fig:mll} $r_0 m^2_{ll}/\bar{m}_l$ in function of $r_0 \bar{m}_l$: NLO curvatures are clearly visible and cut-off effects are smaller than 10\% at the lattice spacing $\sim$ 0.085 fm.
$\hat{m}$ is deduced by imposing that $[m_{ll}/f_{ll}](\hat{m})=m_\pi/f_\pi$. $r_0$ is then determined by imposing $r_0 f_{ll} (\hat{m})=r_0 f_\pi$ and the lattice spacing can be set through $a f_{ll}(\hat{m}) = a f_\pi$.\\
In a first work, analysing the data at $\beta=3.9$ only, we had obtained \cite{BlossierVV}
\begin{equation}
\hat{m}^{\overline{\rm MS}}(2\,{\rm GeV})=3.85 \pm 0.12 \pm 0.40\,{\rm MeV},
\end{equation}
with a first error of statistical origin while the second error included the spread among different chiral analyses: $\chi$ PT and polynomial fits. However no cut-off effects could be systematically kept under control.
The complete set of available ensembles were analysed in a second work and the result reads \cite{BlossierCR}
\begin{equation}
\hat{m}^{\overline{\rm MS}}(2\,{\rm GeV})=3.6(1)(2)\, {\rm MeV},
\end{equation}
where the first error is statistical and the second error includes\\ 
-- the discrepancies between NLO and NNLO chiral fits (the latter is done by adding a term in $ m^2_l$ and imposing priors on the NLO LEC's) (4\%),\\ 
-- taking into account or not the cut-off effects in the fit (3\%),\\ 
-- the truncation error in the perturbative conversion from RI-MOM to $\overline{\rm MS}$ schemes (estimated as large as the ${\cal O}(\alpha^3_s)$ term of the perturbation theory) (2\%).\\
We passed the quality criteria by FLAG so that our result is included in their ${\rm N_f}=2$ average \cite{AokiLDR}.\\
The strange quark mass has been measured by studying the data in the kaon sector. The analysis is based
either on an NLO SU(2)-$\chi$PT \cite{AlltonPN}, without any chiral logs,
\bea
m_K^2(m_s, m_l, a)&=&Q_1(m_s)+Q_2(m_s)\,m_l+Q_3(m_s)\,a^2\,,\quad\forall\, m_s\,,
\label{eq:mK2SU2_2}\\
m_K^2(m_s, \hat{m}, a=0)&\equiv& Q_1(m_s)+Q_2(m_s)\,\hat{m}=Q_4+Q_5\, m_s\,,
\eea
or on the SU(3)-$\chi$PT~\cite{SharpeBY} formula:
\bea
m_K^2(m_s, m_l, a)&=&B_0 (m_s+m_l) (1+
Q_6(m_s)+Q_7(m_s)\, m_l+Q_8(m_s)\,a^2)\,,\,\forall\, m_s\,,
\label{eq:mK2SU3_3}\\
m_K^2(m_s, \hat{m}, a=0)&\equiv& B_0 (m_s+\hat{m}) (1+
Q_6(m_s)+Q_7(m_s)\, \hat{m})\nn\\
&=&B_0 (m_s+\hat{m}) \left(1+\frac{2\,B_0\,m_s}{(4\pi f_0)^2}\,\ln\frac{2\,B_0\,m_s}{(4\pi f_0)^2}+
Q_9\,m_s\right)\,,
\eea
where $B_0$ and $f_0$ come from the pion sector fit. 
We have followed a strategy of, first, fixing the strange quark mass to three "reference" masses, close to the physical point, interpolating the lattice data at those points by a quadratic spline fit, and finally extrapolating to the chiral and continuum limits using either the SU(2) or the SU(3) formulae (\ref{eq:mK2SU2_2}) and (\ref{eq:mK2SU3_3}).\\
We have also studied how the mass of a fictitious $\eta_s$ meson, made of two valence strange quarks with "up" and "down" isospin types, depends on the strange and light quark masses: a very weak dependence on $m_l$ is expected, because it is a sea effect.
The mass of the ficticious $\eta_s$ meson is related to $m_K$ and $m_\pi$. Again, one can refer to a SU(2) $\chi$PT expression
\beq
m_{\eta_s}^2=R_1+R_2\,(2\,m_K^2-m_\pi^2)+R_3\,m_\pi^2+R_4\,a^2\,,
\label{eq:etaKpiSU2}
\eeq
or an SU(3) $\chi$PT one
\bea
m_{\eta_s}^2&=&(2\,m_K^2-m_\pi^2)\cdot\left[1 + (\xi_s-\xi_l)\,\log(2\,\xi_s)+
(R_7+1)\,(\xi_s-\xi_l)+R_8\,a^2\right]\nn\\
&&- m_\pi^2\,\left[-\xi_l\,\log(2\,\xi_l)+\xi_s\,\log(2\,\xi_s)+R_7\,(\xi_s-\xi_l)\right]\,
\label{eq:etaKpiSU3},
\eea
with $\xi_l=m^2_\pi/(4\pi f_0)^2$ and $\xi_s=(2m^2_K - m^2_\pi)/(4\pi f_0)^2$.
\begin{figure}[t]
    \begin{center}
\begin{tabular}{cc}      
\includegraphics*[draft=false,angle=-90,width=0.5\textwidth]{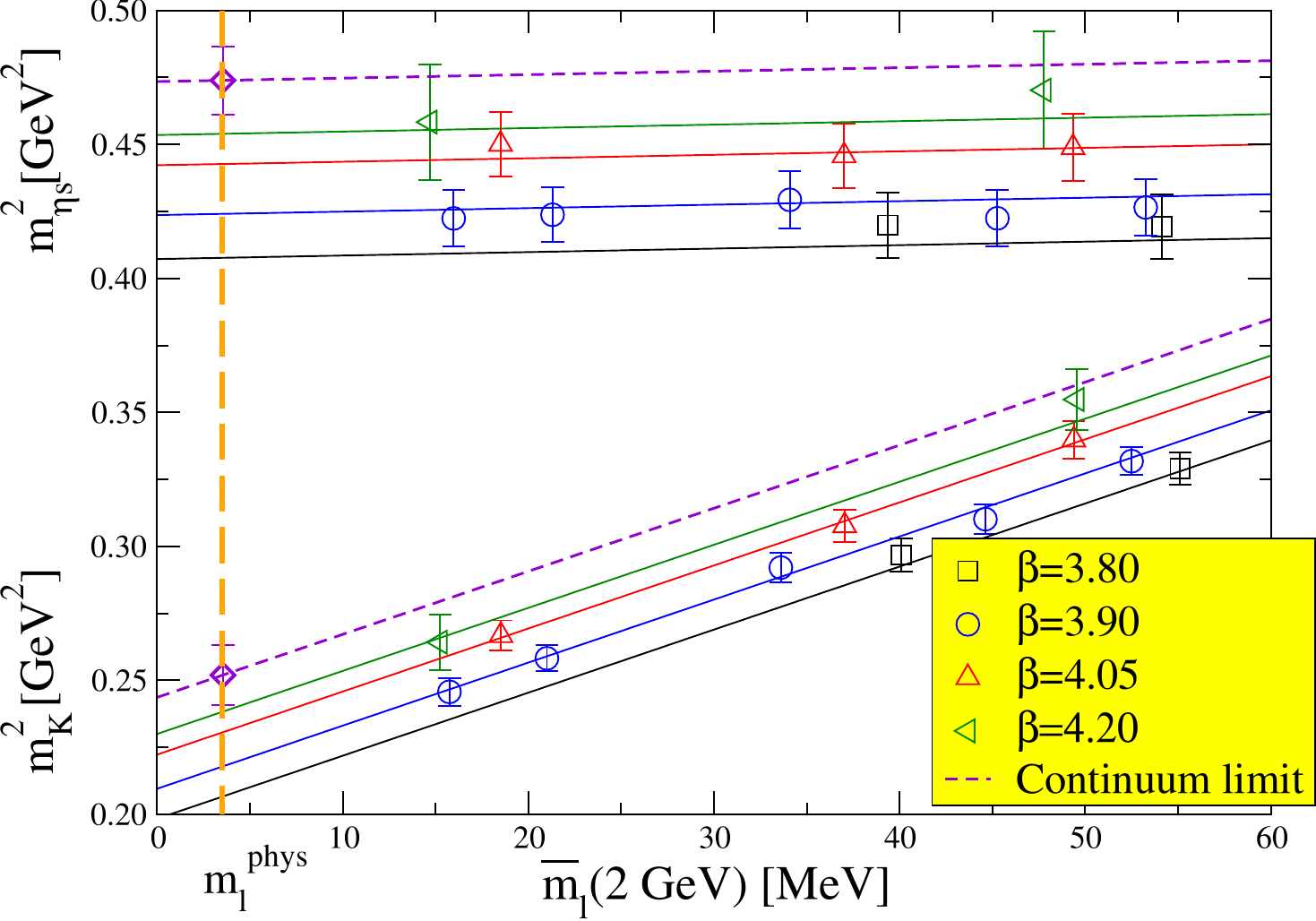}
&
\includegraphics*[draft=false,angle=-90,width=0.5\textwidth]{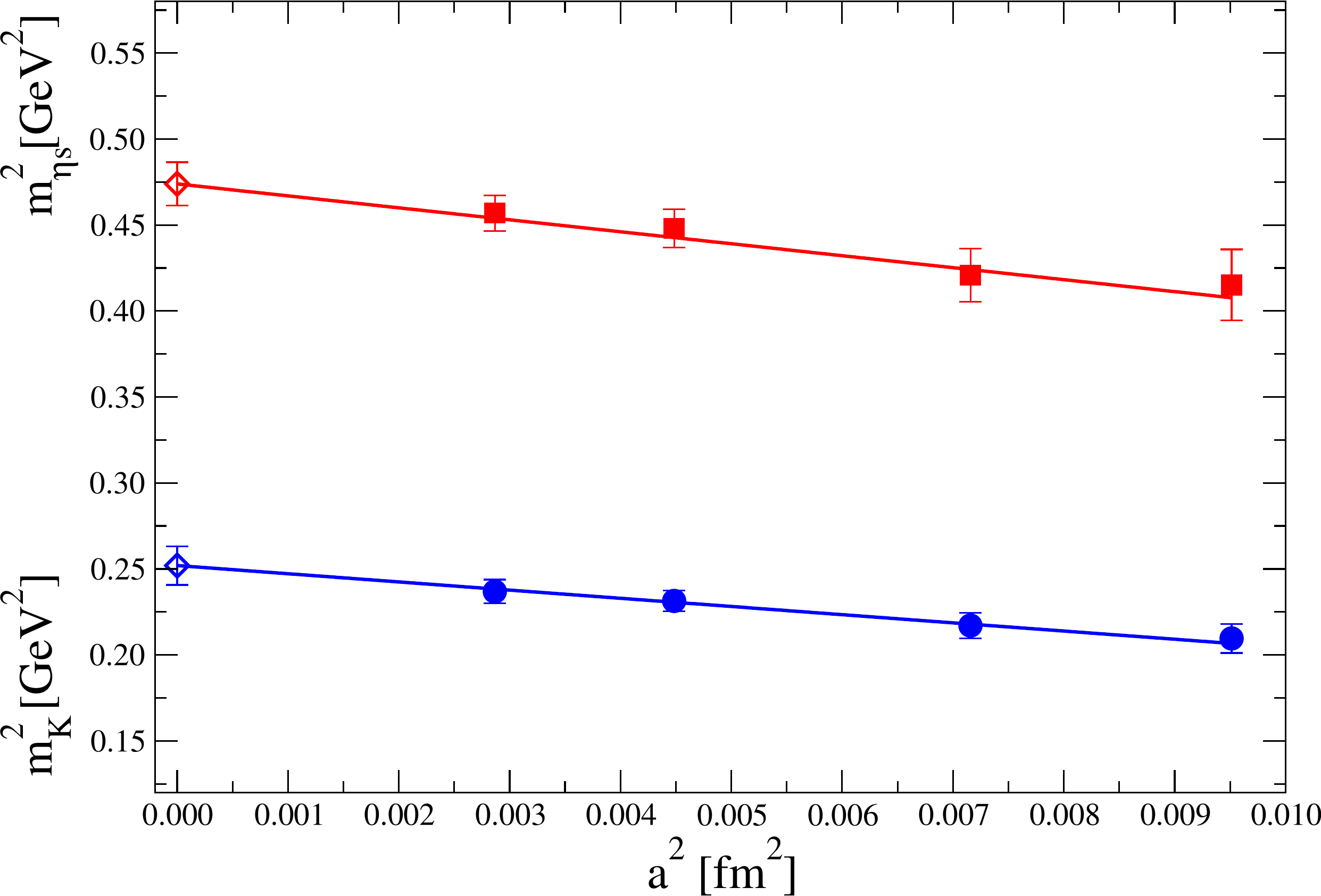}
\end{tabular}
\end{center}
\caption{Left panel: dependence of $m^2_K$ and $m^2_{\eta_s}$ on the renormalized light quark mass, for a fixed reference strange quark mass ($m^{\rm ref}_s = 95\, {\rm MeV}$) and at the four lattice spacings considered by ETMC; the orange vertical line corresponds to the physical $\hat{m}$ light mass. Right panel: dependence of $m^2_K$ and $m^2_{\eta_s}$ on the squared lattice spacing, for a fixed reference strange quark mass ($m^{\rm ref}_s = 95\, {\rm MeV}$) 
and at the physical $\hat{m}$ light mass; empty diamonds represent continuum limit results.
\label{fig:ms}}
\end{figure}
Our finding is that, within SU(2), $m_{\eta_s}=692(1)$ MeV and, within SU(3), $m_{\eta_s}=689(2)$ MeV, to be compared with the LO SU(3) prediction $(m_{\eta_s})^{\rm LO}=\sqrt{2m^2_K-m^2_\pi}=686$ MeV. Then, we have analysed $m_{\eta_{s}}$ data with the following fit functions of the light quark mass, the lattice spacing, and the strange quark mass from SU(2) and SU(3) $\chi$PT: 
\bea
\label{eq:mssSU2_1}
m_{\eta_s}^2(m_s, m_l, a)&=&T_1(m_s)+T_2(m_s)\,m_l+T_3(m_s)\,a^2\,,\quad \forall\,m_s\,,\\
m_{\eta_s}^2(m_s, \hat{m}, a=0)&\equiv& T_1(m_s)+T_2(m_s)\,\hat{m}=T_4+T_5\,m_s\,
\label{eq:mssSU2_2},
\eea
\bea
\label{eq:mssSU3_1}
m_{\eta_s}^2(m_s, m_l, a)&=&2\,B_0\,m_s\cdot [1+T_6\,(m_s)+T_7(m_s)\,m_l+T_8(m_s)\,a^2]\,,\quad \forall\,m_s\,,\\
\nonumber
m_{\eta_s}^2(m_s, \hat{m}, a=0)&\equiv&2\,B_0\,m_s\cdot[1+T_6\,(m_s)+T_7(m_s)\,\hat{m}]\\
&=&2\,B_0\,m_s\cdot \left[2\,\frac{2\,B_0\,m_s}{(4\pi f_0)^2}\,\log\left(2\,\frac{2\,B_0\,m_s}{(4\pi f_0)^2}\right)+T_9+T_{10}\,m_s \right]\,.
\label{eq:mssSU3_2}
\eea
$T_2$ and $T_7$ are found to be independent of the light quark mass. We have shown in Figure \ref{fig:ms} that our analysis could keep under control the cut-off effects and the mild light quark mass dependence of our results. We have collected in Table \ref{tab:ms} the strange mass estimates from $m_K$ and $m_{\eta_s}$: the consistency among several fits is reassuring. Taking the weighted average between the different determinations (\emph{via} $m_K/m_{\eta_s}$ and SU(2)/SU(3) fits), we obtain 
\beq
m_s(\overline{\rm MS},\, 2\,{\rm GeV})=95(2)(6)\,{\rm MeV},
\eeq
where the first error is statistical and the second error includes the discrepancy between:\\ 
-- $m_s(m_K)$ and $m_s(m_\eta)$ (3\%),\\
-- SU(2) and SU(3) fits in $m_s(m_K)$ (3\%),\\
-- different fits in the light sector (3\%),\\
-- exclusion or not of the $\beta=3.8$ data in the analysis, to assess an uncertainty on the continuum extrapolation (2\%),\\
-- inclusion or not of the ${\cal O}(\alpha^3_s)$ term in the conversion from RI-MOM to $\overline{\rm MS}$ schemes, to get an idea of the truncation error on perturbation theory (2\%).\\
Having a look to the right panel of Figure \ref{fig:ms}, we evaluate the cut-off effects to be around 15\% 
\begin{table}[t]
\begin{center}
\begin{tabular}{|c|c|c|c|c|}
\hline
$m_s(\overline{\rm MS},\, 2\,{\rm GeV})\, [{\rm MeV}]$& $K$-SU(2) & $K$-SU(3) & $\eta_s$-SU(2) & $\eta_s$-SU(3) \\ \hline
L1 & $92.1(3.8)$ & $94.7(2.2)$ & $96.0(2.6)$ & $95.5(2.1)$\\\hline
L2 & $91.6(3.9)$ & $94.6(2.3)$ & $95.4(2.6)$ & $95.3(1.9)$\\\hline
L3 & $95.4(3.8)$ & $94.7(2.1)$ & $99.4(2.9)$ & $97.7(2.2)$\\
\hline
\end{tabular}
\end{center}
\vspace{-0.4cm}
\caption{Results for the strange quark mass in the $\overline{\rm MS}$ scheme at 2 GeV, as obtained from the different fits within the light and strange quark sectors: $\hat{m} (\overline{\rm MS},\,2\,{\rm GeV})$ is extracted by a chiral fit at NLO
with ${\cal O}(a^2)$ discretization effects (L1), a chiral fit at NLO without including discretization effects (L2) or a chiral fit at NNLO
with ${\cal O}(a^2)$ discretization effects (L3).}
\label{tab:ms}
\end{table}
on $m^2_K$ at $a=0.085$ fm: they were affecting the first estimate of $m_s$ performed by ETMC at $\beta=3.9$.
We obtain for the scale and scheme-independent ratio $m_s/\hat{m}=27.3(5)(7)=27.3(9)$.\\
Finally the charm quark has been extracted by studying the $D$, $D_s$ and "charged" $\eta_c$ sectors 
\begin{figure}[t]
    \begin{center}
\begin{tabular}{cc}      
\includegraphics*[draft=false,angle=-90,width=0.5\textwidth]{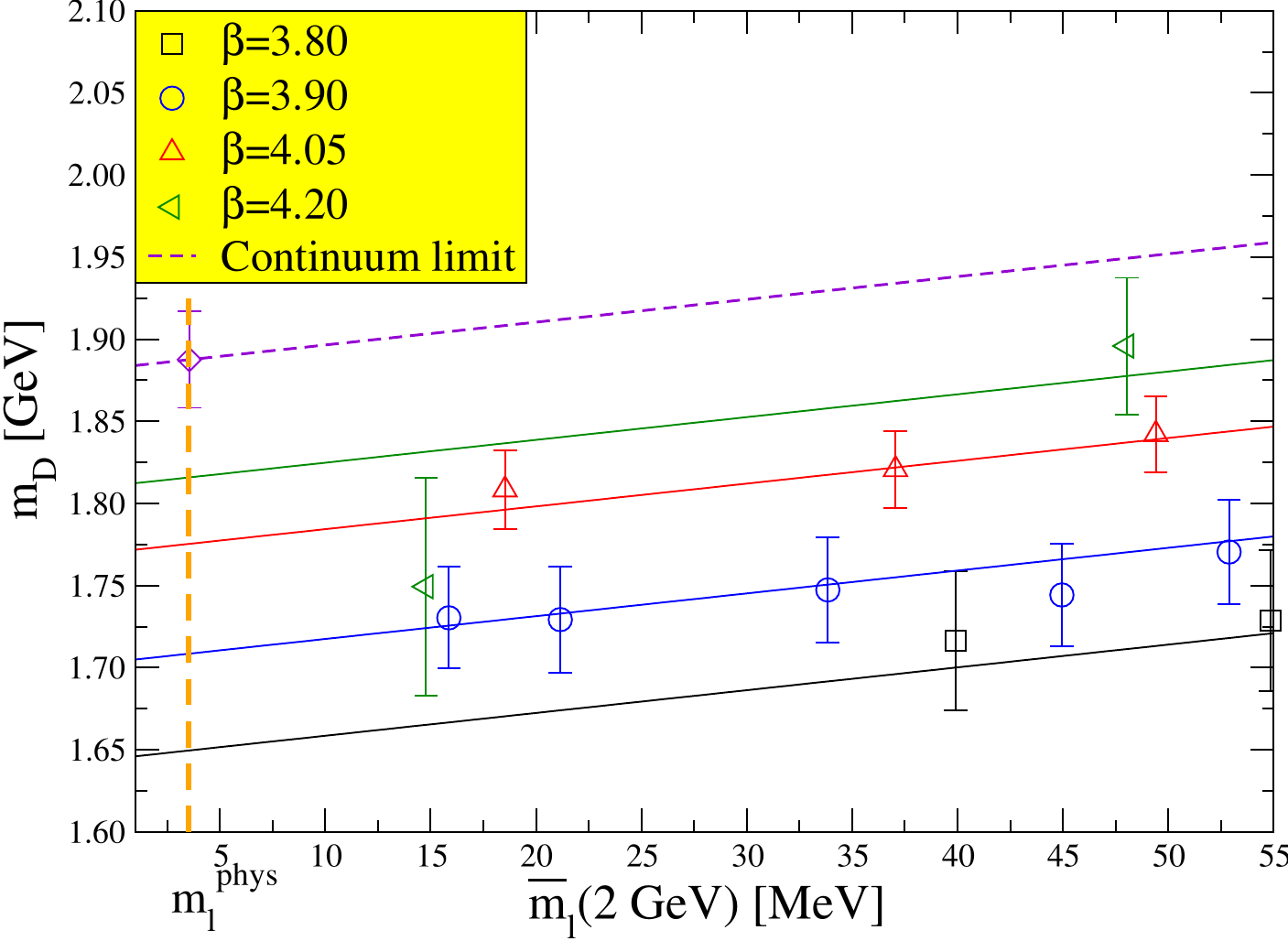}
&
\includegraphics*[draft=false,angle=-90,width=0.5\textwidth]{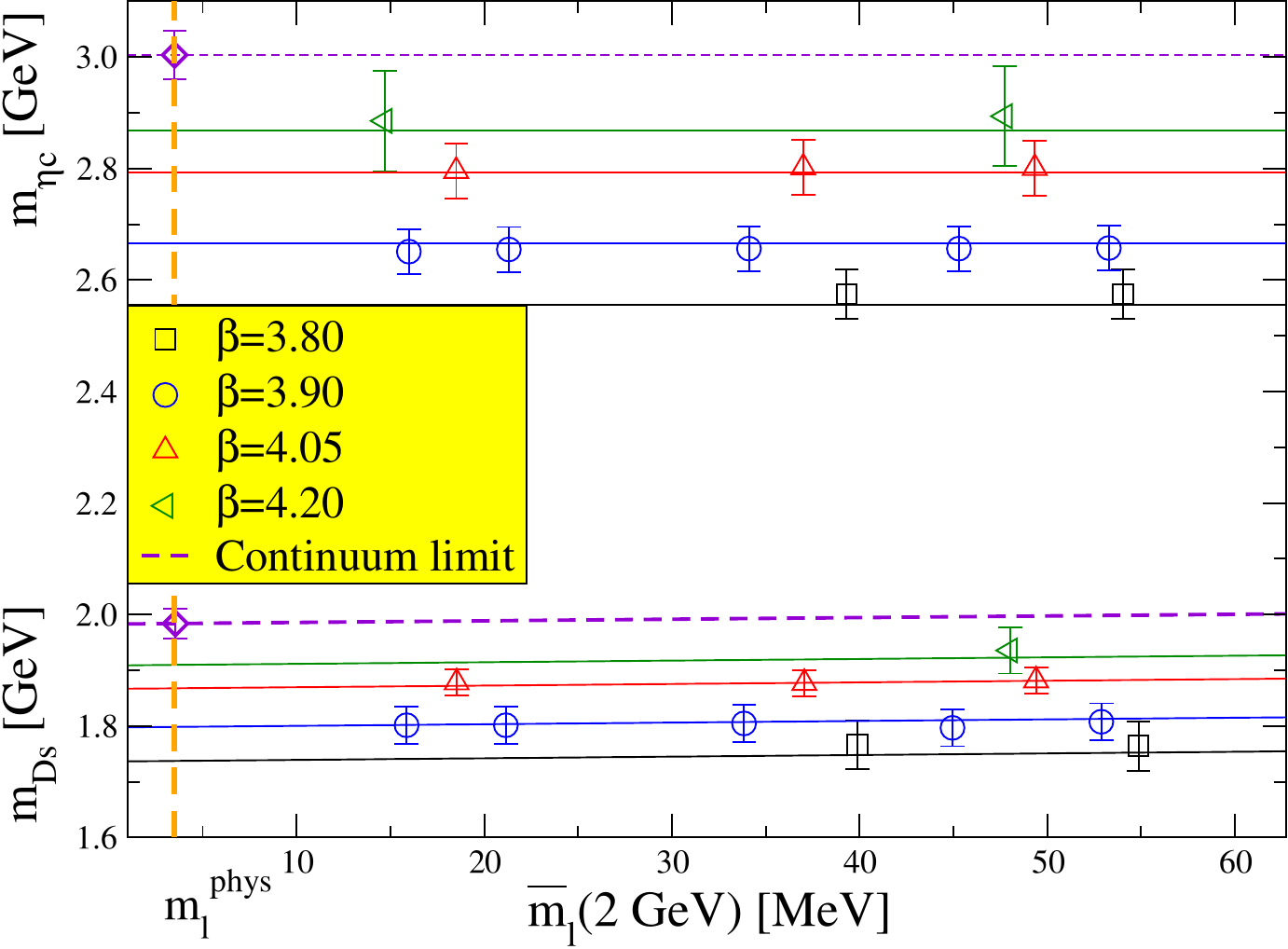}
\end{tabular}
\end{center}
\caption{Left panel: dependence of $m_{D}$ on the renormalized light quark mass, at fixed reference charm quark mass ($m_c^{\rm ref}=1.16\, {\rm GeV}$) and at the four lattice spacings considered by ETMC; the orange vertical line indicates the physical $\hat{m}$ light mass. Right panel: dependence of $m_{D_s}$ and $m_{\eta_c}$ on the renormalized light quark mass, at fixed reference strange quark mass ($m_s^{\rm ref} = 95\, {\rm MeV}$) and charm quark mass ($m_c^{\rm ref}=1.16\, {\rm GeV}$) and at the four lattice spacings considered by ETMC; the orange vertical line indicates the physical $\hat{m}$ light mass.
\label{fig:mcml}}
\end{figure}
\begin{figure}[t]
    \begin{center}
\includegraphics*[draft=false,angle=-90,width=0.5\textwidth]{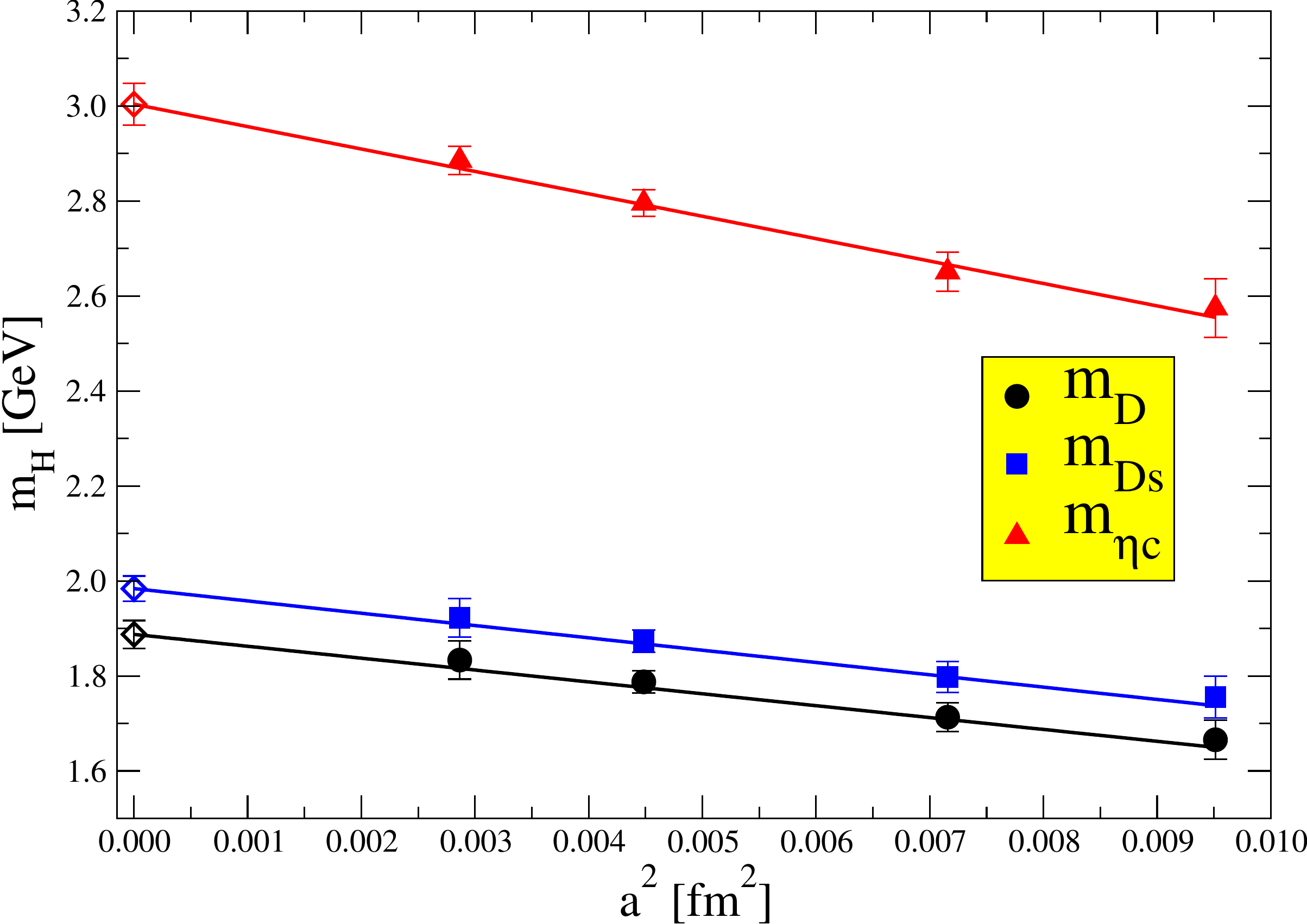}
\end{center}
\caption{Dependence of $m_D$, $m_{D_s}$ and $m_{\eta_c}$, at fixed reference charm quark mass ($m_c^{\rm ref} = 1.16\, {\rm GeV}$) and at physical light and strange quark masses, on $a^2$; empty diamonds represent continuum limit results.
\label{fig:mccutoff}}
\end{figure}
-- it means that we have neglected any disconnected contribution to the $\eta_c$ correlator. Again, we have first interpolated heavy-light and charmonium meson masses to three reference charm masses. Then the dependence on $m_l$, $m_s$ and $a^2$ is settled. $m_c$ is given from $m^{\rm exp}_D=1.87$ GeV, 
$m^{\rm exp}_{D_s}=1.969$ GeV and $m^{\rm exp}_{\eta_c}=2.981$ GeV. We have considered polynomial fits to deal with cut-off effects and $m_{l,s}$ dependences:
\beq
m_{H}(m_c,m_s,m_l,a)=C_1^{H}(m_c)+C_2^{H}(m_c)\,m_l+C_3^{H}(m_c)\,m_s+C_4^{H}(m_c)\,a^2\,,\quad \forall m_c\,,
\label{eq:DDsetac_ref}
\eeq
where $H\equiv D$, $D_s$ and $\eta_c$. Inspired by scaling laws of Heavy Quark Effective Theory \cite{NeubertMB}, we have fitted the $m_c$ behaviour of $m_H$ using the following function:
\bea
m_{H}(m_c,m_s^{phys},\hat{m},a=0)&\equiv& C_1^{H}(m_c)+C_2^{H}(m_c)\,\hat{m}
+C_3^{H}(m_c)\,m_s^{phys}\nn\\
&=&C_5^{H}+\frac{C_6^{H}}{m_c}+C_7^{H}\,m_c\,.
\label{eq:mhq_mc}
\eea
We have shown in Figure \ref{fig:mcml} the $m_l$ dependence of the heavy-light meson masses, at a given 
$m_c^{\rm ref}$. We observe that it is almost invisible for $m_{D_s}$ and $m_{\eta_c}$, as expected because it reflects the weak influence of the sea quarks for those systems. In Figure \ref{fig:mccutoff}, we note that the cut-off effects on $m_{\eta_c}$ are $\sim 15\%$ at the coarsest lattice spacing, while they are $\sim 12\%$ for $m_{D_{(s)}}$. The different values of $m_c(\overline{\rm MS},\, 2\,{\rm GeV})$ are collected in Table \ref{tab:mc}.
\begin{table}[t]
\begin{center}
\begin{tabular}{|c|c|c|c|}
\hline
& $D$ & $D_s$ & $\eta_c$ \\ \hline
$m_c(\overline{\rm MS},\, 2\,{\rm GeV})\, [{\rm GeV}]$ & $1.14(3)$ & $1.14(3)$ & $1.15(2)$ \\
\hline
\end{tabular}
\end{center}
\vspace{-0.4cm}
\caption{Results for the charm quark mass in the $\msb$ scheme at 2 GeV, as obtained from the different fits within the charm quark sector.}
\label{tab:mc}
\end{table}
We obtain
\bea
\label{eq:rismc}
m_c(\overline{\rm MS},\,2\,{\rm GeV})&=&1.14(3)(3)\, {\rm GeV}=1.14(4)\, {\rm GeV},\\
m_c(\overline{\rm MS},m_c)&=&1.28(4)\,{\rm GeV},
\eea
where the first error is statistical and the second error counts for 
the discrepancy between:\\ 
-- $m_c(m_D)$, $m_c(m_{D_s})$ and $m_c(m_{\eta_c})$ (1\%),\\
-- exclusion or not of the $\beta=3.8$ data in the analysis, to assess an uncertainty on the continuum extrapolation (1.5\%),\\
-- inclusion or not of the ${\cal O}(\alpha^3_s)$ term in the conversion from RI-MOM to $\overline{\rm MS}$ schemes, to get an idea of the truncation error on perturbation theory (2\%).\\
We also obtained $m_c/m_s=12.0(3)$.\\
We have shown in Figure \ref{fig:massesFLAG} the collection made by FLAG for the quark masses. Results at ${\rm N_f}=2$ and $2+1$ are quoted. \emph{We notice a small variation $\simeq 10$\% of the results with the number of flavours}: this is in our opinion the main message of the series of works we discussed. Very recently a first estimate with ${\rm N_f}=2+1+1$ was presented \cite{CarrascoLAA} which confirms this tendency of a weak dependence on the number of flavours: $m^{{\rm N_f}=2+1+1}_s(\overline{\rm MS},\,2\,{\rm GeV})=99.2(3.9)\,{\rm MeV}$. Concerning the charm quark mass, apart the other recent estimate at ${\rm N_f}=2$ $m_c(\overline{\rm MS},m_c)=1.273(6)\,{\rm GeV}$ \cite{McNeileJI} included in the world average by PDG \cite{BeringerZZ}, a result at ${\rm N_f}=2+1+1$ was also publised in \cite{CarrascoLAA}: $m_c(\overline{\rm MS},m_c)=1.350(49)\,{\rm GeV}$.
We point out also the fact that taking carefully into account the cut-off effects in analyses is crucial.
\begin{figure}[t]
    \begin{center}
\begin{tabular}{cc}
\includegraphics*[draft=false,width=0.5\textwidth]{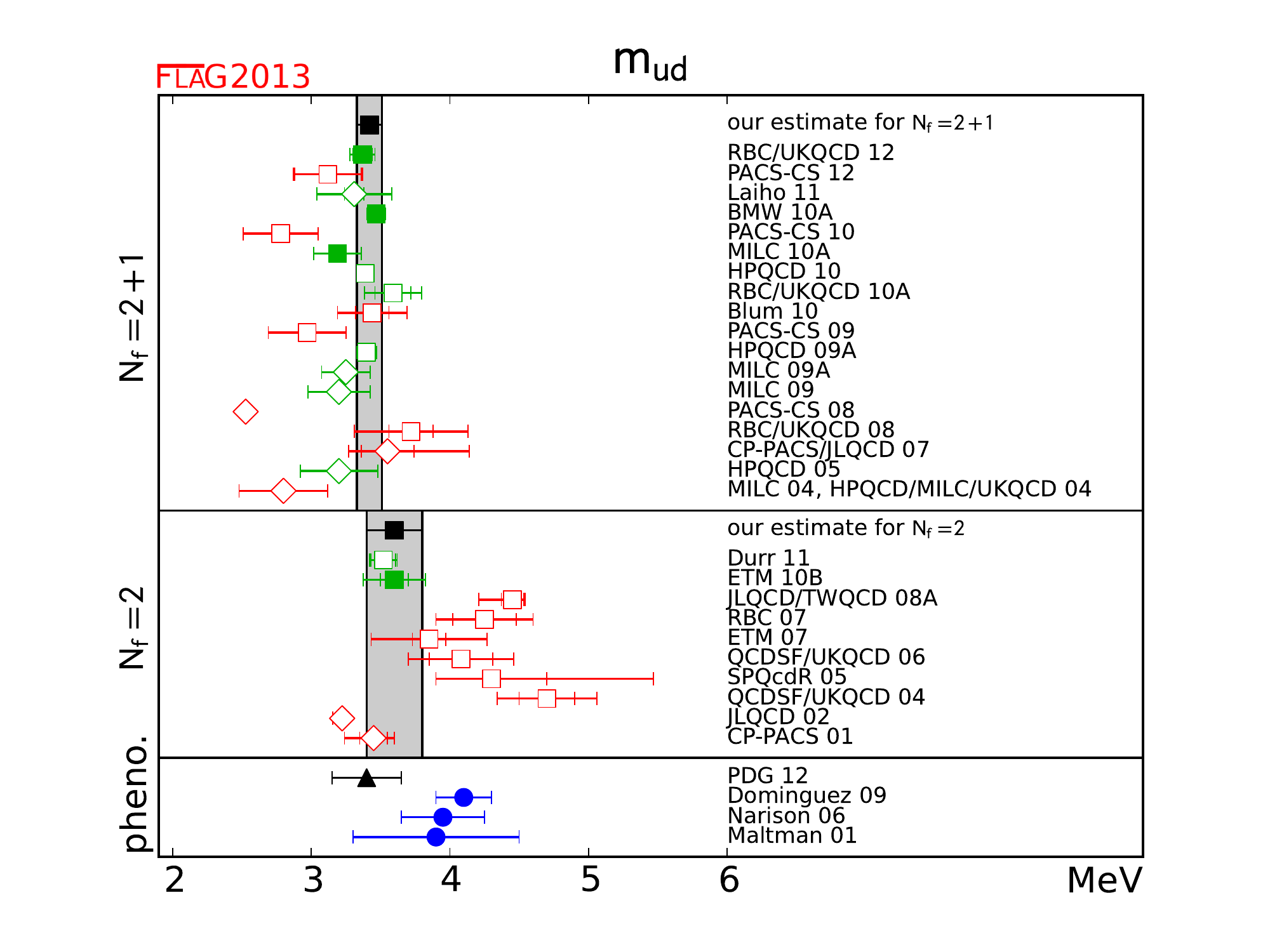}
&
\includegraphics*[draft=false,width=0.5\textwidth]{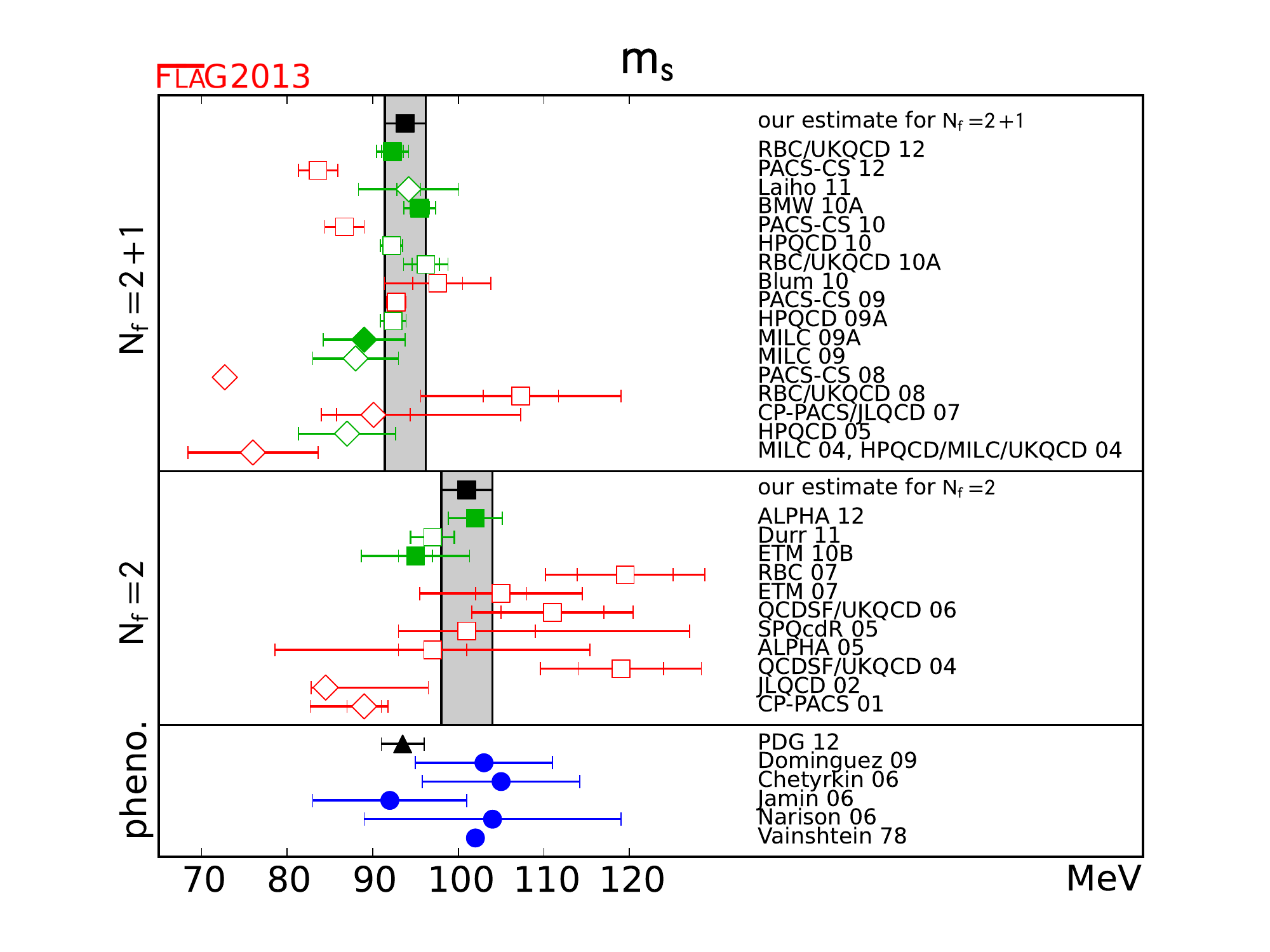}
\end{tabular}
\end{center}
\caption{Collection of lattice and analytical results for light (left panel) and strange (right panel) quark masses made by FLAG \cite{AokiLDR}.
\label{fig:massesFLAG}}
\end{figure}

\section{Back to phenomenology}

Our value of $m_{s}/\hat{m}$ is, as other lattice results that are rather obtained at ${\rm N_{f}}=2+1$, in good agreement with the LO formula of SU(3) $\chi$PT \cite{WeinbergHB}
\beq\nonumber
m_{s}/\hat{m}\stackrel{{\rm LO}}{=} \frac{\hat{M}^{2}_{K^{+}}+\hat{M}^{2}_{K^{0}} - \hat{M}^{2}_{\pi^{+}}}
{\hat{M}^{2}_{\pi^{+}}}\stackrel{{\rm isospin}}{\equiv}25.9,
\eeq
where $\hat{M}$ is a meson mass obtained by neglecting electromagnetic effects. It tells that, though the 
SU(3)-flavour symmetry breaking effects are large within the 8 Goldstone bosons, the corresponding mass spectrum obeys formulae dictated by SU(3)$\times$SU(3) chiral symmetry.

\noindent Precision Higgs physics cannot be studied in such a ``dirty'' environment as the LHC. $e^{+}e^{-}$ linear colliders are appropriate. At the moment we do not know whether ILC will be planned but a lot of investigations have been led to predict which accuracy can be reached in the measurement of, for instance, Higgs couplings to quarks. The process $H \to c \bar{c}$ is quite favoured at a Higgs mass of 126 GeV and the partial width is given by 
\beq\nonumber
\Gamma(H \to c \bar{c})=\frac{3G_{F}}{4\sqrt{2}\pi}m_{H}m_{c}^{2}(\overline{{\rm MS}},m_{H})\left[1+\Delta_{cc} + \Delta^{2}_{H}\right],
\eeq 
where $\Delta_{cc}$ and $\Delta^{2}_{H}$ are QCD correction \cite{qcd1higgs},  \cite{qcd2higgs}. The error of $\sim$ 2\% on $m_{c}(\overline{{\rm MS}},m_{c})$ quoted by PDG propagates onto an error of $\sim$ 3\% on $m_{c}(\overline{{\rm MS}},\, 126\,{\rm GeV})$. The uncertainty on the branching ratio of $H \to c \bar{c}$ experimentally measured at ILC with $\sqrt{s}=500$ GeV is estimated to be $\sim$ 8\% \cite{DeschXQ}. The uncertainty on $m_{c}$ clearly dominates this error.



\chapter{The bottom mass}

\fancyhead[LO]{\bfseries \leftmark}
\fancyhead[RE]{\bfseries \rightmark}

The recent experimental observation of a new particle with a mass $\sim$ 126 GeV is the kick-end of the
tests to validate the Standard Model, that describes the subatomic physics down to the fermi scale; if its interpretation
as the SM Brout-Englert-Higgs boson is correct, the most probable hypothesis, it still remains to study carefully its various couplings, in particular those to quark-antiquark pairs. It is well known that the dominant channel decay at this mass scale is $H \to b \bar{b}$ \cite{DjouadiGI} and the width reads, as for $H \to c \bar{c}$,
\beq\nonumber
\Gamma(H \to b \bar{b})=\frac{3G_{F}}{4\sqrt{2}\pi}m_{H}m_{b}^{2}(\overline{{\rm MS}},m_{H})\left[1+\Delta_{bb} + \Delta^{2}_{H}\right].
\eeq
The uncertainty on the branching ratio measured at ILC is expected to be $\sim$ 2.5\% \cite{DeschXQ}. Again, the largest part comes from the uncertainty on $m_b$.\\
In the meanwhile, though an intense activity takes place in the search of exotic 
particles up to the TeV scale, low energy processes offer nice complementary
test benches of New Physics scenarios, especially if they are highly suppressed
in the Standard Model. The $B$ sector is potentially
among the most promising ones to detect in that way effects beyond the SM. 
Typically, the inclusive $B \to X_s \gamma$ radiative decay has received a lot of
attention \cite{PazSA}: for dimensional reasons its analytical expression scales as the
fifth power of the $b$ quark mass $m_b$. In practice, normalising by the inclusive  
$B \to X_c l \nu$ semileptonic decay width does not help much because $\sim$ 10\% of the
final component in $X_c$ is not fully understood theoretically \cite{BigiQP}. Studying asymmetries is often beneficial because a
lot of theoretical uncertainties (Cabibbo-Kobayashi-Maskawa matrix elements, hadronic form factors)
cancel in ratios such that describing experimental data by models in Effective
Field Theory approaches puts stringent constraints on dynamics at high energy,
by fixing bounds on the Wilson coefficients.
For instance it has been advocated that comparing the experimental $CP$ asymmetry in
$B \to X_s \gamma$ to the SM prediction still leaves room for NP \cite{AmhisBH}. However
a subtantial uncertainty remains from $m_c/m_b$ \cite{HurthDK}, in particular as far
as the renormalization scale and scheme are concerned.
On another hand, the process $B \to D^* l \nu$ appeared recently as a possible probe of NP
as well because an excess was experimentally observed with respect to SM prediction
\cite{LeesXJ}.
In one of the proposed NP scenarios, based on the exchange of scalar currents
mediated by charged Higgs, running the effective couplings $\propto m_b$ from $m_b$ scale
to TeV scale is a key ingredient \cite{FajferVX}. An important point to outline is that it
is theoretically acceptable
to choose the starting point of the renormalization Group Equation (RGE) at a much higher
scale, for instance 100 GeV. It is thus of great help to compute $m_b$ in the
Schr\"odinger Function renormalization scheme \cite{CapitaniMQ}, well defined both in the
asymptotically
free region and at the ${\cal O}(\Lambda_{\rm QCD})$ scale, and then convert it into
$\overline{\rm MS}$ \emph{at that scale of 100 GeV}, or at the Higgs mass scale of 126 GeV.

\noindent Determining from first principles of Quantum Field Theory the value of $m_{b}$ needs some
care: as other estimates, based on $e^{+}e^{-} \to Q\bar{Q}$ total cross section analysed by QCD sum rules
\cite{HoangXM}, \cite{PinedaGX}, \cite{BoughezalPX}, \cite{ChetyrkinFV}, \cite{NarisonRN} - \cite{LuchaGTA}, quarkonia spectrum modelised by the $Q \bar{Q}$ static potential in
perturbation theory \cite{PeninZV}, \cite{BrambillaQK}, \cite{LaschkaZR}, fitted moments of inclusive $B$ decays \cite{MahmoodTT}, \cite{BauerVE}, \cite{BuchmullerZV}, \cite{SchwandaKW}, \cite{AubertQDA},
or jet-events expressed in perturbation theory \cite{AbdallahCV}, it has its own
systematics. Indeed, lattice regularisation in a finite volume assumes that one
can control border size effects on a wide range of hadronic scales: there is roughly a factor 30 between the
pion mass and the $B$ meson mass. For the moment it is unrealistic to get rid simultaneously of the IR and UV
cut-off effects because the typical parameters of numerical simulations, the lattice spacing $a$ and lattice
extent $L$, are such that $L m_{\pi} \sim 4$ and $a m_{b} \sim 1$.
In the literature, methods based on Non Relativistic QCD \cite{GrayUR} and extrapolation up to the $b$ region results obtained in the charm region \cite{deDivitiisIY}, \cite{McNeileJI}, \cite{DimopoulosGX}, \cite{CarrascoZTA} have been investigated. ALPHA Collaboration has followed an effective field theory approach, i.e. extracting hadronic quantities in the
Heavy Quark Effective Theory (HQET) framework \cite{EichtenZV} with couplings of the effective Lagrangian and
currents derived through a non perturbative matching with QCD. In this chapter, particularly rich in theoretical concepts, 
we will introduce the Schr\"odinger Functional formalism, recall the basics of HQET, present its regularisation on the lattice in the context of Schr\"odinger Functional and sketch the different ingredients of the whole strategy we have applied: non-perturbative matching between QCD and HQET, extraction
of hadronic matrix elements using a powerful numerical tool to improve the statistics and paying a lot of attention to remove carefully the contribution from excited states to the correlators, in order to get $m_b$ in the quenched approximation and at ${\rm N_f}=2$.

\section{Quarks in a can of lattice}

Historically, very important conceptual steps in lattice gauge field theories have been made in the 1990's with the idea of performing numerically the running of physical quantities. First, it was tried to compute the $\Lambda$ parameter of QCD \cite{LuscherWU}. It means to introduce a scale, $L$, through the definition of a renormalized coupling $\bar{g}^2(L)$ in an appropriate scheme, i.e. a scheme that is not intimately related to perturbation theory, unlike the $\overline{\rm MS}$ scheme. A regularisation and mass independent scheme is particularly appealing. Then, the notion of \emph{step scaling function} has been developed: it allows, for instance, to apply the running $\bar{g}^2(L) \longrightarrow \bar{g}^2(sL) \equiv \sigma(s, \bar{g}^2(L))$ up to a very high scale where perturbative expressions can be used thanks to the asymptotic free behaviour of the theory. Computing the step scaling functions by a lattice regularisation needs a careful analysis of cut-off effects, the extrapolation to the continuum limit is a key ingredient.
Those pioneering studies led to the proposal of employing the Schr\"odinger Functional scheme, whose the main characteristics are the following:\\
-- The basic object is a partition function ${\cal Z}[C,C'] =\elematrice{C}{e^{-HT}}{C'}$ \cite{SymanzikWD} where $C(x_0=0)$ and $C'(x_0=T)$ are two given field configurations. In other words, the framework is the background field method.\\
-- It has been proved that ${\cal Z}$ is renormalizable with Yang-Mills theories \cite{LuscherAN}, i.e. there is no counterterm to add to cancel ultraviolet divergences.\\
-- The effective potential $\Gamma(\Phi_{\rm cl})\equiv - \ln {\cal Z}[C,C']$, obtained at the minimum field configuration of the action $S[\Phi]$, $\left.\frac{\delta S}{\delta \Phi}\right|_{\Phi=\Phi_{\rm cl}}=0$, can be expanded in function of the bare coupling $g_0$ and the classical field solution $\Phi_{\rm cl}$: $\Gamma(\Phi_{\rm cl})=g^{-2}_0\Gamma_0(\Phi_{\rm cl}) + \Gamma_1(\Phi_{\rm cl}) + g^2_0 \Gamma_2(\Phi_{\rm cl})+ \cdots$.\\ 
Letting depend the boundary fields $C$ and $C'$ on a generic parameter $\eta$ (for instance, a phase), the renormalized coupling in the Schr\"odinger Functional is then defined:
\beq\nonumber
\bar{g}^2(L) = \left.\left[\frac{\partial \Gamma_0(\Phi_{\rm cl})}{\partial \eta}\right] / \left[\frac{\partial \Gamma(\Phi_{\rm cl})}{\partial \eta}\right]\right|_{\eta=0}, \quad \bar{g}^2(L)=\left.\left\langle \frac{\partial S}{\partial \eta}\right\rangle\right|_{\eta=0}.
\eeq
So, it can be estimated by Monte-Carlo estimates \cite{LuscherZX}. The goal is then to tune the bare coupling $g_0$ to get a target value for the renormalized coupling $\bar{g}^2(L_{\rm max})$ from which the product $s^{n}L_{\rm max} \Lambda^{\rm SF}$ can be deduced after applying $n$ scaling steps:
\beq\nonumber
\Lambda^{\rm SF}=\frac{1}{s^n L_{\rm max}} [\beta_0 \bar{g}^2(s^nL_{\rm max})]^{-\frac{\beta_1}{2\beta^2_0}}\exp 
\left(-\frac{1}{2\beta_0 \bar{g}^2(s^n L_{\rm max})}\right)
\exp \left[ -\int^{\bar{g}(s^n L_{\rm max})}_0 dg \left(\frac{1}{\beta(g)}+\frac{1}{\beta_0 g^3}
-\frac{\beta_1}{\beta^2_0 g} \right)\right],
\eeq
\beq\nonumber
\beta(g)=-g^3(\beta_0 + \beta_1 g^2 + \beta_2 g^4 + \cdots).
\eeq
-- Adding matter fields is a bit of work, in particular to maintain the invariance under $CPT$ of the theory. However it has been proved that the Schr\"odinger Functional in QCD is renormalizable as well \cite{SintUN}, the fermion fields obey inhomogeneous Dirichlet boundary conditions in time and twisted boundary conditions in space: 
\begin{align*}
P_+\psi(x)|_{x_0=0}= \rho(\vec{x}), \quad P_-\psi(x)|_{x_0=T}= \rho'(\vec{x}),\\
\bar{\psi}(x)|_{x_0=0}\; P_- = \bar{\rho}(\vec{x}), \quad \bar{\psi}(x)|_{x_0=T}\; P_+ = \bar{\rho}'(\vec{x}),\\
\psi(x+L \hat{k})=e^{i\theta_k} \psi(x), \quad \bar{\psi}(x+L \hat{k})=e^{-i\theta_k} \bar{\psi}(x),
\end{align*}
where $P_\pm = \frac{1\pm \gamma^0}{2}$. The transfer matrix is still mathematically well defined.
An artistic view of the Schr\"odinger Functional is depicted in Figure \ref{fig:SF}: the lattice is put in a can.\\
The expectation value $\langle O \rangle$ of an operator is then given by
\begin{figure}[t]
\begin{center}
\includegraphics*[width=0.2\textwidth]{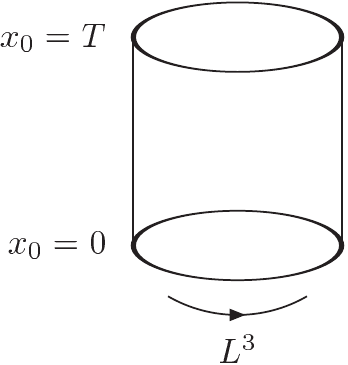}
\end{center}
\caption{\label{fig:SF} Pictorial view of lattice QCD in the Schr\"odinger Functional framework.}
\end{figure}
\beq\nonumber
 \lgl O \rgl = \left.\left(\frac{1}{\cal Z} \int [{\cal D} U][{\cal D} \psi][{\cal D} \bar{\psi}]
O e^{-S(U,\psi,\bar{\psi})}\right)\right|_{\rho=\bar{\rho}=\rho'=\bar{\rho}'=0}, 
\eeq
where $S\equiv S_g + S_F$ includes the Yang-Mills and Dirac actions regularised on the lattice. The formulation
used hereafter is the following:
\bea\nonumber
S_g&=&\frac{1}{g^2_0} \sum_{\vec{x}}\sum_{x_0=0}^T\sum_{\mu,\nu} w(P){\rm Tr}[1-P_{\mu\nu}(x)],\\
\nonumber
S_f&=&\sum_{\vec{x},\vec{y}} \sum_{x_0,y_0=1}^{T-1} \bar{\psi}(x) M^{\rm Wilson\,\theta}_{xy} \psi(y)\\
\nonumber&+& \sum_{\vec{x}}
\bar{\psi}(x) P_-\{\sum_{k=1}^3 \gamma_k [e^{i\theta_k/L}U_k(x) \psi(x+\hat{k}) - U^\dag_k(x-\hat{k})e^{-i\theta_k/L}\psi(x-\hat{k})]
-U_0(x) \psi(x+\hat{0})\}_{x_0=0}\\
\nonumber
&+&\sum_{\vec{x}}
\bar{\psi}(x) P_+\{\sum_{k=1}^3 \gamma_k [e^{i\theta_k/L}U_k(x) \psi(x+\hat{k}) - e^{-i\theta_k/L}
U^\dag_k(x-\hat{k})\psi(x-\hat{k})]
-U^\dag_0(x-\hat{0}) \psi(x)-\hat{0})\}_{x_0=T},
\eea
where $P_{\mu\nu}(x)$ is a plaquette in the plane $\{\mu,\nu\}$ and the weight factor $w$ is equal to 1 except on the boundaries where it is set to 1/2 for space-like plaquettes.\\
${\cal O}(a)$ improvement of the Schr\"odinger Functional can be realised. In order to assure that, one adds for the bulk contribution to the Wilson-Dirac action the Clover term, while the fermionic boundary terms are multiplied by a coefficient $\tilde{c}_t$ and the weight factor $w(P)$ at the boundaries is a coefficient noted $c_t$ \cite{LuscherSC}. $c_t$ and $\tilde{c}_t$ are known from perturbation theory at 1-loop \cite{LuscherVW} and 2-loop order \cite{BodeSM}. Different $\theta$-boundary conditions on the $\psi$ fields were used to determine non perturbatively the improvement coefficient $c_{SW}$ \cite{LuscherSC}.\\
Even in the chiral limit the Dirac operator with inhomogeneous Dirichlet boundary conditions does not have any zero mode, the spectrum has a lower bound which is $\sim$ $1/T$. It means that one can work in a mass independent scheme.
As sketched in Figure \ref{fig:correlSF}, 2 types of correlators are computed on the lattice: boundary to bulk and boundary to boundary correlation functions. In the Schr\"odinger Functional framework, the "boundary fields" are defined in a peculiar way with respect to the usual formulation of lattice QCD with anti-periodic boundary conditions in time for the fermions:
\bea\nonumber
\zeta(\vec{x})\equiv \frac{\delta}{\delta \bar{\rho}(\vec{x})}, \quad \zeta'(\vec{x})\equiv \frac{\delta}{\delta \bar{\rho}'(\vec{x})},\\
\nonumber
\bar{\zeta}(\vec{x})\equiv -\frac{\delta}{\delta \rho(\vec{x})}, \quad \bar{\zeta}'(\vec{x})\equiv -\frac{\delta}{\delta \rho'(\vec{x})}.
\eea
Boundary to bulk correlation functions will take the form 
\beq\nonumber
C_{\Gamma_1\Gamma_2}(x_0)\propto \sum_{\vec{y},\vec{z}} \langle \bar{\psi}(x)\Gamma_2\psi(x) \bar{\zeta}(\vec{y})\Gamma_1 \zeta(\vec{z})\rangle,
\eeq 
and the boundary to boundary correlation functions read 
\beq\nonumber
C'_{\Gamma_1\Gamma_2}(x_0)\propto \sum_{\vec{u},\vec{v},\vec{y},\vec{z}} \langle \bar{\zeta}'(\vec{u})\Gamma_2\zeta'(\vec{v}) \bar{\zeta}(\vec{y})\Gamma_1 \zeta(\vec{z})\rangle.
\eeq
The time dependence of boundary to bulk correlators can be fitted to extract hadron masses and matrix elements \cite{GuagnelliZF} while a technique to compute the quark mass renormalization constant
\beq\nonumber
Z_m(L)\equiv Z_A/Z_P(L)
\eeq 
(from PCAC relation) was developed in \cite{JansenCK}: in particular, one imposes the renormalization condition
\beq\label{eq:SFrenorm}
Z^2_P(L)\langle P(x_0=T/2) \Pi \rangle^2 = - (9/L^6) \langle \Pi' \Pi \rangle, \quad
\Pi^{(')}=\sum_{\vec{x},\vec{y}}\bar{\zeta}^{(')}(\vec{x})\gamma^5\zeta^{(')}(\vec{y}), \quad T=2L,
\eeq
and lets $Z_P$ run up to high energy scales through step scaling functions. The normalisation factor has been introduced to eliminate the first order term in perturbation theory. We have sketched graphically what represents that renormalization condition for the Schr\"odinger Function scheme in Figure \ref{fig:SFrenorm}.
\begin{figure}[t]
\begin{center}
\includegraphics*[width=0.4\textwidth]{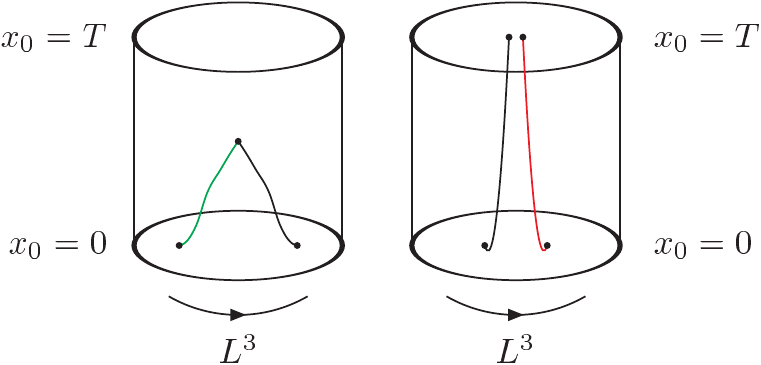}
\end{center}
\caption{\label{fig:correlSF} Boundary to bulk and boundary to boundary correlation functions in the Schr\"odinger Functional formalism.}
\end{figure}
\begin{figure}[t]
\begin{center}
\includegraphics*[width=0.6\textwidth]{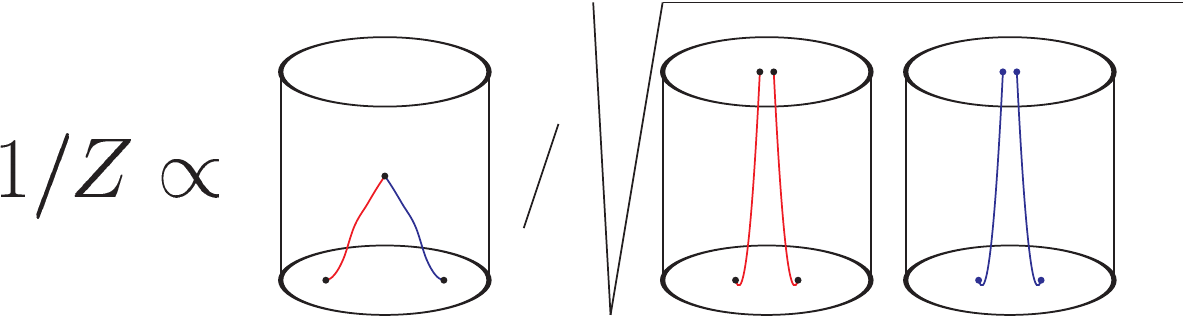}
\end{center}
\caption{Pictorial definition of the Schr\"odinger Functional renormalization condition of bilinear of quarks.
\label{fig:SFrenorm}}
\end{figure}
As we do not consider an automatically ${\cal O}(a)$ improved regularisation of QCD, dimension 4 operators and coefficients of improvement have to be introduced to guarantee extrapolations to the continuum limits in $a^2$ of the physical results. We will be concerned by a couple of them:
\begin{align*}
A^I_\mu(x)=& A_\mu(x) + a c_A \frac{\partial_\mu + \partial^*_\mu}{2} P(x),\\
\tilde{m}_q=& m_q(1 + b_m a m_q),\\
A^R_\mu(x) =& Z_A (1 + b_A a m_q) A^I_\mu(x),\\
P^R(x) =&Z_P(1 + b_P a m_q)P(x).
\end{align*}

\section{Heavy Quark Effective Theory}

From the complete theory of QCD one can derive an effective theory if one is interested by a physics for which 
a characteristic scale $\mu \gg \Lambda_{\rm QCD}$ is relevant, $\Lambda_{\rm QCD}/\mu$ and 
$\alpha_s(\mu)$ will constitute the parameters of the expansion. For instance it is the case for $B$ physics and heavy-light hadron physics, in general, where the hard scale is the mass $m_Q$ of the heavy quark of the hadrons. Thus, those systems are described by Heavy Quark Effective Theory (HQET) \cite{NeubertMB}, \cite{HQETref}, in which one considers only the degrees of freedom of ${\cal O}(\Lambda_{\rm QCD}) \ll m_Q$.
Formally it is still a field theory governed by the confinement but its prediction power is a bit larger than the QCD one because analytical expressions of physical quantities involving the heavy quark are simplified thanks to more symmetries than in the original theory. Heavy-light hadrons are, from the strong interaction point of view, like the hydrogen atom with respect to electromagnetism: here, the proton is replaced by the heavy quark, other quarks and gluons play the same role as electrons and photons. One writes the heavy quark momentum as 
$p_Q=m_Q\,v+k$ in a "classical part" $m_Q\,v$ and a "quantum fluctuation" part $k \sim {\cal O}(\Lambda_{\rm QCD})$. $v$ is the quadri-velocity of the heavy quark ($v^2=1$) and is a constant of motion in momentum exchanges  $\sim$ ${\cal O}(\Lambda_{\rm QCD})$ between light quarks, gluons and the heavy quark. Hence, $v$ is a vector 
parameterizing Hilbert space of the effective theory that one tries to build and one defines the effective fields
$h_v(x)$ and $H_v(x)$ in function of the heavy quark field $Q(x)$ 
($Q=b$ or $c$) by:
\beq\nonumber
Q(x)=e^{-im_Q v\cdot x}[h_v(x) + H_v(x)],\quad h_v(x) = e^{i m_Q
v\cdot x} \Prvp Q(x), \quad H_v(x) = e^{i m_Q v\cdot x} \Prvm  Q(x). 
\eeq
One factorises the term of fast oscillations $e^{-im_Q v\cdot x}$, to keep only the soft degrees of freedom, and one projects the heavy quark field on its large and small components. They verify $\not\!v h_v = h_v$ and $\not\!v H_v = -H_v$. Small components $H_v$ are related to the creation of heavy antiquarks with the same velocity $v$ and correspond to ${\cal O}(1/m_Q)$ terms of the expansion. To describe the physics of heavy antiquarks, one defines $Q$ in function of the fields $\bar{h}_{-v}$ and $\bar{H}_{-v}$ by   
\beq\nonumber
Q(x)=e^{im_Q v\cdot x}[\bar{h}_{-v}(x) + \bar{H}_{-v}(x)],\quad \bar{h}_{-v}(x) = 
e^{-i m_Q v\cdot x} \Prvm Q(x), \quad \bar{H}_{-v}(x) = e^{-i m_Q v\cdot x} \Prvp  Q(x).
\eeq
From the projectors $\frac{1\pm \,\not v}{2}$ it is easy to understand that the creation of a $Q\bar{Q}$ pair is forbidden in HQET: it is a phenomenon whose the characteristic scale is $m_Q$ and has been integrated in Wilson coefficents of the expansion. The expression of $Q$ induces a content of the effective theory in degrees of freedom of the heavy quark or of the heavy antiquark.\\
The HQET Lagrangian can be written from the QCD counterpart\footnote{Here we are back in Minkowski space.} ${\cal L}_{\rm QCD} = \bar{Q} (i\!\not\!\! D -m_Q)Q$, where $X^\m_\perp = X^\m - x\cdot v v^\m$:
\bea\nonumber
{\cal L}_{\rm QCD}&=&[\bar{h}_v+\bar{H}_v][m_Q(\not\!v -1) +i\!\not\!\! D][h_v+H_v]\\
\nonumber
&=&[\bar{h}_v+\bar{H}_v][i\!\not\!\! D h_v + (i\!\not\!\! D - 2m_Q) H_v]\\
\nonumber
&=&\left[\bar{h}_v \left(\Prvp\right) + \bar{H}_v \left(\Prvm\right)\right] 
\left[i\!\not\!\! D \left(\Prvp\right) h_v + (i\!\not\!\! D - 2m_Q) 
\left(\Prvm\right) H_v\right]\\
\nonumber
&=&\bar{h}_v (iv\cdot D) h_v + \bar{H}_v i\!\not\!\! D h_v + \bar{h}_v i\!\not\!\! D H_v 
- \bar{H}_v (iv\cdot D + 2m_Q) H_v\\,
&=&\bar{h}_v (iv\cdot D) h_v + \bar{H}_v i\!\not\!\! D_\perp h_v + 
\bar{h}_v i\!\not\!\! D_\perp H_v - \bar{H}_v (iv\cdot D + 2m_Q) H_v,
\eea
\beq
{\cal L}_{\rm HQET}=\bar{h}_v (iv\cdot D) h_v + {\cal O}(1/m_Q) \equiv {\cal
L}_0 + {\cal L}_1.
\eeq
The leading order of HQET is the limit $m_Q \to \infty$ in which one neglects ${\cal L}_1$.
At this order, the theory has a flavour symmetry, because ${\cal L}_0$ is independent of the heavy quark mass,
and a spin symmetry, because the interaction with gluons is of the chromoelectric kind. One gathers them under the name of "\emph{Heavy Quark Symmetry}" (HQS) and they form a group $U(2N_h)$ where $N_h$ is the number of heavy flavours under investigation. Physically, those symmetries reflect the fact that the exchange of soft gluons between the heavy quark and the light quark(s) does not probe the flavour nor the spin of the heavy quark. 
Effects of the symmetry breaking between heavy quarks $Q_i$ and $Q_j$ are proportional to $1/m_i-1/m_j$, one can trace them back in the discrepancy of $B$ and $D$ physics, while spin symmetry breaking effects enter at the order $1/m_Q$: they come from the chromomagnetic interaction that involves the heavy quark spin. HQS is similar to
the symmetry of hydrogeno\"\i d atom, in which the proton mass and spin decouple from the dynamics.\\
Feynman rules are easily found from ${\cal L}_0$:
\bea\nonumber
{\rm HQET\; quark\; propagator:}&\frac{i}{k\cdot v + i\epsilon} \frac{1+\not\, v}{2}\\
\nonumber
{\rm HQET\; quark-quark-gluon\; vertex:}&-ig_s T_a v^\mu
\eea
One gets the expression of ${\cal L}_1$ by writing the equation of motion $H_v = \frac{1}{iv\cdot D + 2m_Q} i\not\!\!D_\perp h_v$. Then,
\bea\nonumber
{\cal L}_1&=&\bar{h}_v i\!\not\!\! D_\perp\frac{1}{2m_Q+iv\cdot D} i\!\not\!\! D_\perp h_v \\
\nonumber
&=&\bar{h}_v \frac{(i\!\not\!\! D_\perp)^2}{2m_Q} h_v + {\cal O}(1/m^2_Q)\\
&=&-\bar{h}_v \frac{D^2_\perp}{2m_Q} h_v 
- g_s\bar{h}_v \frac{\sigma_{\alpha \beta} G^{\alpha \beta}}
{4m_Q} h_v + {\cal O}(1/m^2_Q)\label{Lagr1},
\eea
where $\sigma_{\alpha \beta}=i[\gamma_\m, \gamma_\n]/2$. ${\cal L}_1$ has two contributions: a kinetic term ${\cal L}_{\rm kin}=\frac{1}{2m_Q} \bar{h}_v (iD_\perp)^2 h_v$ and a chromomagnetic term ${\cal L}_{\rm mag}=-\frac{1}{2m_Q} \bar{h}_v \frac{g_s}{2} 
\sigma_{\alpha\beta} G^{\alpha \beta}h_v$\footnote{Actually, taking into account radiative corrections, ${\cal L}_{\rm mag}$ is proportional to a coefficient $a(\mu)$ whose the dependence on the effective scale $\mu$ cancels with the dependence of operator $\bar{h}_v \frac{g_s}{2} \sigma_{\alpha\beta} G^{\alpha \beta}h_v$. 
In dimensional regularization, the invariance under reparameterization $v \to v+\epsilon/m_Q$, $k\to k-\epsilon$, $h_v\to e^{i\epsilon \cdot
x}\left(1+\frac{\not\,\epsilon}{m_Q}\right) h_v$ \cite{Luke} protects the kinetic term: the "invariant quadri-velocity" reads ${\cal V}=v+\frac{iD}{m_Q}$. The full Lagrangian that is invariant under this reparameterization has the following expression: $\frac{1}{2}m_Q \bar{h}_v \{({\cal V}^2-1) -\frac{i}{2} \sigma_{\alpha\beta}[{\cal V}^\alpha, {\cal
V}^\beta]\}h_v$. That form remains unchanged after adding quantum corrections if the leading term ${\cal L}_{\rm stat}$ and the kinetic term renormalize in the same way. To give an idea of what it corresponds to, changing the quadri-velocitity by a factor ${\cal O}(1/m_Q)$ does not change the physics of the heavy quark described by HQET.}.
The former breaks only the flavour symmetry while the latter breaks also the spin symmetry.

\section{HQET on the lattice}

The lattice regularisation of HQET led to very rich developments as "lattice QCD in a can"\footnote{Metaphor made by A. Kronfeld.}. since a first attempt at the turning of 1990's \cite{EichtenZV}. Perturbative matching to QCD was performed, in particular to extract the $B$ meson decay constant \cite{BoucaudGA} - \cite{HernandezBX}, improvement coefficients of static-light bilinears were computed in perturbation theory by several authors \cite{FlynnQZ} -  \cite{GimenezMW}.\\ The basis of a non perturbative renormalization was then established in the context of Schr\"odinger Functional \cite{KurthKI}.
In the Monte Carlo sampling, one considers, in addition to the Yang-Mills and light quark actions, the static HQET action
\beq\nonumber
S_h = a^4 \frac{1}{1+a\delta m}\sum_x \bar{\psi}_h(\vec{x},x_0)[(1+a\delta m) \psi_h(\vec{x},x_0) - U^\dag_0(\vec{x},x_0 - \hat{0}) \psi_h(\vec{x},x_0-\hat{0})],\quad  P_+\psi_h(x)=\psi_h(x),
\eeq
where $\delta m$ is an additive mass counterterm that absorbs the linear divergence of the static quark self energy.
Boundary conditions for the heavy quark fields are
\beq\nonumber
\left.\psi_h(x)\right|_{x_0=0} = \rho_h(\vec{x}), \quad \left.\bar{\psi}_h(x)\right|_{x_0=T} = \bar{\rho}'_h(\vec{x}).
\eeq
"Boundary fields" are defined by:
\beq\nonumber
\zeta'_h(\vec{x}) = \frac{\delta}{\delta \bar{\rho}'_h(\vec{x})}, \quad \bar{\zeta}_h(\vec{x})=-\frac{\delta}{\delta \rho_h(\vec{x})}.
\eeq
At the static order, the theory regularised on the lattice verifies the flavour and spin symmetries:
\beq\nonumber
\psi_h(x) \longrightarrow e^{i \phi(\vec{x})} \psi_h(x), \quad \bar{\psi}_h(x) \longrightarrow \bar{\psi}_h(x) e^{-i \phi(\vec{x})}, \quad \psi_h \longrightarrow e^{-i \alpha \cdot \sigma} \psi_h(x), \quad \bar{\psi}_h \longrightarrow \bar{\psi}_h e^{i \alpha \cdot \sigma}.
\eeq
At this order there is no improvement term in the HQET action thanks to the equations of motion $\nabla_0 \psi_h(x)=0$, $\bar{\psi}_h \stackrel{\leftarrow}{\nabla_0}=0$, because the projector $P_+$ forbids the possible contribution $\bar{\psi}\sigma_{0i} G_{0i} \psi_h$, the spin symmetry protects against the counterterm $\bar{\psi}_h \sigma_{ij}G_{ij} \psi_h$ while the flavour symmetry does not allow the counterterm $\bar{\psi}_h \vec{\nabla}^2 \psi_h + \bar{\psi}_h \stackrel{\leftarrow}{\nabla}^2 \psi_h(x)$.
In the following the HQET improvement program will concern the axial static-light operator:
\bea\nonumber
A^{\rm stat}_0(x)&=&\bar{\psi}_l \gamma_0 \gamma^5 \psi_h(x),\\ 
\nonumber
(A^{\rm stat}_I)_0(x)&=&A^{\rm stat}_0(x) + \sum_j a c^{\rm stat}_A \bar{\psi}_l(x) \frac{\nabla_j + \stackrel{\leftarrow}{\nabla}^*}{2}\gamma_j\gamma^5\psi_h(x),\\
\nonumber
(A^{\rm stat}_R)_0 &=& Z^{\rm stat}_A (1+ b^{\rm stat}_A am_q) (A^{\rm stat}_I)_0.
\eea
With the correlators $f^{\rm stat}_A$, $f^{\rm stat}_1$, $f^{\rm stat}_{\delta A}$,
\bea\nonumber
f^{\rm stat}_A(x_0,\theta)&=&-\frac{a^6}{2} \sum_{\vec{y},\vec{z}} \langle A^{\rm stat}_0(x) \bar{\zeta}_h(\vec{y})\gamma^5\zeta_l(\vec{z})\rangle ,\\
\nonumber
f^{\rm stat}_1(\theta)&=&-\frac{a^{12}}{2L^6} \sum_{\vec{u},\vec{v},\vec{y},\vec{z}} \langle \bar{\zeta}'_l(\vec{u})\gamma^5\zeta'_h(\vec{v}) \bar{\zeta}_h(\vec{x})\gamma^5\zeta_l(\vec{y}),\\
f^{\rm stat}_{\delta A}(x_0,\theta)&=&-\frac{a^6}{2}\sum_{\vec{y},\vec{z}}\sum_j \left\langle \bar{\psi}_l(x) \frac{\nabla_j +\stackrel{\leftarrow}{\nabla}^*_j}{2}\gamma_j\gamma^5\psi_h(x) \bar{\zeta}_h(\vec{y})\gamma^5\zeta_l(\vec{z})\right \rangle
\label{correlSFHQET},
\eea
one builds the ratio $X_I(g_0,L/a)$:
\beq\nonumber
X_I(g_0,L/a)=\frac{f^{\rm stat}_A(T/2,\theta)+ac^{\rm stat}_A f^{\rm stat}_{\delta A}(T/2,\theta)}
{\sqrt{f^{\rm stat}_1(\theta)}}.
\eeq
We recall that, here, the $\theta$ variable refers to the twisted boundary conditions in space 
verified by all quark fields, i.e. those described by QCD and those described by HQET:
\beq\nonumber
\psi(x+L\hat{k})=e^{i\theta_k} \psi(x),\quad \bar{\psi}(x+L\hat{k})=e^{-i\theta_k} \bar{\psi}(x).
\eeq
The renormalization condition is then \cite{KurthKI}
\beq\nonumber
\left.Z^{\rm stat}_A(g_0,\mu=1/L) X_I(g_0,L/a)\right|_{\kappa=\kappa_c}=X^{(0)}(L/a),
\eeq
where $X^{(0)}$ is the tree-level estimates of $X_I$. In \cite{HeitgerXG} another renormalization condition, more favorable from the statistical point of view, has been proposed:
\bea\nonumber
f_1(\theta)&=&-\frac{a^{12}}{2L^6} \sum_{\vec{u},\vec{v},\vec{y},\vec{z}} \langle \bar{\zeta}'_{l,2}(\vec{u}) \gamma^5\zeta'_{l,1}(\vec{v}) \bar{\zeta}_{l,2}(\vec{x})\gamma^5\zeta_{l,1}(\vec{y}),\\
f^{hh}_1(x_3,\theta)&=&-\frac{a^{8}}{2L^2} \sum_{x_1,x_2,\vec{y},\vec{z}} \langle \bar{\zeta}'_h(\vec{x}) \gamma^5\zeta'_h(\vec{0}) \bar{\zeta}_{l,2}(\vec{x})\gamma^5\zeta_{l,1}(\vec{y}),\label{correlSFHQET2}\\
\nonumber
\Xi(L)&=&\frac{f^{\rm stat}_A(T/2,\theta)+ac^{\rm stat}_A f^{\rm stat}_{\delta A}(T/2,\theta)}
{\sqrt{f_1(\theta) f^{hh}_1(L/2,\theta)}},\\
\nonumber
\left.Z^{\rm stat}_A(g_0,\mu=1/L) \Xi(g_0,L/a)\right|_{\kappa=\kappa_c}&=&\Xi^{(0)}(L/a).
\eea
A renormalized matrix element $\Phi(\mu=1/L) \equiv Z^{\rm stat}_A(g_0,L) \Phi^{\rm bare}(g_0)$ can be run up to a scale $s^{-n} \mu$ using the step scaling function $\Sigma_{Z_A} (g_0,L) = Z^{\rm stat}_A(g_0,sL)/Z^{\rm stat}_A(g_0,L)$ and converted into RGI by the relation
\beq\nonumber
\frac{\Phi^{\rm RGI}}{\Phi(1/L_{\rm max})}=\Pi_{i=1}^n \Sigma_{Z_A}(g_0,s^{i-1}/L)[2b_0 \bar{g}^2(\mu)]^{-\gamma_0/2b_0}\exp\left[-\int_0^{\bar{g}(\mu)}dg \left(\frac{\gamma(g)}{\beta(g)} - \frac{\gamma_0}{b_0g}\right)\right],
\eeq
with $b_0$ and $\gamma_0$ the first universal coefficients of the $\beta$ function and the anomalous dimension of the HQET axial current.

\subsubsection{Matching QCD with HQET}

A very important step forward was realised in \cite{HeitgerNJ}, where the authors showed that the couplings of the effective action and operators can be computed non perturbatively, order by order in the $1/m$ expansion, with the lattice regularisation. Generically, one writes the HQET action and operators expanded up to ${\cal O}(1/m^n)$ as
\bea\nonumber
S^{\rm HQET} &=& a^4 \sum_x {\cal L}^{\rm stat}(x) + \sum_{i=1}^n \sum_{j=1}^{n_i} \omega_{i,j} {\cal L}^{(i,j)}(x),\\
J^{\rm HQET}(x)&=&J^{\rm stat}(x) + \sum_{i=1}^n \sum_{j=1}^{n'_i} \omega'_{i,j} J^{(i,j)}(x),
\eea
where ${\cal L}^{(i,j)}$ and $J^{(i,j)}$ are $n_i$ and $n'_i$ operators of dimension $4+i$ and $3+i$, respectively. The path integral is only evaluated over the static Lagrangian, in addition to the gluonic and light fermions parts:
\bea\nonumber
\langle O \rangle_{\rm HQET}&\equiv&\frac{1}{Z}\int [{\cal D} \Phi] O(\Phi) e^{-S_{g,l} - S^{\rm HQET}},\\
\nonumber
Z&=&\int  [{\cal D} \Phi] e^{-S_{g,l} - S^{\rm HQET}},\\
\nonumber
\langle O \rangle_{\rm HQET}&\propto&\frac{1}{Z_{\rm stat}}\int [{\cal D} \Phi] O(\Phi) 
\left(1-a^4 \sum_x \sum_{i=1}^n \sum_{j=1}^{n_i} \omega_{i,j} {\cal L}^{(i,j)}(x) + \cdots\right) e^{-S_{g,l} - S^{\rm stat}}\\
\nonumber
&\propto&\frac{1}{Z_{\rm stat}}\int [{\cal D} \Phi] O(\Phi)  \left(1-a^4 \sum_x \omega_{\rm kin} {\cal L}^{\rm kin}(x) - a^4 \sum_x \omega_{\rm mag} {\cal L}^{\rm mag}(x)\right.\\ 
&&\left.+ \frac{1}{2} [a^4 \sum_x \omega_{\rm kin} {\cal L}^{\rm kin}(x) + a^4 \sum_x \omega_{\rm mag} {\cal L}^{\rm mag}(x)]^2 + \cdots\right) e^{-S_{g,l} - S^{\rm stat}}.
\eea
\begin{figure}[t]
    \begin{center}
      \includegraphics*[width=0.5\textwidth]{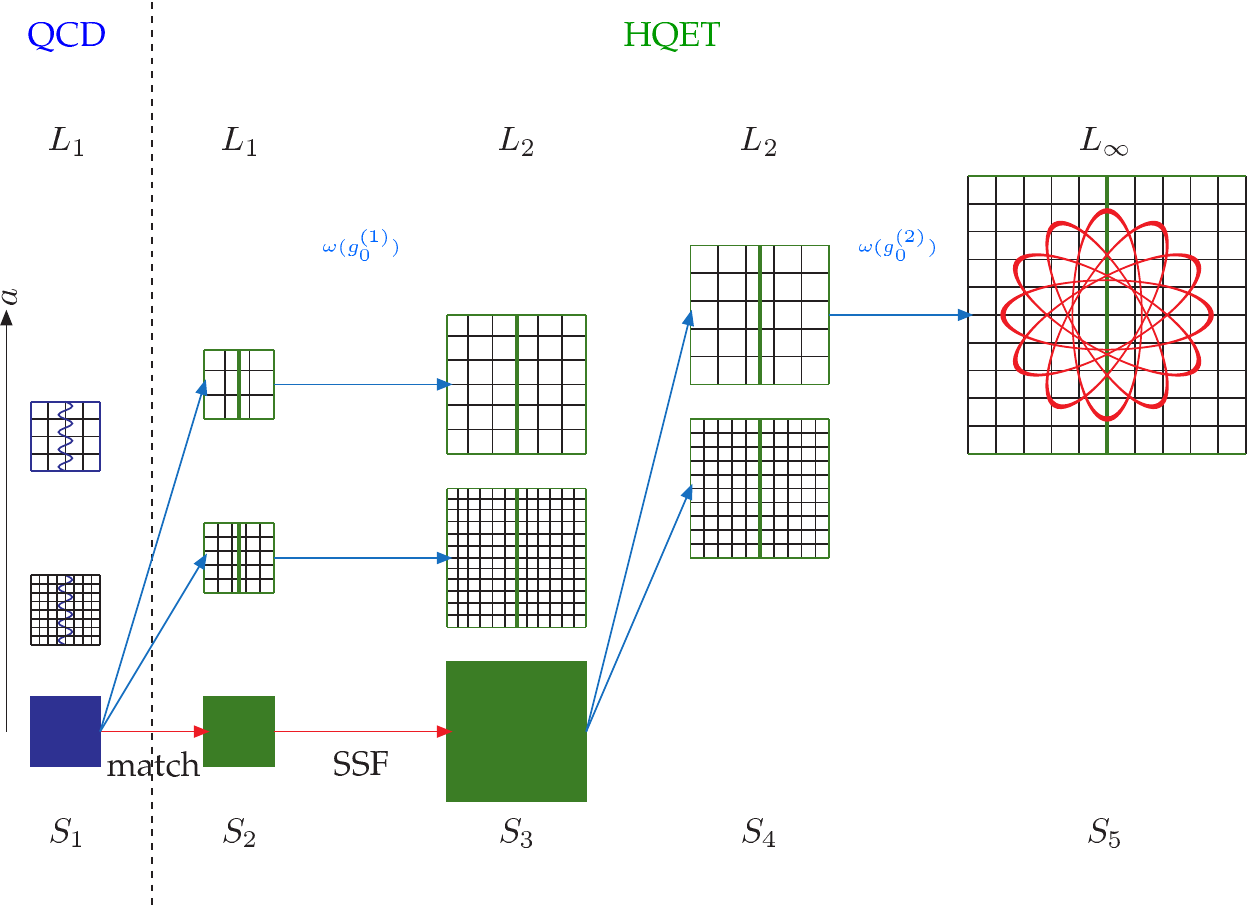}
    \end{center}
\caption{Strategy followed by the ALPHA Collaboration to deal with heavy quarks on the lattice.
\label{fig:programHQET}}
\end{figure}
It means that the ${\cal O}(1/m)$ terms of the HQET Lagrangian are included as operator insertions.\\ 
The program proposed in \cite{HeitgerNJ} consists in determining the parameters $\delta m$ (entering ${\cal L}^{\rm stat}$), $\omega_{i,j}$ and $\omega'_{i,j}$, at a given bare coupling $g_0$: it is sketched in Figure \ref{fig:programHQET}. The idea is to measure, in full QCD, with a physical $b$ quark, as many correlators $\Phi^{\rm QCD}_\alpha$ as the number of HQET parameters to know: they must have a proper continuum limit. One expresses their HQET counterparts
\beq\nonumber
\Phi^{\rm QCD,\; cont}_\alpha=f_{\alpha \beta}[\omega(g_0^{(1)})] \Phi^{\rm HQET}_\beta(g_0^{(1)}).
\eeq
HQET parameters are obtained, for a first series $\{g_0\}^{(1)}$ of bare couplings.
To have a good control on extrapolation to the continuum limit in QCD, the computation is necessarily done in a small physical volume $\sim$ 0.2 fm so that $am_b\ll 1$ without $L/a$ $\gtrsim$ 100. It generates huge finite size effects but, at this stage, the goal is only to absorb ultraviolet divergences of the theory. Hadronic quantities are measured on larger volumes, $L$ $\gtrsim$ 2 fm, with bare couplings larger than $\{g_0\}^{(1)}$ as well and, overall, \emph{with $L/a \in [10, 20]$ perfectly handable with computers that were available in the 2000's}. The trick is, again, to let the observables $\Phi^{\rm QCD}_\alpha(L_i)$ run from a volume $L_i$, with step scaling functions (SSF's) $\sigma^{\rm HQET}$, to hadronic volumes $L_f$ to get $\Phi^{\rm QCD}_\alpha (L_f)$. Those SSF's are defined such that a continuum limit exists. Then, a second matching procedure is applied:
\bea\nonumber
\Phi^{\rm QCD,\; cont}_\alpha(L_f) &=&\sigma^{\rm HQET}_{\alpha \beta}(L_i,L_f) \Phi^{\rm QCD,\; cont}_\beta(L_i),\\
\Phi^{\rm QCD,\; cont}_\alpha(L_f)&=&f'_{\alpha \beta}[\omega(g_0^{(2)})] \Phi^{\rm HQET}_\beta(g_0^{(2)},L_f),
\eea
in order to get HQET parameters for a second series $\{g_0\}^{(2)}$ of bare couplings. Let's illustrate this on a very simple example, assuming a single parameter:
\bea\nonumber
\Phi^{\rm QCD,\; cont}(L_f)& =& \frac{\Phi^{\rm QCD,\; cont}(L_f)}{\Phi^{\rm QCD,\;cont}(L_i)} 
\Phi^{\rm QCD,\; cont}(L_i)\\
\nonumber
&=& \lim_{a^{(1)} \to 0} \frac{ \omega(g_0(a^{(1)})) \Phi^{\rm HQET}(L_f, g_0(a^{(1)}))}
{ \omega(g_0(a^{(1)})) \Phi^{\rm HQET}(L_i, g_0(a^{(1)}))}\Phi^{\rm QCD,\; cont}(L_i)\\
\nonumber
&=&\lim_{a^{(1)} \to 0} \Sigma^{\rm HQET}(g_0(a^{(1)}),L_i,L_f) \Phi^{\rm QCD,\; cont}(L_i)\\
&=&\sigma^{\rm HQET}(L_i,L_f)\Phi^{\rm QCD,\; cont}(L_i),
\eea
\beq
\Phi^{\rm QCD,\; cont}(L_f) = \omega(g_0(a^{(2)})) \Phi^{\rm HQET}(L_f, g_0(a^{(2)}))
\eeq
One obtains the HQET parameter $\omega(g_0(a^{(2)}))$ (from the simulations $S_4$) after having computed $\sigma^{\rm HQET}(L_i,L_f)$ (from the simulations $S_2$ and $S_3$) and $\Phi^{\rm QCD,\; cont}(L_i)$ (from the simulations $S_1$).\\
At a given order in $1/m$, what we have just described is a closed procedure, once known the fact that the expansion in $1/m^n$ mixes with the expansion in $a^n$: symmetries, considering the effective theory on-shell and the corresponding equations of motion help to reduce the number of effective operators.
Following a first work reported in \cite{DellaMorteCB} where the ideas were investigated in details, this long-term research program continued and led to a series of publications about HQET on the lattice, expanded at ${\cal O}(1/m)$, in the quenched approximation \cite{BlossierKD} - \cite{BlossierMK} and at ${\rm N_f}=2$ \cite{BlossierQU}, \cite{BernardoniXBA}.
Adopting the notations of those papers, the HQET Lagrangian reads:
\beq
{\cal L}_{\rm HQET}(x) =  {\cal L}_{\rm stat}(x) - \omega_{\rm kin} O_{\rm kin}(x)
        - \omega_{\rm spin}O_{\rm spin}(x)  \,,
\eeq
where 
\beq\nonumber
{\cal L}_{\rm stat}(x) =
\bar{\psi}_h(x) \,\nabla^*_0\, \psi_h(x),\quad
  O_{\rm kin}(x) = \bar{\psi}_h\vec{\nabla}^2\psi_h(x), \quad
  O_{\rm spin}(x) = \bar{\psi}_h(x){\boldsymbol\sigma}\!\cdot\!{\bf B}\psi_h(x)\,.
\eeq
and ${\bf B}$ is the chromomagnetic field. At tree level, one has $\omega_{\rm kin}=\omega_{\rm spin}=1/2m_b$. In the following we will write a couple of words about the decay constant $f_B$, so we need to define the heavy-light axial current in HQET:
\bea\nonumber
 A_0^{\rm HQET}(x)&=& Z_A^{\rm HQET}\,[A^{\rm stat}_0(x)+  \sum_{i=1}^2c_A^{(i)}A_0^{(i)}(x)]\,, \\
\nonumber
 A_0^{(1)}(x)\equiv \delta A^{\rm stat}_0(x) &=& \sum_j \bar{\psi}_l(x){1\over2}
            \gamma_5\gamma_j(\nabla^S_j-\stackrel{\leftarrow}{\nabla}^S_{j})\psi_h(x)\,,\\
A_0^{(2)}(x) &=& -\sum_j \partial^S_{j}\,A^{\rm stat}_j(x)\,,\quad A^{\rm stat}_j(x)=\bar{\psi}_l(x)\gamma_j\gamma_5
\psi_h(x)\,,
\eea
where the derivatives are symmetrized:
\beq\nonumber
  \partial^S_{j} = \frac12(\partial_{j}+\partial^*_{j}),\quad
  \stackrel{\leftarrow}{\nabla}^S_{j} = \frac12(\stackrel{\leftarrow}{\nabla}_{j}+
  \stackrel{\leftarrow}{\nabla}^*_{j}),\quad
  \nabla^S_{j} = \frac12(\nabla_{j}+\nabla^*_{j}).
\eeq
However, as we are only interested by $B$-meson at rest, the contribution $A_0^{(2)}$ is not included at all
in the analysis. In principle $Z^{\rm HQET}_A$ depends slightly on the light quark mass $m_l$ but we will neglect it afterwards. In QCD we consider the axial-axial two-pt correlation function
\beq\nonumber
C_{AA}(x_0) = Z_A^2 a^3\sum_{\vec{x}} \Big\langle A_0(x)  (A_0(0))^{\dagger}
              \Big\rangle,
\eeq
where $A_\mu=\bar{\psi}\gamma_\mu\gamma^5 \psi_b$ is the axial current. We expand formally the correlator $C_{AA}$ in $1/m_b$:
\beq\nonumber
C_{AA}(x_0) = C^{\rm stat}_{AA}(x_0) + C^{1/m}_{AA}(x_0).
\eeq
Including the light quark and gluonic actions and developing the integral path in $1/m_b$, we get
\bea\nonumber
  C_{AA}(x_0) &=&  e^{-m^{\rm bare} x_0} (Z^{\rm HQET}_A)^2 \,\Big[
 C_{AA}^{\rm stat}(x_0)+\omega_{\rm kin}C_{AA}^{\rm kin}(x_0)+
 \omega_{\rm spin}C_{AA}^{\rm spin}(x_0)\\
\nonumber && \qquad \qquad \qquad\qquad\;+ 
  c^{(1)}_A[C_{\delta A A}^{\rm stat}(x_0)+C_{A\delta A}^{\rm stat}(x_0)]
   \Big]\,\\
 \nonumber  &=&  e^{-m^{\rm bare} x_0} (Z^{\rm HQET}_A)^2 C_{AA}^{\rm stat}(x_0)\,\Big[
 1+\omega_{\rm kin}R_{AA}^{\rm kin}(x_0)+
 \omega_{\rm spin}R_{AA}^{\rm spin}(x_0)\\
 && \qquad \qquad \qquad\qquad\;+ 
  c^{(1)}_A[R_{\delta A A}^{\rm stat}(x_0)+R_{A\delta A}^{\rm stat}(x_0)]
   \Big]\,.
     \label{e:caahqet}
        \eea
 At large time, the asymptotic behaviour of the correlators is the following:
 \bea\nonumber
C_{AA}(x_0) &=&  {\cal A}^2 
    e^{-m_Bx_0 }\, [\,1 + {\cal O}(e^{-\Delta\,x_0 })\,]\\
& \sim & ({\cal A}^{\rm stat})^2\, e^{-m_{B}^{\rm stat} x_0 } 
    [1 + 2 {\cal A}^{1/m} 
    - m_{B}^{1/m} x_0]\,,\\
C_{AA}^{\rm stat}(x_0) &=& ({\cal A}^{\rm stat})^2\,e^{-E^{\rm stat} x_0 }\,
   [\,1 + {\cal O}(e^{-\Delta^{\rm stat}\,x_0 })\,],\\
 R_{AA}^{\rm kin,\,spin}(x_0) &=& 2{\cal A}^{\rm kin,\,spin} -x_0 E^{\rm kin,\,spin} + 
   {\cal O}(x_0\,e^{-\Delta^{\rm stat}\,x_0 }),\\
R_{\delta AA}^{\rm stat}(x_0) &=& {\cal A}^{\delta A}\,
   [\,1 + {\cal O}(e^{-\Delta^{\rm stat}\,x_0 })\,],
\eea
with ${\cal A}=f_B\sqrt{m_B/2}$. Finally the expansion of the $B$-meson mass and decay constant read
\footnote{The ${\cal O}(a)$ improvement contribution $b^{\rm stat} a m_{l}$ is also included.}
\bea\nonumber
m_B &=& m^{\rm bare} + E^{\rm stat} + \omega_{\rm kin} E^{\rm kin} +  \omega_{\rm spin} E^{\rm spin} \,,
   \\ 
   \label{e:expandfb}
   \ln({\cal A}r_0^{3/2})  &=&  \ln(Z^{\rm HQET}_A) + \ln({\cal A}^{\rm stat}\,r_0^{3/2}) 
   + c^{(1)}_A  {\cal A}^{\delta A} + \omega_{\rm kin} {\cal A}^{\rm kin} 
 + \omega_{\rm spin} {\cal A}^{\rm spin}\,.
 \eea
One needs 5 QCD observables to determine the 5 couplings 
\beq\{\omega\}\equiv (m^{\rm bare},\omega_{\rm kin},\omega_{\rm spin}, c^{(1)}_A, \ln(Z^{\rm HQET}_A)).
\eeq 
Restricting ourselves to the (${\cal O}(a)$ improved) static order, we get
\bea\nonumber
m^{\rm stat}_B &=& m^{\rm bare}_{\rm stat} + E^{\rm stat} \,,\\ 
\ln({\cal A}r_0^{3/2})^{\rm stat}  &=&  \ln(Z^{\rm stat}_A) + b^{\rm stat} a m_{l} + \ln({\cal A}^{\rm stat}\,r_0^{3/2}) 
   + c^{\rm stat}_A  {\cal A}^{\delta A}\,,
 \eea
with 2 other couplings $\{\omega^{\rm stat}\} \equiv (m^{\rm bare}_{\rm stat}, \ln(Z^{\rm stat}_{A}))$.
We use the Schr\"odinger Functional framework, where the physical volume $L$ is a natural scale of the problem. Heavy-light correlators are involved and there is an implicit dependence on the heavy quark mass of our results: precisely it helps to compute $m_b$, as we will discuss hereafter; in that purpose we introduce the RGI mass $M_h$:
\beq\nonumber
M_h=h(\mu)\,Z_m(\mu) a\,m_{q,h}(1+b_m a m_{q,h}),\quad Z_m(\mu) = \frac{Z_A}{Z_P(\mu)} \frac{1+b_A am_{q,h}}{1+b_P a m_{q,h}}, \quad h(\mu)=\frac{M}{m^{\rm SF}(\mu)}.
\eeq
In addition to the correlators of eq. (\ref{correlSFHQET}) and (\ref{correlSFHQET2}), we introduce the following ones:
\bea
f_A(x_0,\theta) &=& -{a^6 \over 2}\sum_{\vec{y},\vec{z}}\,
  \left\langle
  (A_I)_0(x)\,\bar{\zeta}_b(\vec{y})\gamma_5\zeta_l(\vec{z})
  \right\rangle  \,,\\
  k_V(x_0,\theta) &=& -{a^6 \over 6}\sum_{\vec{y},\vec{z},k}\,
  \left\langle
  (V_I)_k(x)\,\bar{\zeta}_b(\vec{y})\gamma_k\zeta_s(\vec{z})
  \right\rangle  \,,\\
 k_1(\theta) &=&
  -{a^{12} \over 6L^6}\sum_{\vec{u},\vec{v},\vec{y},\vec{z},k}
  \left\langle
  \bar{\zeta}'_l(\vec{u})\gamma_k\zeta'_b(\vec{v})\,
  \bar{\zeta}_b(\vec{y})\gamma_k\zeta_l(\vec{z})
  \right\rangle\,.
  \eea
3 different twisted boundary conditions in space, defined through $\theta_0$, $\theta_1$ and $\theta_2$ 
in the following, are necessary to perform a closed matching. With 
\beq\nonumber
\Gamma^P(\theta_0) = -\frac{\partial_0+\partial^*_0}{2}\left.\ln(-f_A(x_0,\theta_0))\right|_{(x_0=T/2,T=L)}, 
R_A(\theta_1,\theta_2)=\left.\ln(f_A(T/2,\theta_1)/f_A(T/2,\theta_2))\right|_{T=L}
\eeq and 
\beq\nonumber
R_1(\theta_1,\theta_2)=\frac14\left[\ln(f_1(\theta_1)k_1(\theta_1)^3)-\ln(f_1(\theta_2)k_1(\theta_2)^3)\right],
\eeq
the observables read:
\beq\nonumber
\Phi_1(\theta_0) =     L\Gamma^P(\theta_0),\quad
\Phi_2(\theta_0)=\ln\left({-f_A(T/2,\theta_0)\over\sqrt{f_1(\theta_0)}}\right), \quad
\Phi_3(\theta_1,\theta_2)=R_A(\theta_1,\theta_2),
\eeq
\beq\nonumber 
\Phi_4(\theta_1,\theta_2)=R_1(\theta_1,\theta_2),\quad
\Phi_5(\theta_0)= {3 \over 4}\ln\left({f_1(\theta_0)}\over{k_1(\theta_0)}\right).
\eeq
The observables that are computed on the HQET side are given in papers attached to that report and their explicit definition is not illuminating.

\section{$b$ quark mass in the quenched approximation}

In the quenched approximation, 4 series of simulations have been realised, whose main characteristics are given in Tables \ref{tab:simulSFnf0}, with 3 different twisted boundary conditions for the fermion fields $(1,1,1)\theta/L$: $\theta=\{0,0.5,1\}$. The light quark is regularised by the nonperturbatively ${\cal O}(a)$ improved Wilson-Clover fermion,
two different discretizations of the static quark are studied, HYP1 and HYP2 \cite{HasenfratzHP}, \cite{DellaMorteYC}: they are pretty similar to the Eichten-Hill action \cite{EichtenZV} except that the Wilson line of the static propagator looks more like a flux tube. Use of that smearing is beneficial to improve the signal to noise ratio because the linear divergent part of the quark self energy is much reduced \cite{DellaMorteMN} - \cite{GrimbachUY}.
\begin{table}[t]
  \begin{center} 
    \begin{tabular}{|cccc|}
      \hline
      Simulation & $L$ & Theory & $L/a$ \\
      \hline
      $S_1$ & $\tilde{L}_1$ & QCD  & $40, 32, 24, 20$     \\
      $S_2$ & $L_1$ & HQET & $16, 12, 10, 8, 6$   \\
      $S_3$ & $L_2$ & HQET & $32, 24, 20, 16, 12$ \\
      $S_4$ & $L_2$ & HQET & $16, 12, 8$          \\
      \hline
    \end{tabular}
  \end{center}
\caption{\label{tab:simulSFnf0}
\footnotesize Summary of the simulations used in this work. Note that, for $S_2$ and $S_3$, 
$L_1/a=16,\;L_2/a=32$ are in addition to those of \cite{DellaMorteCB}.}
\end{table}
For $S_1$, the renormalized coupling is $\bar{g}^2(\tilde{L}_1/4)=1.8811$ and three heavy quark masses $z_h\equiv \tilde{L}_1 M_h$ are $z_h=10.4$, $12.1$ and $13.3$. We collect the parameters of the simulations, extracted from \cite{DellaMorteCB}, in Table \ref{tab:simulQCD}.
\begin{table}[t]
\begin{center}
\begin{tabular}{|c|c|c|c|c|c|c|c|}
\hline
$L\over a$   &   $\beta$  &  $\kappa_{\rm l}$  & $\bar{g}^2({L\over4})$  & $Z_{\rm P}(g_0,{L\over2})$ & $b_{\rm m}$
& $\frac{1+b_A am_{q,h}}{1+b_P am_{q,h}}$  & $\kappa_h$ \\
\hline
$20$ & $7.2611$ & $0.134145$ & $1.8811(19)$ & $0.6826(3)$ & $-0.621$ & $1.0955$ & $0.124195$ \\
     &          &               &              &             &               &             & $0.122119$ \\
     &          &               &              &             &               &             & $0.120483$ \\
\hline
$24$ & $7.4082$ & $0.133961$ & $1.8811(22)$ & $0.6764(6)$ & $-0.622$ & $1.0941$ & $0.126055$ \\
     &          &               &              &             &               &             & $0.124528$ \\
     &          &               &              &             &               &             & $0.123383$ \\
\hline
$32$ & $7.6547$ & $0.133632$ & $1.8811(28)$ & $0.6713(8)$ & $-0.622$ & $1.0916$ & $0.127991$ \\
     &          &               &              &             &               &             & $0.126967$ \\
     &          &               &              &             &               &             & $0.126222$ \\
\hline
$40$ & $7.8439$ & $0.133373$ & $1.8811(22)$ & $0.6679(8)$ & $-0.623$ & $1.0900$ & $0.128989$ \\
     &          &               &              &             &               &             & $0.128214$ \\
     &          &               &              &             &               &             & $0.127656$ \\
\hline
\end{tabular}
\end{center}
\caption{Parameters of the simulation $S_1$ at $L=\tilde L_1$.}\label{tab:simulQCD}
\end{table}
Concerning $S_2$ and $S_3$, they are set by the renormalized coupling $\bar{g}^2(L_1)=3.48$, with $\bar{g}^2(\tilde{L}_1)-\bar{g}^2(L_1)=-0.17(5)$
\footnote{This mismatch arose because, at that time, numerical simulations with $\beta \geq 7.25$, $L/a \geq 20$, were particularly expensive to reach a statistical target of $\sim$ 1000 measurements; it was more appropriate to tune the $\beta$ parameters on simulations defined with the coupling $\bar{g}^2(L_1/4)$ and applying twice the step scaling function of the running coupling. The price to pay was a slight mismatch between $S_1$ and $S_2$ series of simulations, introduced by errors on the step scaling function.}, their parameters are given in Table \ref{tab:simulHQET} (left) and taken from \cite{HeitgerXG}. Finally the parameters of $S_4$, tuned from $L_2=1.436 r_0$ and the empirical formula \cite{NeccoXG}
\beq
\ln (a/r_{0})=-1.6804 - 1.7331(\beta-6.0) + 0.7849(\beta-6.0)^{2}-0.4428(\beta-6.0)^{3}, \quad 5.7 \leq \beta \leq 6.92,
\eeq
are collected in Table \ref{tab:simulHQET} (right) and taken from \cite{DellaMorteYC}. We show in Figure \ref{fig:matchingCL} examples of continuum limit extrapolations for the observables $\Phi^{\rm QCD}_1(\tilde{L}_1)$, proportional to the (finite volume) $B$ meson mass, and $\Phi^{\rm QCD}_5(\tilde{L}_1)$, related to the (finite volume) $B^* - B$ mass splitting, and in Figure \ref{fig:matchingCL2} the static and the $1/m$ parts of $\Phi^{\rm HQET}_1(L_2)$. As the $1/m$ order in HQET regularised on the lattice is formally equivalent to the first order in the Symanzik expansion, one extrapolates ${\cal O}(1/m)$ quantities in $a/L$ instead of $(a/L)^2$.\\
The formula (\ref{e:expandfb}) gives an information about the $m_B$ dependence on the $b$ quark mass $m_b$, through the dependence of the HQET parameters on $z$. Hence, $z_b$ is obtained by asking that $m_B(z_b)=m_B^{\rm exp}$, once the hadronic quantities $E^{\rm stat,\,kin,\, spin}$ are known for a given light quark mass, residual cut-off are properly removed, as in \cite{DellaMorteCB}, \emph{and excited states are really under control}. From investigations reported in \cite{DellaMorteIJ}, \cite{DellaMorteSB}, it appeared numerically obvious that boundary fields of the Schr\"odinger Functional couple strongly to excitations. The use of boundary-to-boundary correlators to extract $f_B$ does not seems so great from the statistical point of view. That is why we performed quenched simulations with standard boundary conditions for the fields, with the same parameters as for $S_4$ simulations. We considered $B_s$ correlators, $\kappa_s$ being set from \cite{GardenFG}. Parameters of $S_5$ are collected in Table \ref{tab:simulhadronHQET}. 
\begin{table}[t]
\begin{center}
\begin{tabular}{cc}
\begin{tabular}{|c|c|c|}
\hline
$\beta$ &{$\kappa$} & $L/a$ \\
\hline
  6.2204 &0.135470 &  6 \\
  6.4527 &0.135543 &  8 \\
  6.6350 &0.135340 & 10 \\
  6.7750 &0.135121 & 12 \\
  7.0203 &0.134707 & 16 \\
\hline
\end{tabular}
&
\begin{tabular}{|c|c|c|}
\hline
$\beta$ &{$\kappa$} & $L/a$ \\
\hline
  6.0219 &0.135081 &  8\\
  6.2885 &0.135750 &  12 \\
  6.4956 &0.135593 & 16 \\
\hline
\end{tabular}
\end{tabular}
\end{center}
\caption{Parameters of the HQET simulations $S_2$, $S_3$ (left table) and $S_4$ (right table). \label{tab:simulHQET}}
\end{table}

\begin{figure}[t]
\begin{center}
\begin{tabular}{cc} 
\includegraphics*[width=0.5\textwidth]{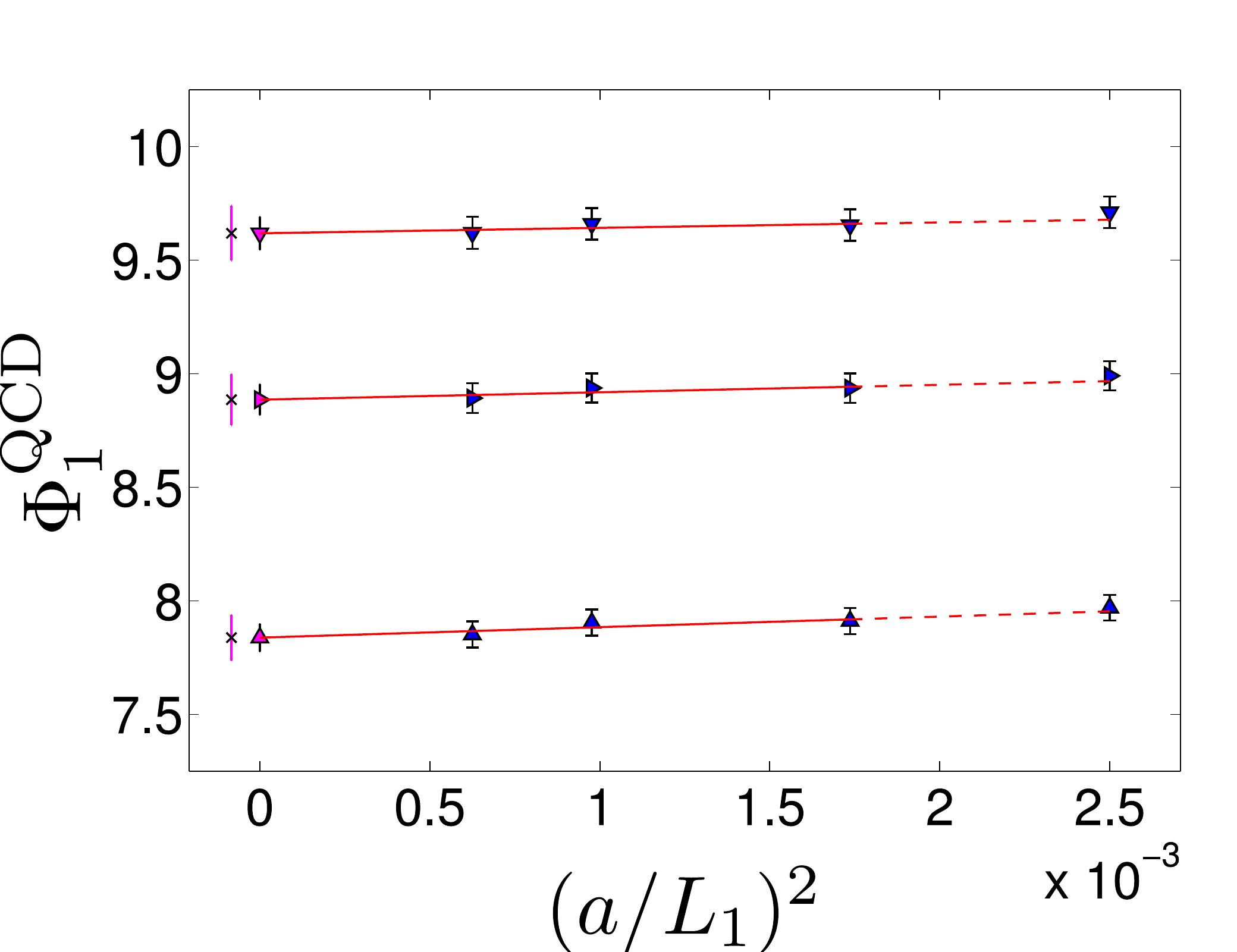}
&   
\includegraphics*[width=0.5\textwidth]{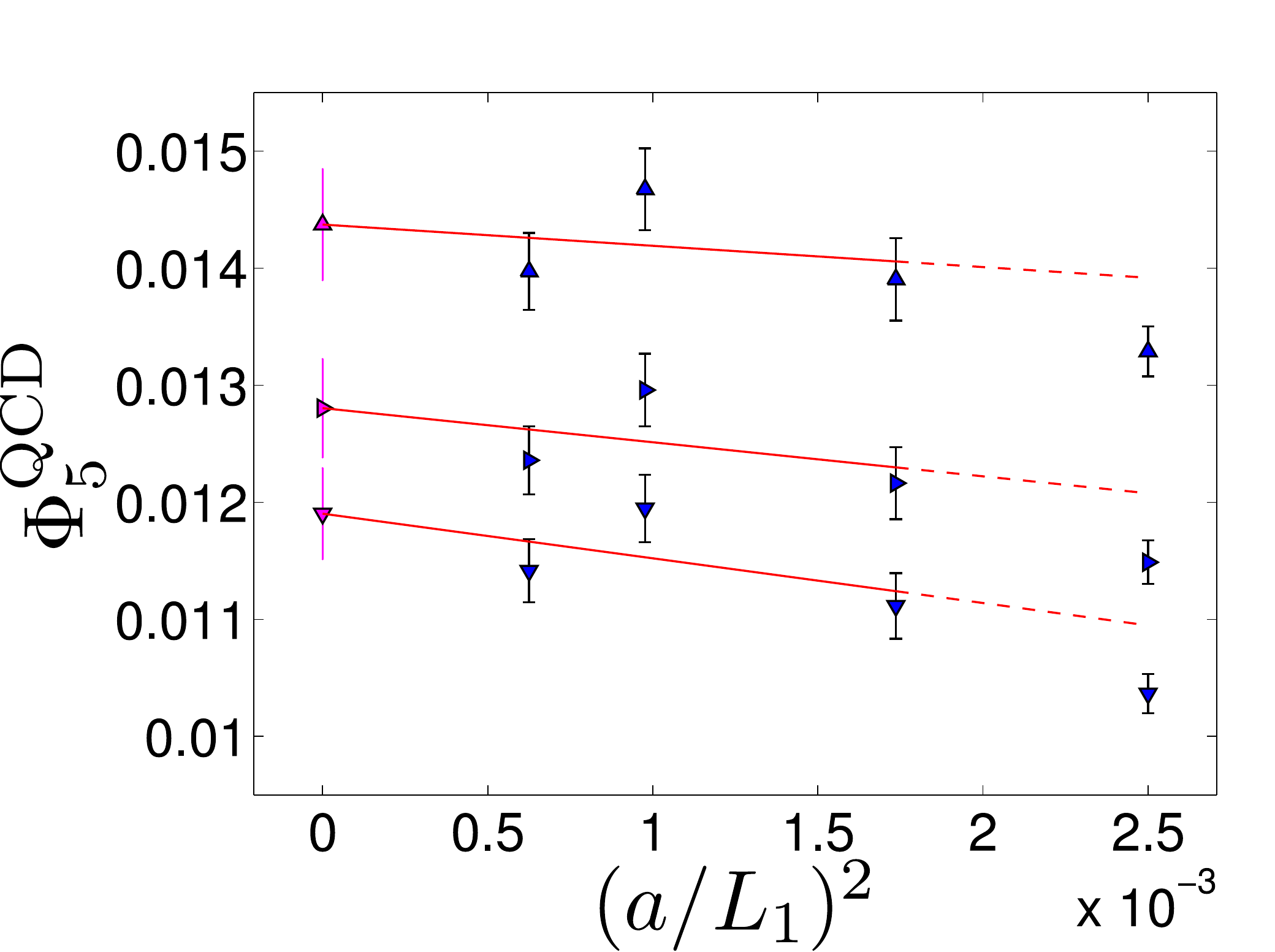}
\end{tabular}
\end{center}
\caption{Extrapolation to the continuum limit of the QCD observables $\Phi_1(L_1,\theta_0)$ and $\Phi_5(L_1,\theta_0)$ that give access to 
the HQET parameters $m^{\rm bare}$ and $\omega_{\rm spin}$: $\Phi_1$ is related to the $B$ meson mass and
$\Phi_5$ is related to the $B^*$-$B$ mass splitting; $\theta_0=0.5$ and the three curves correspond to the three heavy masses $z$.
\label{fig:matchingCL}}
\end{figure}

\begin{figure}[t]
\begin{center}
\begin{tabular}{cc}
\includegraphics*[width=0.5\textwidth]{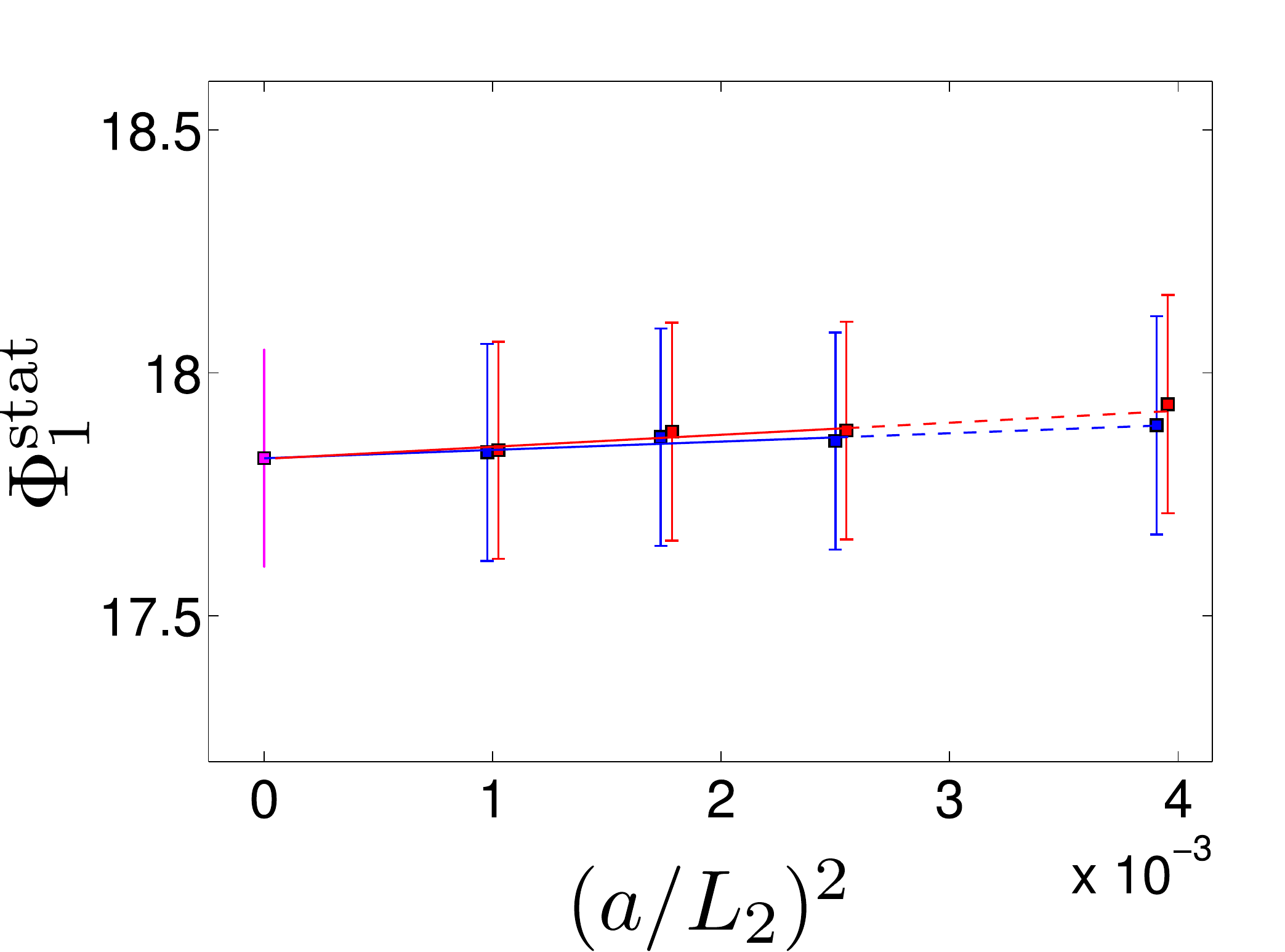}
&
\includegraphics*[width=0.5\textwidth]{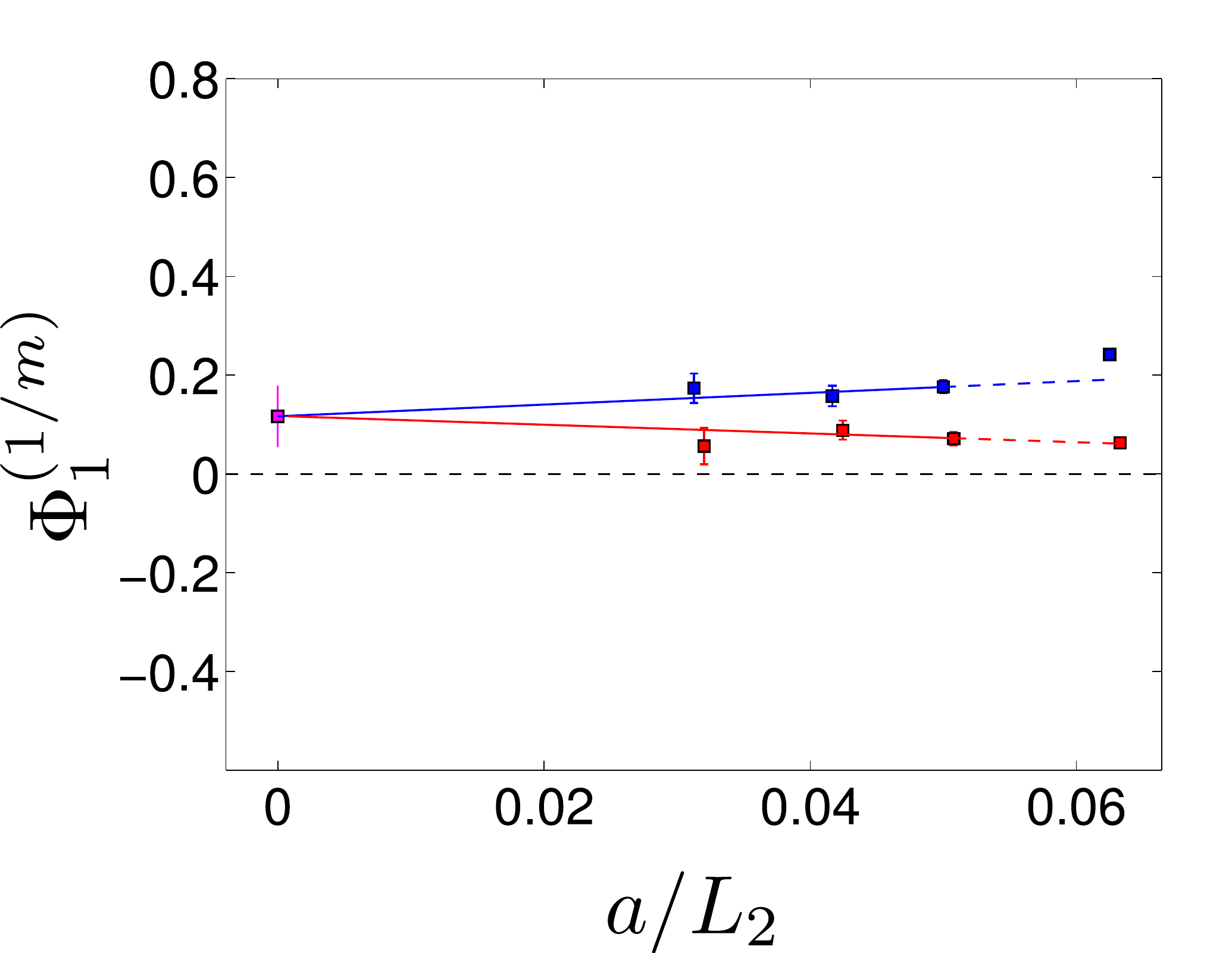}
\end{tabular}
\end{center}
\caption{Extrapolation to the continuum limit of the observables $\Phi^{\rm stat}_1(L_2,\theta_0)$ and $\Phi^{1/m}_1(L_2,\theta_0,\theta_1,\theta_2)\equiv [\Phi_1 - \Phi^{\rm stat}_1](L_2,\theta_0,\theta_1,\theta_2)$: $\theta_0=0.5$, $(\theta_1,\theta_2)=(0.5,1)$ and $z=12.1$; red and blue points correspond to the HQET actions HYP1 and HYP2, respectively.
\label{fig:matchingCL2}}
\end{figure}

\begin{table}[t]
\begin{center}
\begin{tabular}{|c|c|c|c|c|c|}
\hline
$\beta$ &{$\kappa_s$} & $L/a$&$N_{\rm ev}$&$N_{\rm stoch}$&\{$R_{i}$\} \\
\hline
  6.0219 &0.133849 &  16 &50&2&0, 10, 20, 30, 40, 60, 80, 100\\
  6.2885 &0.1349798 &  24 &50&2&0, 22, 45, 67, 90, 135, 180, 225\\
  6.4956 &0.1350299 & 32 &0&4&0, 40, 80, 120, 160, 240, 320, 400\\
\hline
\end{tabular}
\end{center}
\caption{Parameters of the HQET simulations $S_5$ with periodic boundary conditions in space and time; we have also indicated the parameters to compute all to all propagators and those of the Gaussian smearing for the $B_{s}$ interpolating fields. 
\label{tab:simulhadronHQET}}
\end{table}

\subsubsection{All to all propagators}

In this subsection we will present our numerical implementation of the strange quark "all to all" propagator and discuss the interest of time dilution to reduce the variance on static-light correlators. We have adopted the (slightly modified) techniques of \cite{FoleyAC} to improve the statistics by using a low-mode averaging and/or a stochastic estimator for the strange quark propagator:
\beq\nonumber
Q^{-1}(y,x) = \sum_{i=1}^{N_{\rm ev}}
\frac{1}{\lambda_i}u_i(y) u^\dag_i(x) + 
\lgl\lgl \psi(y) \eta^\dag(x) \rgl \rgl_{\eta}, \quad
\lgl\lgl \eta(y) \eta^\dag(x) \rgl \rgl_{\eta} = \delta_{xy}, \quad Q \psi =  P^\perp_{\rm N_{ev}}\eta,
\eeq
where
\beq\nonumber
Q = 2 \kappa_{s} \gamma^5 D, \quad\quad P^\perp_{\rm N_{ev}}\eta(x) = \eta(x) - \sum_{y,i} u^\dag_i(y) \eta(y) u_i(x)\,.
\eeq
The changing we have made are that we have used approximate low modes and the even-odd preconditioning: it reduces the computational effort and the storage requirement. 
Indeed, we have the following block structure:
\beq\nonumber
   Q = \gamma_5 
       \left(\begin{array}{cc}
          M_{ee} & M_{eo} \\
          M_{oe} & M_{oo}
       \end{array}\right)
  \,,
\eeq
where $M_{ee}$ ($M_{oo}$) is different from ${\mathbbm 1}$ by the Clover term on the even (odd) 
sites, and $M_{oe}$ ($M_{eo}$) is the hopping term.  Defining
\beq\nonumber
   J \equiv \left(\begin{array}{cc}
       {\mathbbm 1}_e  & - M_{ee}^{-1}M_{eo} \\
       0         & {\mathbbm 1}_o  \end{array}\right) 
   \,, 
\eeq
the preconditioned matrix $J^\dagger Q J$ is block-diagonal 
and the propagator can be written as
\beq
    Q^{-1} = J
             \left(\begin{array}{cc}
                 \hat{Q}_{ee}^{-1} & 0 \\
                 0            & \hat{Q}_{oo}^{-1} 
             \end{array}\right)
             J^\dagger
    \,,
    \label{invQ}
\eeq
where $\hat{Q}_{ee} = \gamma_5 M_{ee}$ is diagonal in space-time, and
$\hat{Q}_{oo} = \gamma_5(M_{oo} - M_{oe} M_{ee}^{-1}M_{eo}) = \hat{Q}_{oo}^\dagger$.
Concerning the approximate low modes, 
we take an orthonormal basis  $\{ v_i: i = 1, \ldots, N_{\rm ev} \}$ 
of an $N_{\rm ev}$ dimensional subspace of all fermion fields
which live only on odd sites. Defining the projectors
\beq\nonumber
    P_{IR} \equiv \sum_{i=1}^{N_{\rm ev}} v_i  v_i^\dagger, \quad
P_{UV} \equiv {\mathbbm 1}_{o} - P_{IR}
    \,,
\eeq
we write 
\bea\nonumber
   \hat{Q}_{oo}^{-1} 
   &=& \hat{Q}_{oo}^{-1} (P_{IR}+P_{UV})\\
&   =& \sum_{i=1}^{N_{\rm ev}} (\hat{Q}_{oo}^{-1} v_i)  {v_i}^\dagger + 
   \hat{Q}_{oo}^{-1} P_{UV}
   \,.
   \label{invQhatHL}
\eea
It reveals appropriate to choose for $v_i$ approximate eigenvectors of the low-lying 
eigenvalues of $\hat{Q}_{oo}$
\beq
   \hat{Q}_{oo} v_i = \lambda_i v_i + \hat{r}_i
    \,.
   \label{evQhat}
\eeq
with $|v_i| = 1$ and ${v_i}^\dagger \hat{r}_k = 0$.
Then, the part $\hat{Q}_{oo}^{-1}P_{IR}$  is expected to 
approximate the long-distance behaviour of the propagator \cite{FoleyAC},\cite{NeffZR}, 
and the inversions needed in $(\hat{Q}_{oo}^{-1}v_i)$ are cheap. The fact that we do not explicitly use $\hat{Q}_{oo}^{-1} v_i \simeq \lambda_i^{-1} v_i$ allows us to impose a weak constraint on the errors $\|\hat{r}_i\|$.
In practice, we require $\|\hat{r}_i\|\le 0.001 \cdot \vert\lambda_i\vert$,
    and take $\lambda_i^{-1} v_i$ only as starting vectors for the inversion.\\
 On the other hand, we introduce a stochastic estimator for $P_{UV}$. 
We take random vectors $\eta_i$ with
\bea\nonumber
  \lgl \lgl \eta_{i,\alpha} \rgl \rgl_\eta &=& 0\,, \label{eta1}\\
  \lgl \lgl \eta_{i,\alpha}\,\eta_{j,\beta}^\ast\rgl \rgl_\eta &=& \delta_{ij} \delta_{\alpha\beta} \,, \\
  \lgl \lgl\eta_{i,\alpha}\,\eta_{j,\beta}\rgl \rgl_\eta &=& 0 \,, \label{eta2}
\eea
where $\alpha, \beta$ are combined (colour, spinor, and site) indices.
Thus, the second part of (\ref{invQhatHL}) reads
\beq
   \hat{Q}_{oo}^{-1}\,P_{UV} 
     = {1 \over N_{\rm stoch}} \sum_{i=1}^{N_{\rm stoch}} \big \lgl \big\langle  
             \hat{Q}_{oo}^{-1}\, P_{UV} \, \eta_i \cdot \eta_i^\dagger 
       \big \rgl \big\rangle_{\eta}
   \,,
   \label{invQhatH}
\eeq
and the estimator of $\hat{Q}_{oo}^{-1}$ can be written as a sum of products
\beq
    \hat{Q}_{oo}^{-1} = \sum_{i=1}^{N_{\rm ev}+N_{\rm stoch}} \big \lgl \big\langle \hat{w}_i  \hat{u}_i^\dagger \big \rgl \big\rangle_{\eta}
    \,,
    \label{invQhat}
\eeq
with
\beq\nonumber
   \begin{array}{rclrcll}
       \hat{w}_i & = & \hat{Q}_{oo}^{-1}\,\hat{u}_i \ , &
       \hat{u}_i & = & v_i &                      
       (i=1, \ldots, N_{\rm ev}), \\
       \hat{w}_i & = & \hat{Q}_{oo}^{-1}\,P_{UV}\,\hat{u}_i \ , \ &
       \hat{u}_i & = & N_{\rm stoch}^{-1/2}\,\eta_i &                      
       (i=N_{\rm ev}\!+\!1, \ldots, N_{\rm ev}\!+\!N_{\rm stoch}). \\
   \end{array}
\eeq
The full propagator $Q^{-1}$ is then obtained from (\ref{invQ}).
Since $J$ connects only adjacent time slices,  the block $\hat{Q}_{ee}^{-1}$ 
does not contribute to the propagator between sites with time
separation $\vert x_0-y_0\vert > 2a$. In this case, we can simply write
\beq
    Q^{-1} = \sum_{i=1}^{N_{\rm ev}+N_{\rm stoch}} \big \lgl \big\langle w_i  u^\dagger_i \big \rgl \big\rangle_{\eta}    \,,
    \eeq
with 
\beq\nonumber
  w_i \equiv J\left(\begin{array}{c}0\\\hat{w}_i\end{array}\right), \quad
 u_i \equiv J\left(\begin{array}{c}0\\\hat{u}_i\end{array}\right)
  \,.
  \label{WfromWhat}
\eeq
In addition, we use the time dilution scheme.
It is implemented by replacing $P_{UV}$ with
$P_{UV} \sum_t P_{t}$ where $P_{t}$ projects on the components corresponding 
to odd sites with time coordinate $t$.
Then, an independent stochastic estimator is introduced for each term
\beq\nonumber
  P_{UV} P_{t} = {1 \over N_{\rm stoch}} \sum_{i=1}^{N_{\rm stoch}}
            \big\langle (P_{UV}\,\eta_{ti})  \eta_{ti}^\dagger \big\rangle_{\eta}
  \,,
\eeq
where the noise vectors $\eta_{ti}$ have non-vanishing components only for 
odd sites on time-slice $t$. Note that due to the hopping term in $J$ the full propagator (\ref{invQ}) from time slice $x_0$ to $y_0$ receives contributions 
from noise vectors $\eta_{ti}$ on three time slices, $t=x_0, x_0\pm a$,
i.e. three inversions are required for the propagator from one time slice $x_0$. 
However, a total of $T$ inversions is sufficient, and hence no 
extra effort is required, if one computes the propagator for 
all $x_0$, as we do in our measurements.\\ 
Analysing the variance of a heavy-light two-point correlator
as described in \cite{LuscherAE}, one sees that the
variance with time dilution decays roughly as $e^{-(x_0-y_0) m_\pi}$,
while the expression without time dilution contains
pieces independent of $x_0-y_0$. Indeed, we estimate a static-light 
correlation function $C$ as
\bea\nonumber
  C &=& C_{IR} + \lgl \langle {\rm Tr} \Upsilon K_1\rangle  \rgl + \lgl \langle {\rm Tr} \Upsilon K_2\rangle\rgl \,,\; 
\lgl  \langle  \Upsilon \rangle \rgl ={\mathbbm 1}\,, \;
  K_i={\cal Q}_i\bar F(x_0) {\cal S}^{\rm HQET}(x_{0},y_{0})\bar G(y_0) \,,
\eea
where we have 
\bea\nonumber
\Upsilon_{xy}=\eta(x)\eta^*(y), \quad {\cal Q}_1=P_{UV}  A(Q), \quad  {\cal Q}_2=P_{UV}  [Q^{-1}-A(Q)]\,;
\eea
${\cal S}^{\rm HQET}$ is the (exactly known) static propagator, $A(Q)$ is an approximation of the light quark propagator $Q^{-1}$ (for instance, a polynomial) and $\bar{F}$ and $\bar{G}$ are ``wave functions'' entering the $B$-meson interpolating fields. For simplicity we have restricted our discussion to the time variable.
$C_{IR}$ is the contribution to the correlator where the low mode part of $Q^{-1}$
is entering. The statistical uncertainty over the noise estimator
is
\bea\nonumber
  \Delta C &=&
  \sqrt{{{\cal V}_1^2 \over  N_1} + {{\cal V}_2^2 \over  N_2}} \\
  {\cal V} &=&  \frac12 \big\{
	\langle \langle [{\rm Tr} \Upsilon K + {\rm Tr} \Upsilon ^\dagger K^\dagger]^2 \rangle \rgl - 2 \lgl \langle {\rm Tr} \Upsilon K \rgl \rangle^2 
	\big\}^{1/2}\,,
\eea
where we assumed that 
\bea\nonumber
\lgl  \langle {\rm Tr} \Upsilon K\rangle \rgl = \lgl \langle {\rm Tr} \Upsilon K \rgl \rangle^*\; \leftrightarrow \; {\rm Tr} K = {\rm Tr} K^\dagger \,.
\eea 
The simple rules are ($I=(x,c,\alpha)$, we do not use a $Z(2)$ noise)
\bea\nonumber
 \mbox{no time dilution} && \Upsilon_{IJ}=\eta_I\eta_J^* \,,\quad \lgl \langle \eta_I\eta_J^*\rangle \rgl = \delta_{IJ}, \\
\nonumber  && \langle\lgl \eta_I\eta_J\rangle \rgl = 0\,, 
  \quad \lgl \langle \eta_I\eta_J^* \eta_K\eta_L^*\rangle \rgl = \delta_{IJ}\delta_{KL} + \delta_{IL}\delta_{KJ}\,.
\eea
\bea\nonumber
 \mbox{time dilution} && \Upsilon_{IJ}=\eta_I\eta_J^* \,,\quad \langle \lgl \eta_I\eta_J^*\rangle \rgl= \delta_{IJ}, \\
\nonumber  && \lgl \langle \eta_I\eta_J\rangle \rgl = 0\,, 
  \quad \lgl \langle \eta_I\eta_J^* \eta_K\eta_L^*\rangle \rgl = 
  \delta_{IJ}\delta_{KL}\delta(x_0^J-x_0^K) + 
  \delta_{IL}\delta_{KJ}\delta(x_0^I-x_0^K)\,.
\eea
Without time dilution we get
\bea\nonumber
 \lgl \langle [{\rm Tr} \Upsilon K + {\rm Tr} \Upsilon ^\dagger K^\dagger]^2 \rgl \rangle& =& 
 \lgl \langle  \eta_I\eta_J^* K_{IJ} \eta_K\eta_L^*  K_{KL} +
           \eta_I\eta_J^* K_{IJ} \eta_K^*\eta_L  K_{KL}^* \rgl \rangle +c.c. \\
\nonumber  &=& 2({\rm Tr} K)^2 + {\rm Tr} K^2 + {\rm Tr} K K^\dagger +c.c. \\
 {\cal V}^2&=&  \big\{{\rm Tr} K^2  + 2 {\rm Tr} K K^\dagger +  {\rm Tr} K^\dagger K^\dagger \big\} / 4,
\eea
and with time dilution we have
\bea
 {\cal V}^2&=&  \delta(x_0^I-x_0^J) \,\big\{K_{IJ}K_{JI}  
                 + 2 K_{IJ}K^*_{JI} +  K^*_{IJ}K^*_{JI} \big\} / 4 \,.
\eea
Graphically, we have
\bea\nonumber
  K(z_0',z_0) &=& {\cal Q}(z_0',y_0) \gamma_5 F(y_0) M^{-1}(y_0,x_0)G(x_0)
  \,\delta_{x_0z_0}\,\,, \\
\nonumber
  {\rm Tr} K^2 &=& {\rm tr} \big[{\cal Q}(x_0,y_0) \gamma_5 F(y_0) M^{-1}(y_0,x_0) G(x_0) \\
  \nonumber &&
                   \quad {\cal Q}(x_0,y_0) \gamma_5 F(y_0) M^{-1}(y_0,x_0) G(x_0)\big] \\
                    &&
                   =\vbb{0.15mm}  \label{e:vbb}\quad\quad\quad\quad,
                            \\
 \nonumber {\rm Tr}  K K^\dagger &=& \sum_{z_0'} 
                      {\rm tr} \big[{\cal Q}(z_0',y_0) \gamma_5 F(y_0) M^{-1}(y_0,x_0)
                      G(x_0) \\
                      \nonumber &&
                      \quad G^\dagger(x_0) [M^{-1}(x_0,y_0)]^\dagger F^\dagger(y_0)
                      \gamma_5 [{\cal Q}(y_0,z_0')]^\dagger\big] \\ &&
                   =\vbbdag{0.15mm} \label{e:vbbdag}\quad\quad\quad
                   \quad\quad\quad\quad\quad\quad\quad.
\eea
Time runs horizontally in those graphs, the different vertices are
fixed in time, while all space positions in the graphs are fully summed
over. Blue lines are heavy quark propagators, red lines are light quark propagators and
black lines are wave functions. All the vertices are summed over all
space positions. 
The second contribution (\ref{e:vbbdag}) will be rather large, since 
$z_0'\approx y_0$ contributes, where ${\cal Q}(y_0,z_0')$ is large.
Furthermore, as long as the wave functions have some support
at short distances, there will be terms where the static
propagators (almost) cancel. In the case of time dilution, the region $z_0'\approx y_0$ is removed.
(\ref{e:vbb}) is unchanged; however, for the dominating part
of the variance we get
\bea\nonumber
  {\rm Tr}  K K^\dagger &=& 
                      {\rm tr} \big[{\cal Q}(x_0,y_0) \gamma_5 F(y_0) M^{-1}(y_0,x_0)
                      G(x_0) \\
                      \nonumber &&
                      G^\dagger(x_0) [M^{-1}(x_0,y_0)]^\dagger F^\dagger(y_0)
                      \gamma_5 [{\cal Q}(y_0,x_0)]^\dagger\big] \\
                      &&
                   =\vbbdagt{0.15mm} \label{e:vbbdagt}\quad\quad\quad
                   \quad\quad\quad\quad\quad\quad\quad.
\eea
Since ${\cal Q}(x_0,y_0)$ will decay with $|x_0-y_0|$, (\ref{e:vbbdagt}) will be much
smaller than  (\ref{e:vbbdag}). That is why time dilution works for static-light correlators.

\subsubsection{Generalized Eigenvalue Problem}

We have followed a strategy discussed a long time ago in the literature \cite{GriffithsAH} - \cite{LuscherCK} to isolate efficiently the ground state contribution to correlation functions.
One considers a basis of $N$ interpolating fields $O_k$, $k=1,\cdots,N$, coupled to states with the same quantum numbers $J^{PC}$. One computes a matrix of those correlators $C_{ij}(t)\equiv \sum_{\vec{x},\vec{y}} \langle O_i(\vec{x},t)O^{\dag}_j(\vec{y},0)\rangle$ and solves the generalized eigenvalue problem (GEVP)
\beq
\sum_j C_{ij}(t) v^{(n)}_j(t,t_0)=\lambda^{(n)}(t,t_0) \sum_j C_{ij}(t_0) v^{(n)}_j(t,t_0).
\eeq
From the spectral decomposition $C_{ij}(t)=\sum_n \psi^{(n)}_i \psi^{(n)\star}_j e^{-E_n\,t}$, $\psi^{(n)}_i = \elematrice{n}{O_j}{0}$, we deduce that $\lambda^{(n)}(t,t_0) \longrightarrow_{t \to \infty} e^{-E_n (t-t_0)}$. In \cite{LuscherCK} it was proved actually that 
\beq\nonumber
\lambda^{(n)}(t,t_0) = e^{-E_n(t-t_0)}(1+ {\cal O}(e^{-\tilde{\Delta} E_n t})),\; \tilde{\Delta}E_n = {\rm min}_{m \leq N, m \neq n} |E_m - E_n|.
\eeq
Furthermore, and that is the main result of this subsection, \emph{the convergence rate is even better when $t_0$ is large enough, $t_0\geq t/2$}: 
\beq
\lambda^{(n)}(t,t_0) = e^{-E_n(t-t_0)}(1 + {\cal O}(e^{-\Delta E_{N+1,n} t}),\; \Delta E_{m,n}=E_m-E_n, 
\eeq
as we will discuss quickly using perturbation theory. With
\beq\nonumber
A v^{(n)} = \lambda^{(n)} B v^{(n)},\quad A=A^{(0)} + \epsilon A^{(1)},\quad B=B^{(0)}+ \epsilon B^{(1)},
\eeq
\begin{align*}
A^{(0)}= C^{(0)}(t)\,,\quad B^{(0)}= C^{(0)}(t_0)\,,\quad 
C_{ij}^{(0)}(t) =\sum_{n=1}^N \psi^{(n)}_{i}\psi_{j}^{(n)\star} e^{-E_{n}t}\,,\\
\epsilon A^{(1)}= C^{(1)}(t), \quad \epsilon B^{(1)}= C^{(1)}(t_0)\,, \quad
C_{ij}^{(1)}(t) =\sum_{n=N+1}^\infty \psi^{(n)}_{i}\psi_{j}^{(n)\star} e^{-E_{n}t}\,,
\end{align*}
one has at lowest order
\beq\nonumber
  A^{(0)}{v}^{(n,0)} = \lambda^{(n,0)} B^{(0)}\, {v}^{(n,0)}\,,
\eeq
satisfying the orthogonality condition
\beq\nonumber
({v}^{(n,0)},B^{(0)}{v}^{(m,0)}) = \rho_n \, \delta_{nm}\,,
\eeq
where $(a, M b) \equiv \sum_{i,j} a_i M_{ij} b_j$ is the scalar product. Expanding $\lambda^{(n)} = \sum_k \epsilon^k \lambda^{(n,k)}$, $v^{(n)} = \sum_k \epsilon^k v^{(n,k)}$, one has at the order $k$
\bea\nonumber
 0 &=& (A^{(0)}-\lambda^{(n,0)} B^{(0)})v^{(n,k)} 
  + (\Delta_n -\lambda^{(n,1)} B^{(0)}) v^{(n,k-1)} \\
\nonumber  &&+(-\lambda^{(n,1)} B^{(1)} -\lambda^{(n,2)} B^{(0)}) v^{(n,k-2)} 
+(-\lambda^{(n,2)} B^{(1)} -\lambda^{(n,3)} B^{(0)}) v^{(n,k-3)} \\
  &&+ \cdots
  +(-\lambda^{(n,k-1)} B^{(1)} -\lambda^{(n,k)} B^{(0)}) v^{(n,0)}\,,
\eea
where $\Delta_n = A^{(1)} - \lambda^{(n,0)} B^{(1)}$. A projection with $v^{(n,0)}$ gives
\beq\label{eq:lambdan}
  \lambda^{(n,k)} \,\rho_n = (v^{(n,0)}, \Delta_n v^{(n,k-1)} ) 
  - \sum_{l=1}^{k-1}\lambda^{(n,l)}\, (v^{(n,0)}\,,\, B^{(1)} v^{(n,k-1-l)} )\,,
\eeq
and a projection with $v^{(m,0)}\,,\;m\neq n$, gives
\bea\nonumber
v^{(n,k)} &=& \sum_{m\neq n} \alpha_{mn}^{(k)} v^{(m,0)} \\
\nonumber
\alpha_{mn}^{(k)} \,\rho_m &=& (v^{(m,0)},B^{(0)} v^{(n,k)})\\ 
&=&
\nonumber
  {1 \over \lambda^{(n,0)} - \lambda^{(m,0)}} \left\{ (v^{(m,0)},\Delta_n v^{(n,k-1)}) -
   \sum_{l=1}^{k-1}\lambda^{(n,l)}\, (v^{(m,0)}\,,\, B^{(1)} v^{(n,k-1-l)} ) \right. \\ 
\label{eq:alphamn}   && \left. \qquad\qquad\quad 
          - \sum_{l=1}^{k-1}\lambda^{(n,l)}\, (v^{(m,0)}\,,\, B^{(0)} v^{(n,k-l)} )
          \right\} \,.
\eea
Combining the recursion formulae
\beq
\alpha_{mn}^{(k)} =  \rho_m^{-1}\,  {1 \over \lambda^{(n,0)} - \lambda^{(m,0)}} \left\{ (v^{(m,0)},\Delta_n v^{(n,k-1)}) -
 \sum_{l=1}^{k-1}\lambda^{(n,l)}\, \left[ (v^{(m,0)}\,,\, B^{(1)} v^{(n,k-1-l)} )
          + \rho_m \alpha_{mn}^{(k-l)}\right] \right\}\,, 
\eeq
and those on $\lambda^{(n,k)}$, one determines the solution to arbitrary order
in the perturbations starting from the initial values 
$\alpha_{mn}^{(0)}=\delta_{mn}$, $\lambda^{(n,0)}(t,t_0)=e^{-E_n\,(t-t_0)}$. With 
\beq\nonumber
\lambda^{(n)}(t,t_0)=\lambda^{(n,0)}(t,t_0) \left(1 + \sum_{k\geq 1} \frac{\epsilon^k \lambda^{(n,k)}(t,t_0)}{\lambda^{(n,0)}(t,t_0)}\right), 
\eeq
we define
\bea\nonumber
  \epsilon_n(t,t_0) &\equiv& - \partial_t \ln \left(1+\sum_{k\geq1} \epsilon^k {\lambda^{(n,k)}(t,t_0) 
                                                \over \lambda^{(n,0)}(t,t_0) }\right)\\
   &=& - \sum_{l=1}^{\infty} {(-1)^{l+1} \over l} \sum_{k_1,\cdots,k_l\ge 1} \epsilon^{\sum_{i}k_i} \partial_t
      \left\{   {\lambda^{(n,k_1)}(t,t_0) \over \lambda^{(n,0)}(t,t_0) }
   \cdots {\lambda^{(n,k_l)}(t,t_0) \over \lambda^{(n,0)}(t,t_0) }
                \right\}\,,
\eea
If $\partial_t {\lambda^{(n,k)}(t,t_0) \over \lambda^{(n,0)}(t,t_0) } = 
    {\cal O}(e^{-\Delta E_{N+1,n}\, t})$ and ${\lambda^{(n,k)}(t,t_0) \over \lambda^{(n,0)}(t,t_0) } = 
 {\cal O}(1)$, then $\epsilon_n(t,t_0)$ will behave like $e^{\Delta E_{N+1,n}\, t}$, and our statement
 at the beginning of the subsection will be true.
Now, let's examine the asymptotic behaviour on the various contributions in ${\lambda^{(n,k)}(t,t_0) \over \lambda^{(n,0)}(t,t_0) }$ and $\partial_t {\lambda^{(n,k)}(t,t_0) \over \lambda^{(n,0)}(t,t_0) }$. At large time, 
we have
\bea\nonumber
 v_n^{(0)} &=& {\cal O}(1)\,,    \\ 
\nonumber
  \epsilon B^{(1)}&=& {\cal O}(e^{-E_{N+1} t_0 })\,, \\
\nonumber  \epsilon A^{(1)} &=& {\cal O}(e^{-E_{N+1} t })\,, \\ 
\nonumber
  \epsilon \Delta_n &=& {\cal O}(e^{-E_{N+1} t_0 }\,
          e^{-E_n (t-t_0) })  + {\cal O}(e^{-E_{N+1} t })\,, \\  
\nonumber
  \epsilon\rho_m^{-1}B^{(1)} &=& {\cal O}(e^{-(E_{N+1}-E_m) t_0 } )\,, \\ 
  {\epsilon\rho_m^{-1}\Delta_n\over\lambda^{(n,0)}} &=& {\cal O}(e^{-(E_{N+1}-E_m) t_0 }) 
          + {\cal O}(e^{-(E_{N+1}-E_m) t_0 }\, e^{-(E_{N+1}-E_n)(t-t_0) } )\,.
\eea
It is useful to introduce the notations
\bea\nonumber
  \eta_{Nm}(t) &=& {\cal O}(e^{-(E_{N+1}-E_m) t})\,,\\
 \nonumber \gamma_{nm}(t) &=& {e^{-E_n t}\over e^{-E_n t} - e^{-E_m t}}\\
\nonumber
&=& \left\{ \begin{array}{ll} \sum_{j=0}^\infty e^{-j(E_m-E_n) t} \quad &\mbox{when }m>n    \\ 
-\sum_{j=1}^\infty e^{-j(E_n-E_m) t} \quad &\mbox{when }n>m      \end{array} \right. \,.
\eea
Then, we have at large time
\bea\nonumber
   {\lambda^{(n,0)} \over \lambda^{(n,0)} - \lambda^{(m,0)}} 
   &=& \gamma_{nm}(t-t_0)  
  = \left\{ \begin{array}{ll} 1 + {\cal O}(e^{-(E_m-E_n) t} )\quad &\mbox{when }m>n   \\  
{\cal O}( e^{-(E_n-E_m) t}) \quad &\mbox{when }n>m    \end{array} \right. \,, \\
 \nonumber
  {\rho_{m'}^{-1}\,\epsilon\Delta_n \over \lambda^{(n,0)} - \lambda^{(m,0)}} &=& 
  \eta_{Nm'}(t_0)\,[1+\eta_{Nn}(t-t_0)]\, \gamma_{nm}(t-t_0),\\
   {\rho_{m'}^{-1}\,\epsilon B^{(1)}\lambda^{(n,0)} \over \lambda^{(n,0)} - \lambda^{(m,0)}} &=& 
  \eta_{Nm'}(t_0)\, \gamma_{nm}(t-t_0)\,.
\eea
From eq. (\ref{eq:lambdan}) and (\ref{eq:alphamn}), at first order, 
\bea\nonumber
  {\epsilon\lambda^{(n,1)}\over \lambda^{(n,0)}} &=& \eta_{Nn}(t_0)\,[1+\eta_{Nn}(t-t_0)]
                 \,,\quad \\
  \epsilon\alpha_{mn}^{(1)}  &=&  \eta_{Nm}(t_0)\,[1+\eta_{Nn}(t-t_0)]\, \gamma_{nm}(t-t_0)\,,
\eea
while the recursions are 
\bea\nonumber
  \epsilon^k\,{\lambda^{(n,k)}\over \lambda^{(n,0)}} &=& 
                    \epsilon^k\,{\lambda^{(n,1)}\over \lambda^{(n,0)}}\,\alpha^{(k-1)}
            + \epsilon^{k-1}\,\sum_{l=1}^{k-1}  {\lambda^{(n,l)}\over \lambda^{(n,0)}}
                     \,\eta_{Nn}(t_0)\,\alpha^{(k-1-l)} \\
\nonumber
   &=& {\lambda^{(n,1)}\over \lambda^{(n,0)}} \,
       \left\{\epsilon^{k}\,\alpha^{(k-1)} + 
         \epsilon^{k-1}\,\sum_{l=1}^{k-1}  {\lambda^{(n,l)}\over \lambda^{(n,1)}}
                     \,\eta_{Nn}(t_0)\,\alpha^{(k-1-l)}\right\}\,,
\eea
\bea\nonumber
   \epsilon^k\,\alpha_{mn}^{(k)}
   &=& \epsilon^k\,\alpha_{mn}^{(1)}\,\alpha^{(k-1)} + 
          \epsilon^{k-1}\,\sum_{l=1}^{k-1} {\lambda^{(n,l)}\over \lambda^{(n,0)}}
               \, \eta_{Nm}(t_0)\, \alpha^{(k-1-l)} \, \gamma_{nm}(t-t_0)\,\\
\nonumber
   && \quad +  \epsilon^k\,\sum_{l=1}^{k-1} {\lambda^{(n,l)}\over \lambda^{(n,0)}}
               \, \alpha_{mn}^{(k-l)} \, \gamma_{nm}(t-t_0)\,\\
  &=& \alpha_{mn}^{(1)}\,\left\{ \epsilon^k\,\alpha^{(k-1)} + 
          \sum_{l=1}^{k-1} {\lambda^{(n,l)}\over \lambda^{(n,1)}}
               \, \left[\epsilon^{k-1}\,\alpha^{(k-1-l)} \, \eta_{Nn}(t_0)
                +  \epsilon^k\,\alpha_{mn}^{(k-l)}
              \right]\right\}\, , 
\eea
where $\alpha^{(k)} = \max_m \alpha_{mn}^{(k)}$. We also need the starting values of the derivatives
\bea\nonumber
   \epsilon\,\partial_t{\lambda^{(n,1)}\over \lambda^{(n,0)}} &=& 
   \eta_{Nn}(t_0)\,\eta_{Nn}(t-t_0)
   = \eta_{Nn}(t)
                 \,,\quad \\
\nonumber
  \epsilon\,\partial_t\alpha_{mn}^{(1)}  &=&  \eta_{Nm}(t_0)\,
     [\partial_t\gamma_{nm}(t-t_0)\,+\,\partial_t \eta_{Nn}(t-t_0) 
     \, \gamma_{nm}(t-t_0) \\
 \nonumber     
 && \qquad\qquad +\,  \eta_{Nn}(t-t_0) \, \partial_t \gamma_{nm}(t-t_0)\,]\,\\
   &=&   \eta_{Nm}(t_0) \,e^{-|E_m-E_n|(t-t_0)} \,.
\eea
In particular, with
\bea\nonumber
\epsilon\,\eta_{Nn}(t_0)\partial_t\alpha_{mn}^{(1)} &=&e^{-(E_{N+1}-E_n) 2t_0}
e^{-(E_m-E_n)t},\quad E_m\geq E_n,\\
\nonumber
\epsilon\,\eta_{Nn}(t_0)\partial_t\alpha_{mn}^{(1)} &=&e^{-(E_{N+1}-E_m) 2t_0}
e^{-(E_n-E_m)t},\quad E_m< E_n,
\eea
we have \emph{in the case $t_0 \geq t/2$}
\beq\nonumber
\epsilon\,\eta_{Nn}(t_0)\partial_t\alpha_{mn}^{(1)} \leq \eta_{Nn}(t)
\eeq
or, as an equivalent,
\beq\nonumber
\epsilon\,\partial_t\alpha_{mn}^{(1)} \leq \eta_{Nn}(t-t_0).
\eeq 
With the other notations
\beq\nonumber
  r_{k,n} =  \epsilon^{k-1}\,{\lambda^{(n,k)}\over \lambda^{(n,1)}}\,,\quad
  x_{k,mn} =  \epsilon^k\,\alpha_{mn}^{(k)} \,,\quad
  X_k =  \epsilon^k\,\alpha^{(k)} = \epsilon^k\,\max_m \alpha_{mn}^{(k)},
\eeq
we can show that
\bea\nonumber
  x_{k,mn} &=& \eta_{Nm}(t_0)\,\gamma_{nm}(t-t_0)\,, \quad 
  r_{k,n} = {\cal O}(1) \,, \\
\partial_t x_{k,mn} &=& \eta_{Nn}(t-t_0)  \,, \quad 
  \partial_t r_{k,n} = \eta_{Nn}(t-t_0)\,.
\eea
Here ${\cal O}(1)$ means that there no $t$ dependence. In the case $k=1$, it is obvious from the behaviour of the derivatives on the starting values $\lambda^{(n,1)}$ and $\alpha_{mn}^{(1)}$. For the induction steps $k-1$ to $k$,  $k\geq2$, we have:
\bea\nonumber
  r_{k,n} &=&  \epsilon^{k-1}\,\alpha^{(k-1)} + 
        \sum_{l=1}^{k-1}  \epsilon^{k-1-l+l-1}{\lambda^{(n,l)}\over \lambda^{(n,1)}}
                     \,\eta_{Nn}(t_0)\,\alpha^{(k-1-l)}\\
\nonumber
&=&X_{k-1} +  \eta_{Nn}(t_0) \sum_{l=1}^{k-1} r_{l,n} X_{k-1-l};
\eea
As we are only interested by the behaviour in $t$, we can write
\bea\nonumber
r_{k,n}& = & \max_m  \eta_{Nm}(t_0)\,\gamma_{nm}(t-t_0).
\eea
This gives
\bea\nonumber
r_{k,n}&=& \max_m e^{-(E_{N+1}-E_m)t_0} [1 + {\cal O}(e^{-(E_m - E_n)(t-t_0)})]\,, 
E_m\geq E_n,\\
\nonumber
r_{k,n}&=& \max_m e^{-(E_{N+1}-E_m)t_0} {\cal O}(e^{-(E_n - E_m)(t-t_0)})\,,
E_m<E_n.\\
\eea
We obtain $r_{k,n}={\cal O}(1)$. We have also
\beq
x_{k,mn} = x_{1,mn}\,\left\{r_{k,n}+  \sum_{l=1}^{k-1} r_{l,n} x_{k-l,mn}\right\}.
\eeq
The second term in the bracket is subleading compared to the first one and
we obtain
\beq\nonumber
x_{k,mn}=  \eta_{Nm}(t_0) \,\gamma_{nm}(t-t_0)\,.
\eeq
The derivative $\partial_t r_{k,n}$ reads
\bea\nonumber
  \partial_t r_{k,n} &=&  \partial_t X_{k-1} +  \eta_{Nn}(t_0) \sum_{l=1}^{k-1} \partial_t[r_{l,n}\, X_{k-1-l}]\\
  \nonumber
&=&  \partial_t X_{k-1} +  \eta_{Nn}(t_0) \sum_{l=1}^{k-1}\left( r_{l,n} \partial_t X_{k-1-l} 
+ \partial_t [r_{l,n}] X_{k-1-l}\right)\\
 \nonumber
   &=&\eta_{Nn}(t-t_0) \,,
  \eea
  where, again, we have neglected the second term in the sum over $l$.
 Keeping the leading term only, the derivative $\partial_t x_{k,mn}$ reads
 \bea\nonumber
 \partial_t x_k &=& \partial_t [x_1\,r_k]+  \sum_{l=1}^{k-1}  \partial_t [x_1\,r_l\, x_{k-l}] \\
  & =&\eta_{Nn}(t-t_0)  \,.
\eea
The proof is achieved by noticing that
\bea\nonumber
\partial_t \frac{\epsilon^k \lambda^{(n,k)}}{\lambda^{(n,0)}}&=&
r_{k,n} \partial_t \frac{\epsilon \lambda^{(n,1)}}{\lambda^{(n,0)}} + \partial_t r_{k,n}\,  
\frac{\epsilon \lambda^{(n,1)}}{\lambda^{(n,0)}}\\
\nonumber
&=&\eta_{Nn}(t) + \eta_{Nn}(t-t_0)\eta_{Nn}(t_0)[1 + \eta_{Nn}(t-t_0)]\\
\nonumber
&=&\eta_{Nn}(t) + \eta_{Nn}(t)\eta_{Nn}(t-t_0),\\
\nonumber
 \frac{\epsilon^k \lambda^{(n,k)}}{\lambda^{(n,0)}}&=& r_{k,n} 
 \frac{\epsilon \lambda^{(n,1)}}{\lambda^{(n,0)}}\\
\nonumber
 &=&\eta_{Nn}(t_0)+ \eta_{Nn}(t).
\eea
Indeed, we get that $\partial_t \frac{\epsilon^k \lambda^{(n,k)}}{\lambda^{(n,0)}} =
{\cal O}(e^{-(E_N-E_n)t})$ and $\frac{\epsilon^k \lambda^{(n,k)}}{\lambda^{(n,0)}}={\cal O}(1)$:
it means that the correction $\epsilon_n(t,t_0)$ goes like $e^{-(E_{N+1}-E_n)t}$ and that 
$E^{\rm eff}_n(t,t_0) \equiv \ln \left(\frac{\lambda^{(n)}(t,t_0)}{\lambda^{(n)}(t+1,t_0)}\right)$ has the asymptotic behaviour $E_n + {\cal O}(e^{-\Delta E_{N+1,n} t})$.\\
In the HQET framework, a further comment is in order. It is required to solve the GEVP only on the static system. Indeed, in analogy with perturbation theory in quantum mechanics and exploiting the orthogonality property
\beq\nonumber
(v^{(n),{\rm stat}}(t,t_{0}), C^{\rm stat}(t) v^{(m), {\rm stat}}(t,t_{0}))\propto \delta_{mn},
\eeq
we have
\bea\nonumber
E^{{\rm eff}, 1/m}&=&\frac{\lambda^{(n),1/m}(t,t_{0})}{\lambda^{(n),{\rm stat}}(t,t_{0})}
-\frac{\lambda^{(n),1/m}(t+a,t_{0})}{\lambda^{(n),{\rm stat}}(t+a,t_{0})},\\
\frac{\lambda^{(n),1/m}(t,t_{0})}{\lambda^{(n),{\rm stat}}(t,t_{0})}&=&
(v^{(n),{\rm stat}}(t,t_{0}),\left[\frac{C^{1/m}(t)}{\lambda^{(n),{\rm stat}}(t,t_{0})} - C^{1/m}(t_{0})\right]
v^{(n),{\rm stat}}(t,t_{0})),
\eea
$1/m \in [\rm kin, spin]$. The asymptotic behaviour is expected to be
\bea\nonumber
E^{\rm eff, stat}_{n}(t,t_{0})&=&E^{\rm stat}_{n} + \beta^{\rm stat}_{n} e^{-\Delta E^{\rm stat}_{N+1,n}t},\\
\label{eq:energytimedependence} 
E^{{\rm eff}, 1/m}_{n}(t,t_{0})&=&E^{1/m}_{n} + [\beta^{1/m}_{n} - \beta^{\rm stat}_{n} t\Delta E^{1/m}_{N+1,n}] 
e^{-\Delta E^{\rm stat}_{N+1,n}t},
\eea
where $E^{\rm stat}_n$, $E^{1/m}_n$, $\beta^{\rm stat}_n$ and $\beta^{1/m}_n$ are
fit coefficients. This has been numerically checked in \cite{BlossierKD} where we have analysed, on the quenched set-up $\beta=6.2885$, an $8 \times 8$ matrix of correlators, using interpolating fields for static-light mesons of the form~\cite{GuskenAD}
\beq\nonumber
O_i = \bar{\psi}_h \gamma^{5}\,(1 + \kappa_G a^2 \Delta)^{R_i} \psi_l\,,
\eeq
where $\kappa_G=0.1$, $r_i\equiv 2a \sqrt{\kappa_G R_i} \leq 0.6\mathrm{fm}$, 
and $\Delta$ is a gauge covariant Laplacian made of 3 times APE-blocked 
links~\cite{AlbaneseDS}. The $R_{i}$ radii $R_{i}$ are collected in Table \ref{tab:simulhadronHQET}.
In Figures \ref{fig:Estat} and \ref{fig:E1om} we observe a good agreement for the time dependence of energies with their analytical expressions (\ref{eq:energytimedependence}).
\begin{figure}[t]
\begin{center}
\begin{tabular}{cc}
\includegraphics[angle=270,totalheight=5cm]{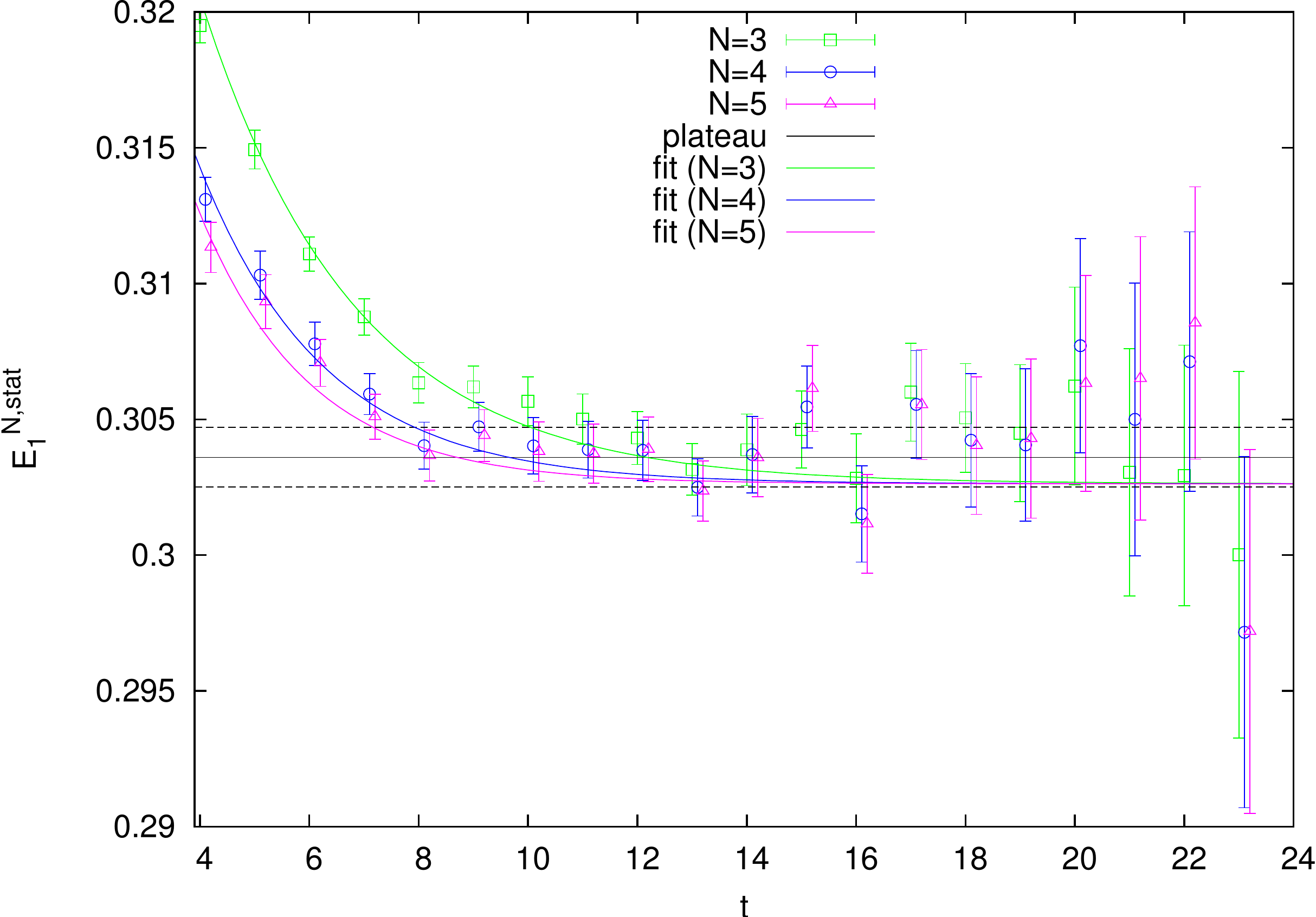}
&
\includegraphics*[angle=270, totalheight=5cm]{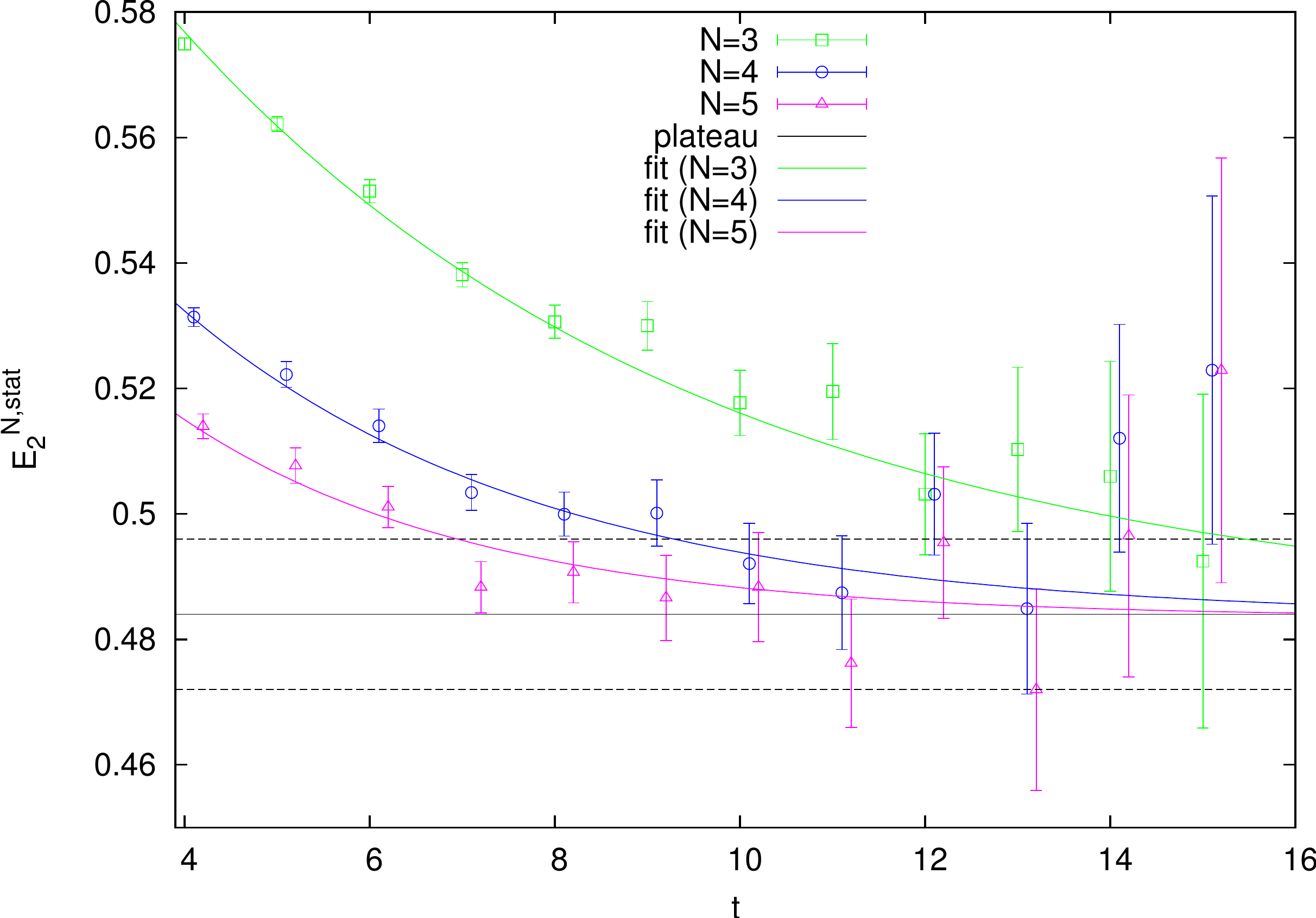}
\end{tabular}
\end{center}
\caption{Time dependence of the static effective energies $E^{\rm stat}_{1}$ (left panel) and $E^{\rm stat}_{2}$ (right panel) extracted from GEVP. The convergence rate is improved when the basis of interpolating fields is enlarged.
\label{fig:Estat}}
\end{figure}
\begin{figure}[t]
\begin{center}
\begin{tabular}{cc}
\includegraphics[angle=270,totalheight=5cm]{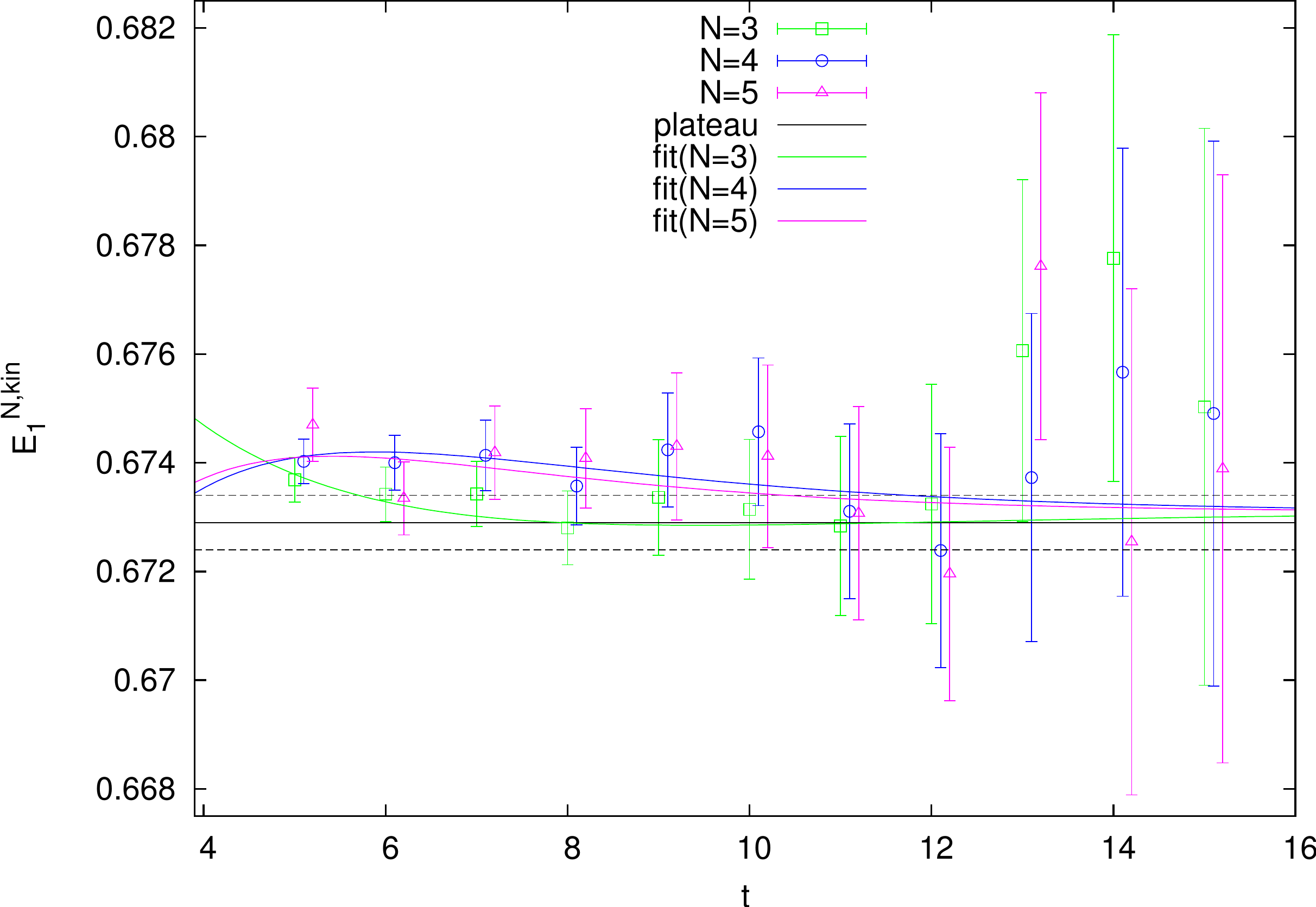}
&
\includegraphics*[angle=270,totalheight=5cm]{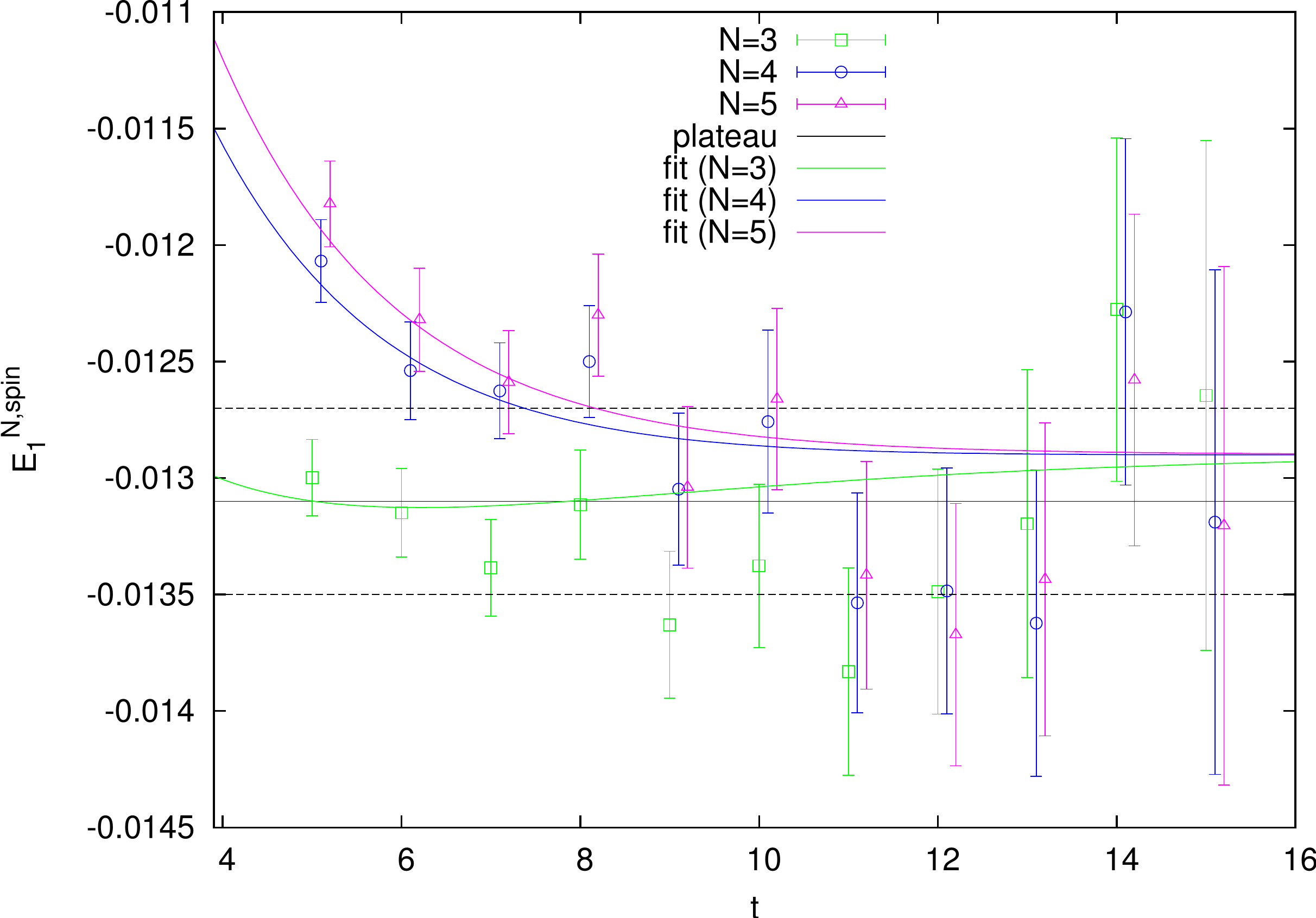}
\end{tabular}
\end{center}
\caption{Time dependence of the effective energies $E^{\rm kin}_{1}$ (left panel) and $E^{\rm spin}_{1}$ (right panel) extracted from GEVP. There is a good agreement between numerical data and the analytical expression (\ref{eq:energytimedependence}).
\label{fig:E1om}}
\end{figure}

\subsubsection{Concluding remarks on the quenched estimate of $m_{b}$}

Having everything in our hands, i.e. the HQET hadronic matrix elements $E^{\rm stat,kin,spin}$ got from a GEVP analysis and the HQET parameters obtained by a matching with QCD, and following \cite{DellaMorteCB} to interpolate the $m_{B_{s}}$ values to find $z_{b}$, as we illustrate in Figure \ref{fig:zbquenched}, our estimates of the RGI $b$ quark mass in the quenched approximation reads
\bea\nonumber
z^{\rm stat}_{b}=12.30(19),&& r_{0} M^{\rm stat}_{b}=17.12(26),\\
z^{\rm HQET}_{b}=12.48(20),&& r_{0} M^{\rm HQET}_{b}=17.38(28),
\eea
where the error is statistical, includes the uncertainty on the quark mass renormalization constant in QCD, the $\sim$ 2\% uncertainty on $L_{2}/r_{0}$ \cite{NeccoXG} 
and the small mismatch $\tilde{L}_{1}-L_{1}$ as described in the appendix of \cite{DellaMorteCB}: it consists in measuring the changing in $z_{b}$ when, at fixed $g^{2}_{0}$, one computes the observables in 2 volumes $L/a$=10 and $12$, with $\left.\bar{g}^{2}\right|_{L/a=12} = 3.48$:
\begin{align*}
\frac{\Delta M_{b}}{m_{B_s}} = |\tilde{u} - u| \rho(u) K'(u),\quad \tilde{u}-u = \bar{g}^{2}(\tilde{L}_{1}) - \bar{g}^{2}(L_{1})
=-0.17(5),\\
 \rho(u) = \frac{z}{L_1 \Gamma^{\rm av}(L_1)}\sim1.44,\quad K(u)=\frac{\Gamma^{\rm stat}(L_{1})-E^{\rm stat}}{m_{B_s}},
\end{align*}
where $\Gamma^{\rm av}$ is the (finite volume) averaged $B_s$ meson mass $(m_{B_s} +
3 m_{B^*_s})/4$.
\begin{figure}[t]
\begin{center}
\includegraphics[width=0.5\textwidth]{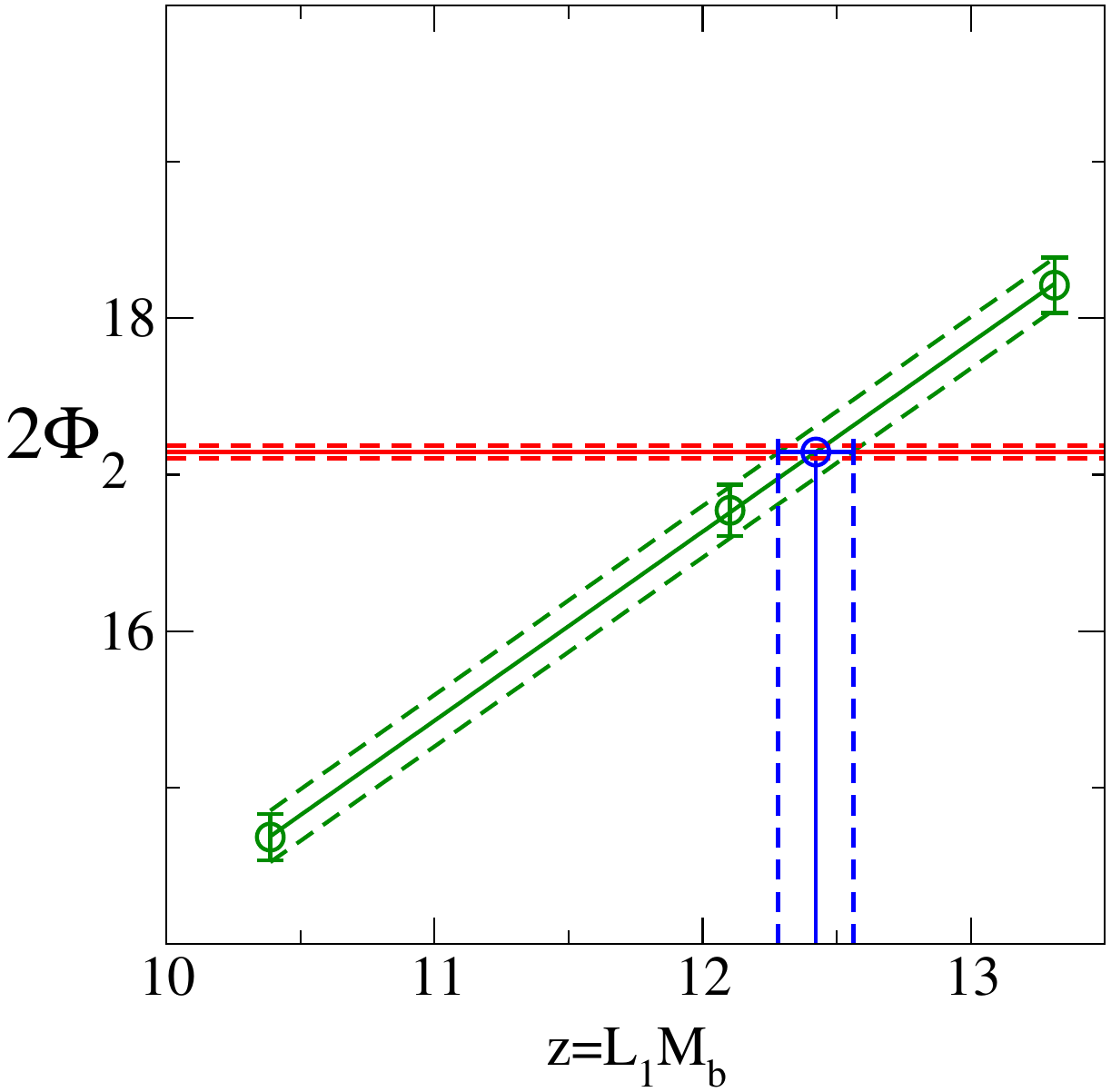}
\end{center}
\caption{Interpolation to $z_{b}$ of the spin average $B_s$ meson mass $m^{\rm av}_{B_s}$ related to a dimensionless observable $\Phi_2=L(m^{\rm bare}+\Gamma_1^{\rm stat}(L,\theta_0)
+\omega_{\rm kin} \Gamma^{\rm kin}_1(L,\theta_0)$, $\Gamma^{\rm stat,kin}_1 
\stackrel{L \to \infty}{\rightarrow} E^{\rm stat, kin}$, analysed in \cite{DellaMorteCB}, from which that plot is taken; the red line corresponds to the experimental value $m^{\rm av}_{B_s}=5.405$ GeV.
\label{fig:zbquenched}}
\end{figure}
\\
Then we have interpolated HQET parameters \{$\omega$\} quadratically in $z$ to $z_{b}$ to extract the hyperfine splitting
$\Delta E_{P-V} \equiv \frac{4}{3} \omega_{\rm spin}E_{\rm spin}$ \cite{BlossierVZ} and the decay constant $f_{B_{s}}$ \cite{BlossierMK}:
\beq
\Delta E_{P-V}=-29.8(3.2)\, {\rm MeV}, \quad f^{\rm stat}_{B_{s}}=229(3)\, {\rm MeV}, \quad f^{\rm HQET}_{B_{s}}=216(5)\,{\rm MeV},
\eeq
where we have used $r_{0}=0.5$ fm in the conversion of dimensionless results into physical units and we show in Figure \ref{fig:spinsplitting} the continuum limit extrapolation of $r_{0} \Delta E_{P-V}$ and $r^{3/2}_{0} f_{B_{s}}\sqrt{m_{B_{s}}/2}$. At this stage the quenching error, although unknown, is certainly large because, experimentally, $m_{B_{s}}-m_{B^{*}_{s}}=-46.1(1.5)$ MeV \cite{BeringerZZ}. To extract the decay constant $f_{B_s}={\cal A}/\sqrt{m_{B_s}/2}$ we have considered the couplings
\bea\nonumber
{\cal A}^{{\rm eff}}_{n}(t,t_{0}) &= &{\cal A}^{{\rm eff\,stat}}_{n}(t,t_{0})(1+\omega {\cal A}^{{\rm eff\,1/m}}_{n}(t,t_{0})),\\
\nonumber
{\cal A}^{\rm eff\, stat}_{n}(t,t_{0})&=&\frac{\sum_{j} C_{0j}(t)v^{(n)\,\rm stat}(t,t_{0})}{\sqrt{(v^{(n)\,\rm stat}(t,t_{0}), C^{\rm stat}(t) v^{(n)\,\rm stat}(t,t_{0}))}}\left(\frac{\lambda^{(n)}(t_{0}+a,t_{0})}{\lambda(t_{0}+2a,t_{0})}\right)^{t/2a},\\
\nonumber
{\cal A}_n^{{\rm eff},1/m}(t,t_0) &=& {R_n^{1/m} \over R_n^{\rm stat}}\,+\,
   \frac{\sum_{j} C_{0j}^{1/m}(t) v^{(n)\,{\rm stat}}(t,t_{0})}{\sum_{j} C_{0j}^{\rm stat}(t) v^{(n)\,{\rm stat}}(t,t_{0})} \,+\,
  \frac{\sum_{j} C_{0j}^{{\rm stat}}(t) v^{(n)\,1/m}(t,t_{0})}{\sum_{j} C_{0j}^{\rm stat}(t) v^{(n)\,{\rm stat}}(t,t_{0})},
\eea
\bea\nonumber
{R^{1/m}_n \over R^{\rm stat}_n} &=& -{1 \over 2}
\frac{(v^{(n)\,\rm stat}(t,t_{0}), C^{1/m}(t) v^{(n)\,\rm stat}(t,t_{0}))}
{(v^{(n)\,\rm stat}(t, t_{0}),C^{\rm stat}(t) v^{(n)\,\rm stat}(t,t_{0}))} + {t \over 2a}
\left({\lambda^{(n)\,1/m}(t_0+a,t_0)\over\lambda^{(n)\,\rm stat}(t_0+a,t_0)}
- {\lambda^{(n)\,1/m}(t_0+2a,t_0)\over\lambda^{(n)\,\rm stat}(t_0+2a,t_0)}\right)\,,\\
v^{(n)\,1/m} &=& \sum_{k=1, k\neq n}^N v^{(k)\,\rm stat}(t,t_{0})
      {\left({v}^{(k)\,\rm stat}(t,t_{0}), [C^{1/m}(t)-\lambda^{(n)\,\rm stat}(t,t_0)C^{1/m}(t_0)]\,
         {v}^{(n)\,\rm stat}(t,t_{0})\right) \over
               \lambda^{(n)\,\rm stat}(t,t_0)-\lambda^{(k)\,\rm stat}(t,t_0) }\,,
\eea  
with, at large time,
\bea\nonumber
{\cal A}_n^{N,\rm stat}(t,t_0) &=& {\cal A}_n^{\rm stat}
+ \gamma_{n,N}^{\rm stat}\,e^{-\Delta E^{\rm stat}_{N+1,n}t_0} \,, \nonumber \\
{\cal A}_{n}^{N,\rm x}(t,t_0) &=& {\cal A}_n^{\rm x} +\left[ \gamma_{n,N}^{\rm x}
-{\gamma_{n,N}^{\rm stat}\over {\cal A}_n^{\rm stat}}
\,t_0\,(\Delta E^{\rm x}_{N+1,n})
\right] e^{-\Delta E^{\rm stat}_{N+1,n}t_0}, 
\eea
where ${\cal A}^{\rm stat}_n$, ${\cal A}^{\rm x}_n$, $\gamma^{\rm stat}_{n,N}$ and $\gamma^{\rm x}_{n,N}$ are fit coefficients.
\begin{figure}[t]
\begin{center}
\begin{tabular}{cc}
\includegraphics[width=0.5\textwidth]{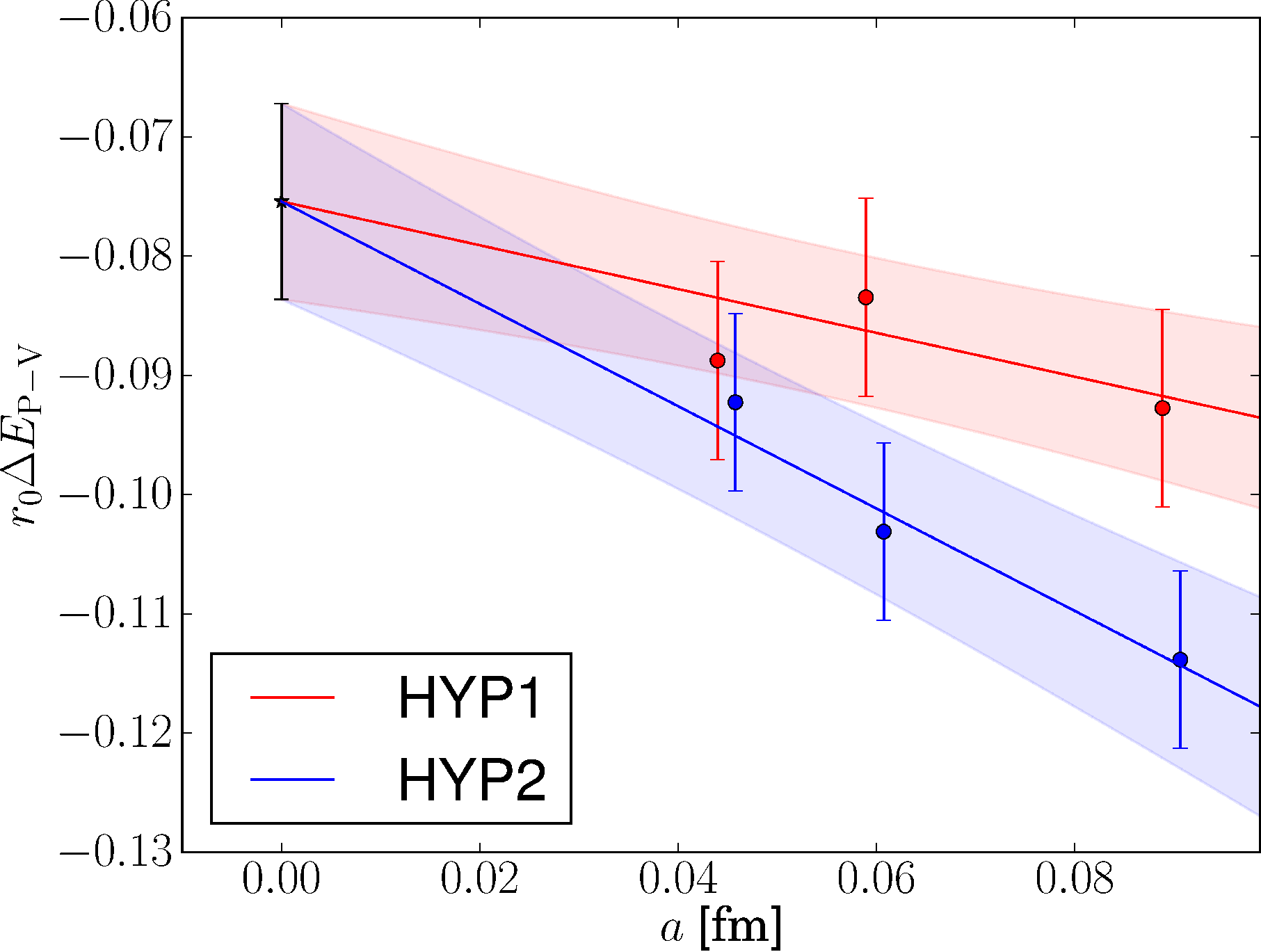}
&
\includegraphics[width=0.5\textwidth]{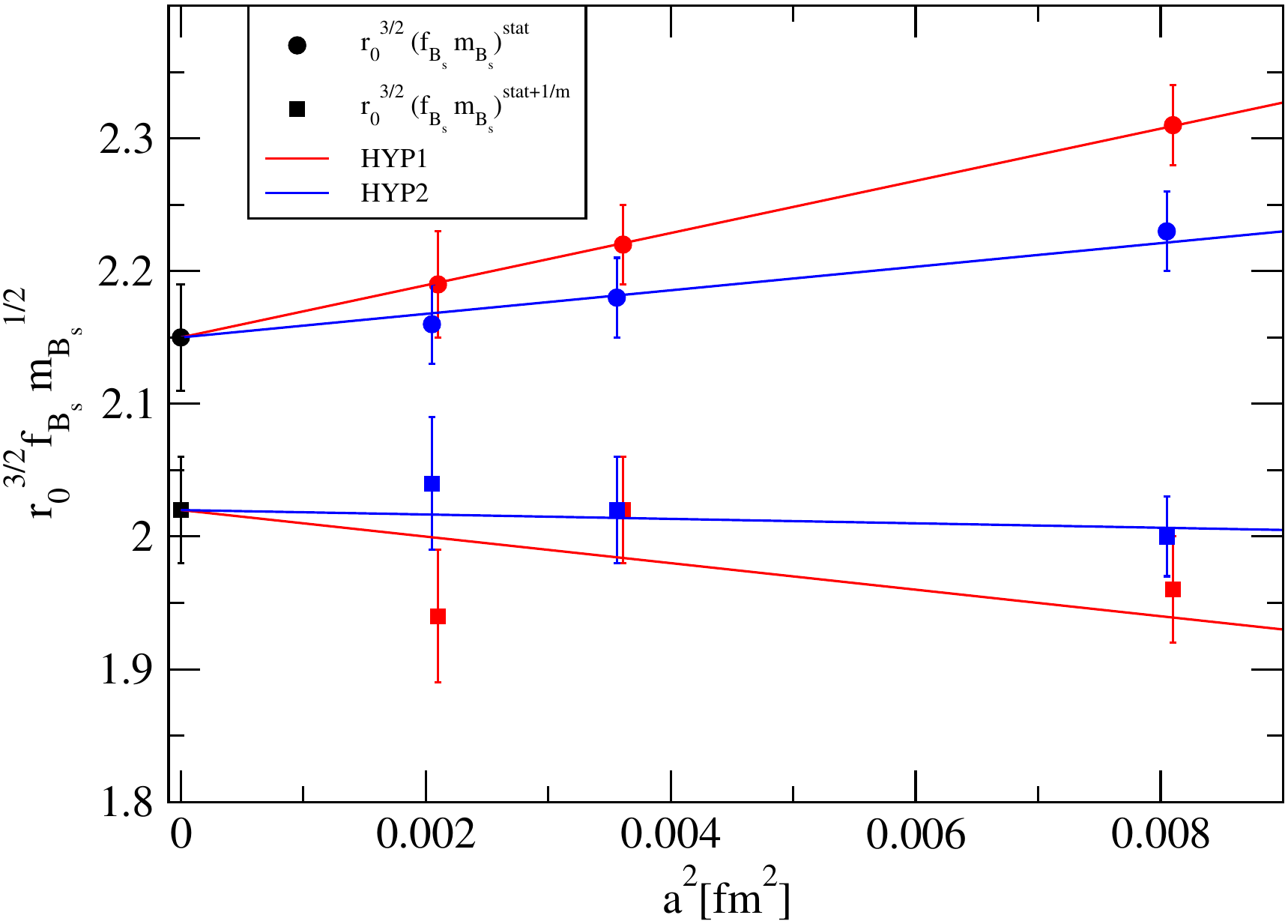}
\end{tabular}
\end{center}
\caption{Continuum limit extrapolations of the $m_{B_{s}}-m_{B^{*}_{s}}$ mass splitting (left panel) and the 
decay constant $f_{B_{s}}$ in the static limit and up to ${\cal O}(1/m_{b})$ of HQET (right panel).
\label{fig:spinsplitting}}
\end{figure}

\section{$b$ quark mass at ${\rm N_{f}=2}$}

We have roughly followed the same strategy as described in the previous section. In QCD, 4 simulations have been realised in a volume $L_{1}\sim 0.4$ fm defined by $\bar{g}^{2}(L_{1}/2)=2.989$. 9 RGI heavy masses $z=4,\,6,\,7,\,9,\,11,\,13,\,15,\,18,\,21$ have been simulated while the PCAC light quark mass is set to zero. Parameters of the 
simulations are collected in Table \ref{tab:simulQCDnf2}.
We recall that
\beq
M=h(L)Z_{m}(g_{0},L/a) (1+b_{m}am_{q,h}) m_{q,h} + {\cal O}(a^{2}),
\eeq
with
\beq\nonumber
Z_{m}(g_{0},L/a) = \frac{Z(g_{0}) Z_{A}(g_{0})}{Z_{P}(g_{0},L/a)}, \quad am_{q,h}=\frac{1}{2}
\left(\frac{1}{\kappa_{h}} - \frac{1}{\kappa_{c}}\right),\quad L=L_{1}/2, \quad h(L)=\frac{M}{m^{\rm SF}(\mu=1/L)}=1.521(14).
\eeq
$Z_{A}$ is known from \cite{DellaMorteXB}, $h(L_{1}/2)$ from \cite{DellaMorteKG} and $Z$ and $b_{m}$ have been determined in \cite{FritzschAW}. In HQET, 2 series of simulations, $S_2$ and $S_3$, have been realised. They are defined by $\bar{g}^{2}(L_{1})=4.484(48)$ from the step scaling function of the coupling that was determined in \cite{DellaMorteBC}. Parameters of the simulations are collected in Table \ref{tab:simulSFHQETnf2}.
\begin{table}[t]
\begin{center}
\begin{tabular}{|c|c|c|c|c|c|c|c|c|} 
\hline
$L/a$ & $\beta$ & $\kappa$ & $\bar{g}^{2}(L/2)$ 
& $Z_{P}\big(g_0,\frac{L}{2a}\big)$ & $b_m$ & $Z$&
$z$ & $\kappa_h$ \\
\hline
  $20$ & $6.1569$ & $0.1360536$ & $2.989(36)$ & $0.6065(9)$  & $-0.6633(12)$ &  $1.10443(17)$ & $4$  
  & $0.1327094$  \\
        &&&&&&& $6$  & $0.1309180$  \\
        &&&&&&& $7$  & $0.1299824$  \\
        &&&&&&& $9$  & $0.1280093$  \\
        &&&&&&& $11$ & $0.1258524$  \\
        &&&&&&& $13$ & $0.1234098$  \\
        &&&&&&& $15$ & $0.1204339$  \\
        &&&&&&& $18$ & $\text{---}$ \\
        &&&&&&& $21$ & $\text{---}$ \\
\hline
  $24$ & $6.2483$ & $0.1359104$ & $2.989(30)$ & $0.5995(8)$  & $-0.6661(9)$  &  $1.10475(12)$ &$4$  
  & $0.1331966$  \\
        &&&&&&& $6$  & $0.1317649$  \\
        &&&&&&& $7$  & $0.1310257$  \\
        &&&&&&& $9$  & $0.1294907$  \\
        &&&&&&& $11$ & $0.1278628$  \\
        &&&&&&& $13$ & $0.1261106$  \\
        &&&&&&& $15$ & $0.1241815$  \\
        &&&&&&& $18$ & $0.1206988$  \\
        &&&&&&& $21$ & $0.1140810$  \\
\hline
  $32$ & $6.4574$ & $0.1355210$ & $2.989(35)$ & $0.5941(10)$ & $-0.6674(23)$ & $1.10455(17)$ & $4$  & $0.1335537$  \\
        &&&&&&& $6$  & $0.1325329$  \\
        &&&&&&& $7$  & $0.1320117$  \\
        &&&&&&& $9$  & $0.1309446$  \\
        &&&&&&& $11$ & $0.1298401$  \\
        &&&&&&& $13$ & $0.1286909$  \\
        &&&&&&& $15$ & $0.1274876$  \\
        &&&&&&& $18$ & $0.1255509$  \\
        &&&&&&& $21$ & $0.1233865$  \\
\hline
  $40$ & $6.6380$ & $0.1351923$ & $2.989(43)$ & $0.5949(12)$ & $-0.6692(27)$ &  $1.10379(17)$ & $4$  & $0.1336432$  \\
        &&&&&&& $6$  & $0.1328462$  \\
        &&&&&&& $7$  & $0.1324413$ \\
                &&&&&&& $9$  & $0.1316178$  \\
        &&&&&&& $11$ & $0.1307738$  \\
        &&&&&&& $13$ & $0.1299065$  \\
        &&&&&&& $15$ & $0.1290126$  \\
        &&&&&&& $18$ & $0.1276125$  \\
        &&&&&&& $21$ & $0.1261232$  \\
\hline
\end{tabular}
\end{center}
\caption{Bare parameters $(L/a,\beta,\kappa_l,\kappa_h)$ used in the computation of the 
heavy-light QCD observables for $L=L_1$.
The entering renormalization constants $Z_{P}$ and $Z$ and the improvement 
coefficient $b_m$ are known from \cite{FritzschAW}.\label{tab:simulQCDnf2}}
\end{table}
\begin{table}[t]
\begin{center}
\begin{tabular}{|c|c|c|c|}
\hline
    $L/a$ & $\beta $ & $\kappa_{l}$ & $\bar{g}^{2}$  \\ 
\hline
    $ 6 $        & $5.2638$ & $0.135985 $ & $4.423(75)$          \\
    $ 8 $        & $5.4689$ & $0.136700 $ & $4.473(83)$          \\
    $ 10$        & $5.6190$ & $0.136785 $ & $4.49(10) $\hphantom{0}    \\
    $ 12$        & $5.7580$ & $0.136623 $ & $4.501(91)$      \\
    $ 16$        & $5.9631$ & $0.136422 $ & $4.40(10) $\hphantom{0}  \\
   \hline
    $8^{\ast}$   & $5.4689$ & $0.13564  $ & $4.873(99)$            \\
    $12^{\ast}$  & $5.8120$ & $0.136617 $ & $4.218(49)$          \\
\hline
\end{tabular}
\end{center}
\caption{Bare parameters and results of the tuning to $\bar{g}^{2}(L)=4.484$ for the HQET simulations $S_2$ and $S_3$. The additional lattices $L/a=8^{\ast},12^{\ast}$ are used
to estimate and propagate a potential error, resulting from not meeting the line of
constant physics condition exactly. \label{tab:simulSFHQETnf2}}
\end{table}
The hadronic volumes we will consider in simulations $S_5$ are made at $\beta$ that are in the range [5.2, 5,5]. It means that the HQET parameters are easily obtained at the set of bare couplings 
$\{g_{0}\}^{(2)}$ by interpolating those extracted from the HQET SSF's in $S_3$. So, here, we have not realised a fourth series $S_4$ as in the quenched case.\\
We show in Figure \ref{fig:phi1QCDnf2} the continuum limit extrapolation of $\Phi^{\rm QCD}_{1}$ for each heavy quark mass $z$ and the dependence on $z$ of $\Phi^{\rm QCD,\,cont}_{1}$. A difference with the quenched study is that, as we have simulated quite heavy quarks, inducing potentially large cut-off effects, but that depend smoothly on $a/L_{1}$ and $z$, we have performed a global fit with the following formula:
\beq\nonumber
\Phi^{\rm QCD}(L,M,a)=\Phi^{\rm QCD,\, cont}(L,M)\left[1+(a/L_{1})^{2}(A+Bz + C z^{2})\right].
\eeq
We have included in the fit only data points such that $a M \le 0.7$; above that threshold the ${\cal O}(a)$ improvement of observables seems to break down \cite{KurthYR}, \cite{HeitgerUE}. We show in Figure \ref{fig:PhiL2} the continuum limit extrapolations of the static observables $\Phi^{\rm stat}_{1}(L_{2})$ and the ${\cal O}(1/m)$ observable $\Phi_{5}(L_{2})$ for the 9 heavy quark masses; those extrapolations are smooth.
\begin{figure}[t]
\begin{center}
\begin{tabular}{cc}
\includegraphics[width=0.5\textwidth]{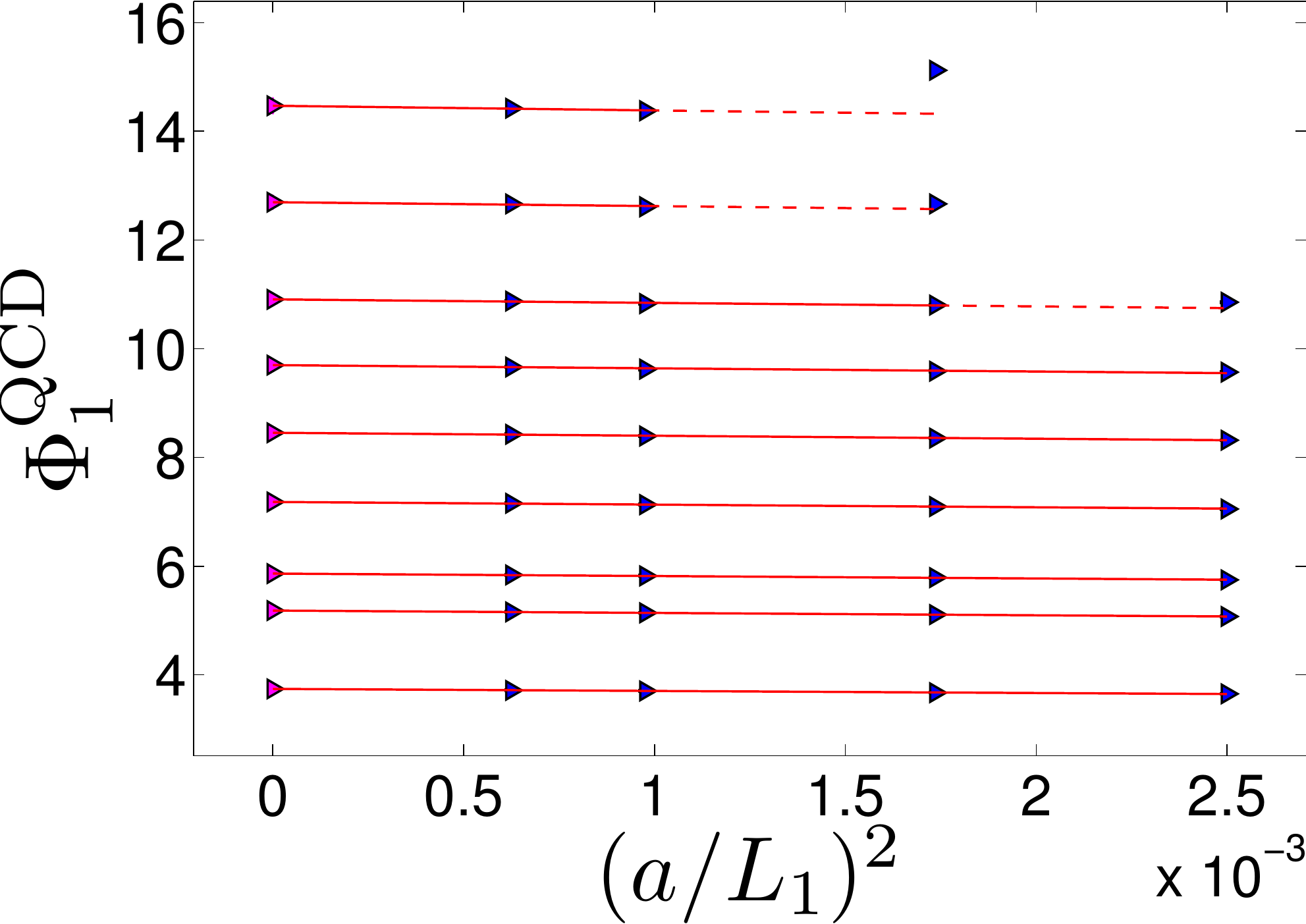}
&
\includegraphics[width=0.5\textwidth]{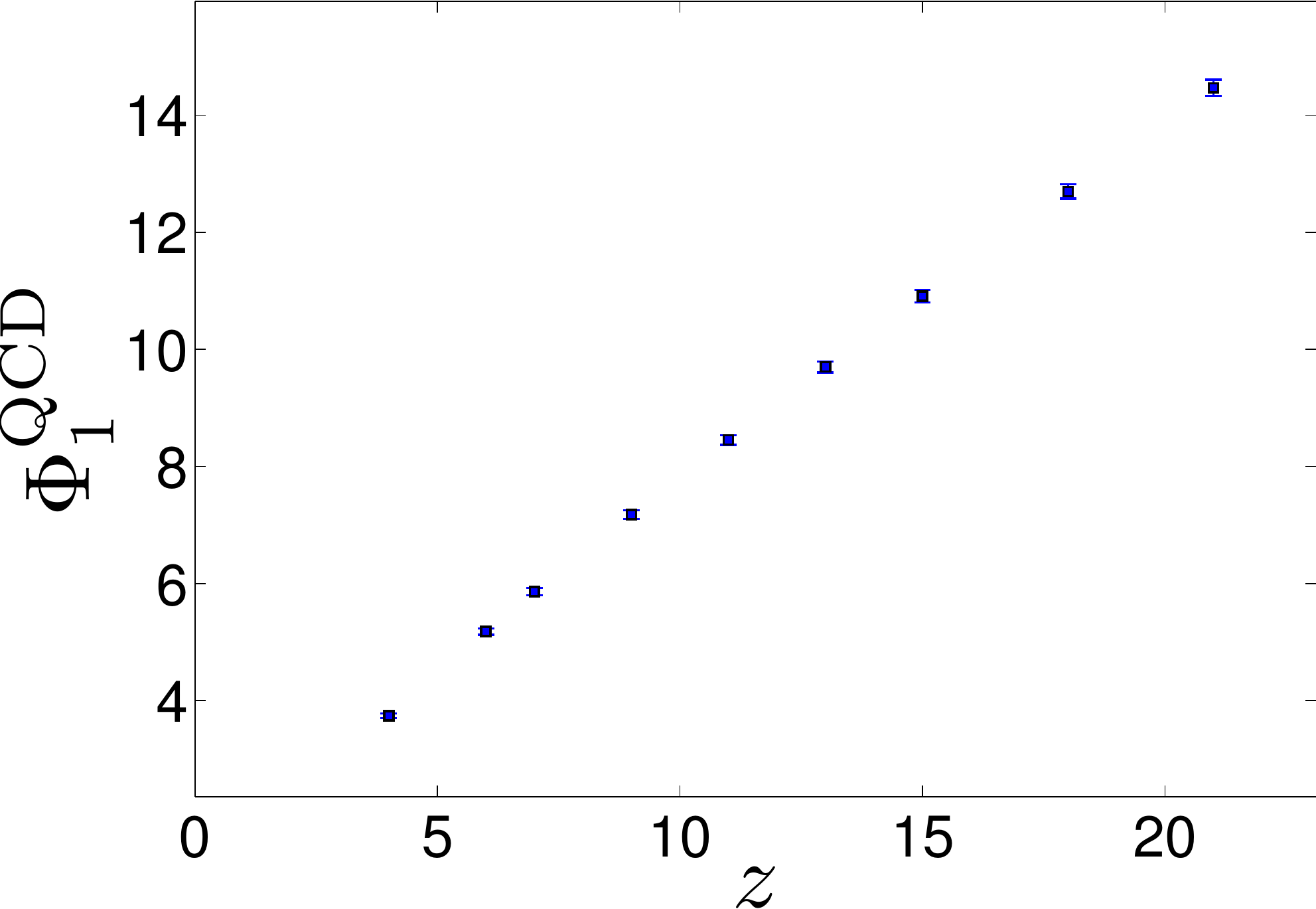}
\end{tabular}
\end{center}
\caption{Continuum limit extrapolation of the QCD observable $\Phi^{\rm QCD}_{1}(L_1,\theta_0)$ for each heavy quark mass (left panel) and dependence on $z$ of $\Phi^{\rm QCD,\,cont}_{1}(L_1,\theta_0)$ (right panel); $\theta_{0}=0.5$.
\label{fig:phi1QCDnf2}}
\end{figure}
\begin{figure}[t]
\begin{center}
\begin{tabular}{cc}
\includegraphics[width=0.5\textwidth]{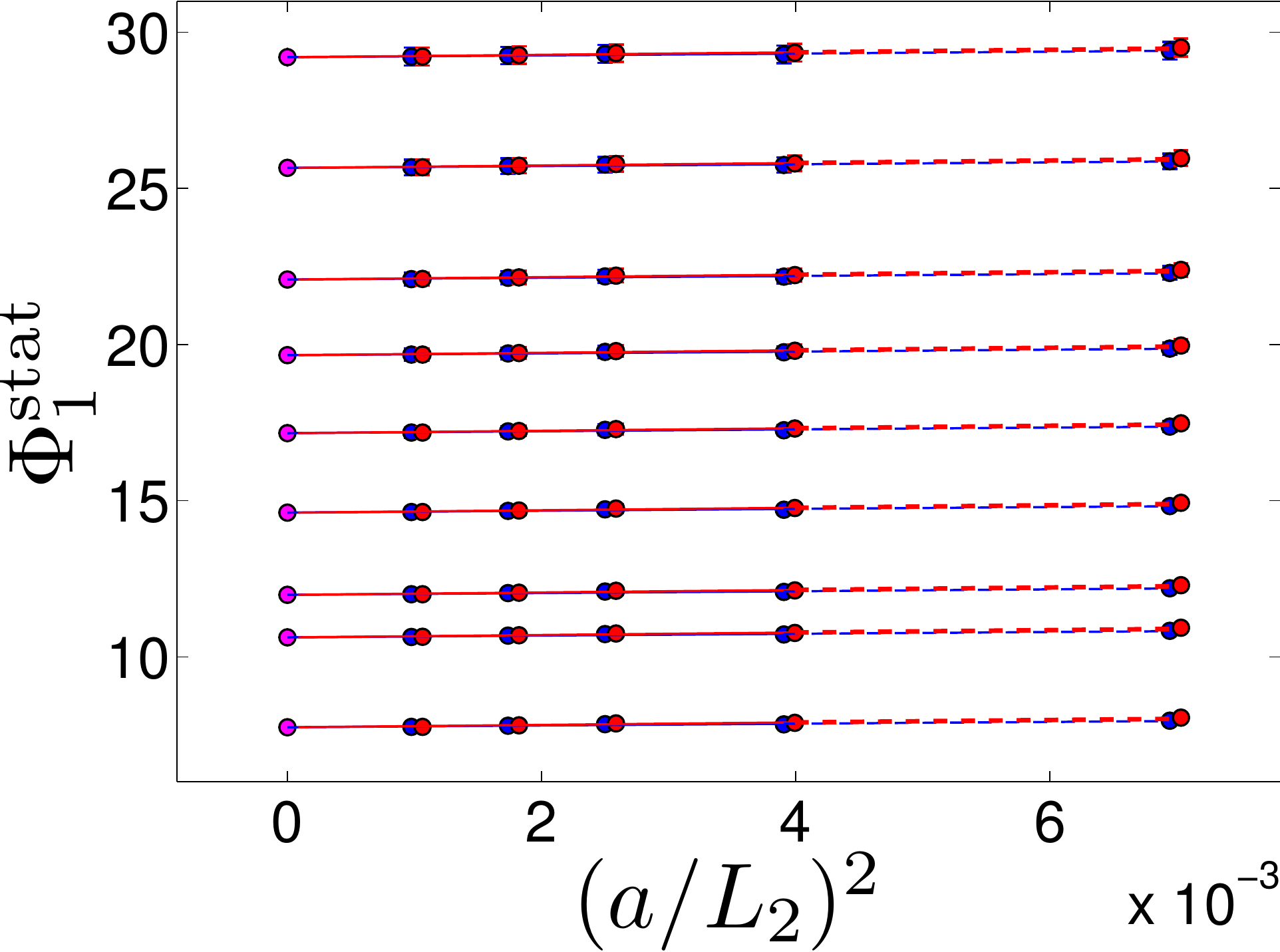}
&
\includegraphics[width=0.5\textwidth]{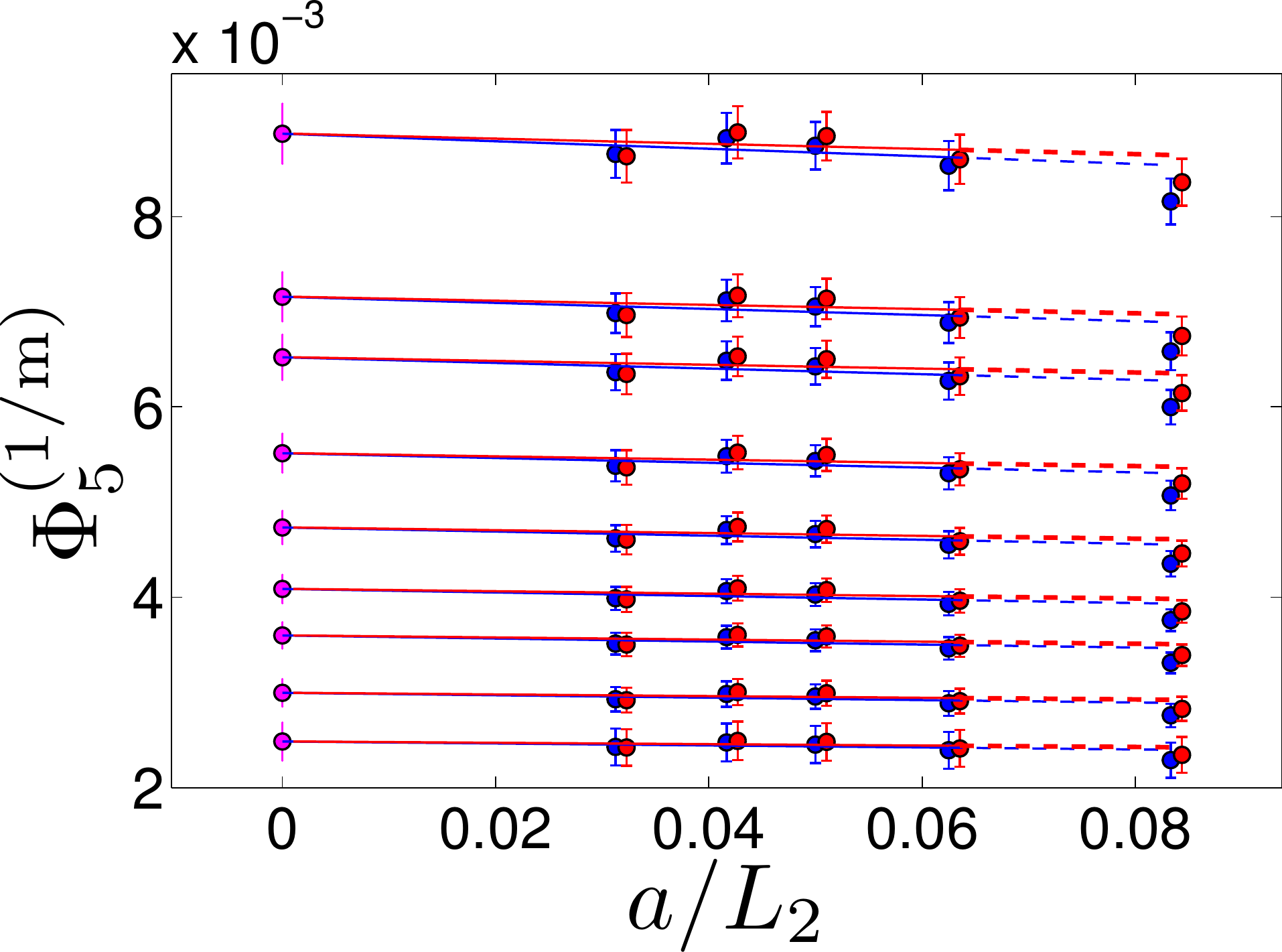}
\end{tabular}
\end{center}
\caption{Continuum limit extrapolations of the observables $\Phi^{\rm stat}_{1}(L_{2},\theta_0)$ (left panel) and $\Phi_{5}(L_{2},\theta_0)$ (right panel) for each heavy quark mass; $\theta_{0}=0.5$. \label{fig:PhiL2}}
\end{figure}
\begin{table}[t]
\begin{center}
 \begin{tabular}{|c|c|c|c|c|c|c|c|c|}
\hline
 $\beta$ & $a$[fm] & $L/a$ & $m_\pi$[MeV] & $m_\pi L$ & \#cfgs &  $\dfrac{\#\rm cfgs}{\tau_{\rm exp}}$  & id  & $\{R_1,R_2,R_3\}$ \\    
\hline
     $5.2$ &$0.075$ & $32$  & $380$        & $4.7$     &  1012  &  122  & A4 & $\{15,  60, 155\}$ \\
           &        & $32$  & $330$        & $4.0$     &  1001  &  164  & A5  & \\
           &        & $48$  & $280$        & $5.2$     &   636  &  52   & B6   & \\ 
\hline
     $5.3$ &$0.065$ & $32$  & $440$        & $4.7$     &  1000  &  120  & E5 & $\{22,  90, 225\}$ \\
           &        & $48$  & $310$        & $5.0$     &   500  &  30   & F6   & \\
           &        & $48$  & $270$        & $4.3$     &   602  &  36   & F7  &  \\
           &        & $64$  & $190$        & $4.1$     &   410  &  17   & G8  & \\
\hline
     $5.5$ &$0.048$ & $48$  & $440$        & $5.2$     &   477  &  4.2  & N5 & $\{33, 135, 338\}$ \\
           &        & $48$  & $340$        & $4.0$     &   950  &  38   & N6  & \\
           &        & $64$  & $270$        & $4.2$     &   980  &  20   & O7   & \\
\hline
  \end{tabular}
\end{center}
  \caption{Details of the CLS ensembles used to extract the HQET hadronic matrix elements: bare coupling 
  $\beta=6/g_0^2$, 
           lattice spacing $a$, spatial extent $L$ in lattice units ($T=2L$),
           pion mass $m_\pi$, $m_\pi L$, number of configurations employed,
           and number of configurations employed normalised in units of
           the exponential autocorrelation time $\tau_{\rm exp}$ as 
           estimated in \cite{SchaeferHU}. Additionally, we specify
           the CLS label id and the Gaussian smearing parameters $R_k$ used 
           to build different interpolating fields of the $B$ meson.
           } 
  \label{tab:simulCLSnf2}
\end{table}
As described in details in \cite{DellaMorteAD} the ensembles have been generated using Hybrid Monte Carlo algorithm \cite{DuaneDE} with a trajectory length of the molecular dynamics $\tau=2$: it is expected to reduce the autocorrelation time \cite{MeyerTY}. Concerning the integration scheme, we have used a multiple time scales one with a leap-frog integrator \emph{\`a la} Sexton-Weingarten \cite{SextonWein}. A mass preconditioner \emph{\`a la} Hasenbusch \cite{HasenbuschNE} is introduced to make faster the convergence of the inversion algorithm. The acceptance rate is $\gtrsim$ 85 \%, something totally safe. Tuning the bare parameters to target values (for instance $g_{0}$) is not always perfect: we have to take into account the mismatch in the error. For $z$, the uncertainty on $Z_{A}$, $Z_{P}$, $Z$ (entering in $Z_{m}$) and $b_{m}$ are combined in quadrature, leading to a relative error on $z$ of $0.38 - 0.41$\%; the uncertainty $\Delta h /h=0.92$\% on the universal factor $h(L_{1}/2)$, that finally dominates the error $\sim$ 1\% on $z$, has to be propagated into the QCD $\Phi$ observables only once they are extrapolated in the continuum limit. On the HQET side, the effect of a mistuning in $\bar{g}^{2}$ is studied with the run ``$12^{*}\,''$; it is the dominant error on static observables $R^{\rm stat}_{A}$ and $R^{\rm stat}_{1}$, that are statistically very precise.\\
FInally the HQET hadronic matrix elements have been extracted on a large set of ensembles built within the Coordinate Lattice Simulations consortium (CLS) \cite{CLS}, using the Domain Decomposition Hybrid Monte Carlo algorithm \cite{LuscherQA} - \cite{LuscherES} and, for the most recently ensembles produced, the mass preconditioning HMC \cite{HasenbuschNE}. Parameters of the runs are collected in Table \ref{tab:simulCLSnf2}. Lattice spacings have been extracted from $f_{K}$ \cite{FritzschWQ}, with an update reported in \cite{LottiniRFA}. $\tau_{\rm exp}$ is used to estimate the long-term effects of autocorrelations induced by the coupling of our observables of interest to the slow modes of the Markov chain, that decay in $\exp^{-\tau/\tau_{\rm exp}}$, as discussed in \cite{SchaeferHU}, where the method proposed in \cite{WolffSM} to include autocorrelation in the statistical error was improved. For a given series of observables $O_{\alpha}$, one considers the auto-correlation function $\Gamma_{\alpha\beta}(t)$
\bea\nonumber
\Gamma_{\alpha\beta}(t) &=& \frac{1}{N_{\rm traj}}\sum_{i=1}^{N_{\rm traj}}
[O_{\alpha}(i+t) - \langle O_{\alpha} \rangle][O_{\beta}(i) - \langle O_{\beta} \rangle ],\\
\nonumber
\langle O_{\alpha} \rangle&=&\frac{1}{N_{\rm meas}}\sum_{i=1}^{N_{\rm meas}}
O_{\alpha}(i)\,.
\eea
An error $\delta \bar{F}$ of a function $F(O_{\{\alpha\}})$ is given by:
\bea\nonumber
(\delta \bar{F})^{2}=\frac{\sigma^{2}_{F}}{N_{\rm meas}} 2 \tau_{\rm int}(F), \quad \sigma^{2}_{F}=\Gamma_{F}(0), \quad \Gamma_{F}(t) = \sum_{\alpha, \beta} F_{\alpha} \Gamma_{\alpha\beta}(t)
F_{\beta},\\
\nonumber
\tau_{\rm int}(F) = \frac{1}{2} + \sum_{t=1}^{\infty} \rho_{F}(t), \quad 
\rho_{F}(t)=\frac{\Gamma_{F}(t)}{\Gamma_{F}(0)}, \quad F_{\alpha}=\frac{\partial F}{\partial O_{\alpha}}.
\eea
In practice, $\tau_{\rm int}(F)$ is estimated from a summation over a window $W$:
\beq
\tau_{\rm int}(F,W) = \frac{1}{2} + \sum_{t=1}^{W-1} \rho_{F}(t),
\eeq
where $W$ is fixed by minimizing the evaluator \cite{WolffSM}
\beq\nonumber
E(W)=e^{-W/(2S\tau_{\rm int}(F,W))} + \sqrt{W/N_{\rm meas}}
\eeq
and $S$ is typically set to 1.5. However, due to the slow modes of Markov chain and the fact that, at large time of the molecular dynamics, $\Gamma_{F}(t) \sim A_{F} e^{-t/\tau_{\rm exp}}$, the integrated auto-correlation time $\tau_{\rm int}$ has to be modified further:
\beq
\tau'_{\rm int}(F,W) = \tau_{\rm int}(F,W) + \tau_{\rm exp} \rho_{F}(W), \quad  
\tau_{\rm exp} = \lim_{t\to \infty} \frac{t}{2\ln \{{\rm max}_{\beta} [\rho_{\beta}(t/2)/\rho_{\beta}(t)]\} }.
\eeq
It has been found by CLS that the square of the topological charge is the most affected observable
by the tail of Markov chain: then, it helps to set $\tau_{\rm exp}$, whose an empirical formula has been proposed in \cite{SchaeferHU}:
\beq\nonumber
\tau_{\rm exp} \sim 200 e^{7(\beta-5.5)} \tau.
\eeq
As in our quenched study, we have solved a generalized eigenvalue problem to eliminate in an appropriate way the contribution from excited states to HQET correlation functions. It revealed enough to analyse a $3\times 3$ system. The smearing parameter is given in Table \ref{tab:simulCLSnf2}. To extract our energies 
$E_1^{{\rm stat},\,{\rm kin},\, {\rm spin}}$, we choose the time intervals 
$[t_{\rm min}, t_{\rm max}]$ over which we fit the plateaux such that 
\begin{equation}
  r(t_{\rm min}) = \frac{|A(t_{\rm min})-A(t_{\rm min}-\delta)|}
                  {\sqrt{\sigma^2(t_{\rm min})+\sigma^2(t_{\rm min}-\delta)}} 
                   \le 3\,,
\end{equation}
where $A$ is the plateau average over the window $[t_{\rm min},t_{\rm max}]$, 
$\sigma$ is the statistical error, 
\beq\nonumber
\delta = 2/(E^{{\rm stat}}_{N+1}-E^{{\rm stat}}_1)\sim 0.3\,\,{\rm fm},
\eeq 
and $t_{\rm max}$ is 
set to $\sim 0.9$ fm.  This assures that our selection criterion 
$\sigma_{\rm sys}\leq \sigma/3$ is satisfied \cite{BlossierVZ}, where 
$\sigma_{\rm sys}\propto e^{-(E_{N+1}-E_1) t_{\rm min}}$.  An illustration 
of two typical plateaux of $E_1^{\rm stat}$ and $E_1^{\rm spin}$ is shown in
Figure \ref{fig:plateauxEstatEspin}. We have performed a combined chiral and continuum extrapolation, using the NLO HM$\chi$PT fit formula \cite{GoityTP}, \cite{BernardoniSX}
\beq\label{chimB}
   m_{B}^{\rm sub}\left(z,y,a\right) 
        = B(z) + C \left(y-y^{\rm exp} \right) + D_{\iota} a^2\,,
        \eeq
where $\iota=1$ or $2$ stands for the HYP discretisation of HQET, 
$y \equiv \frac{m^2_\pi}{8\pi^2f_\pi^2}$ and 
\beq\nonumber
m_{B}^{\rm sub}\left(z,y,a\right) 
  \equiv m_{B}\left(z,m_\pi,a\right)+ \frac{3\hat{g}^2}{16\pi }
                \left( \frac{m^3_{\pi}}{f_{\pi}^2} -  \frac{(m_\pi^{\rm exp})^3}{(f_\pi^{\rm exp})^2} \right).
                \eeq
$\hat{g}=0.489(32)$ is taken from a recent estimate by the ALPHA Collaboration \cite{BulavaEJ} and we have used the convention where $f^{\rm exp}_{\pi}=130.4$ MeV. The quality of the fit is shown on Figure \ref{fig:chiralfitmBnf2} (left panel) for $z=11$, $13$ and $15$, that are close to $z_{b}$. The $z$ dependence of $m_{B}$, as shown in the right panel of Figure \ref{fig:chiralfitmBnf2}, is linear: it is an indication that the HQET expansion we have employed is sufficiently precise for that observable. However we have made a crosscheck by adding a quadratic term in the $z$ interpolation. Imposing the condition $m_{B}(z_{b})=m^{\rm exp}_{B}$, where $m^{\rm exp}_{B}=5.279$ GeV, we obtain
\beq
z_{b}=13.25(22)(13)_{z},
\eeq
where the first error is statistical and contains the uncertainty on the chiral and continuum extrapolation, while the second error comes from the uncertainty on the conversion factor in QCD $h(L_{1}/2)$. Adding a $F_\delta a/z$ factor in the fit formula, not forbidden because the ${\cal O}(1/m)$ correction to lattice HQET in the static limit is not ${\cal O}(a)$ improved, does not change the
unnormalised $\chi^2$.\\
The RGI $b$ mass $M_{b}$ is got in physical units with $M_{b}=z_{b}/[L_{1} f_{K}] f_{K}$, where $L_{1}f_{K}$ is extracted by extrapolating to the continuum limit the product $(L_{1}/a) (a f_{K})$: $L_{1}/a$ is computed by extrapolating to $\beta=$5.2, 5.3 and 5.5 those of the simulations $S_2$ and $S_3$. We quote
\beq
M_{b}=6.58(17)\, {\rm GeV},
\eeq
with the convention \cite{GasserAP} 
$  M = \lim_{\mu \to \infty} \left(2 b_0 \bar{g}^2(\mu)\right)^{-d_0/(2b_0)} \,\bar{m}(\mu) $, 
where $b_0=(11 - 2{\rm N_{f}}/3)(4\pi)^{-2} $ and $d_0=8 (4\pi)^{-2}$.
To convert in the $\overline{\rm MS}$ scheme, we defined
\beq 
 m^{\overline{\rm MS}}_{b}(m^{\overline{\rm MS}}_{b}) = M_{b} \nu(M_{b}/\Lambda_{\overline{\rm MS}}) \,, \quad  m^{\overline{\rm MS}}_{b}(2\,{\rm GeV}) = M_{b} \nu'(M_{b}/\Lambda_{\overline{\rm MS}}),
 \eeq
with conversion functions $\nu(r)$ and $\nu'(r)$ that can be evaluated accurately
using the known 4-loop anomalous dimensions of quark masses and coupling \cite{ChetyrkinPQ}, \cite{MelnikovQH}. Indeed, one considers the RGI quantities
\bea\nonumber
\frac{\Lambda}{\mu} &= & 
  [b_{0}\bar{g}^{2}(\mu)]^{-\frac{b_1}{2b_0^2}}\,e^{-\frac{1}{2 b_{0}\bar{g}^{2}(\mu)}}
  \exp\bigg\{ \!-\! \int_{0}^{\bar{g}(\mu)}\!\! {\rm d}g \left[ \frac{1}{\beta(g)}+\frac{1}{b_0 g^3}
  -\frac{b_1}{b_0^2 g}  \right]
                          \bigg\}\equiv \varphi_g(\bar{g})  \;,\\
\frac{M}{\bar{m}(\mu)} &=& [2b_{0}\bar{g}^{2}(\mu)]^{-\frac{d_0}{2b_0}}
\exp\bigg\{\!-\! \int_{0}^{\bar{g}(\mu)}\!\! {\rm d}g \left[ \frac{\tau(g)}{\beta(g)}-\frac{d_0}{b_0 g}  \right]
                          \bigg\} \equiv \varphi_m(\bar{g})\,,
\eea
where $b_{0}$, $b_1=(102 - 38{\rm N_{f}}/3)(4\pi)^{-4} $ and $d_0$ are universal coefficients.
Taking their ratio one obtains
\beq
r \equiv \frac{M}{\Lambda} = \frac{\bar{m}(\mu)}{\mu}\times 
 \frac{\varphi_m(\bar{g}(\mu))}{\varphi_g(\bar{g}(\mu))}.
\eeq
For a given ${\bar{m}(\mu)}/{\mu}$, one parameterizes the renormalized
coupling $\bar{g}^{2}(\mu)$ through $r$. For instance, with the scale invariant mass $\mu=m_*$ and $\bar{g}(m_*)=g_*$, one writes the functional dependence
\beq\nonumber
        m_* = M \cdot\nu(r)\; , \quad
        \nu(r) = 1/\varphi_m(g_*)\,.
\eeq
As already mentionned, we evaluate $\nu$ at 4-loop order in the $\overline{\rm MS}$ scheme for ${\rm N_{f}}=2$
flavours and obtain to a very good approximation $\nu(r)=0.6400-0.0043\cdot(r-21)$ 
close to $r=21$. To estimate $m^{\overline{\rm MS}}_{b}(2\,{\rm GeV})$, the function $\nu'$ reads \cite{Gerardinmb}
\beq
\nu'(r)=1.1207 - 0.0900 y + 0.0048 y^{2}, \quad y=\mu/\Lambda_{\overline{\rm MS}}
\in [5.5, 7.0]\,.
\eeq
That procedure is helpful to propagate in a consistant way the errors until the end of the calculation. For example, to get the error on $\bar{m}(\bar{m})$, we write
\beq
r=\frac{L_{1}/2 M}{L_{1}/2 \Lambda_{\overline{\rm MS}}}=\frac{L_{1} \bar{m}_{\rm SF}(L_{1}/2)}{2} \frac{h(L_{1}/2)}{k(L_{1}/2)}, \quad k(L_{1}/2)=\frac{L_{1}}{2} \Lambda_{\rm SF} \left[\frac{\Lambda_{\overline{\rm MS}}}{\Lambda_{\rm SF}}\right], \quad 
\Lambda_{\overline{\rm MS}}/\Lambda_{\rm SF}=2.382035(3).
\eeq
Errors on $h$, $k$ and their correlation are known from \cite{DellaMorteKG}; having an analytical expression of $\nu(r)$, it is then straightforward to propagate the error on $r$, using the derivative of that function. The uncertainty from the perturbative running can safely be neglected because the numbers do not change at the permille level if the recently determined 5-loop term of the quark mass anomalous dimension \cite{Chetyrkin5loop} is included. \\
We obtain finally
\beq
m^{\overline{\rm MS}}_{b}(m^{\overline{\rm MS}}_{b}) =  4.21(11)\, {\rm GeV}\,,\quad
m^{\overline{\rm MS}}_{b}(2\,{\rm GeV}) =  4.88(15)\, {\rm GeV}\,.
\eeq
Restricting ourselves to the static order of HQET, we get
\beq
z_{b}^{\rm stat} = 13.24(21)(13)_z, \quad M_{b}^{\rm stat} = 6.57(17)\,  {\rm GeV},
[m^{\overline{\rm MS}}_{b}(m^{\overline{\rm MS}}_{b})]^{\rm stat}  = 4.21(11)\, {\rm GeV}.
\eeq
The very small difference observed between $z_{b}$ and $z^{\rm stat}_{b}$ makes us confident that ${\cal O}(1/m^{2}_{b})$ corrections are negligible for the observable of interest with respect to the accuracy we could reach in our work. Indeed, $z^{(1/m)}_{b}\equiv z_{b} - z^{\rm stat}_{b} = -0.0008(51)$. Let's add a couple of words concerning the relative error budget on $z_{b}$:\\
-- 65\% comes from the HQET parameters,\\
-- 20\% comes from $Z_{A}$, that helps to fix the scale through $f_{K}$,\\
-- 15\% comes from the HQET hadronic matrix elements.\\
\begin{figure}[t]
  \centering
  \includegraphics[width=\textwidth]{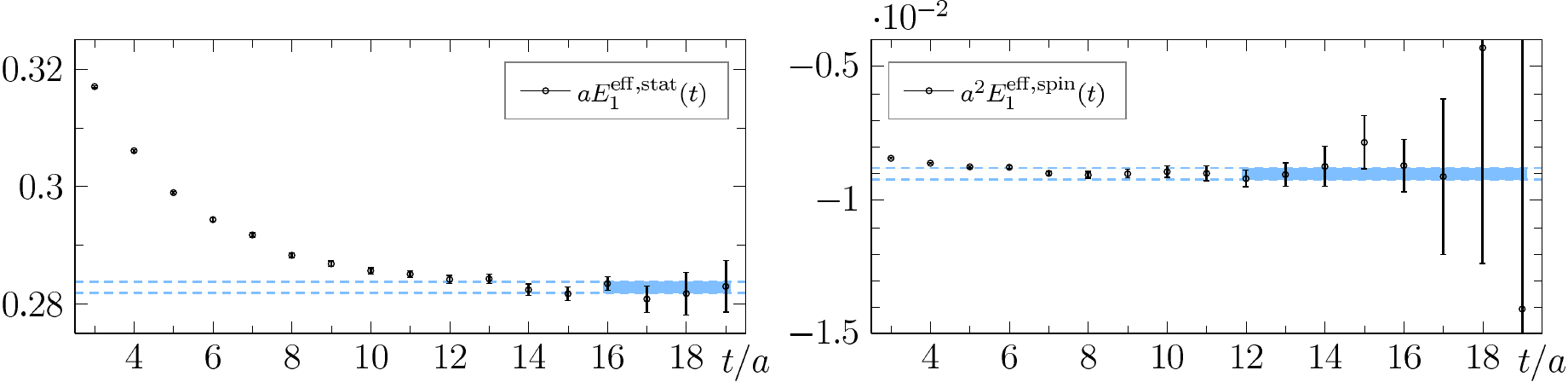}
  \caption{Typical plateaux for $E^{\rm stat}_{1}$ (left panel) and $E^{\rm spin}_{1}$ 
  (right panel); the CLS ensemble shown 
           here is N6 ($a=0.048$\,fm, $m_{\pi}=340$\,MeV).
  \label{fig:plateauxEstatEspin}}
\end{figure}
\begin{figure}[t]
   \centering
   \includegraphics[height=0.4\textwidth]{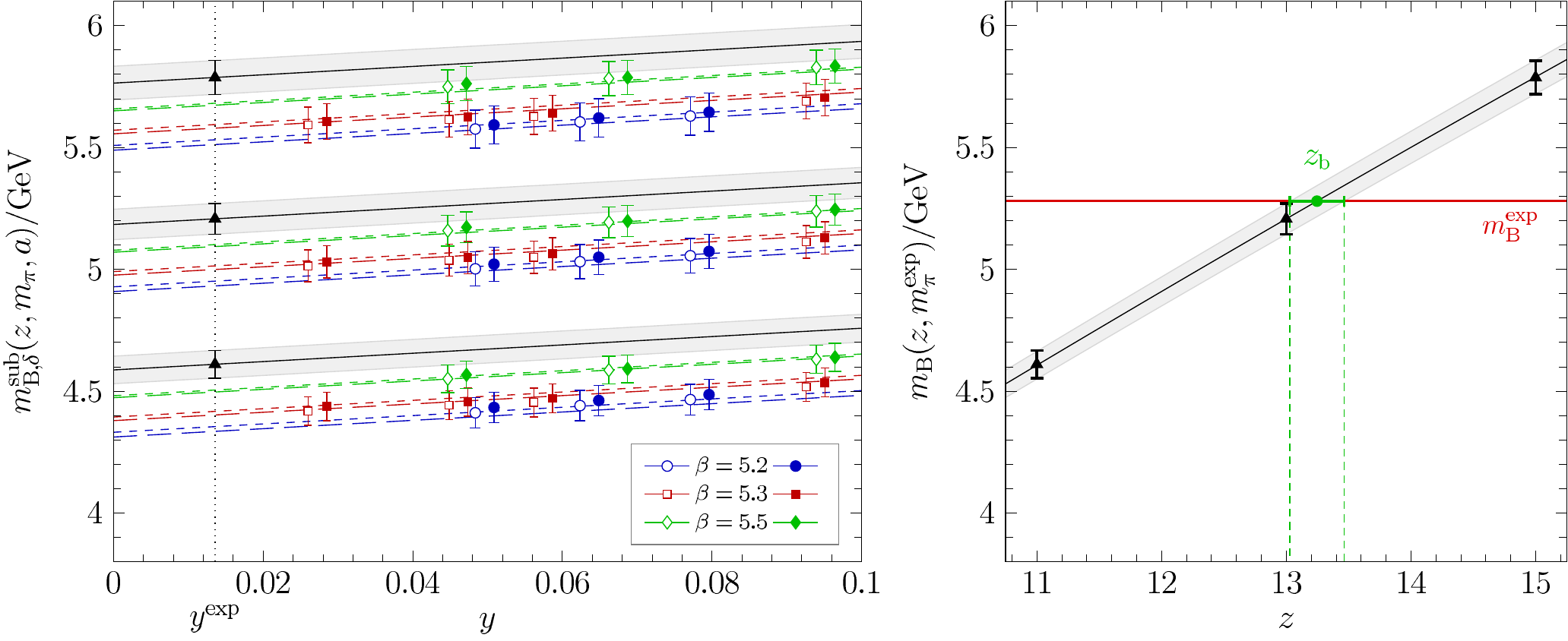}
   \caption{Chiral and continuum extrapolation of 
 $m_{B}^{\rm sub}\left(z,y,a\right)$ for the $z$ used in the 
 determination of $z_{b}$. Open/filled symbols refer to HYP1/HYP2 data 
points as do long/short dashed curves, respectively (left panel). 
 Interpolation to $z_{b}$ by imposing $m_{B}(z_{b}, m^{\rm exp}_{\pi})
 =m^{\rm exp}_{B}$ (right panel).\label{fig:chiralfitmBnf2}}
\end{figure}

\section{Back to phenomenology}

We have reported about the determination of $m_{b}$ at ${\rm N_{f}}=0$, although we have rather performed an update of a work previously done by the ALPHA Collaboration \cite{DellaMorteCB} and taken advantage of the opportunity to explore ideas concerning GEVP analysis, and at ${\rm N_{f}}=2$. We gather our results in Table \ref{tab:resultsmb}, with the PDG value estimated with ${\rm N_{f}}=5$ active flavours \cite{BeringerZZ}. First, our result $m^{\overline{\rm MS}}_{b}(2\,{\rm GeV})\big|_{{\rm N_{f}}=2} = 4.88(15)$  GeV agrees well with the recent ${\rm N_{f}}=2$ result of ETMC \cite{CarrascoZTA}, who indicates a
similar error but uses a pretty different approach to deal with the heavy quark on the lattice, as we briefly described in the first section of this chapter. As for other quark masses, we notice a little
dependence of $m^{\overline{\rm MS}}_{b}(\mu)$ on the number of flavours for ${\rm N_{f}}=0$, $2$, and $5$ and for
typical values of $\mu$ between $m^{\overline{\rm MS}}_{b}$ itself and 2 GeV.  In particular, at the lower scale 
of 2 GeV, where the apparent convergence
of perturbation theory is still quite good, it is difficult to observe any strong dependence on ${\rm N_{f}}$. 
It might be understood by the fact that the effective theories are matched to the real world data at low energies, through $m^{\rm exp}_{B}$ for the $b$-quark mass, $f_{K}$ or $f_{\pi}$ for the lattice 
spacing and $m_{\pi}$ for the $u/d$-quark mass. Having a look at the RGI mass $M_{b}$, it differs significantly between ${\rm N_{f}}=5$ and ${\rm N_{f}}=2$. The discrepancies in $M_{b}$ 
come probably from the ${\rm N_{f}}$ dependence of
both the renormalization group functions and the $\Lambda$ parameters. There is a reinforcement of these two effects between ${\rm N_{f}}=5$ and ${\rm N_{f}}=2$ and a partial compensation between
${\rm N_{f}}=2$ and ${\rm N_{f}}=0$.\\
Let's conclude that part of our report with the 2 following points:\\
-- it is reliable to use the $b$-quark mass at scales around $\mu=2$ GeV when
one attempts to make predictions from theories with a smaller number of
flavours than the physical 5-flavour theory;\\
-- at the present level of accuracy the $b$-quark
mass is appropriately determined from the different approaches; the error budget of our 
computation, as well as the one discussed by other groups, is such that in a future work with ${\rm N_{f}}=2+1$ or $2+1+1$, a competitive number can be obtained, as far as Higgs physics and, in particular, the $h \to b \bar{b}$ decay, are concerned.
\begin{table}[t]
\begin{center}
\begin{tabular}{|c|c|c|c|c|c|}
\hline
    ${\rm N_{f}}$ & $M_{b}$ & $m^{\overline{\rm MS}}_{b}(m^{\overline{\rm MS}}_{b})$ 
    & $m^{\overline{\rm MS}}_{b}(4\,{\rm GeV})$  & $m^{\overline{\rm MS}}_{b}(2\,{\rm GeV})$ & $\Lambda_{\overline{\rm MS}}$ [MeV]\\
\hline
 0  & 6.76(9)  & 4.35(5)  & 4.39(6)  & 4.87(8)  & 0.238(19)\\
 2  & 6.57(17) & 4.21(11) & 4.25(12) & 4.88(15) & 0.310(20) \\
 5 & 7.50(8)  & 4.18(3)  & 4.22(4)  & 4.91(5)  & 0.212(8)\\
\hline
 \end{tabular}
\end{center}
  \caption{\label{tab:resultsmb}
  $b$-quark mass, in GeV, in theories with different 
  quark flavour numbers ${\rm N_{f}}$, with the corresponding $\Lambda_{\overline{\rm MS}}$, 
  and for different schemes and scales, and the  RGI mass $M_{b}$.
           The PDG value of the $b$-quark mass, given in the last row of the table (with ${\rm N_f=5}$) is dominated by 
           \cite{ChetyrkinFV}, \cite{McNeileJI}. The values of $\Lambda_{\overline{\rm MS}}$ we quote are taken at ${\rm N_{f}}=0$ from \cite{CapitaniMQ}, ${\rm N_{f}}=2$ from \cite{FritzschWQ} and ${\rm N_{f}}=5$ from \cite{BeringerZZ}.}
\end{table}



\chapter{The strong coupling constant}

\fancyhead[LO]{\bfseries \leftmark}
\fancyhead[RE]{\bfseries \rightmark}

In the Standard Model, there are mainly four processes to produce the Higgs boson, that we sketched in Figure \ref{fig:prodhiggs}: Higgs strahlung, Higgs production from a vector bosons fusion, gluon-gluon fusion and gluon-gluon fusion that produces, in addition to the Higgs, a $q-\bar{q}$ pair.
\begin{figure}[t]
%
%
%
%
%
\begin{center}
\includegraphics*[width=0.8\textwidth]{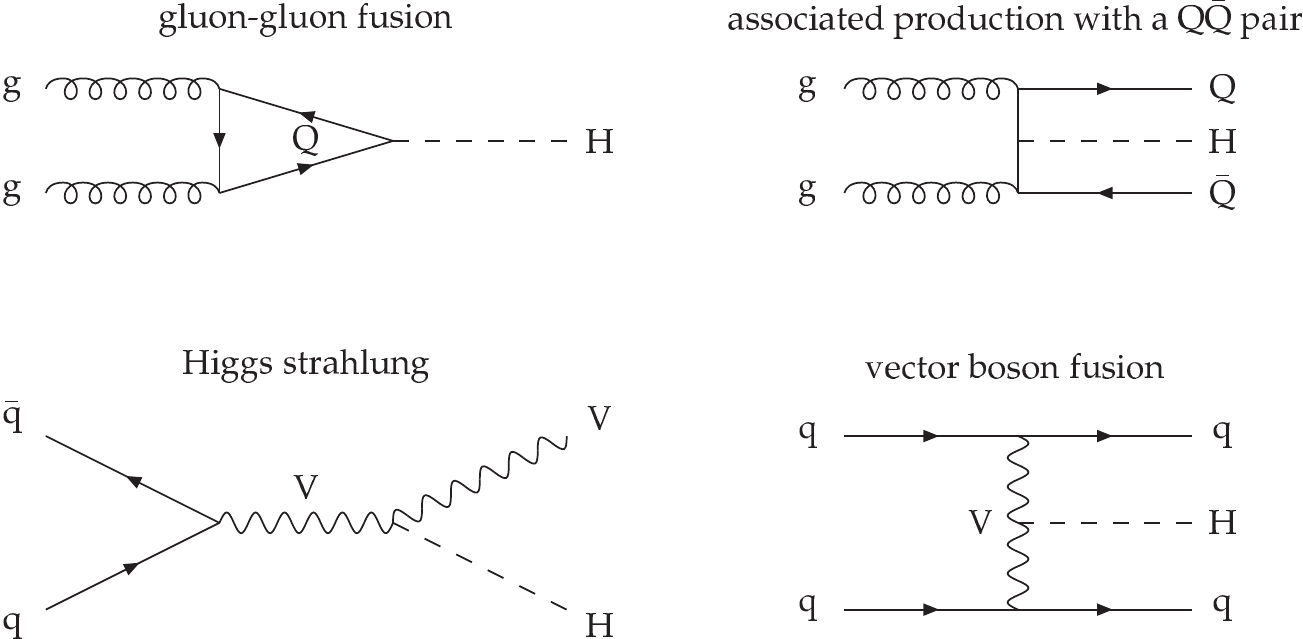}
\end{center}
\caption{Production processes of the Higgs boson. \label{fig:prodhiggs}}
\end{figure}
They are favoured thanks to a strong coupling of the $H$ boson to heavy particles ($W$ and $Z$ bosons, $t$ and $b$ quarks). Among them, the by far dominating one is the gluon-gluon fusion, as we have shown in Figure \ref{fig:TEV4LHC} \cite{TEV4LHC}. At NNLO of QCD, the cross section of $pp \to H + X$ reads
\beq\nonumber
\sigma_{pp \to H + X} = \int_{x_H}^1 \sum_{i,j=g,q,\bar{q}}{\cal F}_{ij} (x,\mu_F,\mu_R) \otimes \hat{\sigma}_{ij}(x,\mu_F,m_Q)\,,
\eeq
where $x_H=m^2_H/s$, $s$ is the center of mass energy of the $pp$ collider, ${\cal F}_{ij}$ are parton distribution functions (PDF's) and $\hat{\sigma}_{ij}$ are Higgs cross sections at the partonic level. 2 scales are introduced: a renormalization scale $\mu_R$, at which the strong coupling constant $\alpha_s$ is defined, and a factorisation scale $\mu_F$ used to separate the partonic contribution and the part that encodes the long-distance physics of QCD. With respect to the LO estimates \cite{WilczekZN} - \cite{GeorgiGS}
\beq\nonumber
\sigma^{LO}_{pp \to H}=\sigma^H_0(\mu_R) x_H \int_{x_H}^1 \frac{dx}{x} g(x,\mu_F)g(x_H/x,\mu_F),
\eeq
NLO QCD corrections induce an increase of $\sim$ 80\% of the cross section \cite{DjouadiTKA} - \cite{SpiraRR}, while including also NNLO QCD corrections (only in an effective field theory approach with $m_t \to \infty$) a further 25\% of enhancement is observed\cite{HarlanderWH} - \cite{RavindranUM}\footnote{Those numbers correspond to $\sqrt{s}=7$ TeV; the convergence is expected to be better at $\sqrt{s}=14$ TeV: all in all, a factor 2 of enhancement between LO and NNLO is expected \cite{BaglioAE}.}. A few percent corrections come from electroweak radiative loops \cite{DjouadiGE} - \cite{ActisTS}. In \cite{BaglioAE} the different sources of theoretical error were discussed:\\
-- changing the factorisation and renormalization scales from $\mu_F=\mu_R=\mu_0$ in a range $[\mu_0/\kappa,\, \kappa \mu_0]$ with $\kappa=2,\,3,\,4,\,\cdots$, helps to evaluate the truncation error; with $\kappa=2$, $m_H=125$ GeV, $\mu_0=m_H/2$, a variation of $\pm$ 10\% was obtained\\
-- comparing the known result of $\sigma^{\rm NLO}(m_t)$ and the result obtained in the infinite mass limit is useful to state the impact of a finite mass $m_b$ in NNLO QCD loops on $\sigma^{\rm NNLO}$: the effect is $\pm$ 3\%; a further $\pm$ 1\% has to be added from the scheme dependence of $m_b$\\
-- 3\% of uncertainty come from unknown NNLO electroweak corrections, set in a conservative way to the NLO number\\
-- the uncertainty on the parameterization of the PDF's, provided by different groups \cite{MartinIQ} - \cite{HERAPDF}, induces $\sim$ 10\% errors on the gluon-gluon fusion cross section\\
-- finally there is also the uncertainty coming from $\alpha_s$ itself: letting $\alpha_s(m_Z)$ vary in the range $[0.107, 0.127]$ with a central value of 0.1171 obtained by analysing deep-inelastic scattering date \cite{MartinBU}, an error of 2-3\% is propagated to $\sigma^{\rm NNLO}_{pp \to H + X}$.\\
The situation does not change with $\sqrt{s}=8$ and $14$ TeV. It is still relevant to study whether theory can help to reduce the uncertainty on $\alpha_s$.
\begin{figure}[t]
\begin{center}
\includegraphics*[angle=90,width=0.5\textwidth]{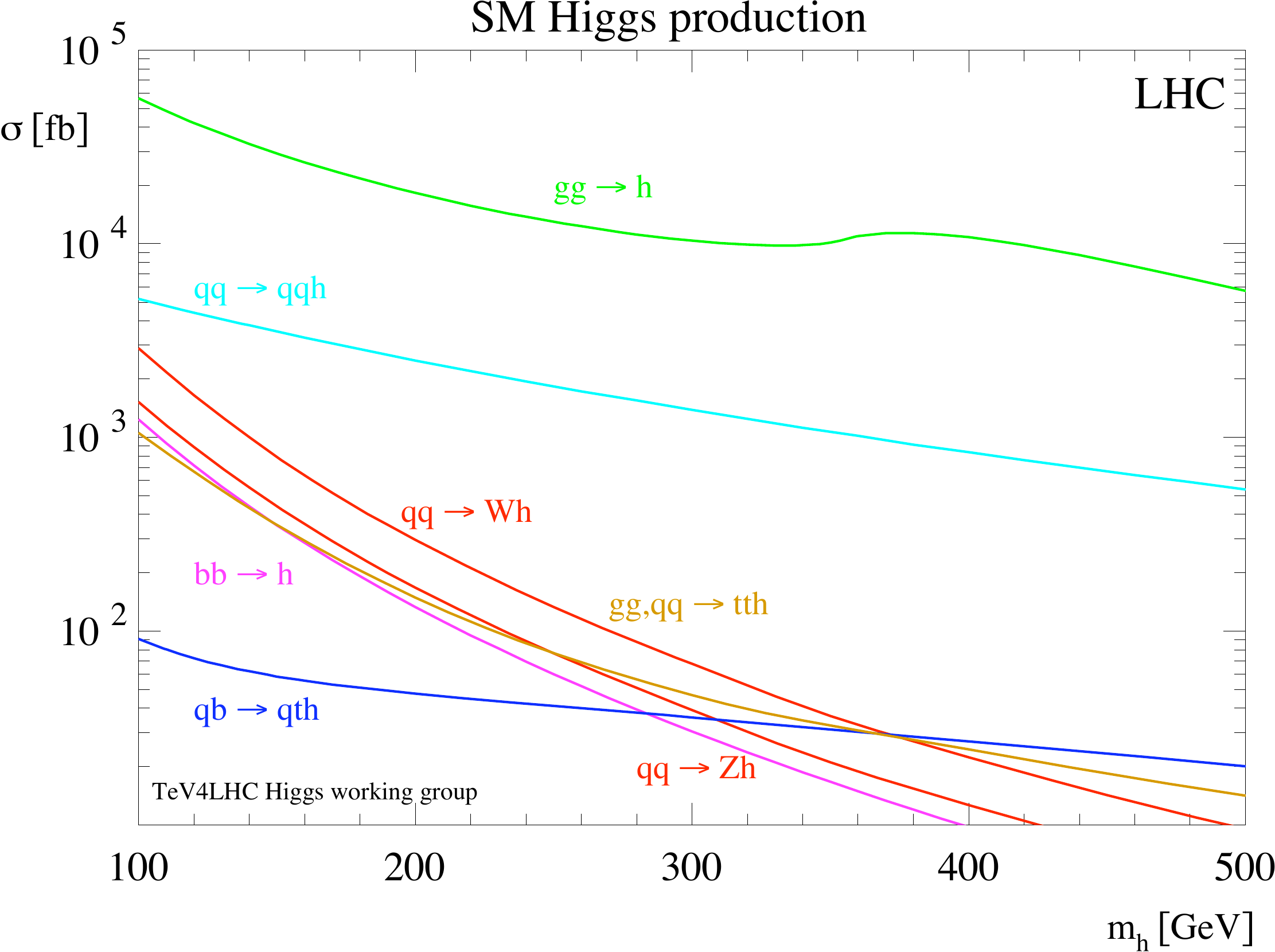}
\end{center}
\caption{Cross section of the Higgs production in function of the Higgs mass; the plot is taken from \cite{TEV4LHC}. \label{fig:TEV4LHC}}
\end{figure}

\section{Phenomenological and experimental determinations of $\alpha_{s}$}

Extracting $\alpha_s$ from deep inelastic scattering processes, schematically depicted on the left panel of Figure \ref{fig:alphaDISES}, is very popular. One compares QCD predictions, made the most often at NNLO, to experimental data in different $Q^2$ (the transferred 4-momentum from the scatter to the final state) and $x$ (the fraction of the longitudinal momentum brought by the parton) regions. Those include structure functions \cite{BlumleinBE}, \cite{JimenezDelgadoTV}, \cite{AlekhinNI} and jet productions at HERA \cite{GlasmanSM}, Tevatron \cite{MartinBU}, \cite{AbazovNC} - \cite{AbazovLUA} and LHC \cite{MalaescuTS} - \cite{ChatrchyanHAA}. A second possibility is the analysis of event shapes, that are depicted on the right panel of Figure \ref{fig:alphaDISES}: $e^+e^-$ annihilation into hadron states with 2 jets, 3 jets, and so on, are matched to QCD computations at NNLO, modelising the hadronisation effects by Monte Carlo methods \cite{DissertoriIK} - \cite{SchieckMP}. Event shapes can also be characterized by the Thrust $\hat{\tau}={\rm max}_{\vec{n}_\tau} \frac{\sum_i |\vec{p}_i \cdot \vec{n}_\tau|}{\sum_i |\vec{p}_i|}$, where $\vec{p}_i$ are the momenta of the hadrons in the final state (they can be those of the jets themselves): a back-to-back $q \bar{q}$ pair will have a Thrust equal to 1, while a process with an isotropic emission of particles will have a Thrust equal to 1/2. Again, QCD predictions at NNLO are used to fit data on Thrust distributions and give an estimate of $\alpha_s$ \cite{BecherCF} - \cite{GehrmannSC}. However, the systematics in the way of incorporating non-perturbative effects in Monte Carlo or in models to describe analytically the Thrust is quite large. Electroweak precision fits, in particular those of the hadronic $Z$ decay, has also been used \cite{AbazovHM}: however a bias exists in that determination of $\alpha_s$ because it assumes the strict validity of the Standard Model in the Higgs sector. QCD analyses at NLO of the heavy quarkonia decay is also a way to obtain $\alpha_s$ \cite{BrambillaCZ}. A further very important approach of measuring the strong coupling constant is the phenomenological analysis of $\tau$ decay, that we sketched in Figure \ref{fig:taudecay}.
\begin{figure}[t]
\begin{center}
\includegraphics*[width=0.8\textwidth]{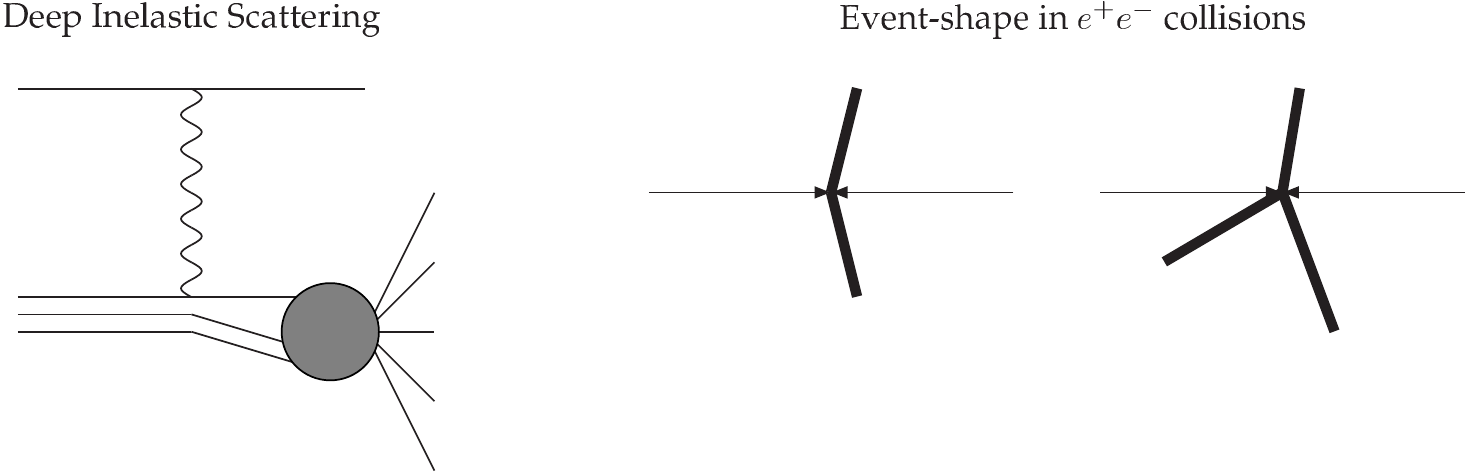}
\end{center}
\caption{Schematic view of processes analysed to extract $\alpha_s$: Deep Inelastic Scattering (left) and event shapes (right). \label{fig:alphaDISES}}
\end{figure}
\begin{figure}[t]
\begin{center}
\includegraphics*[width=0.6\textwidth]{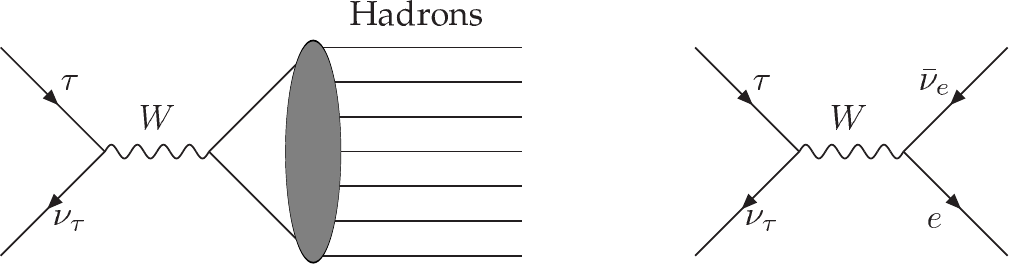}
\end{center}
\caption{Measurement of $\alpha_s$ from $\tau \to$ hadrons decay. \label{fig:taudecay}}
\end{figure}
The decay width, restricted to the $X_{u,d,s}$ final state and normalised by the leptonic process, is given by
\bea\nonumber
R_{\tau}&\equiv&\Gamma[\tau^{-} \to \nu_{\tau}\,\mbox{hadrons}]/\Gamma[\tau^{-} \to
\nu_{\tau} e^{-} \bar{\nu}_{e}]\\
&=&N_{c} |V_{ud}|^{2} S_{\rm EW} (1+\delta_{P} + \delta_{NP}), 
\eea
where $S_{\rm EW}=1.0201(3)$ stands for electroweak corrections, $\delta_{NP}=-0.0059(14)$ \cite{DavierXQ}, \cite{DavierSK} are non perturbative corrections extracted from the invariant mass distribution of the hadronic final state \cite{LeDiberderTE}, \cite{LeDiberderFR} and $\delta_P$ are the dominant QCD corrections \cite{BraatenHC} - \cite{BraatenQM}:
\bea\nonumber
\delta_{P}&=&\sum_{n} K_{n} A^{(n)}(\alpha_{s}) = \sum_{n}(K_{n} + g_{n}) \alpha^{n}(m_{\tau}),\\
\nonumber
A^{(n)} &=& \frac{1}{2\pi i} \oint_{|s|=m^2_\tau} \frac{ds}{s} 
\left(\frac{\alpha_s(-s)}{\pi}\right)^n\left(1 -2\frac{s}{m^2_\tau} 
+ 2 \frac{s^3}{m^6_\tau} - \frac{s^4}{m^8_\tau}\right)\\ 
\nonumber
&=& \alpha^n(m_\tau) + {\cal O}(\alpha^{n+1}(m_\tau)).
\eea
$K_n$ are the coefficients of the Adler function known up to ${\cal O}(\alpha^5_s)$ \cite{BaikovJH} and $A^{(n)}$ depend only on $a_\tau\equiv \alpha_s(m^2_\tau)/\pi$. Expanding the integrals in $a_\tau$ is called Fixed Order Perturbation Theory (FOPT) \cite{BaikovJH} - \cite{BoitoCR} while keeping 
the expression of the integrands as they are and applying a running of $\alpha_s(-s)$ along the integration contour, called Contour-Improved Perturbation Theory (CIPT) \cite{DavierSK}, \cite{BaikovJH}, \cite{BoitoCR} - \cite{NarisonVY} is considered in the literature as giving a better convergence of the results, though strong debates animate experts of the subject \cite{PichSQA}. In Figure \ref{fig:collectalphaspheno} we have collected the various phenomenological determinations of $\alpha_s$ referred by PDG and the corresponding pre-averagings.
\begin{figure}[t]
\begin{center}
\begin{tabular}{ccc}
\includegraphics*[width=0.3\textwidth]{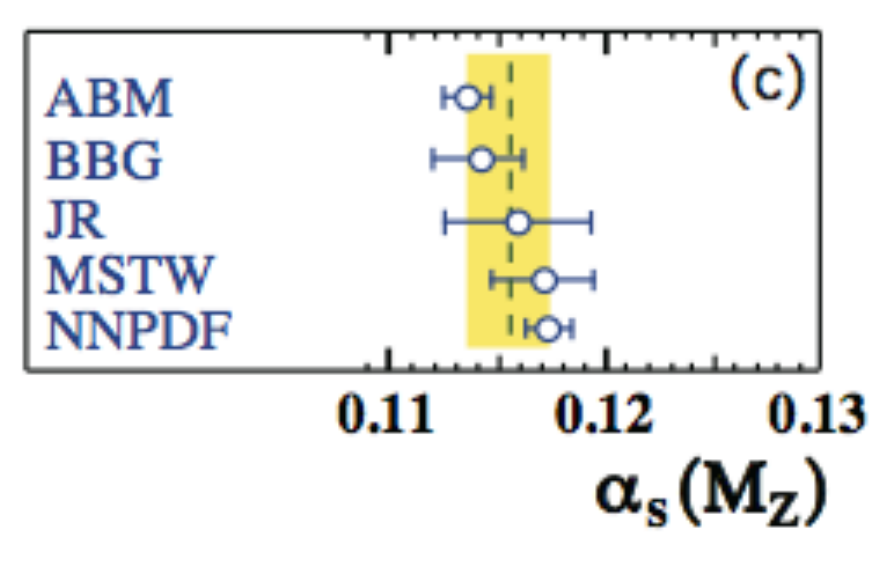}
&
\includegraphics*[width=0.3\textwidth]{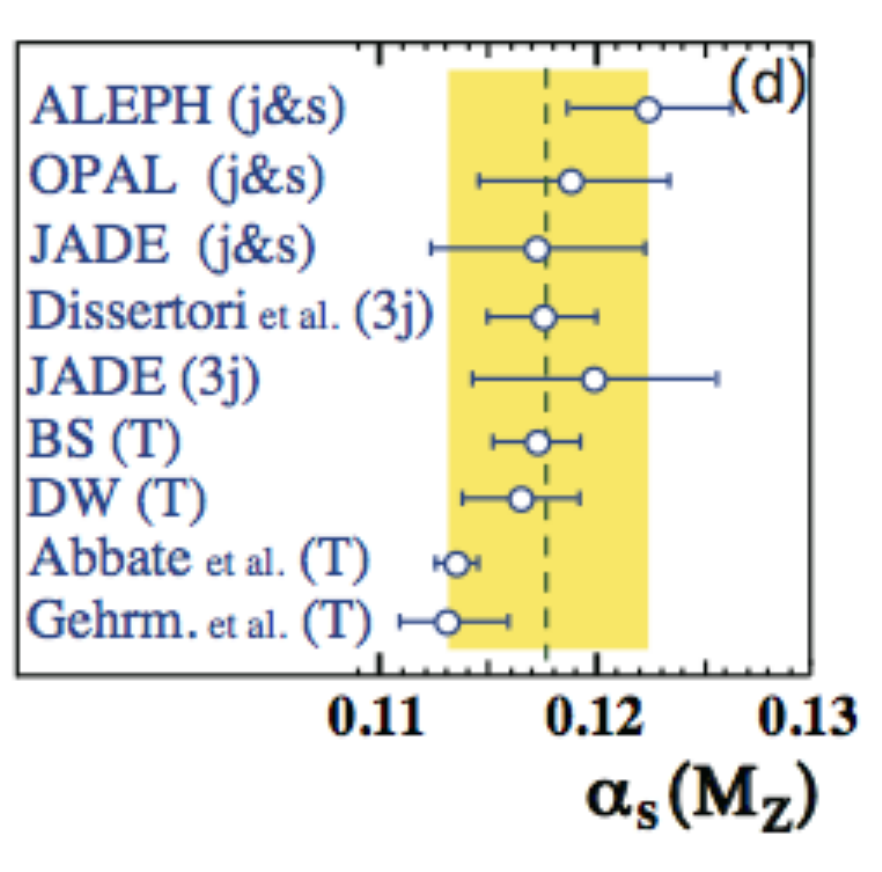}
&
\includegraphics*[width=0.3\textwidth]{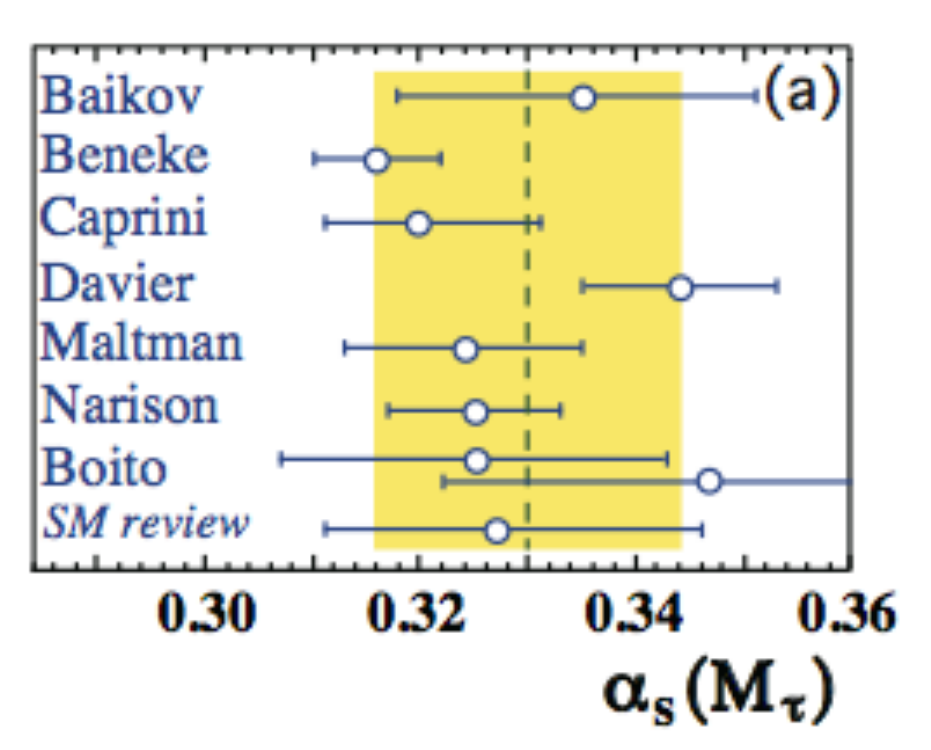}
\end{tabular}
\end{center}
\caption{Phenomenological estimates of $\alpha_s$ included in the PDG average. \label{fig:collectalphaspheno}}
\end{figure}

\section{A ghostly estimate of $\alpha_s$}

The last series of $\alpha_s$ measurements is based on lattice methods. Hadronic schemes are fruitful approaches: a first one, reported for instance in \cite{DaviesSW}, \cite{MaltmanBX} consists in computing Wilson loops ${\cal W}_{m\,n}$ of size $m \times n$ after the tuning of $u/d$, $s$ and $c$ quark masses from $m_\pi$, $2m^2_K - m^2_\pi$ and $m_{\eta_c}$, and extracting the lattice spacing from the $\Upsilon$ spectrum. One develops the Wilson loop at short distance $r\equiv a/d$ in perturbation theory: ${\cal W}_{m\,n}=\sum_i c_i \alpha^i_V(d/a)$, leading to define the $\alpha_s$ coupling in the "potential scheme" \cite{LepageXA}:
\beq
V(q)\equiv \frac{C_F 4 \pi \alpha_V(q)}{q^2};
\eeq
it corresponds to the one-gluon exchange part of the potential. The convergence of the perturbative expansion is faster for $\ln ({\cal W}_{m\,n})$, the tadpole improved term $\ln({\cal W}'_{mn})\equiv \ln [{\cal W}_{mn}/{\cal W}^{(m+n)/2}_{1\,1}]$ and, finally, the Creutz ratios $\ln \left(\frac{{\cal W}_{m\,n+1}}{{\cal W}_{m\,n}} \frac{{\cal W}_{m-1\,n}}{{\cal W}_{m-1\,n+1}}\right)$. The comparison of lattice data with perturbative formulae is possible only after subtracting a gluon-condensate term $-\frac{\pi^2}{36} {\cal A}^2 \langle \alpha_s G^2 / \pi \rangle$, where ${\cal A}$ is the planar area surrounded by the loop, as illustrated in Figure \ref{fig:alphapot}. A running is applied up to the scale of 7.5 GeV and the result is converted to the $\overline{\rm MS}$ scheme and run up to $m_Z$.\\
A second hadronic scheme, exploited by the same group \cite{AllisonXK}, \cite{McNeileJI} as \cite{DaviesSW}, with the same procedure to tune the quark masses and to get the lattice spacing, takes as inputs the moments of an $\eta_c$ correlation function $G(t)=-a^6 \sum_{\vec{x}} (am^{(0)}_h)^2 
\langle [\bar{\psi}_c \gamma^5 \psi_c](\vec{x},t) [\bar{\psi}_c \gamma^5 \psi_c](\vec{0},0)\rangle$, $G_n=\sum_t (t/a)^n G(t)$ and its tree level counterpart $G^{(0)}_n$. Performing the ratios
\beq\nonumber
R_4\equiv \frac{G_4}{G^{(0)}_4}, \quad R_{n\geq 6} \equiv \frac{a m_{\eta_h}}{2am^{(0)}_h} \left(\frac{G_n}{G^{(0)}_n}\right)^{1/(n-4)},
\eeq
one writes their expression in (continuum) perturbation theory \cite{ChetyrkinII} - \cite{MaierYN}, \cite{BoughezalPX}: $R_4 \equiv r_4(\alpha^{\overline{MS}}, \mu/m_h)$, 
$R_{n \geq 6} \equiv \frac{r_n(\alpha^{\overline{MS}}, \mu/m_h)}{2m_h(\mu)/m_{\eta_h}}$. Incorporating a power correction
$[1 + d_n \langle \alpha_s G^2 / \pi \rangle/(2m_h)^4]$ \cite{NovikovDQ}, \cite{BroadhurstQJ}, one matches them to the lattice data, once those are extrapolated to the continuum limit, in order to obtain $\alpha_s^{\overline{\rm MS}}(\mu=3\,{\rm GeV})$ before applying a running up to $m_Z$. We have shown the lattice outputs in Figure \ref{fig:alphamoments}.
\begin{figure}[t]
\begin{center}
\begin{tabular}{cc}
\includegraphics*[width=0.2\textwidth]{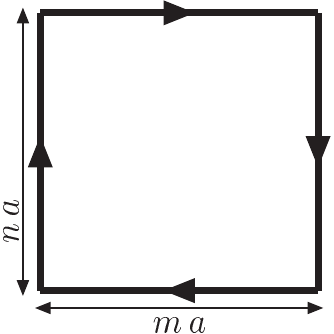}
&
\hspace{1cm}\includegraphics*[width=0.5\textwidth]{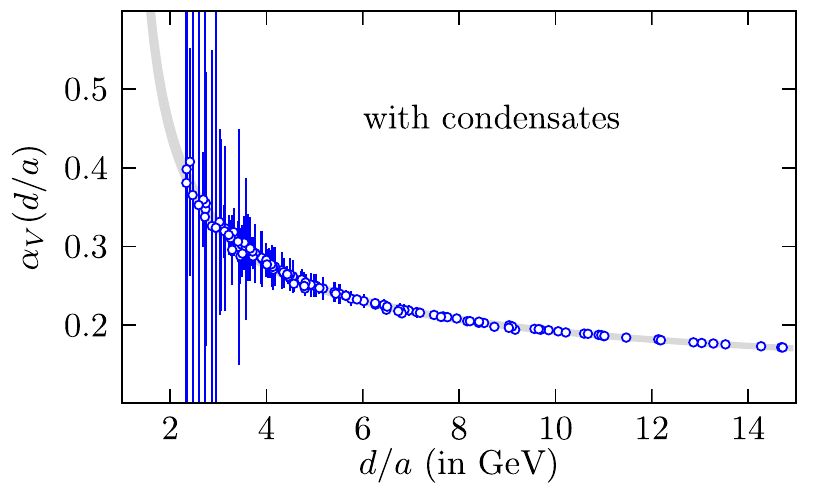}\\
\end{tabular}
\end{center}
\caption{Extraction of $\alpha_s$ from the Wilson loops ${\cal W}_{m\,n}$; the plot of the right panel is taken from \cite{DaviesSW}. \label{fig:alphapot}}
\end{figure}
\begin{figure}[t]
\begin{center}
\begin{tabular}{cc}
\includegraphics*[width=0.5\textwidth]{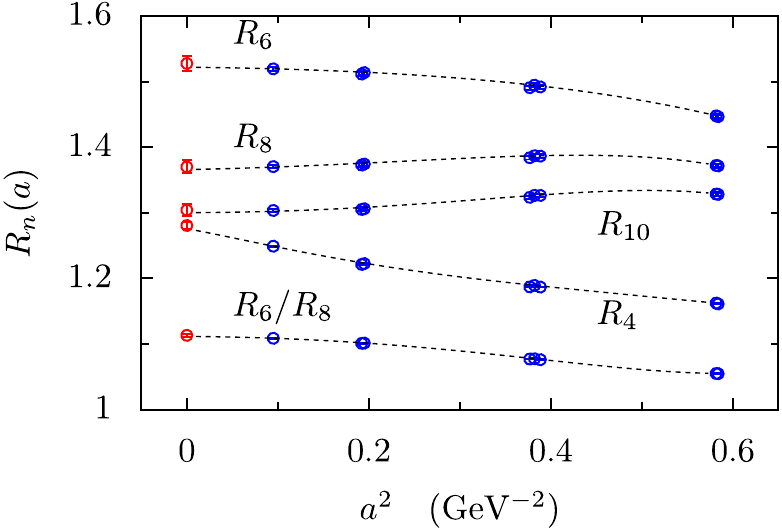}
&
\includegraphics*[width=0.3\textwidth]{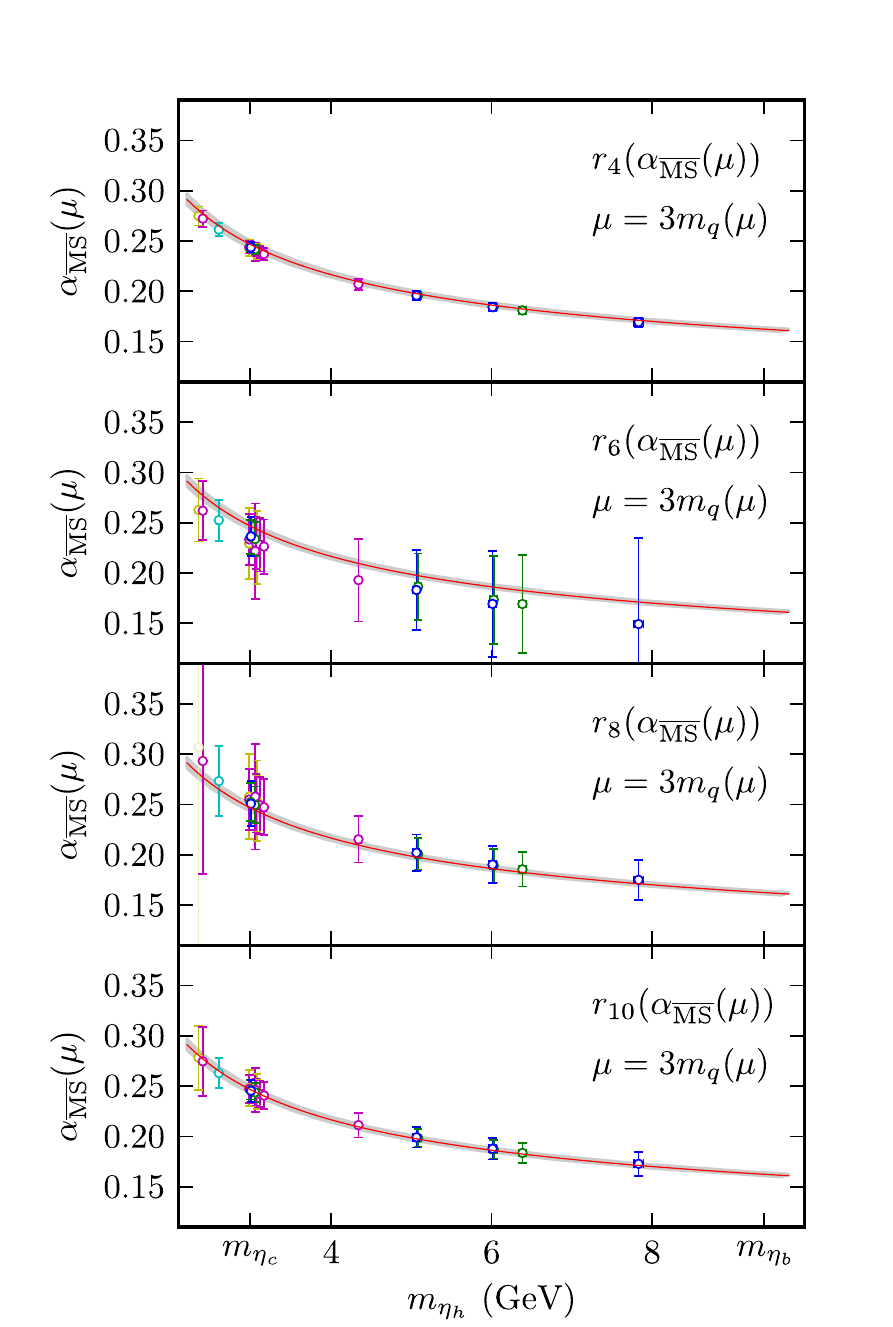}
\end{tabular}
\end{center}
\caption{Extraction of $\alpha_s$ from the moments of $\eta_c$ correlators; plots are taken from \cite{AllisonXK}. \label{fig:alphamoments}}
\end{figure}
\\
\noindent In the same spirit, studying the vacuum polarisation functions 
\beq\nonumber
\langle J_\mu J_\nu \rangle(Q^2)\equiv 
(\delta_{\mu\nu} Q^2 - Q_\mu Q_\nu) \Pi^{(1)}_J(Q^2) + Q_\mu Q_\nu \Pi^{(0)}_J(Q^2),
\eeq 
where $J_\mu$ are light bilinear of quarks, 
is in principle a good way to get $\alpha_s$. The OPE analysis of 
\beq\nonumber
\Pi_{V+A}(Q^2)=\Pi^{(0)}_V(Q^2) + \Pi^{(0)}_V(Q^2) 
+\Pi^{(1)}_A(Q^2) +\Pi^{(1)}_A(Q^2)
\eeq
reads:
\bea\nonumber
\Pi_{V+A}(Q^2, \alpha_s)&=&c+C_0(Q^2,\mu^2,\alpha_s)+C^{V+A}_m(Q^2,\mu^2,\alpha_s)\frac{\bar{m}(Q^2)}{Q^2}\\
\nonumber
&+& \sum_{q=u,d,s}^{V+A} C_{\bar{q}q}(Q^2,\alpha_s) \frac{\lgl m_q \bar{q}q \rgl}{Q^4} 
+ C_{GG}(Q^2,\alpha_s)\frac{\alpha_s / \pi G^2}{Q^4} + {\cal O}(1/Q^6).
\eea
The Wilson coefficients $C_0$ and $C^{V+A}_m$ are known at four-loop order of perturbation theory in $\overline{\rm MS}$ 
scheme \cite{SurguladzeTG}, \cite{GorishniiVF}, \cite{ChetyrkinCF} while $C^{V+A}_{\bar{q}q}$ and $C_{GG}$ are known at three-loop order \cite{ChetyrkinKN}. Lattice data of the Adler function $D(Q^2)=-Q^2 d \Pi(Q^2)/dQ^2$, where the divengent term $c$ is absent, have been analysed in \cite{ShintaniPH} at a single lattice spacing: we have shown in the right panel of Figure \ref{fig:Adler} the result. A preliminary work using CLS ensembles has recently been presented as well \cite{HorchLLA}, \cite{FrancisQKP}.
\begin{figure}[t]
\begin{center}
\begin{tabular}{cc}
\includegraphics*[width=0.2\textwidth]{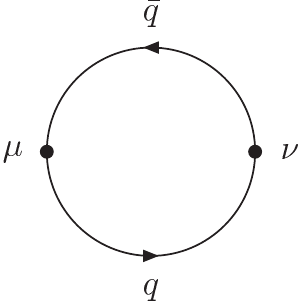}
&
\hspace{2cm}\includegraphics*[width=0.5\textwidth]{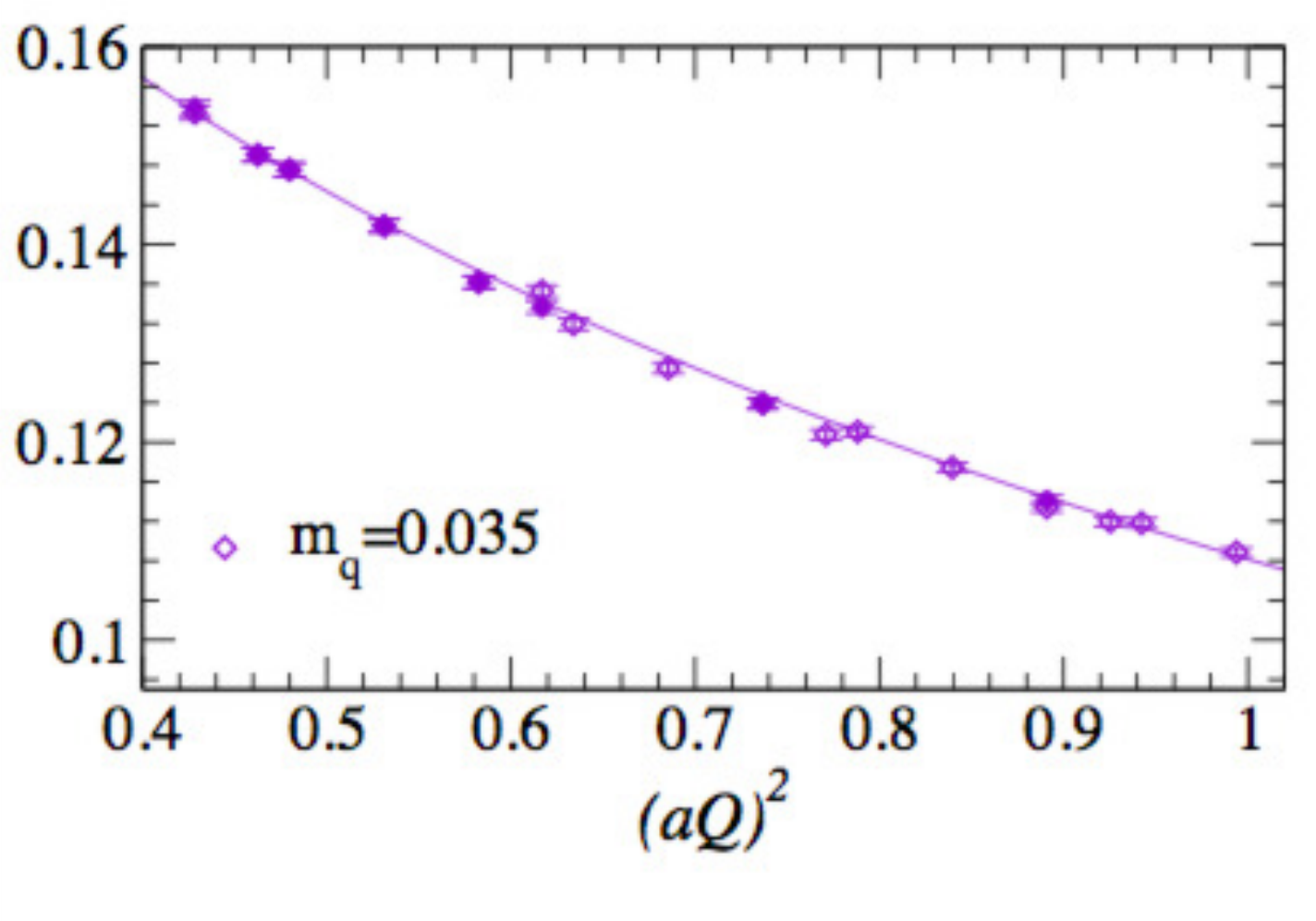}
\end{tabular}
\end{center}
\caption{Lattice data and fit curve of the Adler function $(m_{u/d}\sim m_s/2)$; the plot is taken from \cite{ShintaniPH}. \label{fig:Adler}}
\end{figure}
\\
We have already discussed in full glory the principle of measuring $\alpha_s$ in the Schr\"odinger Functional scheme, a finite volume scheme. 2 groups \cite{DellaMorteBC}, \cite{FritzschWQ}, \cite{AokiTF} have reported on their estimate of $\Lambda^{\overline{\rm MS}}_{\rm QCD}$ and $\alpha_s(m_Z)$ from ${\rm N_f}=2$ and ${\rm N_f}=2+1$ simulations, respectively. The lattice inputs to convert the $\Lambda$ parameter in physical units are $f_K$ (${\rm N_f}=2$) and $m_{\Omega}$ (${\rm N_f}=2+1$).\\
Fixed gauge approaches are also very elegant and offer a complementary input to other lattice calculations. They are MOM scheme kinds and involve Green functions of gluons and ghosts. First, from the gluon field 
\beq\nonumber
A_\mu(x+\hat{\mu}/2) = \left[\frac{U_\mu(x) - U^\dag_\mu(x)}{2iag_0}\right]_{\rm traceless},\quad
A_\mu(p) = \int e^{ipx} A_\mu(x),
\eeq 
one takes the gluon propagator 
\beq\nonumber
G^{(2)}_{\mu\nu}(p) = \lgl A_\mu(p) A_\nu(p) \rgl \equiv G(p^2) 
\left(\delta_{\mu\nu} - \frac{p_\mu p_\nu}{p^2}\right)
\eeq 
and \emph{the 3-gluon vertex} \cite{AllesKA}, \cite{BoucaudBQ}, as depicted
in the left panel of Figure \ref{fig:3gluons}:
\bea\nonumber
G^{(3)}_{\mu\nu\rho}(p_1, p_2, p_3) &=& \lgl A_\mu(p_1) A_\nu(p_2) A_\rho(p_3)\rgl \\
&\equiv &\Gamma_{\alpha\beta\gamma}(p_1,p_2,p_3) 
G^{(2)}_{\alpha\mu}(p_1)G^{(2)}_{\beta\nu}(p_2)G^{(2)}_{\gamma\rho}(p_3).
\eea
The $\widetilde{\rm MOM}$ renormalization scheme conditions read
\beq\nonumber
Z^{-1}_3(\mu) G(p)|_{p^2=\mu^2}=\frac{1}{\mu^2}, \quad \frac{\sum_{\alpha=1}^4 G^{(3)}_{\alpha \beta \alpha}(p,0,-p)}{G^2(p) G(0)}= 6 iZ^{-1}_1(p) g_0 p_\beta,
\eeq
and the renormalized coupling is defined by
\beq
g^{\widetilde{\rm MOM}}_{R}(\mu)= Z^{3/2}_3(\mu)Z^{-1}_{1}(\mu) g_0.
\eeq
Lattice data are fitted by 
\beq\nonumber
\alpha^{\rm Latt}_s(\mu^2) = \alpha_{s, {\rm pert}}(\mu^2) 
\left(1+\frac{c}{\mu^2}\right),
\eeq 
where the running $\alpha_{s, {\rm pert}}(\mu^2)$ is known up to 4 loops in the $\widetilde{\rm MOM}$ scheme \cite{ChetyrkinDQ} and the necessity to introduce the power correction $\left(1+\frac{c}{\mu^2}\right)$ in the analysis was pointed in \cite{BoucaudEY} and related to the presence of a non zero gluon condensate $\langle A^2 \rangle$ \cite{BoucaudND} - \cite{DeSotoQX}. This led to an estimate of $\alpha_s(m_Z)$ from an ${\rm N_f}=2$ simulation and we have shown in the right panel of Figure \ref{fig:3gluons} the behaviour of $\alpha^{\rm Latt}_s(\mu)$.
\begin{figure}[t]
\begin{center}
\begin{tabular}{cc}
%
\includegraphics*[width=0.4\textwidth]{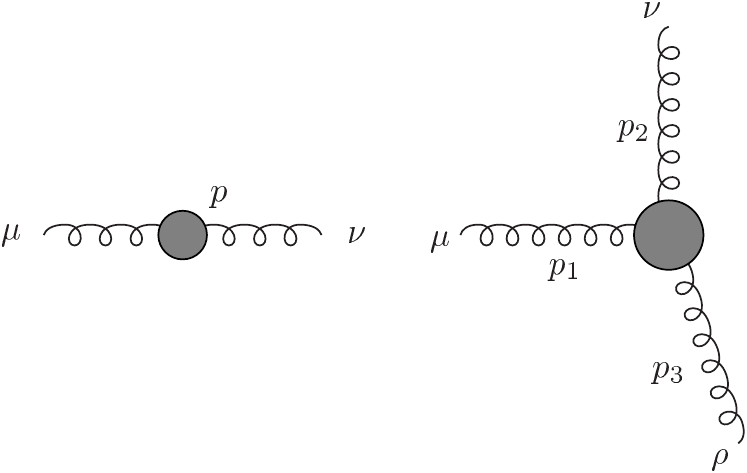}
&
\hspace{2cm}\includegraphics*[width=0.4\textwidth]{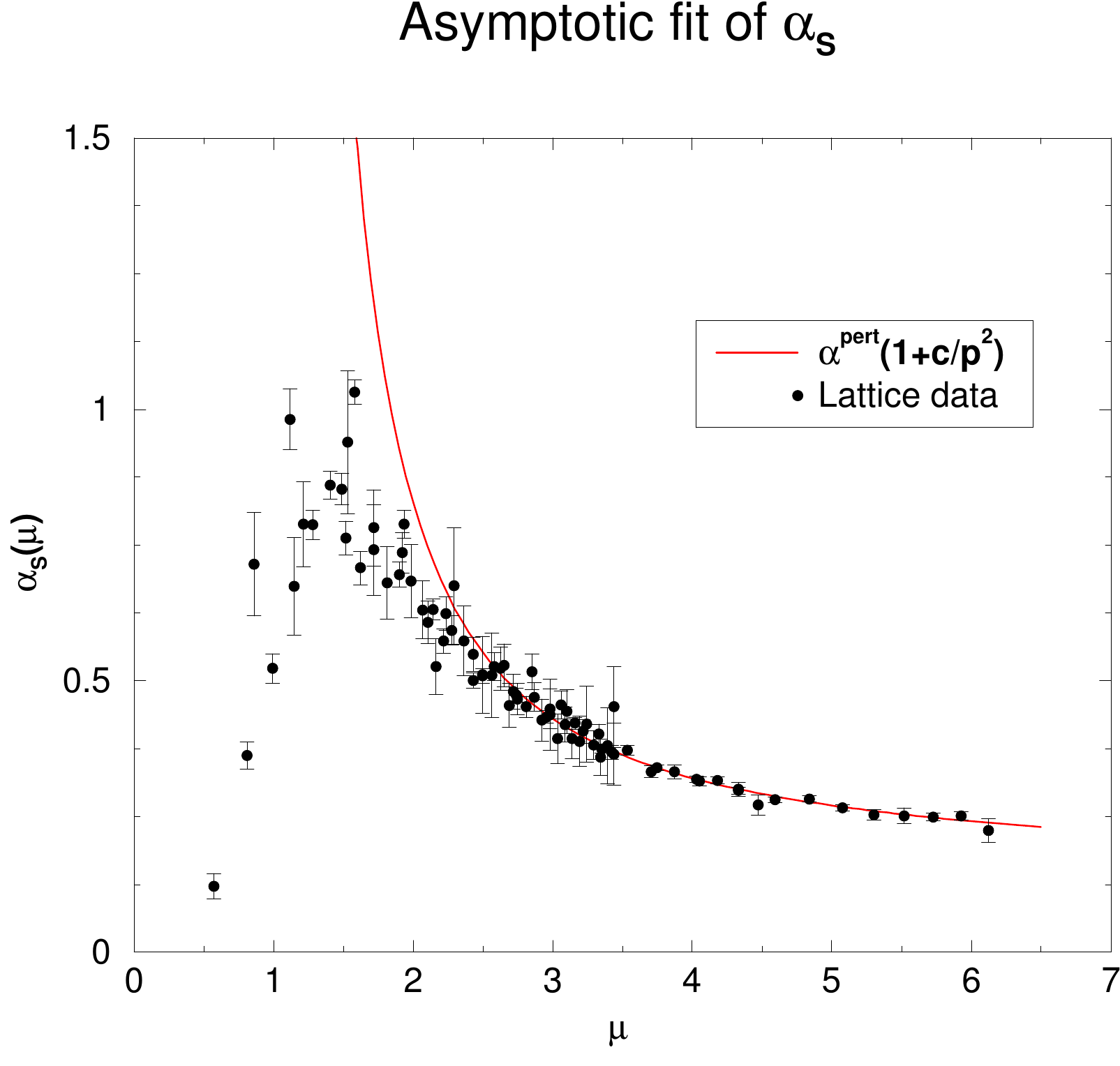}
\end{tabular}
\end{center}
\caption{Extraction of $\alpha_s$ from the 3-gluon vertex; the plot of the right panel is taken from \cite{BoucaudBQ}. \label{fig:3gluons}}
\end{figure}
\noindent The second MOM-like scheme that we will discuss in more details involves the ghost-ghost-gluon vertex. In Landau gauge, the bare gluon and ghost propagators read
\beq\nonumber
\left( G^{(2)} \right)_{\mu \nu}^{a b}(p^2,\Lambda) = \frac{G(p^2,\Lambda)}{p^2} \ \delta_{a b} 
\left( \delta_{\mu \nu}-\frac{p_\mu p_\nu}{p^2} \right), \quad
\left(F^{(2)} \right)^{ab}(p^2,\Lambda) = - \delta_{a b} \ \frac{F(p^2,\Lambda)}{p^2},
\eeq
where $\Lambda$ is the UV cut-off of the theory. Renormalized dressing functions $G_R$ and $F_R$ are defined by
\beq\nonumber
G_R(p^2,\mu^2)\ = \ \lim_{\Lambda \to \infty} Z_3^{-1}(\mu^2,\Lambda) \ G(p^2,\Lambda), \quad
F_R(p^2,\mu^2)\ = \ \lim_{\Lambda \to \infty} \widetilde{Z}_3^{-1}(\mu^2,\Lambda)\ F(p^2,\Lambda).
\eeq
A first set of renomalization conditions are
\beq\nonumber
G_R(\mu^2,\mu^2)=F_R(\mu^2,\mu^2)=1. 
\eeq
The amputated ghost-gluon vertex reads
\beq\nonumber
\widetilde{\Gamma}^{abc}_\nu(-q,k;q-k) =
i g_0 f^{abc} 
\left( q_\nu H_1(q,k) + (q-k)_\nu H_2(q,k) \right), 
\eeq
the renormalized vertex is defined by $\widetilde{\Gamma}_R=\widetilde{Z}_1 \Gamma$ and
$\widetilde{\rm MOM}$ prescriptions are
\beq\nonumber
\left.(H^R_1(q,k) +  H^R_2(q,k))\right\vert_{q^2=\mu^2} = 
\lim_{\Lambda \to \infty}\widetilde{Z}_1(\mu^2,\Lambda)\left.(H_1(q,k;\Lambda) 
+  H_2(q,k;\Lambda))\right\vert_{q^2=\mu^2} =1.
\eeq
The MOM prescription is completed by the kinematical configuration of the incoming momentum $k$: $k^2=(q-k)^2=\mu^2$ (symmetric configuration), $k=0$, $(q-k)^2=\mu^2$ (asymmetric configuration). In terms of  $H_{1}$ and $H_{2}$ scalar form factors, one has
\bea\nonumber
g_R(\mu^2) &=& \lim_{\Lambda \to \infty} \ \widetilde{Z}_3(\mu^2,\Lambda) Z_3^{1/2}(\mu^2,\Lambda) g_0(\Lambda^2)   
\left. \left(  H_1(q,k;\Lambda) + H_2(q,k;\Lambda) 
\rule[0cm]{0cm}{0.5cm}  \right) \right|_{q^2 \equiv \mu^2} \\
\nonumber
&=&  \ \lim_{\Lambda \to \infty} g_0(\Lambda^2) \ 
\frac{Z_3^{1/2}(\mu^2,\Lambda^2)\widetilde{Z}_3(\mu^2,\Lambda^2)}{ \widetilde{Z}_1(\mu^2,\Lambda^2)}.
\eea
The bare amputated vertex can be decomposed as $\widetilde{\Gamma}^{abc}_\nu(-q,k;q-k) = i g_0 f^{abc} q_\nu + {\cal T}_\nu$ as depicted in Figure \ref{fig:vgg}: 
\beq\nonumber
{\cal T}_\nu \sim q^\rho \int d^4 l l^\lambda J_{\rho \nu \lambda}(l,q),
\eeq
where $J$ is the bubble on the right side of the figure. In Landau gauge, the gluon propagator is transverse: $l^\lambda G^{(2)}_{\lambda \nu}(l-k)=
k^\lambda G^{(2)}_{\lambda \nu}(l-k)$. ${\cal T}_\nu$ can finally be expressed as ${\cal T}_\nu = q^\rho k^\lambda {\cal T}'_{\rho \nu \lambda}$. \emph{In the asymmetric configuration $k=0$, the ghost-ghost-gluon vertex keeps its tree level value, $H_1(q,0;\Lambda) 
+  H_2(q,0;\Lambda)=1$, and does not renormalize: $\widetilde{Z}_1(\mu^2,\Lambda^2)=1$}. This non-renormalization theorem was first discussed by Taylor \cite{TaylorFF} and further commented in \cite{MarcianoQCD}. So, in the so-called Taylor scheme, corresponding to ($q^2=\mu^2$, $k=0$), the renormalized strong coupling constant reads
\beq
\alpha_T(\mu^2) \equiv \frac{g^2_T(\mu^2)}{4 \pi}=  \ \lim_{\Lambda \to 
\infty} \frac{g_0^2(\Lambda^2)}{4 \pi} G(\mu^2,\Lambda^2) F^{2}(\mu^2,\Lambda^2).
\eeq
The beauty of this scheme is that the extraction of $\alpha_s$ on the lattice involves only the computation of the gluon and ghost dressing functions \cite{SternbeckBR}, \cite{BoucaudGN}. The relation between Taylor and $\overline{\rm MS}$ scheme is given by
\beq\nonumber
\alpha_T(\mu^2)=\alpha^{\overline{\rm MS}}(\mu^2) \left[1 + \sum_i c_i \left(\frac{\alpha^{\overline{\rm MS}}(\mu)}{4\pi}\right)^i\right],
\eeq 
where the $c_i$ coefficients are known in perturbation theory up to the third order \cite{ChetyrkinMF}, \cite{BoucaudGG}. $\alpha_T$ is related in perturbation theory at four loops to the $\Lambda_T$ parameter by the relation
\bea\nonumber
      \alpha_T(\mu^2) &= &\hspace{-0.2cm}\frac{4 \pi}{\beta_{0}t}
      \left\{1 - \frac{\beta_{1}}{\beta_{0}^{2}}\frac{\log(t)}{t}
     + \frac{\beta_{1}^{2}}{\beta_{0}^{4}}
       \frac{1}{t^{2}}\left[\left(\log(t)-\frac{1}{2}\right)^{2}
     + \frac{\widetilde{\beta}_{2}\beta_{0}}{\beta_{1}^{2}}-\frac{5}{4}\right]\right\} \\
     &+ &\hspace{-0.2cm}\frac{1}{(\beta_{0}t)^{4}}
 \left\{\frac{\widetilde{\beta}_{3}}{2\beta_{0}}+
   \frac{1}{2}\left(\frac{\beta_{1}}{\beta_{0}}\right)^{3}
   \left[-2\log^{3}(t)+5\log^{2}(t)+
\left(4-6\frac{\widetilde{\beta}_{2}\beta_{0}}{\beta_{1}^{2}}\right)\log(t)-1\right]\right\},
\label{eq:alphat}
\eea
wihere $t=\ln \frac{\mu^2}{\Lambda_T^2}$ and $\widetilde{\beta}_{2,3}$ are the NNLO and ${\rm N^3}$LO terms of the $\beta$ function in the Taylor scheme, that one can deduce from the Taylor $\to \overline{\rm MS}$ scheme conversion and the $\beta$ function in the latter scheme \cite{vanRitbergenVA}. $\Lambda_T$ is related to $\Lambda^{\overline{\rm MS}}$ by
\beq\nonumber
\frac{\Lambda_T}{\Lambda^{\overline{\rm MS}}}=e^{-c_1/\beta_0}=
e^{- \frac{507-40 {\rm N_f}}{792 - 48 {\rm N_f}}}.
\eeq
Fitting the $\alpha_T(\mu^2)$ lattice data with the perturbative formula (\ref{eq:alphat}) allows to extract $\Lambda^{\overline{\rm MS}}$.
\begin{figure}[t]
%
%
%
%
\begin{center}
\includegraphics*[width=0.8\textwidth]{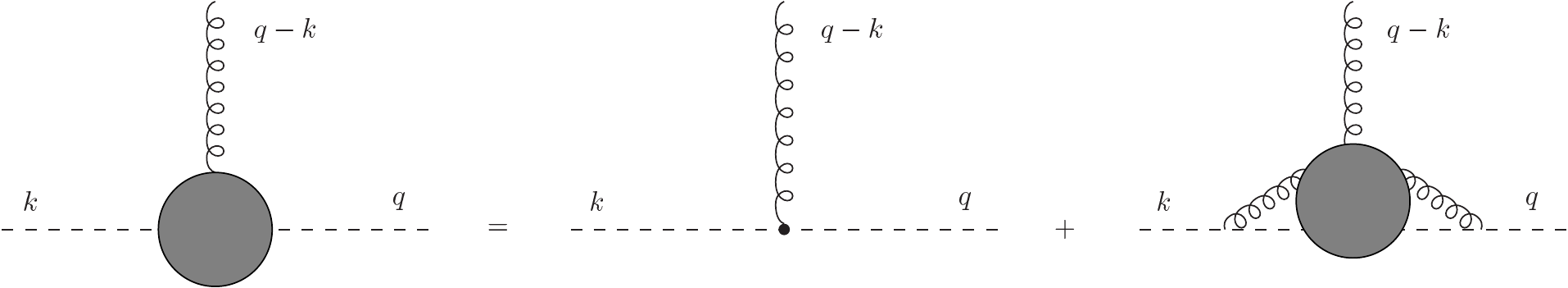}
\end{center}
\caption{Loop corrections to the tree-level part of the ghost-ghost-gluon vertex. \label{fig:vgg}}
\end{figure}
We have used this method to measure $\alpha_s$ from lattice simulations with ${\rm N_f}=2$ \cite{BlossierKY} and ${\rm N_f}=2+1+1$ \cite{BlossierTF} - \cite{BlossierIOA} dynamical flavours regularised by twisted mass fermions at maximal twist.

\subsection{$\alpha_s$ measurement at ${\rm N_f}=2$}

We have used a set of gauge ensembles produced by ETM Collaboration; the parameters are collected in Table \ref{tab:run}. The reader might worry about the low statistics for some simulations but we will see that it was enough for the target level of accuracy. 
\begin{table}[t]
\centering
\begin{tabular}{|c|c|c|c|}
\hline
$\beta$ & $a \mu_q$ & $L^3\times T$ & \#
\\ \hline
$3.9$ &  
\begin{tabular}{c}
0.004 \\
0.0064 \\
0.010
\end{tabular}
& 
$24^3\times48$ & 
\begin{tabular}{c}
$120$ \\
$20$ \\
$20$
\end{tabular}
\\ \hline
$4.05$ &  
\begin{tabular}{c}
0.003 \\
0.006 \\
0.008 \\
0.012
\end{tabular}
& $32^3\times64$ & 
\begin{tabular}{c}
$20$ \\
$20$ \\
$20$ \\
$20$
\end{tabular}
\\ \hline
$4.2$ & 0.0065 & $32^3\times64$ & $200$
\\ \hline
\end{tabular}
\caption{Parameters of the ETMC ensembles analysed in our work. For each ensemble, the time separation between measurements, in molecular dynamics units, is larger than the autocorrelation time of the plaquette.\label{tab:run}}
\end{table}
We have fixed the gauge links into the Landau gauge by minimizing the functional
\beq
F_U[g] = \frac{1}{V}\mbox{\rm Re}\left[ \sum_x \sum_\mu  \hbox{Tr}\left(1-\frac{1}{N_c}g(x)U_\mu(x)g^\dagger(x+\hat{\mu}) \right) \right]
\eeq
with respect to the gauge transform $g$. It has been done by applying a combination of
overrelaxation algorithm \cite{MandulaGF} and (partial) Fourier acceleration \cite{DaviesVS}. Overrelaxation technique consists in using, instead of 
\beq\nonumber
g= \exp \left\{\left[\sum_\nu 2iag_0 \Delta_{\nu} A_\nu\right] \right \}, \quad 
\Delta_\nu U_\mu(x) = U_\mu(x-\hat{\nu}) - U_\mu(x),
\eeq 
the operator 
\beq
g^{(\omega)} = \sum_{n=0}^N \frac{\gamma_n(\omega)}{n!}(g-{\mathbbm 1})^n, \quad 
\gamma_n(\omega)=\frac{\Gamma(\omega+1)}{\Gamma(\omega + 1 -n)};
\eeq
based on past experience, we have set $N=4$ and $\omega=1.72$. The partial Fourier acceleration is realised by
defining the gauge tranformation
\beq
g = \exp \left\{\frac{\alpha}{2} \Box^{-1} \left[\sum_\nu 2iag_0 \Delta_{\nu} A_\nu\right] \right \}.
\eeq
It is claimed that, in Fourier space, all modes converge as fast as the fastest one in the procedure of minimization. Numerically we have to invert the equation 
\beq\nonumber
\Box X = Y, \quad Y=\left[\sum_\nu 2iag_0 \Delta_{\nu} A_\nu\right],
\eeq 
for instance using the Conjugate Gradient algorithm. Referring again to previous studies, we have set $\alpha=0.6$. We have defined convergence criteria to stop the minimization run:
\beq\nonumber
||\partial^\mu A_\mu ||\equiv \frac{1}{V\,(N^2_c-1)} \sum_{x,\mu,a} |A^a_\mu(x+\hat{\mu}/2)- A^a_\mu(x-\hat{\mu}/2)|^2 
 < \delta_1,
 \eeq
\beq\nonumber
||g-\mathbbm{1}||\equiv \frac{1}{V\,N_c}\sum_x ({\rm Re}\,{\rm Tr} [g(x) - \mathbbm{1}])^2 + ({\rm Im}\, {\rm Tr} [g(x)])^2<\delta_2
\eeq 
(the identity matrix is an attractive point in the parameter space),
\beq\nonumber 
\frac{||Q^a(t+a) - Q^a(t)||}{||Q^a(t+a)+Q^a(t)||} 
 \equiv {\rm min}_a \frac{{\rm sup}_t\, {\rm Im}(\sum_{\vec{x}} {\rm Tr} [A(\vec{x}+\hat{0}/2,t) t^a]) - 
 {\rm inf}_t\, {\rm Im}(\sum_{\vec{x}} {\rm Tr} [A(\vec{x}+\hat{0}/2,t) t^a])}{{\rm sup}_t\, {\rm Im}(\sum_{\vec{x}} {\rm Tr} [A(\vec{x}+\hat{0}/2,t) t^a]) + 
 {\rm inf}_t\, {\rm Im}(\sum_{\vec{x}} {\rm Tr} [A(\vec{x}+\hat{0}/2,t) t^a])}
  < \delta_3, 
\eeq
where $t^a = \Lambda^a/2$ 
(in Landau gauge the charge colours are conserved). In practice we have set $\delta_1=10^{-12}$, for which the other criteria are always satisfied. One might worry about Gribov copies, corresponding to a local minimum of the functional $F_U$ but not the absolute one \cite{GribovWM} - \cite{MarinariZV}. We did not implement refined methods, as proposed in \cite{BogolubskyBW}, to get rid of those copies. However, the effect that they induce is felt in the deep infra-red regime of QCD, a region that we did not exploit at all in our analysis. Furthermore we have checked the absence of any systematics by comparing results at different volumes.\\
The Faddeev-Popov operator is obtained by deriving $F_{U}[g]$ twice with respect to the ``trajectory'' parameter $s$ defined by
$g(x) \equiv g^{(0)}(x)e^{s \epsilon(x)}$, where $g^{(0)}(x)$ is the gauge transformation that transforms a given link $U_\mu(x)$ into a link $U^{(0)}_\mu(x)$ at the minimum of $F$:
\bea\nonumber
\frac{dF}{ds}&=& -\frac{1}{V\,N_c}\sum_{x,\mu} {\rm Tr} \left[\epsilon(x) g(x)U_\mu(x)g^\dagger(x+\hat{\mu}) - g(x)U_\mu(x)g^\dagger(x+\hat{\mu}) \epsilon(x+\hat{\mu})\right.\\
\nonumber&&\hspace{2.5cm}\left.-\epsilon(x+\hat{\mu}) g^{\dag}(x+\hat{\mu})U^{\dag}_\mu(x)g(x) + g^{\dag}(x+\hat{\mu})U^{\dag}_\mu(x)g(x) \epsilon(x)\right],\\
\nonumber
\frac{d^{2}F}{ds^{2}}&=& -\frac{1}{V\,N_c}\sum_{x,\mu} {\rm Tr} \left[\epsilon^{2}(x) g(x)U_\mu(x)g^\dagger(x+\hat{\mu}) +g(x)U_\mu(x)g^\dagger(x+\hat{\mu})\epsilon^{2}(x+\hat{\mu})\right.\\
\nonumber&&\hspace{2.5cm}\left.- 2 \epsilon(x)g(x)U_\mu(x)g^\dagger(x+\hat{\mu}) \epsilon(x+\hat{\mu})
-2\epsilon(x+\hat{\mu}) g^{\dag}(x+\hat{\mu})U^{\dag}_\mu(x)g(x)\epsilon(x) \right.\\
\nonumber
&&\hspace{2.5cm}\left.+\epsilon^{2}(x+\hat{\mu}) g^{\dag}(x+\hat{\mu})U^{\dag}_\mu(x)g(x) + g^{\dag}(x+\hat{\mu})U^{\dag}_\mu(x)g(x) \epsilon^{2}(x)\right].
\eea
At the mininum of $F$, one has
\bea\nonumber
\left.\frac{d^{2}F}{ds^{2}}\right|_{s=0}&=& -\frac{1}{V\,N_c} \sum_{x,\mu} {\rm Tr} \left[\epsilon^{2}(x) U^{(0)}_\mu(x) +\epsilon^{2}(x+\hat{\mu})U^{(0)}_\mu(x)- 2 \epsilon(x+\hat{\mu})\epsilon(x)U^{(0)}_\mu(x)\right.\\
\nonumber
&&\left.\hspace{2.5cm}+\epsilon^{2}(x+\hat{\mu}) U^{(0)\dag}_\mu(x)+ \epsilon^{2}(x)U^{(0)\dag}_\mu(x)
-2\epsilon(x)\epsilon(x+\hat{\mu})U^{(0)\dag}_{\mu}(x)\right]\\
\nonumber
&=&-\frac{1}{V\,N_c}\sum_{x,\mu}{\rm Tr} \left[ (\epsilon^{2}(x) + \epsilon^{2}(x+\hat{\mu}))(U^{(0)}_{\mu}(x) + U^{(0)\dag}_{\mu}(x))\right.\\ 
\nonumber
&&\hspace{2.5cm}\left.- 2\epsilon(x+\hat{\mu})\epsilon(x)U^{(0)}_{\mu}(x) - 2\epsilon(x)\epsilon(x+\hat{\mu})U^{(0)\dag}_{\mu}(x)\right]\\
\nonumber
&=&-\frac{1}{V\,N_c}\sum_{x,\mu}{\rm Tr} \left[(\epsilon(x+\hat{\mu})-\epsilon(x))^{2}(U^{(0)}_{\mu}(x) + 
U^{(0)\dag}_{\mu}(x))\right.\\ 
\nonumber
&&\hspace{2.5cm}\left.- [\epsilon(x+\hat{\mu}),\epsilon(x)] (U^{(0)}_{\mu}(x) - U^{(0)\dag}_{\mu}(x))\right]\\
\nonumber
&=&-\frac{1}{V\,N_c} \sum_{x,a}  \epsilon^a(x) ({\cal M}^{\rm lat}_{FP}[U] \epsilon)^a (x),
\eea
where the lattice formulation of the Faddeev-Popov operator reads
\bea\nonumber
{\cal M}^{\rm lat}_{FP}[U] \epsilon^a(x)& =& \frac{1}{V} \sum_{\mu}\left\{G^{ab}_{\mu}(x) \left(\epsilon^{b} ( x+ \hat{\mu}) 
- \epsilon^{b}(x)\right) - (x \leftrightarrow x - \hat{\mu})\right.\\
\nonumber
&&\left.+ \frac{1}{2} f^{abc} \left[\epsilon^{b}( x + \hat{\mu}) A^{c}_{\mu}\left(x + \frac{\hat{\mu}}{2}\right)- \epsilon 
^{b}( x - \hat{\mu}) A^{c}_{\mu}\left(x- \frac{\hat{\mu}}{2}\right)\right]\right\}, 
\eea
and
\beq\nonumber
G^{ab}_{\mu}(x)=-\frac{1}{2} {\rm Tr}\left[\left\{t^{a}, t^{b}\right\}\left(U_{\mu}(x) + U^{\dag}_{\mu}(x)\right)\right],
\quad A^{c}_{\mu}\left(x+\frac{\hat{\mu}}{2}\right) = {\rm Tr}\left[t^{c}\left(U_{\mu}(x) - U^{\dag}_{\mu}(x)\right)\right].
\eeq
The ghost propagator is the inverse of the Faddeev-Popov operator. We have removed its zero mode and solved the equation
\beq
\sum_{x,b} {\cal M}^{{\rm lat}\,ab}_{FP}(x,y) \eta^b(y) = \delta_{a\,a_0}\left(\delta_{x\,0} - \frac{1}{V}\right).
\eeq
The dressing function $F(p)$ is computed from the Fourier transform of ${\cal M}^{-1}$ while the dressing function $G(p^2)$ is obtained from the product of Fourier transforms of the gluon field $A_\mu$:
\bea\nonumber
 \left( G^{(2)}\right)^{a_1 a_2}_{\mu_1\mu_2}(p)&=&
 \langle \tilde{A}_{\mu_1}^{a_1}(p)\tilde{A}_{\mu_2}^{a_2}(-p) \rangle,
\nonumber
\tilde{A}_\mu^a(p)=\hbox{Tr}\left [\sum_x A_\mu(x+ \hat \mu/2)
\exp(i p (x+ \hat \mu/2))t^a\right].
\eea
$F$ and $G$ are dimensionless quantities that depend on the discretized momenta $a p_\mu=2\pi n_\mu/N_\mu$, $n_\mu=0,1,\cdots,N_\mu-1$, $N_4=T$, $N_{1,2,3}=L$, and also on $a\Lambda_{\rm QCD}$. In the computation of the quantity
\beq\label{alphaT}
\widehat{\alpha}_T(\mu)=\lim_{a \to 0} \frac{g^2_0(a)}{4 \pi}  F^2(ap,a\Lambda_{\rm QCD}) G(ap\,,a\Lambda_{\rm QCD})|_{p^2=\mu^2},
\eeq
one has to pay attention to cut-off effects: as we used an ${\cal O}(a)$ improved action, those are quadratic in $a$, but are of different kinds.\\
The first one that has to be removed, following the proposal by \cite{BecirevicUC}, \cite{deSotoHT}, comes from the
breaking of the $O(4)$ rotational symmetry of the Euclidean space-time into the group $H(4)$ restricted to rotation within 
the hypercubic lattice. It is beneficial to perform the average  of any dimensionless lattice quantity $Q(a p_\mu)$ over the orbits 
of the group $H(4)$. Usually several orbits of $H(4)$ correspond to a single value of $p^2$: for instance $(1,1,1,1)$ and $(2,0,0,0)$ have the same $p^2$ but different $p^{[4]}\equiv \sum_\mu p^4_\mu$, 4 versus 16.
Defining the $H(4)$ invariants
 \beq
 p^{[4]}, \quad p^{[6]}=\sum_{\mu=1}^{6} p_\mu^6,\cdots,
 \eeq
helps to label the orbits of $H(4)$. Actually, any $H(4)$-invariant polynomial in $ap$ can be written in terms of the four 
invariants $p^{[2 i]}$ with $i=1,2,3,4$ \cite{deSotoHT}, \cite{Weyl}. Roughly said, indeed, a polynomial $Q(p_1,p_2,p_3,p_4)$ reads
\beq\nonumber
Q(p_1,p_2,p_3,p_4)=\sum_{n_1,n_2,n_3,n_4} c_{n_1\,n_2\,n_3\,n_4} p^{n_1}_1\,p^{n_2}_2\,p^{n_3}_3\,p^{n_4}_4.
\eeq 
The invariance
under $H(4)$ implies that the coefficients $c_{n_1\,n_2\,n_3\,n_4}$ do not vanish for even $n_i$ and that $Q$ is a polynomial in
\bea\nonumber
\varphi_1&=&p^2,\\
\nonumber
\varphi_2&=&p^2_1 p^2_2 + p^2_1 p^2_3 + p^2_1 p^2_4 + p^2_2 p^2_3 + p^2_2 p^2_4+ p^2_3 p^2_4,\\
\nonumber
\varphi_3&=&p^2_1p^2_2p^2_3+p^2_1p^2_2p^2_4+p^2_1p^2_3p^2_4+p^2_2p^2_3p^2_4,\\
\nonumber
\varphi_4&=&p^2_1p^2_2p^2_3p^2_4.
\eea 
Noticing that 
\bea\number
2\varphi_2&=&\varphi^2_1 -p^{[4]},\\ 
\nonumber
3\varphi_3&=&\varphi_2 p^2 - \varphi_1 p^{[4]} + p^{[6]},
\eea
and 
\bea\nonumber
4 \varphi_4 &=& \varphi_3 p^2 - \varphi_2 p^{[4]} + \varphi_1 p^{[6]} - p^{[8]},
\eea 
we conclude that $Q$ depends only on the latter variables. Due to the upper cuts on discretized momenta, it is enough to label all the orbits we deal with by $p^2$, $a^2 p^{[4]}$ and $a^4 p^{[6]}$. We can write the quantity $Q(a p_\mu)$ averaged over 
$H(4)$ $Q(a^2\,p^2, a^4p^{[4]}, a^6 p^{[6]}, a^2\Lambda_{\rm QCD}^2)$.
Developing the function $Q$ in $\epsilon=a^2 p^{[4]}/p^2 \ll 1$, in particular if the lattice spacing is small,
one has
\beq\nonumber
Q(a^2\,{p}^2, a^4p^{[4]}, a^6 p^{[6]},a^2\Lambda_{\rm QCD}^2)
= Q(a^2p^2,a^2\Lambda_{\rm QCD}^2) +
\left.\frac{dQ}{d\epsilon}\right|_{\epsilon=0} a^2
\frac{p^{[4]}}{p^2} + \cdots\,.
\eeq
We have explored a general non-perturbative approach to remove the  $\mathcal{O}(a^2)$
corrections driven by the $p^{[4]}$ term, expanding the perturbative\footnote{and sometimes, controlled in an unsatisfactory manner with respect to the claimed error} computations
reported in the literature, for instance in \cite{AlexandrouME}. The basic method is to fit
from the whole set of orbits that share the same $p^2$ the
coefficient $dQ/d\epsilon$ and get the extrapolated value of $Q$,
free from $H(4)$ artefacts. Assuming that the coefficient 
\beq
\left.R(a^2p^2,a^2\Lambda_{\rm QCD}^2) = 
\frac{dQ\left(a^2p^2 ,0,0,a^2\Lambda_{\rm QCD}^2\right)}{d\epsilon}\right|_{\epsilon=0}
\eeq
has a smooth dependence on $a^2p^2$ over a given momentum window, an alternative consists 
in expanding $R$ as $R=R_0+R_1 a^2p^2$ and making a global fit in a set of
momentum windows centered around a given $p$ over a range $(p-\delta,p+\delta)$. 
In that case the extrapolation does not rely on any particular
assumption for the functional form of $R$. The systematic error
coming from the extrapolation can be estimated by modifying the width of the
fitting window. We have considered in that work anisotropic lattices $L^3 \times 2L$: the full $H(4)$
lattice symmetry is reduced to $H(3)$. We expect deviations from $H(4)$ symmetry rather in the
long-distance physics than in the ultraviolet regime and we have supposed the validity of our treatment of the $H(4)$ artefacts.
\begin{figure}[t]
\begin{center}
\begin{tabular}{cc}
\includegraphics*[width=0.5\textwidth]{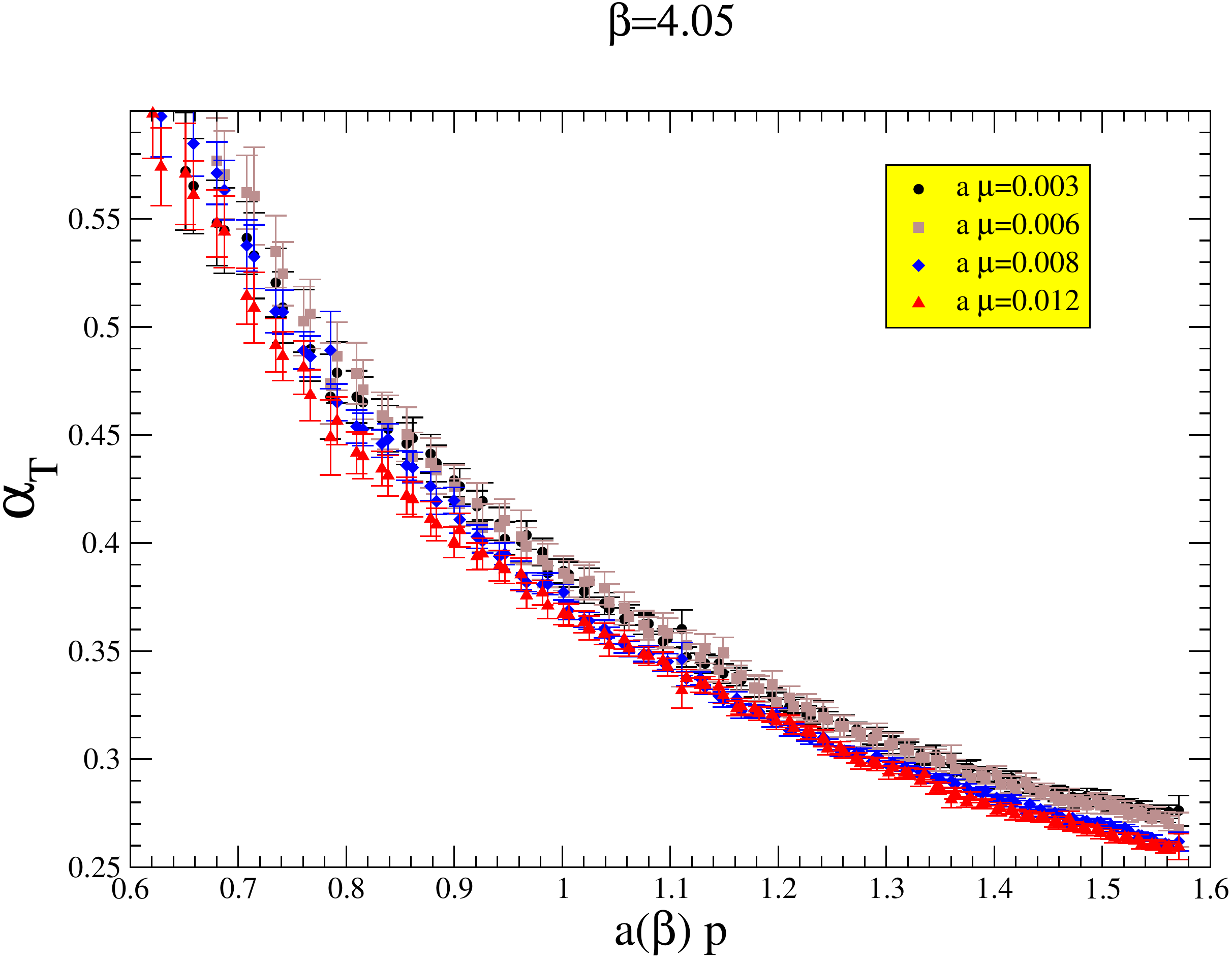}
&
\includegraphics[width=8.2cm]{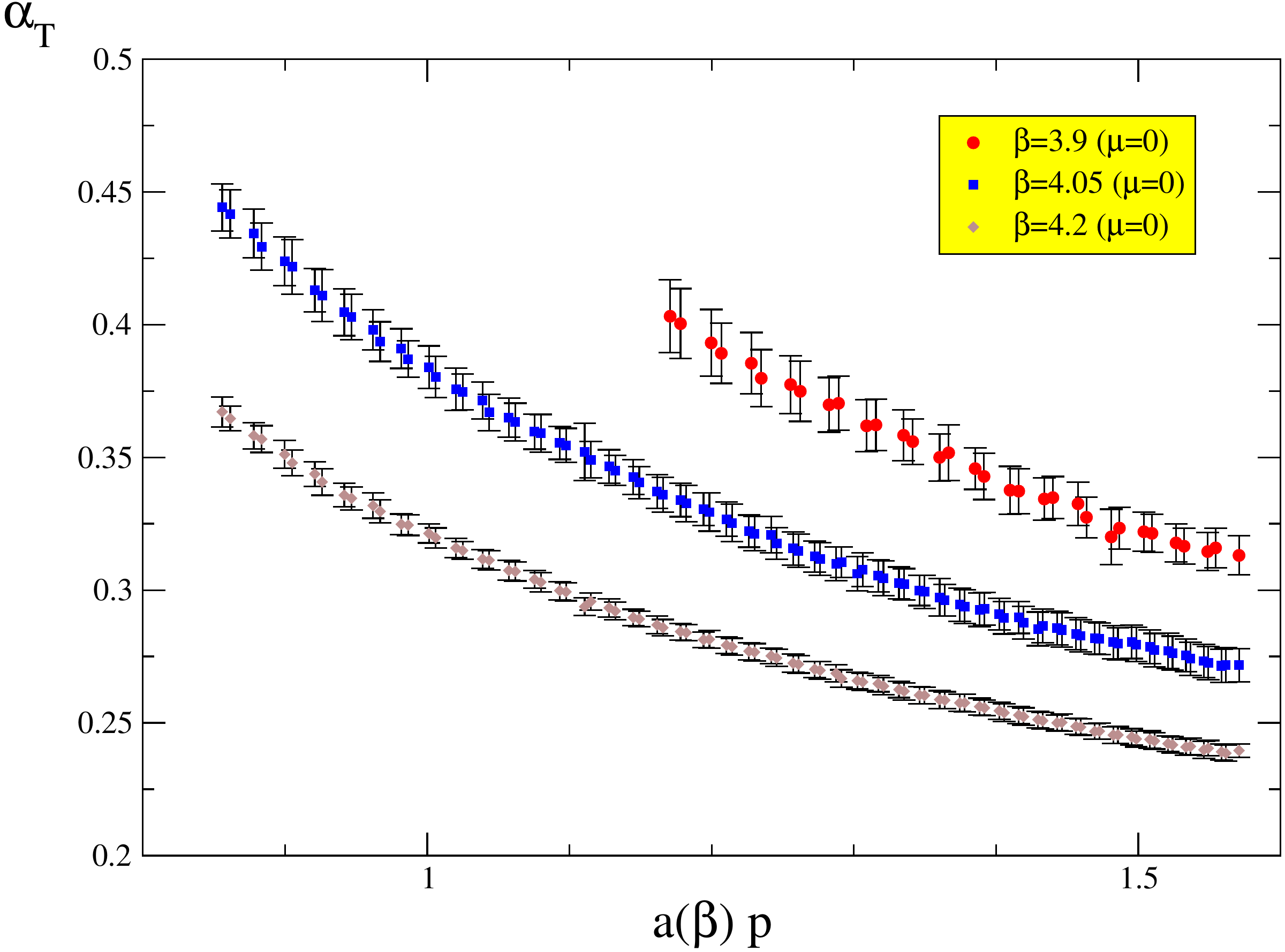} 
\end{tabular}
\end{center}
\caption{The Taylor couplings extracted from eq.(\ref{alphaT}), after an $H(4)$ extrapolation, 
at $\beta=4.05$ (left panel). Chiral extrapolation of the Taylor coupling at our three lattice spacings (right panel).}
\label{fig:brut}
\end{figure}
\\
The second kind of artefact is a residual sea-mass dependence of our results: in Fig.~\ref{fig:brut} (left panel) one can see the Taylor coupling after  hypercubic extrapolation for different $\mu_q$ at fixed $\beta=4.05$. One observes
a visible dependence on $\mu_q$: in case of an artefact,  it should be in $a^2 \mu_q^2$ or $a^2 \mu_q \Lambda_{\rm QCD}$ (from the spontaneous 
breaking of chiral symmetry). In case of a physical dependence, it should scale with some unknown function of the physical mass $\mu^R_q$ renormalized at, say, 2 GeV. 
Trying an $\mathcal{O}(a^2 \mu_q^2)$ dependence, we write the expansion :
\bea\nonumber
\widehat{\alpha}_{T}(a^2p^2,a^2\mu_q^2) &=& \frac{g_0^2(a^2)}{4 \pi}
\widehat{G}(a^2p^2,a^2\mu_q^2) \widehat{F}^{2}(a^2 p^2,a^2\mu_q^2) \\ 
\nonumber &=& \widehat{\alpha}_{T}(a^2p^2,0) + 
\frac{\partial \widehat{\alpha}_{T}}{\partial (a^2\mu_q^2)} \left(a^2 p^2 \right) \ 
a^2 \mu_q^2 \ + \ \cdots\,. 
\eea
If the first-order expansion is reliable, one can expect a linear
behaviour on $a^2 \mu_q^2$  for 
$\widehat{\alpha}_T$ at any fixed lattice momentum, $\beta$, and several  $\mu_q$ masses. This has been checked from our $\beta=3.9$ and $\beta=4.05$
simulations. Neglecting the $\mathcal{O}(a^4)$ contributions, the Taylor-McLaurin expansion reads:
\beq\nonumber
\widehat{\alpha}_T(a^2p^2,a^2\mu_q^2) 
= \alpha_T(p^2) + R'_0(a^2p^2) \ a^2 \mu_q^2 , 
\eeq
where $R'_0(a^2p^2)$ is defined as $R'_0(a^2 p^2) \equiv
\frac{\partial \widehat{\alpha}_{T}}{\partial (a^2\mu_q^2)}$.
We have found that above $p_{\rm min}\simeq 2.8$ GeV, data at both lattice spacings exhibit a fairly constant
 $R'_0(a^2p^2)$ and that a good scaling emerges between both $\beta$'s. The value of $p_{\rm min}$ gives an
 indication for the momentum window
where the chiral extrapolation can be applied at any momentum. The three estimates of the running 
coupling at $\beta=3.9,4.05,4.2$ are plotted in the right panel of Figure \ref{fig:brut}: it shows a very smooth 
running behaviour. Eventually we have to confront those lattice estimates to analytical 
predictions.\\ 
It is well known that perturbation theory for the OPE analysis of Green functions needs the subtraction of power corrections induced by the presence of non-vanishing condensates \cite{ShifmanBX}. It is the case for the gluon and ghost propagators, whose the leading power corrections are given by
\bea
(F^{(2)})^{a b}(q^2) &=& (F_{\rm pert}^{(2)})^{a b}(q^2) \ + \ 
w^{a b} \ \frac{\langle A^2 \rangle}{4 (N_C^2-1)} \ + \ \dots\,,
\nonumber \\
(G^{(2)})_{\mu\nu}^{a b}(q^2) &=& (G_{\rm pert}^{(2)})_{\mu\nu}^{a b}(q^2) \ + \ 
w_{\mu\nu}^{a b} \ \frac{\langle A^2 \rangle}{4 (N_C^2-1)} \ + \ \dots\,, 
\eea
where the Wilson coefficients are diagrammatically represented in Figure \ref{fig:wilsonOPE}.
\begin{figure}[t]
%
%
%
%
\begin{center}
\includegraphics*[width=0.8\textwidth]{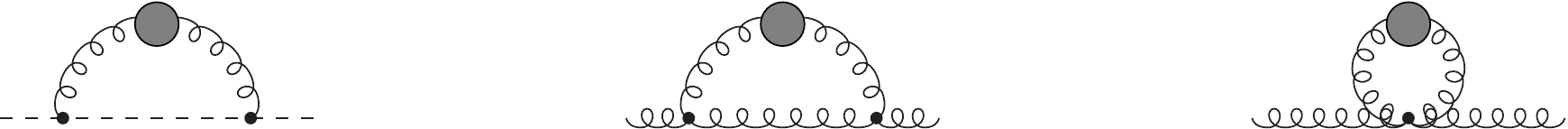}
\end{center}
\caption{Leading power corrections to the ghost and gluon propagators expressed within the OPE framework. \label{fig:wilsonOPE}}
\end{figure}
At tree level one has \cite{BoucaudST}, \cite{DeSotoQX}, \cite{BoucaudXN}, \cite{BoucaudGN}:
\bea\nonumber
F_R(q^2,\mu^2) &= & F_{R, {\rm pert}}(q^2,\mu^2) \
\left(  1 + \frac{3}{q^2} \frac{g^2_R \langle A^2 \rangle_{R,\mu^2}} {4 (N_C^2-1)} \right)\,, \\ 
G_R(q^2,\mu^2) & = & G_{R, {\rm pert}}(q^2,\mu^2) \
\left(  1 + \frac{3}{q^2} \frac{g^2_R \langle A^2 \rangle_{R,\mu^2}} {4 (N_C^2-1)} \right) \,,
\eea
\bea\nonumber
\alpha_T(\mu^2) &=& \lim_{\Lambda \to \infty} 
\frac{g_0^2}{4 \pi} F^2(\mu^2,\Lambda) G(\mu^2,\Lambda) \\
\nonumber
&=&\lim_{\Lambda \to \infty} 
\frac{g_0^2}{4 \pi} F^2(q_0^2,\Lambda) F^2_R(\mu^2,q_0^2) \
G(q_0^2,\Lambda)  G_R(\mu^2,q_0^2)\\
\nonumber
&=& \alpha^{\rm pert}_T(q_0^2) \  
F^2_{R, {\rm pert}}(\mu^2,q_0^2) \ G_{R, {\rm pert}}(\mu^2,q_0^2) 
\ \left(  1 + \frac{9}{\mu^2} \frac{g^2_T(q_0^2) \langle A^2 \rangle_{R,q_0^2}} {4 (N_C^2-1)}
\right)\\
&=& 
\alpha^{\rm pert}_T(\mu^2)
\ \left(  1 + \frac{9}{\mu^2} \frac{g^2_T(q_0^2) \langle A^2 \rangle_{R,q_0^2}} {4 (N_C^2-1)}
\right) \ ,
\eea
where $q_0^2 \gg \Lambda^2_{\rm QCD}$ is some perturbative scale. Including the leading logarithmic correction
to the Wilson coefficient, one gets \cite{BoucaudGN}
\beq\label{alphahNP}
\alpha_T(\mu^2)=
\alpha^{\rm pert}_T(\mu^2)
\left(  1 + \frac{9}{\mu^2} 
\left( \frac{\alpha^{\rm pert}_T(\mu^2)}{\alpha^{\rm pert}_T(q_0^2)}
\right)^{1-\gamma_0^{A^2}/\beta_0}
\frac{g^2_T(q_0^2) \langle A^2 \rangle_{R,q_0^2}} {4 (N_C^2-1)}
\right) \ ,
\eeq
where $\gamma_0^{A^2}$ is the anomalous dimension at leading order of the Wilson coefficient attached to the gluon condensate  \cite{GraceyYT}. One can extend the procedure up to four loops \cite{ChetyrkinKH} and write generically
\beq\label{alphahNPb}
\alpha_T(\mu^2)=
\alpha^{\rm pert}_T(\mu^2)
\left(  1 + \frac{9}{\mu^2} W^{\rm MOM}(\mu,q_0)
\frac{g^2_T(q_0^2) \langle A^2 \rangle_{R,q_0^2}} {4 (N_C^2-1)}
\right)\,.
\eeq
To estimate $\Lambda^{\overline{\rm MS}}$, we fit the data for all momenta inside the window $[p_{\rm min}, p_{\rm max}(a)=1.6 a^{-1}]$, where the upper bound is somehow arbitrary but ${\cal O}(a^4)$ cut-off effects are still negligible: first it is 
done by inverting the four-loop perturbative formula for 
the coupling, giving in principle a constant $\Lambda^{\overline{\rm MS}}$, but they systematically decrease as the lattice momentum increases. It reveals the necessity to incorporate in the analysis some power corrections, with a
non-zero gluon condensate that comes because we work in a fixed gauge.  Doing this, one
requires the best-fit to a constant of the estimates of  $\Lambda^{\overline{\rm
MS}}$, in lattice units, in terms of the lattice momentum, $a p_i$. The procedure is illustrated in the left panel of Figure \ref{fig:fitalphanf2}. We perform the fit for each lattice spacing but, as the running of $\alpha_T$ depends only on the scale in physical units, the various $\alpha_T$ have to sit on the same universal curve, once a proper rescaling of the momenta to a common scale $a_{\beta=3.9}$ is applied. Considering the Wilson coefficient at tree-level, the fit parameters are $a_{\beta=3.9} \Lambda^{\overline{\rm MS}}$, $c\propto a^2_{\beta=3.9} \lgl A^2 \rgl_{R,q^2_0}$ and the ratios of lattice spacings $a_{\beta=4.05}/a_{\beta=3.9}$ and $a_{\beta=4.2}/a_{\beta=3.9}$. We minimize the total $\chi^2$
\beq\nonumber
\chi^2\left(a_{\beta=3.9}\Lambda^{\overline{\rm MS}},c,\frac{a_{\beta=4.05}}{a_{\beta=3.9}},\frac{a_{\beta=4.2}}{a_{\beta=3.9}}\right) \ = \ \sum_{j=0}^2 \sum_{i} \
\frac{\left( \Lambda_i(\beta_j) - \displaystyle \frac{a_{\beta_j}}{a_{\beta=3.9}} a_{\beta=3.9} \Lambda^{\overline{\rm MS}}  
\right)^2}{\delta^2(\Lambda_i)} \,,
\eeq
where the sum over $j$ corresponds to the different lattice spacings and the index $i$ 
runs to cover the fitting window $[p_{\rm min},p_{\rm max}]$. In the case of $\beta=4.2$, as a single simulation has been 
exploited, the extrapolation to the chiral limit has been done by applying the slope $R'_0$ 
computed for $\beta=3.9$ and $\beta=4.05$. $\Lambda_i(\beta_j)$ is again extracted 
for any $\beta_j$ by requiring
\beq\nonumber
\alpha_T^{\rm pert}\left(\log\frac{a^2_{\beta_j} p^2_i}{\Lambda_i^2(\beta_j)}\right) 
\left[ 1+\frac{c}{a^2_{\beta_j} p^2_i} 
\left(\frac{a_{\beta_j}}{a_{\beta=3.9}}\right)^2 W\left(\frac{a^2_{\beta_j} p^2_i}{\Lambda_i^2(\beta_j)},
\frac{a^2_{\beta=3.9} q_0^2}{\Lambda_i^2(\beta_j)} \right)\right]
= \widehat{\alpha}_T(a^2(\beta_j) p^2_i,0)\,,
\eeq 
where now we use the OPE formula including the expression of the Wilson coefficient at different orders in perturbation theory,
with $\alpha_T^{\rm pert}$ given by the perturbative four-loop formula and where $ a_{\beta=3.9} q_0=4.5$, corresponding to $q_0 \approx 10$ GeV. The statistical error on the numerical data has been computed by a jackknife analysis and properly 
propagated through the perturbative inversion to give $\delta(\Lambda_i)$. The numerical values of the fit parameters are collected in Table \ref{tab:fitalphanf2} (top), where we have decided to take the first order of the Wilson coefficient. There are several sources of systematics:\\
-- adding higher order corrections to the Wilson coefficient; the results are collected in the middle of Table \ref{tab:fitalphanf2} and we observe a strong stability of the fits, $\Lambda^{\overline{\rm MS}}a_{\beta=3.9}$ varying less than a 2.5 \%, and a convergent behaviour for the gluon condensate\\
-- including higher power corrections in the OPE; we have tried to add a term in $W'/(p^2)^2$ in the fit but we have obtained unstable results with an anticorrelation between the ${\cal O}(1/p^2)$ and the ${\cal O}(1/(p^2)^2)$ contributions; \emph{it is crucial to take into account a positive non perturbative factor, and we stress that it is well described
by the dominant dimension-two $\lgl A^2 \rgl$ condensate}, as illustrated in the right panel of Figure \ref{fig:fitalphanf2}\\ 
\begin{figure}[t]
\begin{center}
\begin{tabular}{cc}
\includegraphics*[width=0.5\textwidth]{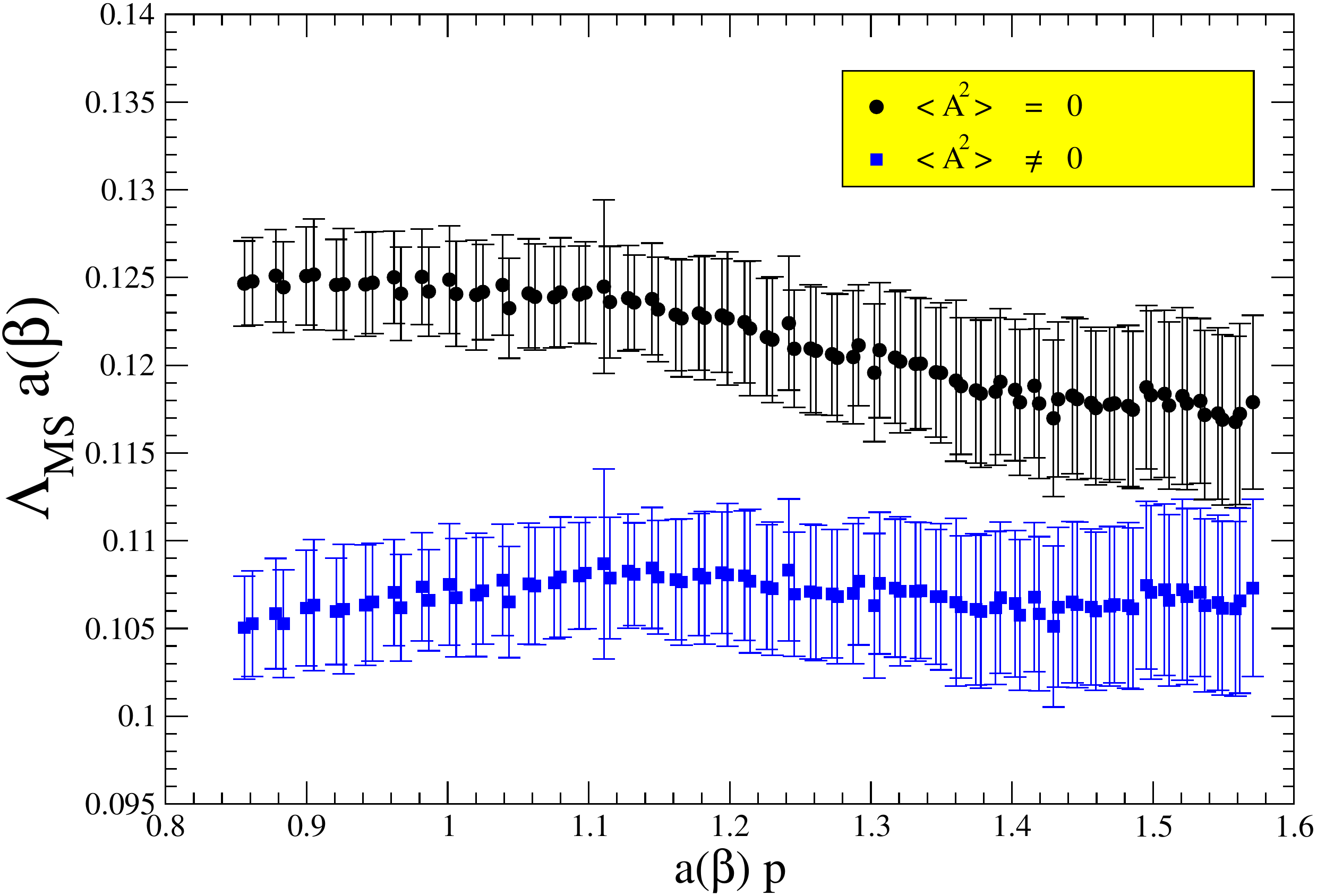}
&
\includegraphics*[width=0.5\textwidth]{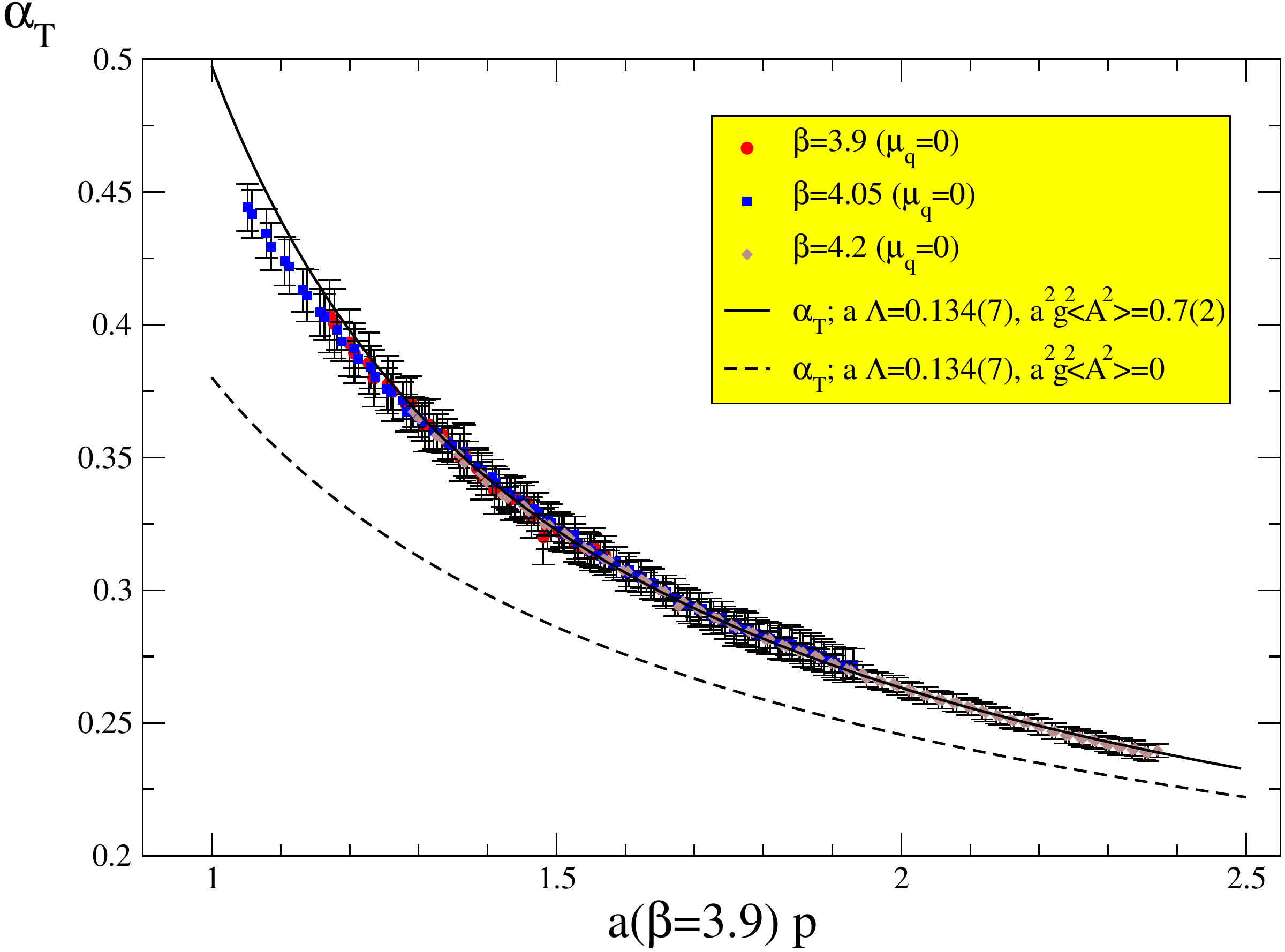}
\end{tabular}
\end{center}
\caption{Extraction of $\Lambda^{\overline{\rm MS}}$ from the analysis of $\alpha_T$ at $\beta=4.05$ (left panel). $\alpha_T$ 
computed for the three $\beta=3.9,4.05,4.2$; the lattice momentum, 
$a_{\beta}p$ in the x-axis, is converted to a physical momentum in units of $a_{\beta=3.9}^{-1}$, by applying the ratios of lattice spacings that are obtained after the fit.The solid curve is the non-perturbative prediction given by
eq.(\ref{alphahNP}) with the best-fit parameters for $\Lambda^{\overline{\rm MS}}$ and the gluon condensate, 
and the dotted one is the same but with zero gluon condensate (right panel).}
\label{fig:fitalphanf2}
\end{figure}
\begin{table}[t]
\begin{center}
\begin{tabular}{c}
\begin{tabular}{|c|c|c|}
\hline
 & $\alpha_T$ & $r_0/a$ \\
\hline
$a_{\beta=3.9}/a_{\beta=4.05}$ & 1.224(23) & 1.255(42)  \\
\hline
$a_{\beta=3.9}/a_{\beta=4.2}$ & 1.510(32)& 1.558(52)  \\
\hline
$a_{\beta=4.05}/a_{\beta=4.2}$ & 1.233(25)& 1.241(39) \\
\hline
$\Lambda^{\overline{\rm MS}} a_{\beta=3.9}$ & 0.134(7) &  \\
\hline
$g^2 \langle A^2 \rangle a^2_{\beta=3.9}$ & 0.70(23) & \\
\hline 
\end{tabular}\\
\\
\\
\begin{tabular}{|c|c|c|c|c|}
\hline
& $W_{\rm one\, loop}$& $W_{\rm two\, loops}$ & $W_{\rm three\, loops}$ & $W_{\rm four\, loops}$ \\
\hline
$\Lambda^{\overline{\rm MS}} a_{\beta=3.9}$ & 0.134(7) & 0.136(7) & 0.137(7) & 0.138(7) \\
\hline
$g^2 \langle A^2 \rangle a^2_{\beta=3.9}$ & 0.70(23) &  0.52(18) & 0.44(14) & 0.39(14) \\
\hline 
\end{tabular}\\
\\
\\
\begin{tabular}{|c|c|c|}
\hline
& $\alpha^{\rm pert}_T$(four loops) & $\alpha^{\rm pert}_T$(three loops) \\
\hline
$a_{\beta=3.9}/a_{\beta=4.05}$ & 1.224(23) & 1.229(23) \\
\hline
$a_{\beta=3.9}/a_{\beta=4.2}$ & 1.510(32)& 1.510(29)  \\
\hline
$a_{\beta=4.05}/a_{\beta=4.2}$ & 1.233(26)& 1.234(25)  \\
\hline
$\Lambda^{\overline{\rm MS}} a_{\beta=3.9}$ & 0.134(7) & 0.125(6) \\
\hline
$g^2 \langle A^2 \rangle a^2_{\beta=3.9}$ & 0.70(23) & 0.80(20) \\
\hline 
\end{tabular}
\\
\end{tabular}
\end{center}
 \caption{Best-fit parameters for the ratios of lattice spacings, $\Lambda^{\overline{\rm MS}}$ 
and the gluon condensate. To compare with another approach to determine the lattice spacings ratios, we quote the results from \cite{BaronWT} (top). Effect from the order in perturbation theory of the Wilson coefficient (middle). 
Effect from the truncation of the perturbative series in $\alpha^{\rm pert}_T$ (bottom).}
\label{tab:fitalphanf2}
\end{table}
-- truncation of the perturbative series for $\alpha^{\rm pert}_T$; we have repeated the whole analysis but limited ourselves to the three loop formula of the $\beta$ function and collected the results in the bottom part of Table \ref{tab:fitalphanf2}; we do not observe any significant impact from the perturbative truncation on the ratios of lattice spacings and, studying the discrepancy, we estimate a systematic uncertainty of roughly 7 \% for $\Lambda^{\overline{\rm MS}}$ and 13 \% for the gluon 
condensate\\
-- finite volume effects and Gribov copies; the lower bound $p_{\rm min}$ imposed on the momenta used in the fits is chosen such that
no uncontrolled finite volume effects are expected and the impressive scaling seen on Figure \ref{fig:fitalphanf2} for $\alpha_T$ makes us confident that our results are not made uncertain by Gribov copies.\\
The conversion into physical units is realised by calibrating $a_{\beta=3.9}$ from $f_\pi$ \cite{BaronWT}; the final results are:
\bea\nonumber
\Lambda^{\overline{\rm MS}} &=& \left(330 \pm 23 \pm 22 _{-33}\right) 
\times \left(\frac{0.0801 \ \mbox{\rm fm} \cdot  130.7 \ \mbox{\rm MeV} \ }{a(3.9) \ f_\pi} 
\right)  \mbox{\rm MeV}\,,\\ 
g^2(q_0^2) \langle A^2 \rangle_{q_0} &=&  \left(4.2 \pm 1.5 \pm 0.7 ^{+ ?}\right) \times 
\left( \frac{0.0801 \ \mbox{\rm fm} \cdot  130.7 \ \mbox{\rm MeV} \ }{a(3.9) \ f_\pi} 
\right)^2 \
\ \mbox{\rm GeV}^2 \ ,
\eea
where the first error is statistical, including that of the determination of $a_{\beta=3.9}$, the second and third error
correspond to the discrepancy observed by changing the order in perturbation theory of $\alpha^{\rm pert}_T$ and of 
the Wilson coefficient and the fourth error is a guestimate of the impact from the contribution of higher 
power corrections in OPE expansion, reminding that, in the case of the gluon condensate, we can only give the sign, positive, of such a correction. Using the Sommer scale $r_0$ taken from \cite{BaronWT}, we obtain $r_0 \Lambda^{\overline{\rm MS}}=0.72(5)$, in  agreement within the error with the more recent estimate by the ALPHA Collaboration (0.79(5)) \cite{FritzschWQ}, which is an update of the old result 0.62(4) \cite{DellaMorteBC} coming mainly from an improved determination of $r_0$. 

\subsection{$\alpha_s$ measurement at ${\rm N_f}=2+1+1$}

Having explored quite successfully the method of analysing the ghost-ghost-gluon vertex to measure $\alpha_s$ on the lattice with ${\rm N_f}=2$ dynamical quarks, we have extended our study to the (almost) real world in the low energy regime of the strong interaction, i.e. with 2 degenerate light quarks, the strange quark and the charm quark in the sea. This allows to compare the numerical findings with phenomenological searches, for instance at the $\tau$ mass scale, without treating charm threshold effects perturbatively. The fact that ETMC was the first collaboration to perform simulations with a dynamical charm close to the physical point is not a coincidence. Twisted-mass fermions regularised on the lattice must come in pairs in order to guarantee the ${\cal O}(a)$ improvement of physical quantities. With $\chi^h \equiv \left(\begin{array}{c} \chi^s\\ \chi^c\end{array}\right)$, the continuum action takes the form \cite{FrezzottiWZ}
\beq
S_F(\bar{\chi}^h,\chi^h) = \int d^4 x\, \bar{\chi}^h [\gamma_\mu D^\mu + m_h + i \gamma^5 \mu_h \tau^3 
+ \epsilon_h \tau^1 ]\chi^h,
\eeq
where $\mu_h$ and $\epsilon_h > 0$. Performing the rotation from the twisted basis to the physical basis
\bea\nonumber
\chi'^h = e^{i\omega_2 \tau^2/2}\chi^h |_{\omega_2=\pi/2} = \left(\frac{1+i\tau^2}{\sqrt{2}}\right)\chi^h,\\
\nonumber
\bar{\chi}'^h = \bar{\chi}'e^{-i\omega_2 \tau^2/2}|_{\omega_2=\pi/2} = \bar{\chi}^h\left(\frac{1-i\tau^2}{\sqrt{2}}\right),\\
\nonumber
\psi^h = e^{-i\omega_1\gamma^5\tau^1/2}\chi'^h, \quad
\bar{\psi}^h = \bar{\chi}'e^{-i\omega_1 \tau^1/2 \gamma^5},
\eea
\begin{table}[t]
\begin{center}
\begin{tabular}{|c|c|c|c|c|c|c|}
\hline
$\beta$ &$a\mu_l$& $m_\pi$ [MeV] & $a \mu_h$ & $a \epsilon_h$ &
$(L/a)^3\times T/a$ & \# \\
\hline
1.9 & \begin{tabular}{c}
     0.005\\
     0.004\\
     0.003\\
     \end{tabular}
    &
     \begin{tabular}{c}
     310 \\
     280\\
     250\\
     \end{tabular}
     & 0.150 & 0.190  & $32^3\times 64$ & \begin{tabular}{c}
     500 \\
     500\\
     500\\
     \end{tabular}\\
    \hline
2.1 & 0.002& 240 & 0.120 & 0.1385 & $48^3\times 96$ & 800 \\  
\hline
\end{tabular}
\end{center}
\caption{Lattice parameters of the ${\rm N_f}=2+1+1$ ETMC ensembles we used in our extraction of $\alpha_s$. $\mu_h$ and 
$\epsilon_h$ are such that the kaon and $D$ meson masses take roughly their physical value. For each ensemble, the time separation between measurements, in molecular dynamics units, is larger than the autocorrelation time of the plaquette.}
\label{tab:set-up}
\end{table}
\begin{figure}[t]
  \begin{center}
\begin{tabular}{cc}
    \includegraphics[width=0.5\textwidth]{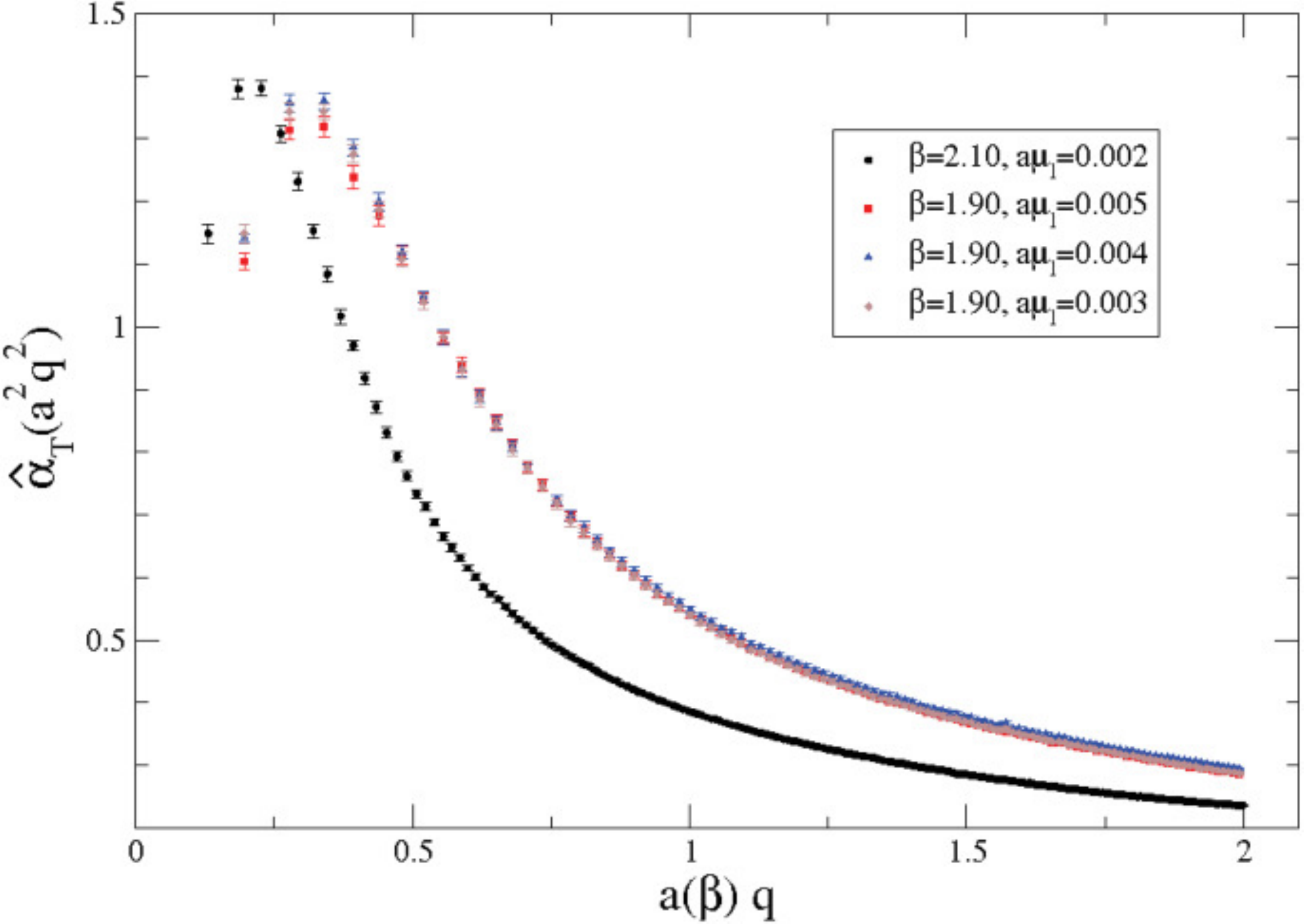}
&
    \includegraphics[width=0.5\textwidth]{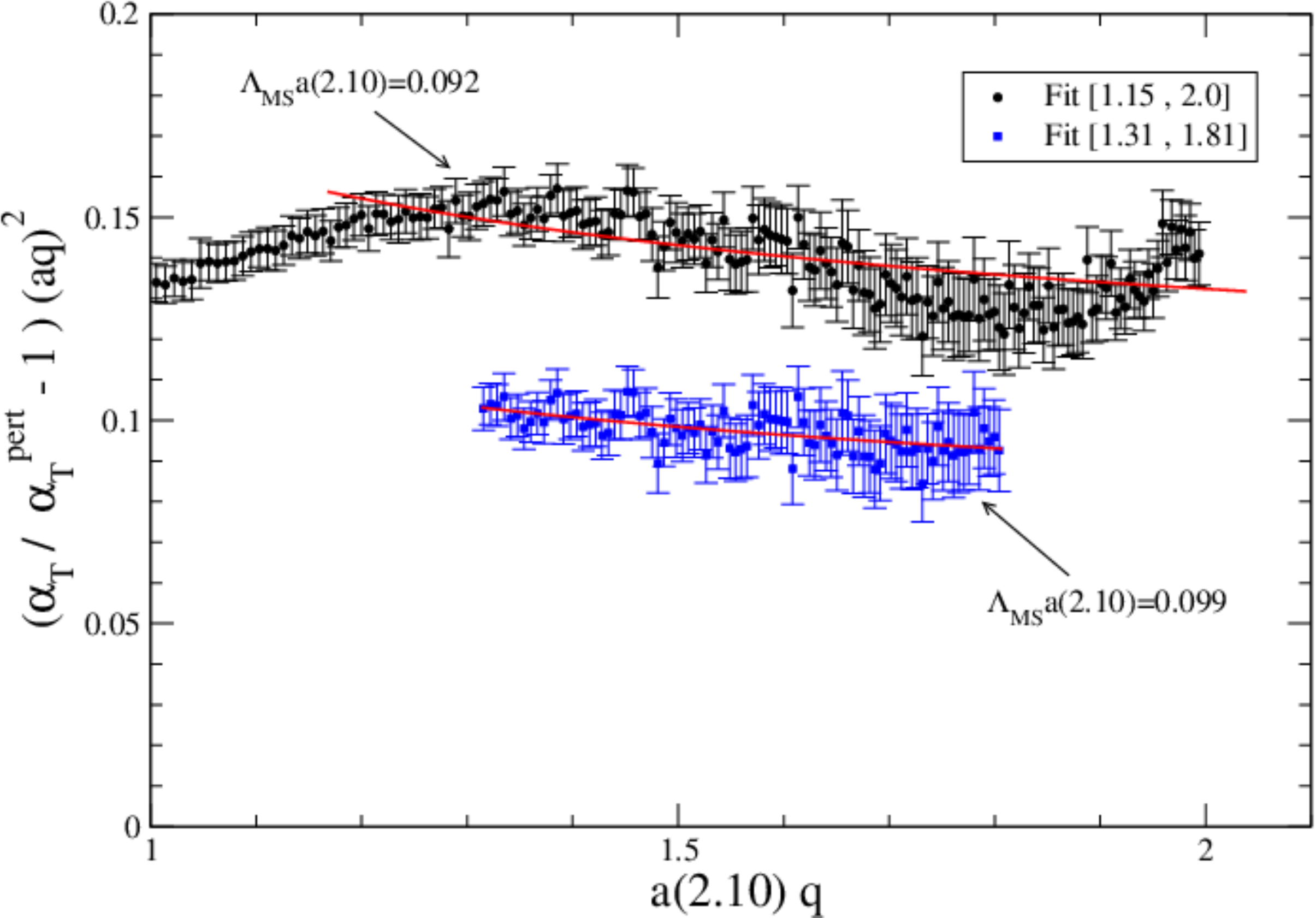}
\end{tabular}
  \end{center}
\caption{Lattice estimates of $\hat{\alpha}_T$ obtained with the ensembles collected in Table \ref{tab:set-up} after 
removing the $H(4)$ artefacts (left panel). $\left( \frac{\alpha_T(q^2)}{\alpha_T^{\rm pert}(q^2)} - 1 \right)  q^2$
estimated  with $\alpha_T$ from the lattice data at $\beta=2.10$ and its four-loop perturbative prediction evaluated with $\Lambda^{\overline{\rm MS}}$ from Table \ref{tab:fitalphanf211}; the solid red line corresponds 
to the evaluation of eq.(\ref{mona}) r.h.s., also with the best fit parameter for $g^2\langle A^2 \rangle$. 
\label{fig:alphat211}}
\end{figure}
the first part of the transformation lets the kinetic term of the action invariant while the mass term takes the form 
\beq\nonumber
m_h + i\mu_h \gamma^5 \tau^3 + \epsilon_h \tau^1 \longrightarrow m_h - i\mu_h \gamma^5 \tau^1 + \epsilon_h \tau^3.
\eeq 
With $\tan \omega_1=\mu_h/m_h$, the action reads 
\beq\nonumber
S_F(\bar{\psi}^h,\psi^h) = \int d^4 x\, \bar{\psi}^h [\gamma_\mu D^\mu + M + \epsilon_h \tau^3 ]\psi^h, 
\quad M=\sqrt{m^2_h + \mu^2_h}.
\eeq 
As we will not be interested by valence fermions, we only make the remark that, with this action, the fermionic determinant is positive under the condition $M > \epsilon_h$. The lattice regularisation is now straightforward, except that the renormalized masses are defined by 
\beq\nonumber
\mu^R_h=Z^{-1}_P \mu_h, \quad \epsilon^R_h=Z^{-1}_S \epsilon_h,
\eeq 
and, with 
\beq\nonumber
\mu^R_c = \mu^R_h + \epsilon^R_h=Z^{-1}_P(\mu_h + Z_P/Z_S \epsilon_h), \quad \mu^R_s=\mu^R_h - \epsilon^R_h=Z^{-1}_P(\mu_h - Z_P/Z_S \epsilon_h),
\eeq 
the positivity of the fermionic determinant occurs if 
\beq\nonumber
\frac{Z_P}{Z_S}> \frac{\mu^R_c - \mu^R_s}{\mu^R_c + \mu^R_s}.
\eeq 
The total lattice action $S=S_g + S^{(2)}_F(\bar{\chi}^l, \chi^l,\mu_l)+S^{(1+1)}_F(\bar{\chi}^h, \chi^h,\mu_h,\epsilon_h)$ includes the discretized Yang-Mills action, the action of a light doublet of degenerate twisted-mass fermions and the action of a non-degenerate doublet of strange and charm twisted mass fermions. The tuning at maximal twist is realised by searching, \emph{for each} $\mu_l$, $\mu_h$ \emph{and} $\epsilon_h$, the critical $\kappa$. This strategy is a bit different from what ETMC had proposed at ${\rm N_f}=2$: it gives more freedom in the exploration of the parameter space, if for instance the conclusion is drawn that lighter masses have to be simulated to get reliable chiral extrapolations. Due to the reinforcement of the Singleton-Sharpe phase transition with the increased number of dynamical quarks \cite{ChiarappaAE}, the choice of the Iwasaki gauge action \cite{IwasakiWE}
\beq\nonumber
S_g = \frac{\beta}{6} \bigg(b_0 \sum_{x,\mu\neq\nu} 
\textrm{Tr}\Big(1 - P^{1 \times 1}(x;\mu,\nu)\Big) + b_1 \sum_{x,\mu\neq\nu} \textrm{Tr}
\Big(1 - P^{1 \times 2}(x;\mu,\nu)\Big)\bigg)\,,
\eeq
$b_1=0.331$, $b_0 = 1 - 8 b_1$, has been made: then, one observes a smoother dependence of phase transition sensitive quantities on the bare quark mass than the tree-level-improved Symanzik gauge action employed at ${\rm N_f}=2$.\\
We have collected the parameters of our runs in Table \ref{tab:set-up}; we have performed the analysis in several steps, with a first milestone put to prove that our method does work on the ${\rm N_f}=2+1+1$ sets \cite{BlossierTF}, \cite{BlossierEF} as well as it did at ${\rm N_f}=2$ \cite{BlossierKY}, before being able to accumulate enough statistics in order to track the systematics \cite{BlossierIOA}. We have plotted in the left panel of Figure \ref{fig:alphat211} the Taylor coupling constant, after the removal of $H(4)$ artefacts that we described in the previous subsection. We obtain pretty smooth curves that we can nicely fit with OPE formulae.  We have decided to exploit the high statistics (800 gauge configurations) of the available data
for the lattice ensemble $\beta=2.1$, where the effect of a finite $a \mu_l$ can be neglected, in order to extract precisely the physical 
running of $\alpha_T(q^2)$. including ${\cal O}(1/q^2)$ higher-power OPE corrections 
and ${\cal O}(a^2q^2)$ lattice artefacts. Then, one needs to identify the appropriate window of momenta, large enough for 
the running behaviour not to be polluted by higher OPE powers but small enough so that it is not affected 
by higher-order discretization effects. At this stage, because we are concentrated to a single set of bare parameters, we can skip 
the question of the absolute lattice calibration: it will be put back to a further stage of the work. So, all the fitted dimensionful 
parameters will be expressed in units of the lattice spacing $a_{\beta=2.1}$ and, in particular, the perturbative running that we express in function of $t=\ln(a^2_{\beta=2.1}q^2/a^2_{\beta=2.1}\Lambda^2_T)$:
\beq
\widehat{\alpha}_T(a^2q^2)=\alpha^{\rm pert}_T(a^2q^2)\left(  1 + \frac{9}{a^2q^2} 
W^{\rm MOM}(q,q_0) \frac{g^2_T(q_0^2) \langle a^2A^2 \rangle_{R,q_0^2}} {4 (N_C^2-1)}
\right) + C_{a^2q^2} a^2q^2\,.
\eeq
With the ${\rm N^3LO}$ formula of the Wilson coefficient $W^{\rm MOM}$, a fit of the data having  lattice 
momenta above $a_{\beta=2.1} q \simeq 1.15$ is done, whose results are collected in Table \ref{tab:fitalphanf211}. The optimal fit
is obtained by shifting the upper and lower bounds of the fitting window, whose the size has been fixed between 
0.4 $a^{-1}_{\beta=2.1}$ and 0.8 $a^{-1}_{\beta=2.1}$: the "optimal" window is $1.31 < a_{\beta=2.1} q < 1.81$. The normalisation
point is $a_{\beta=2.1} q_0=2.92$, $q_0 \sim 10$ GeV. 
\begin{table}[t]
\begin{center}
\begin{tabular}{|c|c|c|c|}
\hline
$[a_{\beta=2.1}q_{\rm min},\, a_{\beta=2.1}q_{\rm max}]$& $\Lambda^{\overline{\rm MS}} a_{\beta=2.1}$ & $g^2 \langle A^2 \rangle a^2_{\beta=2.1}$ & $c_{\rm a2p2}$ \\
\hline
[1.15,2.0] & 0.092(5) & 0.39(11) & -0.0049(8) \\
\hline
[1.31,1.81] & 0.099(3) & 0.26(7) & -0.0066(4)\\
\hline
\end{tabular}
\end{center}
\caption{Fit parameters for the two different fitting windows described in 
the text: $a_{\beta=2.1}q > 1.15$ (first row) and the optimal fit window (second row). The errors that we indicate
are statistical and are obtained by jackknife.\label{tab:fitalphanf211}}
\end{table}
The continuum term $\alpha_T$ can be reexpressed as
\beq\label{mona}
\left( \frac{\alpha_T(q^2)}{\alpha_T^{\rm pert}(q^2)} - 1 \right)  q^2 =  
9 W^{\rm MOM}(\alpha^{\rm pert}_T(q^2),\alpha^{\rm pert}_T(q_0^2)) 
\frac{g^2_T(q_0^2) \langle A^2 \rangle_{R,q_0^2}} {4 (N_C^2-1)}\ .
\eeq
up to terms that vanish at large $q^2$. One can compute the l.h.s. of that equation with $\alpha_T$ extracted numerically 
and its perturbative four-loop prediction with the best fit parameters 
for $\Lambda^{\overline{\rm MS}}$. The r.h.s. of the equation corresponds to the running of the Wilson coefficient.  
This is shown in the right panel of Figure \ref{fig:alphat211}: it is obvious that a nonzero nonperturbative 
contribution appears, as the OPE analysis predicts and as we already discussed in the ${\rm N_f}=2$ case, although 
a systematic deviation between the data and the analytical curves can be noticed from 
the expected behaviour in the case of the fit over every momenta above $a_{\beta=2.1} q > 1.15$. It is an indication that higher order
power corrections are certainly present in the lower part of the momenta values. especially $a_{\beta=2.1}q \lesssim~1.2$. 
Corrections to (\ref{alphahNPb}) should be incorporated through condensates of higher-order local operators and are very
highly suppressed at $a_{\beta=2.1}q\sim 2$. To identify the dominant next-to-leading contribution, 
lattice data deviations from eq.(\ref{alphahNPb}) are shown in Figure \ref{fig:HOPE} in terms of the momenta, using for both 
axes logarithmic scales. We observe that the lattice data follow a logarithmic slope of $\sim 6.15$, after the subtraction of 
eq.(\ref{alphahNPb}) with the best fit parameters corresponding to the optimal window, for momenta
$0.5 \lesssim a_{\beta=2.1}q \lesssim 1.3$. 
\begin{figure}[t]
  \begin{center}
    \includegraphics[width=0.5\textwidth]{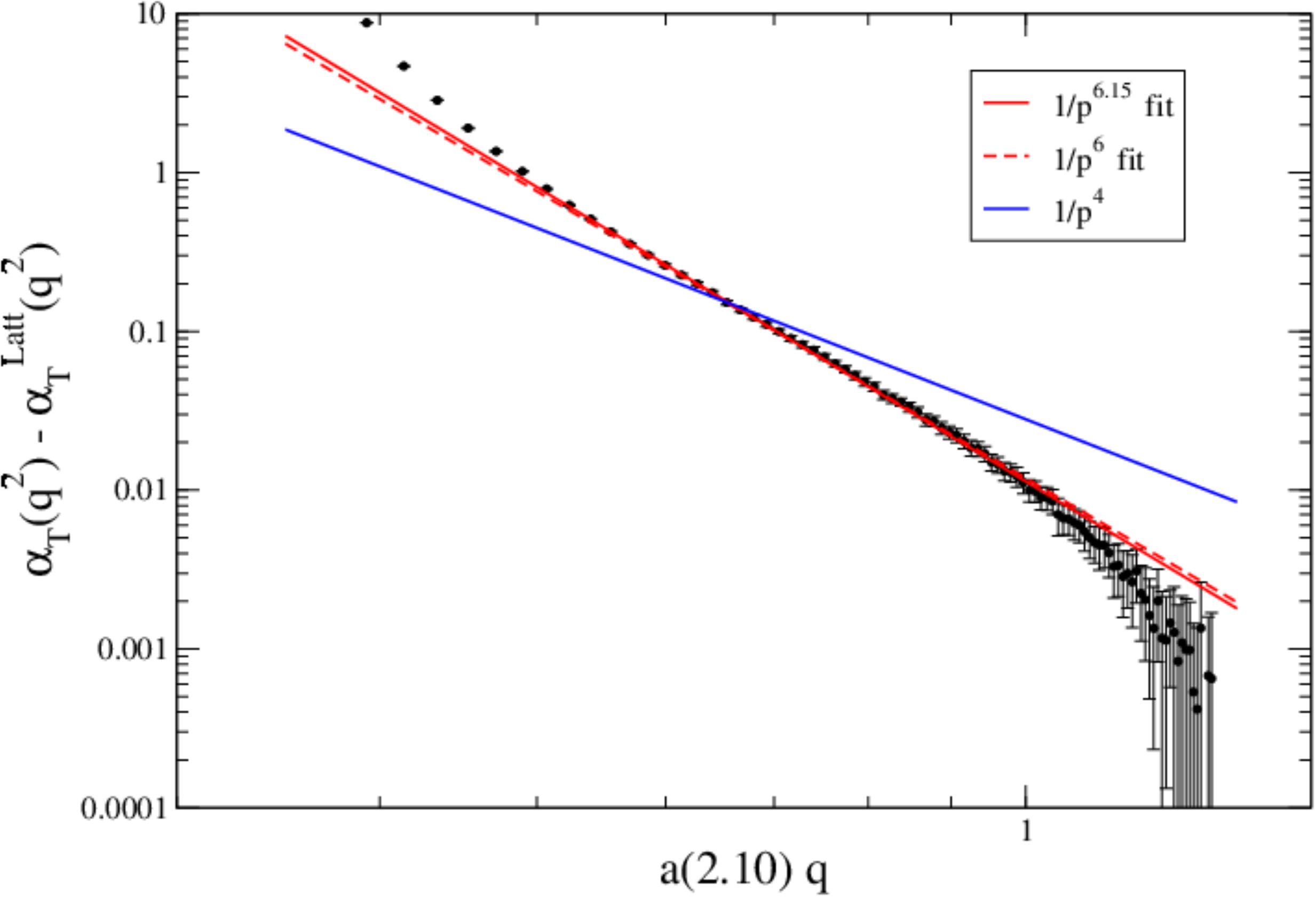}
  \end{center}
\caption{Lattice data deviations from eq.(\ref{alphahNPb}) with the best fit parameters for the optimal window 
in Table \ref{tab:fitalphanf211}. Logarithmic scales are applied for the coordinates in both axes. A best fit for 
the logarithmic slope of 6.15 is shown in red solid 
line as well as fits with slopes fixed to 6 (red dashed) and 4 (blue). \label{fig:HOPE}}
\end{figure}
The logarithmic slope is consistent with what we obtained at a very early stage of the ${\rm N_f=2+1+1}$ program, with a 
much lower statistics: a $1/p^6$-correction was incorporated to describe successfully the 
lattice data all over a large momenta window roughly ranged between 1.75 GeV and 7 GeV. Such a $1/p^6$ correction 
might not be necessarily explained by the dominance of a condensate of dimension six but 
might be rather an effective power, interplay of a next-to-leading power correction and its Wilson coefficient 
or by a different nonperturbative mechanism being dominant at these low-momenta. The main conclusion here is that 
eq.(\ref{alphahNPb}), with the addition of a $1/p^x$ contribution, $x\sim 6$, should account for the running of lattice data 
also at very low momenta. Thus, we can write
\beq\label{eq:1overpx}
\widehat{\alpha}_T(a^2 q^2) = \alpha^{\rm A2}_T(q^2) +  \frac{d_x}{q^x}  +  c_{a2q2}  a^2 q^2  + {\cal O}(a^2) \ ,
\eeq
where the physical running of the coupling is now given by 
\beq\label{eq:atA2}
\alpha_T(q^2) \ = \ \alpha_T^{A2}(q^2) \ + \ \frac{d_x}{q^x},
\eeq
and $\alpha^{A2}_T$ is still given by eq.(\ref{alphahNPb}). Once ${\cal O}(a^2q^2)$ cut-off effects have been taken into account as well as higher order cut-off effects by a completely empirical formula $c^{(0)}_{aq} aq \sin(c^{(1)}_{aq} aq)$, $c^{(0)}_{aq} \sim 0.007$ and $c^{(1)}_{aq}\sim 2.1 \pi$, and the higher power correction to the OPE, we are in a good shape to check the running 
of the leading OPE Wilson coefficient over a large window $a_{\beta=2.1}q > 0.5$. 
So, eq.(\ref{mona})'s l.h.s. can be plotted in terms of $a q$, as in Figure \ref{fig:NPtot}, with $\alpha_T$ replaced by $\alpha^{A2}_T$ obtained from the lattice data and compared with the theoretical expression given in eq.(\ref{mona})'s  r.h.s. Remarkably, we observe a striking agreement between the analytical fomula and the numerical data for momenta $a_{\beta=2.1}q \gtrsim 0.5$.
\begin{figure}[t]
  \begin{center}
    \includegraphics[width=0.5\textwidth]{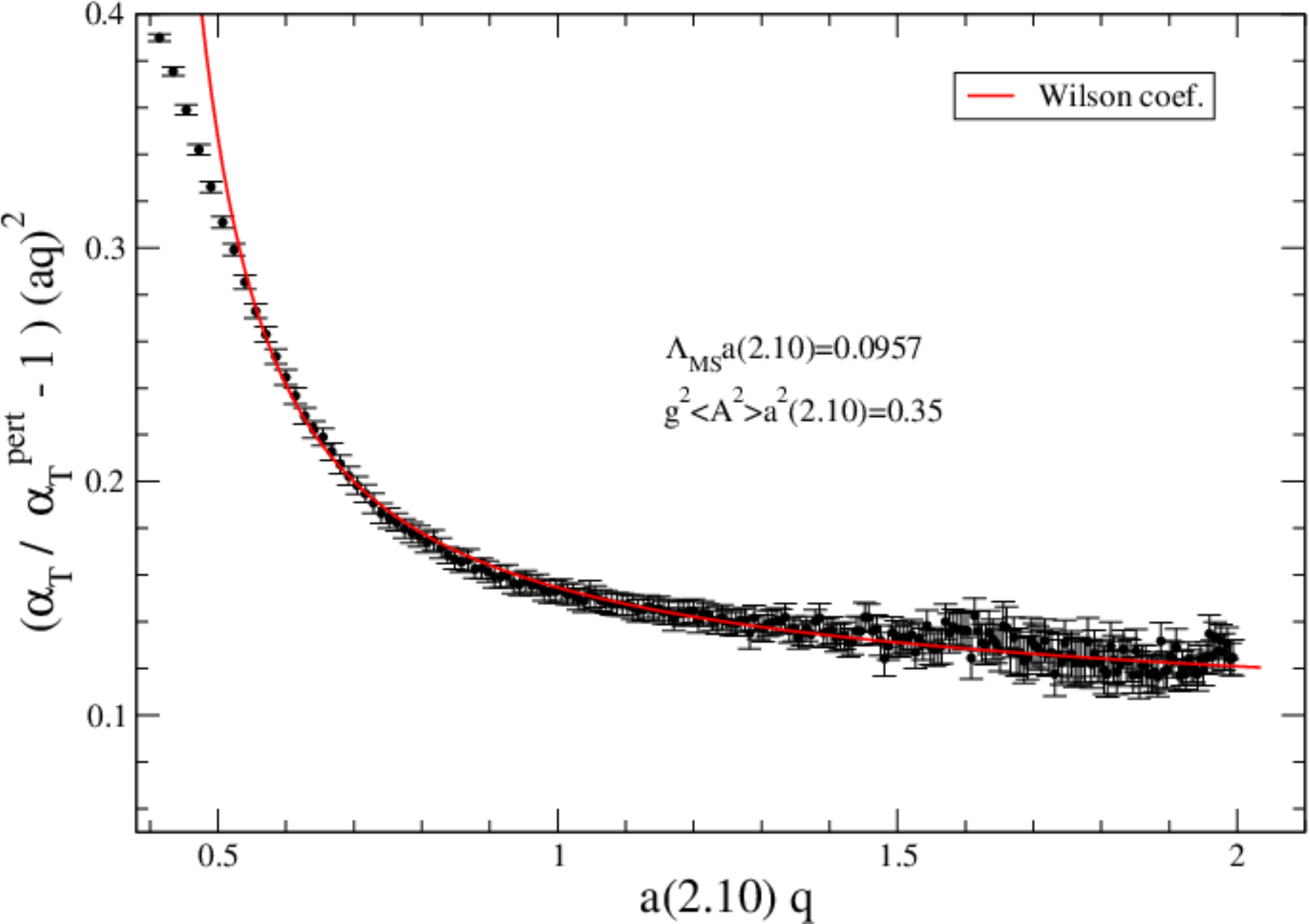}
  \end{center}
\caption{$a^2_{\beta=2.1}q^2 (\alpha_T/\alpha^{\rm pert}_T -1)$ with $\alpha_T$ resulting from the elimination  
of lattice artefacts and the higher-order power correction; the solid red line 
corresponds to the running predicted for the Wilson coefficient of the leading OPE power correction. 
\label{fig:NPtot}}
\end{figure}
\\
\noindent Taking advantage of the ``scaling" for the running of the 
physical coupling at different $\beta$'s, we can analyse the three other 
ensembles of data at $\beta=1.9$. Indeed, the running coupling is a renormalized quantity that does not depend on 
the regularisation scheme nor the regularisation cut-off. Once cured from lattice artefacts, 
the resulting coupling expressed in terms of the physical momentum should be independent of $\beta$ and one expects
a slight dependence on the mass of the active quark flavours, in particular if the momentum is far above the quark mass thresholds. We will assume that any sea mass dependence is properly captured by the lattice spacing, through a small shift in momenta, and we write, as in the previous subsection,
\beq\label{eq:scaling}
\widehat{\alpha}^{(\beta,\mu_l)}_T(a^2_{\beta,\mu_l}q^2) -  c_{a2q2}  a^2_{\beta,\mu_l}q^2 
 \equiv 
\widehat{\alpha}^{(\beta_0,\mu^0_l)}_T(a^2_{\beta_0,\mu^0_l}q^2)  -  c_{a2q2}  a^2_{\beta_0,\mu^0_l}q^2,
\eeq
for any bare coupling parameters, $\beta$ and $\beta_0$, and light flavour masses, $\mu_l$ and $\mu^0_l$. With $\beta_0=2.1$, $\mu^0_l=0.002$ and, replacing eq.(\ref{eq:scaling})'s r.h.s. by the continuum running of $\alpha_T$, we write
\bea\nonumber
\widehat{\alpha}^{(\beta,\mu_l)}_T\left( \left(\frac{a_{\beta,\mu_l}}{a_{\beta_0,\mu^0_l}}\right)^2 k^2_L \right) - c_{a2q2}  \left(\frac{a_{\beta,\mu_l}}{a_{\beta_0,\mu^0_l}}\right)^2 k^2_L
&= &  \alpha_T(k^2_L;\Lambda^{\overline{\rm MS}}a_{\beta_0,\mu^0_l},g^2\langle A^2\rangle a^2_{\beta_0,\mu^0_l}) \\
\label{scaling2}&+& \frac{d_x}{(a_{\beta_0,\mu^0_l} k_L)^x},
\eea
where $k_L=a_{\beta_0,\mu^0_l}q$ is the momentum in $a_{\beta_0, \mu^0_l}$ lattice spacing units. 
In the r.h.s. of this equation the parameters $\Lambda^{\overline{\rm MS}}$ and $g^2\langle A^2 \rangle$ 
used for $\alpha_T$ are taken from the analysis at $\beta=2.1$. 
Concerning its l.h.s., $\widehat{\alpha}_T$ can be obtained from a lattice simulation at any $\beta$ and $\mu_l$, after 
properly shifting the lattice momentum $a q$ to be left with the physical momentum expressed in units of $a^{-1}_{\beta_0,\mu^0_l}$. We perform the same exercice as for the ${\rm N_f}=2$ case of extracting ratios $a_{\beta_0,\mu^0_l}/a_{\beta,\mu_l}$ and  coefficients $c_{a2q2}(a_{\beta,\mu_l})$ with fits done over a region $k_L > 0.5$: the difference with the ${\rm N_f}=2$ study is that we let $c_{a2q2}(a_\beta,\mu_l)$ free.  
We collect the fit results in Table \ref{tab:matching}: we obtain fits of high-quality where the coefficients $c_{a2p2}(a_{\beta, \mu_l})$ are fully compatible with each other, and together with that obtained at $\beta=2.1$. The reasonable consistency of those coefficients, with respect to the fact that they contain higher order cut-off effects than those of ${\cal O}(a^2q^2)$ only, confirms the reliability of our ${\rm N_f}=2$ and ${\rm N_f}=2+1+1$ results.
\begin{table}[t]
\begin{center}
\begin{tabular}{|c|c|c|c|}
\hline 
$\mu_l$ & $a_{\beta=2.1,\mu_l=0.002}/a_{\beta=1.9,\mu_l}$ & $c_{a2q2}$ & $\chi^2/$d.o.f. 
\\
\hline
0.003 &  0.6798(74) & -0.0076(6) & 20.2/82 
\\
\hline 
0.004 & 0.6683(72) & -0.0067(6) & 14.4/82
\\
\hline 
0.005 & 0.6775(73) & -0.0081(6) & 45.7/82
\\ 
\hline
\end{tabular}
\end{center}
\caption{Best fit parameters obtained with eq.(\ref{scaling2}) and the data of our three ensembles at $\beta=1.9$;
the fitting momenta window is defined by $k_L>0.5$. \label{tab:matching}}
\end{table}
Looking more carefully the sea mass dependence of the lattice spacing ratios, we have tried to perform a chiral extrapolation: either
in $a^2 \mu^2_l$ or in $a \mu_l$, the latter being allowed because of the spontaneous breaking of chiral symmetry. We have found that the slopes are compatible with 0 in both cases and we quote, for the quadratic extrapolation in $a \mu_l$, the result
\beq\label{eq:Phy}
a_{\beta=2.1,\mu_l=0.002} =  0.677(13)  a_{\beta=1.9,\mu_l=0}  =  0.0599(27)\,  \mbox{\rm fm}\,,
\eeq
where we used the recent lattice calibration $a_{\beta=1.9,\mu_l=0}=0.0885(36)\ {\rm fm}$ \cite{CarrascoLAA} obtained by ETMC
through chiral fits in the pion sector: $f_\pi$ is again the quantity employed to convert the dimensionful results in physical units.
The lattice estimates of $\widehat{\alpha}_T$ from simulations at $\beta=2.1$ and $\beta=1.9$ are put in the common plot
of Figure \ref{fig:Phy} in function of momentum in physical units: data nicely fall on the same universal curve and we have indicated the fit results in physical units in Table \ref{tab:Phy}.
\begin{table}[t]
\begin{center}
\begin{tabular}{|c|c|c|c|}
\hline
fit window [GeV] & $\Lambda^{\overline{\rm MS}}$ [MeV] & $g^2 \langle A^2 \rangle$ [GeV$^2$]& $(-d_x)^{1/x}$ [GeV] \\
\hline
[4.3,6.0] & 324(18) & 2.8(8) & \\
\hline
[1.7,6.6] & 314(16) & 3.8(6)  & 1.61(7)\\
\hline
\end{tabular}
\end{center}
\caption{Fitted parameters of $\alpha_T$ expressed in physical units. The error estimates include the uncertainty on the lattice spacing determination given in eq.(\ref{eq:Phy}). \label{tab:Phy}}
\end{table}
\begin{figure}[t]
  \begin{center}
  	\begin{tabular}{cc}
    \includegraphics[width=0.5\textwidth]{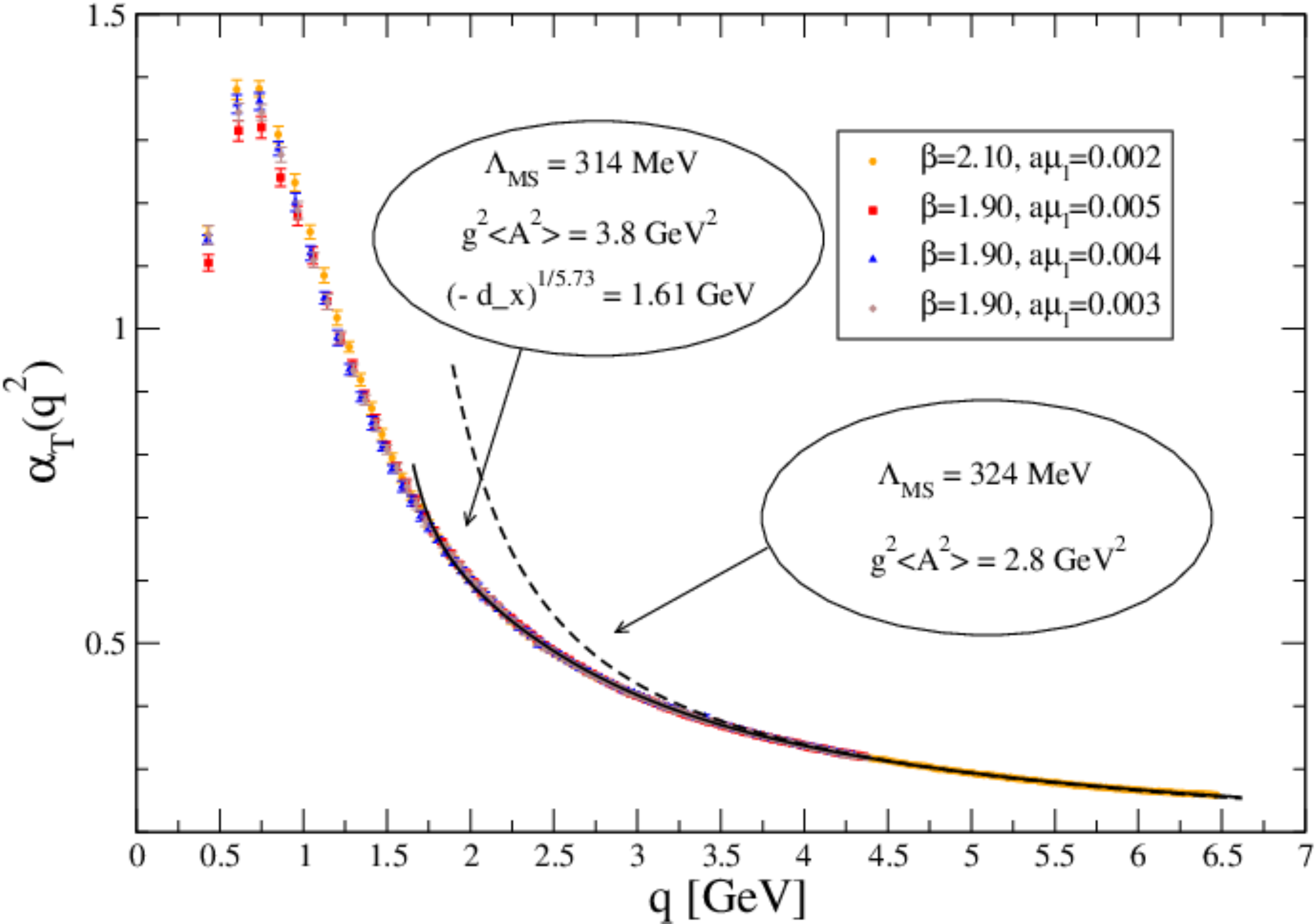} & 
    \includegraphics[width=0.5\textwidth]{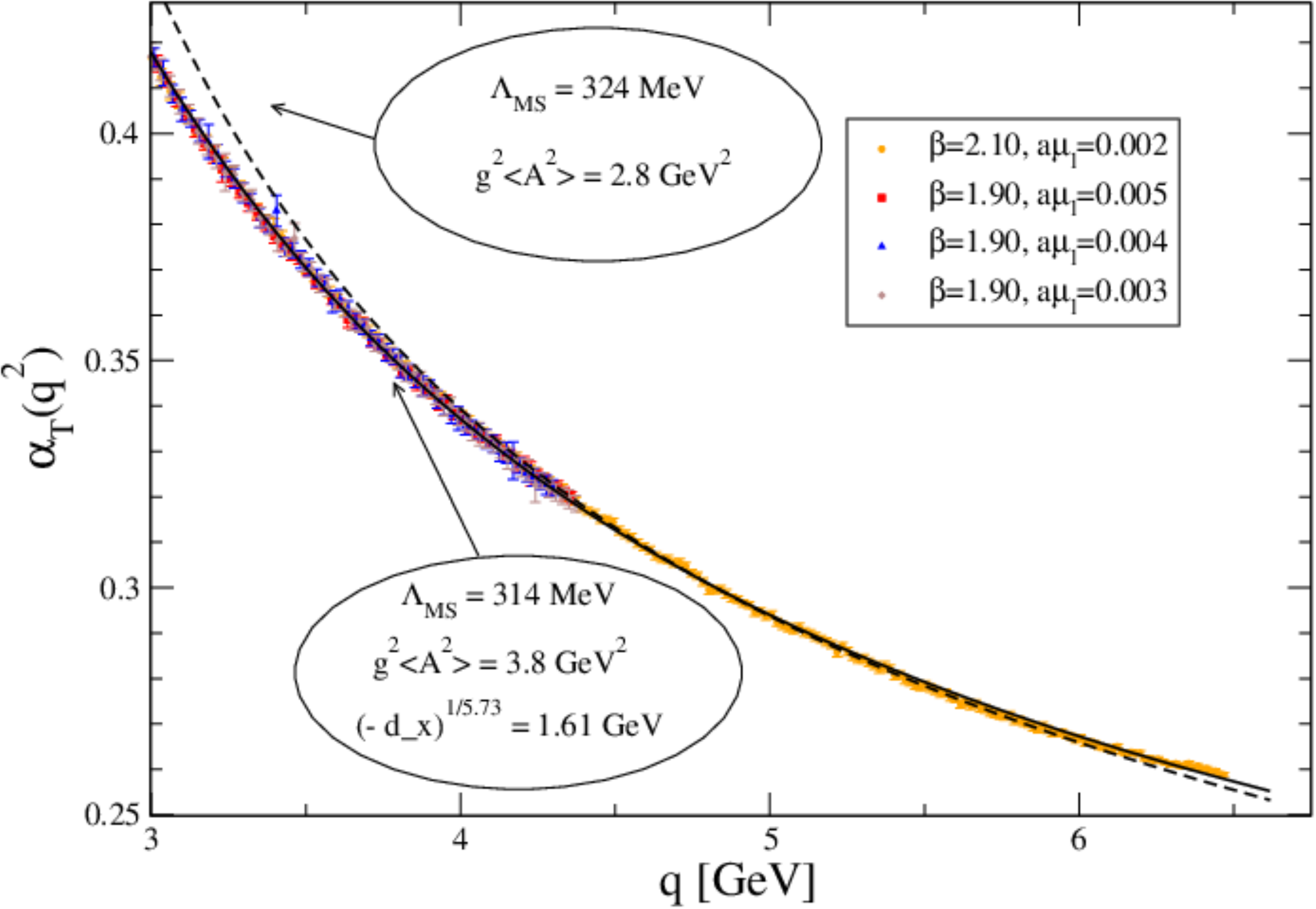}
	\end{tabular}	
  \end{center}
\caption{$\alpha_T$ obtained from the lattice after having removed 
$O(4)$-breaking ($H(4)$-extrapolation) and $O(4)$-invariant artefacts; the curves 
are fitted to the physical running defined by eq.(\ref{eq:atA2}) with the parameters collected in Table \ref{tab:Phy}, 
the solid (dashed) black line refers to the large (small) fitting window (left panel). The right plot is  
a zoom, focussing on the large momenta region, of the left one (right panel).
\label{fig:Phy}}
\end{figure}
We obtain finally
\beq\label{eq:LMS}
\Lambda^{\overline{\rm MS}}_{{\rm N_f}=2+1+1} =  314(7)(14)(10) \times \left( \frac{0.0599 \ \mbox{\rm fm}}{a_{\beta=2.1,\mu_l=0.002}} \right) 
 \ \mbox{\rm MeV},
\eeq
where we take the central value from the fit including a higher-order power correction in the OPE. The error budget is composed as\\
-- 2\% from statistics,\\
-- 4\% from the uncertainty on the lattice spacing calibration given in eq.(\ref{eq:Phy}),\\
-- 3\% of systematics estimated by the difference with the value obtained with the fit over 
the small window, without including the higher-order power correction.\\
We can run the coupling down to the $\tau$ mass scale, 
below the bottom quark mass threshold,  and compare the result with the estimate from 
$\tau$ decays that we have collected in the first section of this chapter: $\alpha_s^{\overline{\rm MS}}(m_\tau^2)=0.330(14)$. 
This gives, propagating the error by 
\beq\nonumber
\sigma^2\left(\alpha_s^{\overline{\rm MS}}(q^2)\right) = 4 \beta^2_{\overline{\rm MS}}\left(\alpha_s^{\overline{\rm MS}}(q^2)\right) \ \frac{\sigma^2\left(\Lambda^{\overline{\rm MS}}\right)}{\Lambda^{2\,\overline{\rm MS}}} 
\eeq
for each error contribution in eq.(\ref{eq:LMS}), the result at the $\tau$-mass scale
\beq\label{eq:amtau}
\alpha_s^{\overline{\rm MS}}(m_\tau^2) = 0.336(4)(8)(6),
\eeq
in good agreement with the one from $\tau$ decays. The three errors are of the same nature as in (\ref{eq:LMS}).
To compute $\alpha_s^{\overline{\rm MS}}(q^2)$ at the $Z^0$ mass scale, we run first 
the coupling up to the $\overline{\rm MS}$ scheme bottom mass $m_b$, 
with $\beta$-coefficients and $\Lambda^{\overline{\rm MS}}$ extracted for 4 quark flavours, 
then apply the matching formula~\cite{BeringerZZ}
\beq\nonumber
\alpha_s^{\overline{\rm MS}\,{\rm N_f}=5}(m_b) = \alpha_s^{\overline{\rm MS}\,{\rm N_f}=4}(m_b) 
\left( 1 + \sum_n c_{n0} \left(\ \alpha_s^{\overline{\rm MS}\,{\rm N_f}=4}(m_b)\right)^n \right),
\eeq
where the coefficients $c_{n0}$ are taken from \cite{ChetyrkinIA}, and 
finally run from the bottom mass up to the $Z^0$ mass scale with $\beta$-coefficients at
${\rm N_f}=5$. We get
\beq\label{eq:amz0}
\alpha_s^{\overline{\rm MS}}(m_Z^2)=0.1196(4)(8)(6)\,,
\eeq
where we have again propagated all the error contributions from eq.(\ref{eq:LMS}). This result is compatible with the last lattice results averaged by PDG \cite{BeringerZZ}, $0.1185(6)$, and with its world average without including lattice results, 
$0.1183(12)$. It is the update of what was reported in an exploratory study, but now with a much higher statistics for our sample of gauge field 
configurations and a calibration of the lattice spacing based on a more complete analysis of the ETMC available data. \emph{Apart 
that output, our method is particularly appropriate because it allows to make a lot of cross-checks: control of cut-off effects, sea mass dependence, power corrections in OPE analysis, and test of the perturbative running of $\alpha_s$ and of a Wilson coefficient}. The Schr\"odinger Functional scheme with 4 degenerate massless quarks offers the same opportunity concerning the running of $\alpha_s$ and the cure of cut-off effects \cite{Tekin2010} but the price to pay is an important numerical effort in tuning the bare parameters, especially to be at the point $\bar{g}^2(L_{\rm max})$, with $1/L_{\rm max}$ above the charm threshold: still, it is necessary to obtain the lattice calibration from an ${\rm N_f}=2+1+1$ simulation.

\section{Back to phenomenology}

We have collected the lattice results of $\alpha_s(m_Z)$ in Figure \ref{fig:latticealpha}, together with partial averages performed by
PDG over the different phenomenological extractions we discussed at the beginning of this chapter. Clearly the numerical simulation computations are much more precise than all the other ones and, in our opinion, it will be tough to reduce much below the \% level the error. Indeed, an input in every lattice calculations is the calibration and, from our own experience with the ALPHA and ETM collaborations and from what is quoted in the literature, a $\sim$ 2\% of uncertainty in $a^{-1}$ through the decay constant $f_\pi$, the nucleon mass $m_N$ or the $\Omega$ mass $m_\Omega$ is almost irreducible \cite{SommerMEA}. Then, from a naive dimensional analysis, it seems to us quite hard to reach a better precision than $\sim$ 1\% on $\alpha_s(m_Z)$: using quite sophisticated statistical tools reduce the statistical error to $\sim$ 0.5\% but, then, one has to deal with systematics as well, which is most probably also $\sim$ 0.5\%. It is expected that in a next future, several groups will put an important effort to come with more ${\rm N_f}=2+1+1$ estimates. Let's outline again that the landscape of lattice evaluations is spanned by complementary approaches (hadronic schemes, Green functions of gluon and ghost) that also shed light on the running. Numerics is in very nice agreement with perturbation theory in the ultraviolet regime and, in the more infrared region, it has confirmed that a proper treatment of correlation functions within OPE framework needs to introduce power corrections. Once efficient tools are used to eliminate lattice artefacts, it is even possible to make visible the running of the corresponding Wilson coefficients.\\ 
In conclusion of that discussion, it seems legitimate to take the spread over lattice values $\Delta \alpha^{\rm latt}_s \sim \pm  0.002$ as the theoretical uncertainty $\Delta \alpha^{\rm th}_s$ on the gluon-gluon fusion Higgs boson production cross section.

\begin{figure}[t]
\begin{center}
\includegraphics[width=0.8\textwidth]{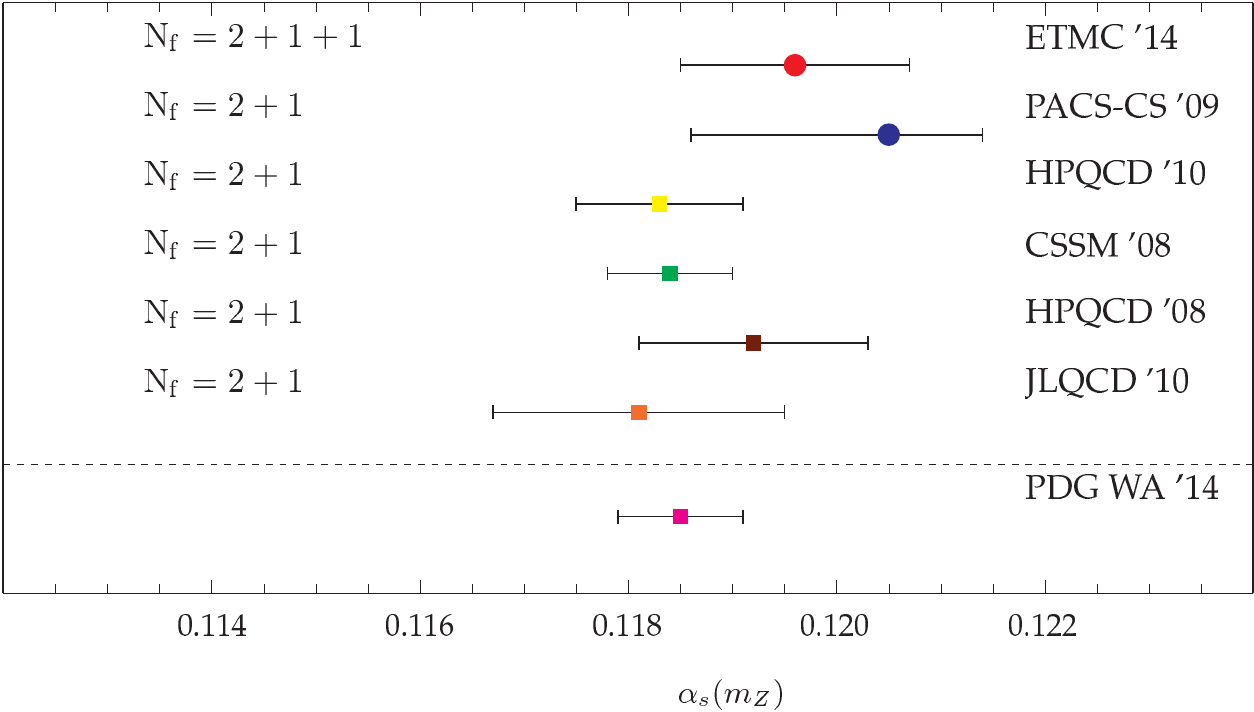}

\vspace{1cm}
\includegraphics[width=0.5\textwidth]{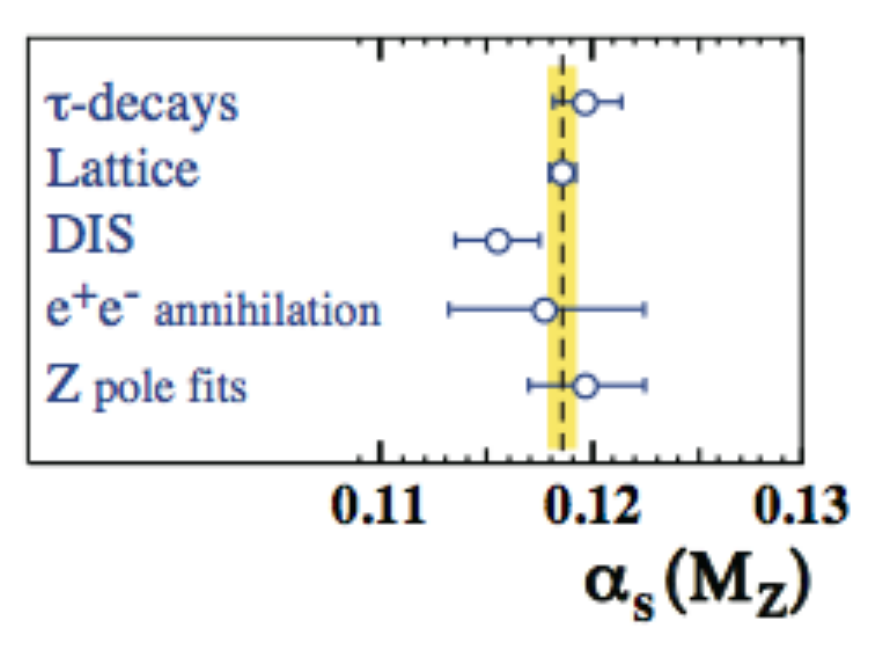}
\end{center}
\caption{Collection of lattice results and PDG average of $\alpha_s(m_Z)$ (top). Partial averages on $\alpha_s$
performed by PDG within each approach discussed in this chapter (bottom). \label{fig:latticealpha}}
\end{figure}



\chapter*{Conclusion and perspectives}

\fancyhead[LO]{\bfseries \leftmark}
\fancyhead[RE]{\bfseries \rightmark}

\noindent We have presented in that report what we have done during the recent past years to provide 
phenomenologists with inputs of the Standard Model in the quark sector: quark masses, strong coupling constant.
They are intimately related to Higgs boson physics, either in the way to produce that particle or in the various decay channels.
Quenched simulations as well as ensembles with ${\rm N_f}=2$ and ${\rm N_f}=2+1+1$ dynamical quarks have been considered.
Different fermionic and gauge regularisations, as well as effective field theory framework, have been used.
Of course the door is still open to some improvements with respect to our effort, for instance measuring the quark masses on configurations
with ${\rm N_f}=2+1+1$ dynamical quarks. Nowadays subleading effects are under investigation by the lattice 
community as well, like the strong and electromagnetic breakings of isospin symmetry: we recall that the stability of the proton
is assured by a large enough difference between down and up quark masses to forbid the capture of an electron by a proton to form a neutron. We had the great opportunity to be involved in the development of tools and techniques that will be precious in future research works, for instance the use of generalized eigenvalue problems to deal with excited states. We did not say any word about heavy flavour phenomenology, that is still, however, one of our favorite research topics. Exploiting the step scaling in masses and twisted boundary conditions are particularly welcome to study heavy-heavy and heavy-light transitions and extract form factors in a wide zone of the physical region. Comparing lattice results with phenomenological models is a deep motivation because it is a way to better understand the confinement mystery. Supporting experimentalists in their task of carefully analysing data and controlling the systematic and theoretical errors, for instance when they have to face with $S$ wave or resonant states, is a source of active thoughts: it is necessary to learn about multihadron states. Let's mention again that lattice QCD is highly relevant in the quest of New Physics in the flavour sector. 
\chapter*{Publications}

\section*{Lattice QCD and flavour physics}

-- D.~Be\'cirevi\'c, B.~Blossier, Ph.~Boucaud, G.~Herdoiza, J.P.~Leroy,A.~Le~Yaouanc,
V.~Mor\'enas, O.~P\`ene, "\emph{Lattice measurement of the Isgur-Wise functions $\tau_{1/2}$ and $\tau_{3/2}$}",
Phys. Lett. B {\bf 609}, 298 (2005) [hep-lat/0406031].\\
-- D.~Be\'cirevi\'c, B.~Blossier, Ph.~Boucaud, V.~Gimenez, V.~Lubicz, F.~Mescia,
S.~Simula and C.~Tarantino, "\emph{Non-perturbatively renormalized light quark masses from a lattice 
simulation at ${\rm N_f}=2$}", Nucl. Phys B {\bf 734}, 138 (2006) [hep-lat/0510014].\\
-- B.~Blossier, A.~Le~Yaouanc, V.~Mor\'enas and O.~P\`ene, 
"\emph{Lattice renormalization of the static quark derivative operator}", 
Phys. Lett. B {\bf 632}, 319 (2006) [hep-lat/0507024].\\
-- B.~Blossier, "\emph{Lattice renormalisation of ${\cal O}(a)$ improved heavy-light operators}", 
\Journal{\PRD}{76}{114513}{2007} [arXiv:0705.4574].\\
-- I.~Bigi, B.~Blossier, A.~Le Yaouanc, L.~Oliver, A.~Oyanguren, O.~P\`ene, P.~Roudeau 
and J.~C.~Raynal, "\emph{Memorino on the `1/2 vs. 3/2 Puzzle' in $\bar B \to l \bar \nu X_c$ -- a Year Later and 
a Bit Wiser}", \emph{Eur. Phys. J} {\bf C 52}, 975 (2007) [arXiv:0708.1621].\\
-- B.~Blossier {\it et al.}  [European Twisted Mass Collaboration],
"\emph{Light quark masses and pseudoscalar decay constants from Nf=2  Lattice QCD
with twisted mass fermions}",  JHEP {\bf 0804}, 020 (2008) [arXiv:0709.4574 [hep-lat]].\\
-- C.~Alexandrou {\it et al.}  [European Twisted Mass Collaboration],
  "\emph{Light baryon masses with dynamical twisted mass fermions},
  Phys.\ Rev.\  D {\bf 78}, 014509 (2008) [arXiv:0803.3190 [hep-lat]].\\
-- B.~Blossier, M.~Della Morte, G.~von Hippel, T.~Mendes and R.~Sommer,
  "\emph{On the generalized eigenvalue method for energies and matrix elements in
  lattice field theory}", JHEP {\bf 0904}, 094 (2009)  [arXiv:0902.1265 [hep-lat]].\\
-- B.~Blossier, M.~Wagner and O.~Pene  [European Twisted Mass Collaboration],
"\emph{Lattice calculation of the Isgur-Wise functions $\tau_{1/2}$ and $\tau_{3/2}$
with dynamical quarks}", JHEP {\bf 0906}, 022 (2009) [arXiv:0903.2298 [hep-lat]].\\
-- B.~Blossier {\it et al.}  [ETM Collaboration],
"\emph{Pseudoscalar decay constants of kaon and D-mesons from Nf=2 twisted mass
  Lattice QCD}", JHEP {\bf 0907}, 043 (2009) [arXiv:0904.0954 [hep-lat]].\\
-- D.~Becirevic, B.~Blossier, E.~Chang and B.~Haas, "\emph{g(B*Bpi)-coupling in the static heavy quark limit}",
  Phys.\ Lett.\  B {\bf 679}, 231 (2009)  [arXiv:0905.3355 [hep-ph]].\\
--  B.~Blossier {\it et al.}  [ETM Collaboration],  "\emph{A proposal for B-physics on current lattices}",
  JHEP {\bf 1004}, 049 (2010)  [arXiv:0909.3187 [hep-lat]].\\
-- B.~Blossier, M.~della Morte, N.~Garron and R.~Sommer,
"\emph{HQET at order 1/m: I. Non-perturbative parameters in the quenched
  approximation}", JHEP {\bf 1006}, 002 (2010)
  [arXiv:1001.4783 [hep-lat]].\\
-- B.~Blossier, M.~Della Morte, N.~Garron, G.~von Hippel, T.~Mendes, H.~Simma
and R.~Sommer [Alpha Collaboration], "\emph{HQET at order $1/m$: II. Spectroscopy in the quenched approximation}",
JHEP {\bf 1005}, 074 (2010)
  [arXiv:1004.2661 [hep-lat]].\\
-- B.~Blossier, Ph.~Boucaud, F.~De Soto, V.~Mor\'enas, M.~Gravina, O.~P\`ene 
and J.~Rodriguez-Quintero [ETM Collaboration],
  "\emph{Ghost-gluon coupling, power corrections and $\Lambda_{\overline {\rm MS}}$
  from twisted-mass lattice QCD at Nf=2}",
  Phys.\ Rev.\  D {\bf 82}, 034510 (2010)  [arXiv:1005.5290 [hep-lat]].\\
-- B.~Blossier, M.~Della Morte, N.~Garron, G.~von Hippel, T.~Mendes, H.~Simma 
and R.~Sommer [ALPHA Collaboration],
  "\emph{HQET at order 1/m: III. Decay constants in the quenched approximation}",
JHEP {\bf 1012}, 039 (2010)   [arXiv:1006.5816 [hep-lat]].\\
-- B.~Blossier {\it et al.}  [ETM Collaboration],
  "\emph{Average up/down, strange and charm quark masses with Nf=2 twisted mass
  lattice QCD}",  Phys.\ Rev.\  D {\bf 82}, 114513 (2010)
  [arXiv:1010.3659 [hep-lat]].\\
-- B.~Blossier, Ph.~.Boucaud, M.~Brinet, F.~De Soto, Z.~Liu, V.~Mor\'enas, 
O.~P\`ene, K.~Petrov and J. Rodriguez-Quintero,
"\emph{Renormalisation of quark propagators from twisted-mass lattice 
QCD at $N_f$=2}",Phys.\ Rev.\ D {\bf 83}, 074506 (2011) [arXiv:1011.2414 [hep-ph]].\\
-- B.~Blossier, "\emph{Lattice renormalisation of O(a) improved heavy-light 
operators: an addendum}",
Phys.\ Rev.\ D {\bf 84}, 097501 (2011) [arXiv:1106.2132 [hep-lat]].\\
-- B.~Blossier, Ph.~Boucaud, M.~Brinet, F.~De Soto, X.~Du, M.~Gravina, 
V.~Mor\'enas, O.~P\`ene,  K.~Petrov, J.~Rodriguez-Quintero,
"\emph{Ghost-gluon coupling, power corrections and $\Lambda_{\bar{\rm MS}}$ 
from lattice QCD with a dynamical charm}", Phys.\ Rev.\ D {\bf 85}, 034503 
(2012) [arXiv:1110.5829 [hep-lat]].\\
-- B.~Blossier, Ph.~Boucaud, M.~Brinet, F.~De Soto, X.~Du, V.~Mor\'enas, 
O.~P\`ene, K.~Petrov, 
J.~Rodriguez-Quintero, "\emph{The Strong running coupling at $\tau$ and $Z_0$ mass scales from lattice QCD}",
Phys.\ Rev.\ Lett.\  {\bf 108}, 262002 (2012) [arXiv:1201.5770 [hep-ph]].\\
-- B.~Blossier {\it et al.}  [ALPHA Collaboration],
"\emph{Parameters of Heavy Quark Effective Theory from Nf=2 lattice QCD}", JHEP {\bf 1209}, 132 (2012) 
[arXiv:1203.6516 [hep-lat]].\\
-- B.~Blossier, P.~Boucaud, M.~Brinet, F.~De Soto, V.~Mor\'enas, O.~P\`ene, K.~Petrov and J.~Rodriguez-Quintero,
"\emph{Testing the OPE Wilson coefficient for $A^2$ from lattice QCD with a dynamical charm}",
Phys.\ Rev.\ D {\bf 87}, 074033 (2013) [arXiv:1301.7593 [hep-ph]].\\
-- D.~Becirevic, B.~Blossier, A.~Gerardin, A.~Le Yaouanc and F.~Sanfilippo,
"\emph{On the significance of B-decays to radially excited D}",
Nucl.\ Phys.\ B {\bf 872}, 313 (2013) [arXiv:1301.7336 [hep-ph]].\\
-- B.~Blossier, J.~Bulava, M.~Donnellan and A.~Gérardin,
"\emph{$B^{*\prime} \to B$ transition}", Phys.\ Rev.\ D {\bf 87}, no. 9, 094518 (2013)
[arXiv:1304.3363 [hep-lat]].\\
-- B.~Blossier {\it et al.}  [ETM Collaboration],
"\emph{High statistics determination of the strong coupling constant in Taylor scheme and its OPE Wilson coefficient from lattice QCD with a dynamical charm}, Phys.\ Rev.\ D {\bf 89}, 014507 (2014)
[arXiv:1310.3763 [hep-ph]].\\
-- F.~Bernardoni, B.~Blossier, J.~Bulava, M.~Della Morte, P.~Fritzsch, N.~Garron, A.~Gerardin and J.~Heitger {\it et al.},
"\emph{The b-quark mass from non-perturbative $N_f=2$ Heavy Quark Effective Theory at $O(1/m_h)$"}
Phys.\ Lett.\ B {\bf 730},  171 (2014) [arXiv:1311.5498 [hep-lat]].

\section*{Dissemination and Education Sciences}

-- B.~Blossier, "\emph{Jouer aux d\'es pour percer les secrets ultimes de la
mati\`ere}", Plein-Sud sp\'ecial Recherche 2010-2011, 96 (2011).\\
-- B.~Blossier, "\emph{De l'apprentissage du Braille au dessin des graphes de
Feynman}", Rep\`eres IREM {\bf 84}, 19 (2011).\\
-- N. Audoin \emph{et al}, "\emph{Adapter sa p\'edagogie au coll\`ege et au lyc\'ee \`a des \'el\`eves en situation
de handicap sensoriel, moteur ou ayant des troubles des apprentissages}", resources coordinated by B. Blossier and published
by the Minist\`ere de l'Education Nationale,\\
{\bf http://eduscol.education.fr/ash-ressources-disciplinaires} (2013).


\end{document}